\documentclass[10pt, aps, prx, twocolumn, nofootinbib, superscriptaddress,floatfix]{revtex4-2}

\usepackage[dvipsnames]{xcolor}
\usepackage{amsmath}
\usepackage{amssymb}
\usepackage{graphicx}
\usepackage{bbold}
\usepackage{mathtools}
\usepackage{makecell}
\usepackage{braket}
\usepackage[export]{adjustbox}
\usepackage{algorithm}
\usepackage{algpseudocode}

\usepackage{hyperref}
\hypersetup{
    colorlinks=true,
    linkcolor=BrickRed,
    citecolor=BrickRed,
    filecolor=VioletRed,      
    urlcolor=Plum
}
\usepackage{physics}

\DeclareMathOperator{\sgn}{sgn}
\def\zz{\mathbb{Z}}
\def\rr{\mathbb{R}}

\def\idop{\mathbb{1}}
\def\mmod{\operatorname{mod}}
\def\ovl{\overline}
\def\col{\mathcal C}
\def\ccol{\mathbf c}
\def\mbd{\text{bd}}
\def\bdin{\text{bdin}}
\def\bdout{\text{bdout}}

\newcommand\myparagraph[1]{\noindent\textbf{#1 ---}}

\makeatletter
\def\l@subsubsection#1#2{} % remove acknowledgments from table of contents
\makeatother

\definecolor{tcol}{rgb}{0.7,0,0.5}
\definecolor{tstarcol}{rgb}{0,0.6,0.2}
\definecolor{colr}{rgb}{1,0,0}
\definecolor{colg}{rgb}{0,1,0}
\definecolor{colb}{rgb}{0,0.2,1}
\definecolor{coly}{rgb}{1,0.5,0}

\usepackage{tikz}
\usetikzlibrary{decorations,decorations.pathmorphing,decorations.markings,calc,snakes,positioning,fixedpointarithmetic,shapes,shapes.geometric,patterns}

\usepackage{etoolbox}
\AtBeginEnvironment{tikzpicture}{\catcode`$=3 } % $ % was important for auctex preview??

\tikzset{alignmid/.style={baseline={([yshift=-.5ex]current bounding box.center)}}} % adjust pictures vertically
\tikzset{every picture/.append style=alignmid}

\tikzset{
bottomzigzag/.style={postaction={draw,decorate, decoration={zigzag,amplitude=1pt,segment length=3pt,raise=1pt}}},
zigzag/.style={draw,decorate, decoration={zigzag,amplitude=1pt,segment length=3pt}},
rc/.style=rounded corners,
}

\tikzset{
    -|/.style={to path={-| (\tikztotarget)}},
    |-/.style={to path={|- (\tikztotarget)}},
}

\usepackage{ifthen}
\usepackage{xparse,pgffor}

\tikzset{
mark/.code={
\tikzset{postaction={/network/mark/.cd,#1,/tikz/.cd,decorate,decoration={name=markings,mark=at position \netmarkpos with{%+\netmarkposoff} with{
\begin{scope}[netmarktrafo]
\netmarkcode
\end{scope}
}}}}
\def\netmarkpos{0.5}%\pgfdecoratedpathlength}
},
}

\def\netmarkpos{0.5}%\pgfdecoratedpathlength}
\def\netmarkposoff{0cm}
\def\netmarkcode{}

\tikzset{
netmarktrafo/.style={},
netmarkstyle/.style={solid,semithick,sharp corners},
}

\pgfkeys{
/network/mark/.cd,
sty/.code={
\tikzset{netmarkstyle/.style={#1}}
},
asty/.code={
\tikzset{netmarkstyle/.append style={#1}}
},
f/.style={asty=fill},
p/.code={ % relative positioning on decorated line
\def\netmarkpos{#1}%\pgfdecoratedpathlength}
},
poff/.code={ % absolute shift on decorated line in cm
\def\netmarkposoff{#1cm}
},
e/.code={ % mark at the End of line
\def\netmarkpos{\pgfdecoratedpathlength-0.005cm-\netmarkposoff}

\tikzset{netmarktrafo/.append style={shift={(-\netmarkwidth,0)}}}
},
s/.code={ % mark at the Start of line
\def\netmarkpos{0.005cm+\netmarkposoff}

\tikzset{netmarktrafo/.append style={shift={(\netmarkwidth,0)},xscale=-1,yscale=-1}}
},
a/.code={ % mark After end of line
\def\netmarkpos{\pgfdecoratedpathlength-0.005cm}

\tikzset{netmarktrafo/.append style={xscale=-1,shift={(-\netmarkwidth,0)}}}
},
b/.code={ % mark Before start of line
\def\netmarkpos{0.005cm}

\tikzset{netmarktrafo/.append style={xscale=-1,shift={(\netmarkwidth,0),yscale=-1}}}
},
-/.code={ % flip horizontally (mark backwards)
\tikzset{netmarktrafo/.append style={xscale=-1}}
},
r/.code={ % flip vertically (mark on the Right instead of left)
\tikzset{netmarktrafo/.append style={yscale=-1}}
},
sideoff/.code={ % shift sideways
\tikzset{netmarktrafo/.append style={shift={(0,#1)}}}
},
slab/.code={ % label next to the line
\def\netmarkwidth{0}
\def\netmarkcode{
\node[inner sep=0.04cm,netmarkstyle,draw=none] (mylabelwidthtest) at (0,0){\phantom{#1}};
\path let \p1=(mylabelwidthtest.north east), \p2=(mylabelwidthtest.south east), \n1 = {max(abs(\y1),abs(\y2))} in node[inner sep=0.04cm,netmarkstyle] at (0,\n1) {#1};
}
},
lab/.code={ % label on the line
\def\netmarkwidth{0}
\def\netmarkcode{
\node[inner sep=0.04cm,anchor=\netmarkanchor] (mylabelwidthtest) at (0,0) {\phantom{#1}};
\draw[white] (mylabelwidthtest.\pgfdecoratedangle)--(mylabelwidthtest.\pgfdecoratedangle+180);
\node[inner sep=0.04cm,anchor=\netmarkanchor,netmarkstyle] at (0,0) {#1};
}
},
rotlab/.code={ % label on the line rotated with the line
\def\netmarkwidth{0}
\def\netmarkcode{
\node[inner sep=0.04cm,fill=white,transform shape,rotate=90,anchor=\netmarkrotanchor,netmarkstyle] (mydecorationnodename) at (0,0) {#1};
}
},
mrotlab/.style={ % small math label on the line rotated with the line
rotlab={$\scriptstyle{#1}$}
},
arr/.code={ % arrow
\def\netmarkwidth{0.04}
\def\netmarkcode{\draw[netmarkstyle] (-0.04,0.08)--(0.04,0)--(-0.04,-0.08);}
},
arrbar/.code={ % arrow
\def\netmarkwidth{0.08}
\def\netmarkcode{\draw[netmarkstyle] (-0.08,0.08)--(0,0)--(-0.08,-0.08) (0.04,0.08)--(0.04,-0.08);}
},
rarr/.code={ % round arrow
\def\netmarkwidth{0.04}
\def\netmarkcode{\draw[netmarkstyle] (-0.04,-0.08)arc(90-180:90:0.08);}
},
dot/.code={ % dot
\def\netmarkwidth{0.08}
\def\netmarkcode{\draw[netmarkstyle] (0,0)circle(0.08);}
},
flag/.code={ % flag to the left
\def\netmarkwidth{0.06}
\def\netmarkcode{\draw[netmarkstyle] (-0.06,0)--(0,0.09)--(0.06,0)--cycle;}
},
darr/.code={ % "double arrow": half-arrow on the left
\def\netmarkwidth{0.08}
\def\netmarkcode{\draw[netmarkstyle] (-0.04,0)--(0.04,0)--(-0.04,0.08)--cycle;}
},
circ/.code={
\def\netmarkwidth{0.1}
\def\netmarkcode{\draw[netmarkstyle] (0,0) circle (0.1);}
},
hcirc/.code={ % half-circle on the left
\def\netmarkwidth{0.1}
\def\netmarkcode{\draw[netmarkstyle] (-0.1,0) arc (180:0:0.1);}
},
diamond/.code={
\def\netmarkwidth{0.1}
\def\netmarkcode{\draw[netmarkstyle] (-0.1,0)--(0,-0.1)--(0.1,0)--(0,0.1)--cycle;}
},
bar/.code={ % tick, short line perpendicular to line
\def\netmarkwidth{0.05}
\def\netmarkcode{
\draw[netmarkstyle] (0,-0.08cm-0.5*\pgflinewidth)--(0,0.08cm+0.5*\pgflinewidth);
}
},
dbar/.code={ % double tick
\def\netmarkwidth{0.13}
\def\netmarkcode{
\draw[netmarkstyle] (-0.04cm,-0.08cm-0.5*\pgflinewidth)--(-0.04cm,0.08cm+0.5*\pgflinewidth) (0.04cm,-0.08cm-0.5*\pgflinewidth)--(0.04cm,0.08cm+0.5*\pgflinewidth);
}
},
hbar/.code={ % tick, short perpendicular line towards the left
\def\netmarkwidth{0.05}
\def\netmarkcode{
\draw[netmarkstyle] (0, 0.5*\pgflinewidth)--++(0,0.12);
}
},
mill/.code={
\def\netmarkwidth{0.16}
\def\netmarkcode{
\draw[netmarkstyle] (0,-0.5*\pgflinewidth)--++(-0.08,-0.08)--++(0,0.08);
\draw[netmarkstyle] (0,0.5*\pgflinewidth)--++(0.08,0.08)--++(0,-0.08);
}
},
three dots/.code={
\def\netmarkwidth{0.2}
\def\netmarkcode{
\fill (-0.12,0) circle (0.5*0.05) (0,0) circle (0.5*0.05) (0.12,0) circle (0.5*0.05);
}
},
}

\tikzset{wid/.style={minimum width=#1cm}}
\tikzset{hei/.style={minimum height=#1cm}}

\tikzset{sx/.style={xshift=#1cm}}
\tikzset{sy/.style={yshift=#1cm}}

\tikzset{box/.style={draw,rectangle}}
\tikzset{fbox/.style={draw,rectangle, line width=1.1}}
\tikzset{roundbox/.style={draw,rectangle,rounded corners}}
\tikzset{froundbox/.style={draw,rectangle, rounded corners, line width=1.1}}
\tikzset{rounddiamond/.style={draw,diamond,rounded corners}}
\tikzset{dot/.style={draw, shape=circle, fill=black, scale=0.5}}

\tikzset{
netbox/.code={
\node[draw,netbdstyle] (\atomname) at (0,0) {#1};
\coordinate (\atomname-r) at (\atomname.east);
\coordinate (\atomname-l) at (\atomname.west);
\coordinate (\atomname-t) at (\atomname.north);
\coordinate (\atomname-b) at (\atomname.south);
\coordinate (\atomname-tr) at (\atomname.north east);
\coordinate (\atomname-br) at (\atomname.south east);
\coordinate (\atomname-tl) at (\atomname.north west);
\coordinate (\atomname-bl) at (\atomname.south west);
},
}

\tikzset{bdlw/.code={\tikzset{mybdstyle/.style={draw, line width=#1}}}}
\tikzset{bdcol/.code={\tikzset{mybdstyle/.append style={#1}}}}

\newcommand\setelements[1]{
\pgfkeys{/network/atom/.cd,#1}
}

\newcommand\atoms[2]{
\foreach \name/\keys in {#2}{
\expandafter\atom\expandafter{\keys,#1}{\name}
}
}

\newcommand\atom[2]{
\def\atomname{#2}
\tikzset{
nettrafo/.style={},
netatompos/.style={},
netdeco/.style={},
netpostdeco/.style={},
}

\pgfkeys{/network/atom/.cd,#1}

\begin{scope}[netatompos] % shift to atom position
\begin{scope}[nettrafo] % rotate, flip and scale
\netshapecoords % set the anchor coordinates
\fill[netbackstyle] \netshapepath;
\clip \netshapepath;
\tikzset{netdeco}
\draw[netbdstyle] \netshapepath;
\end{scope}
\tikzset{netpostdeco} % draw post-decorations, not rotated, flipped, or scaled
\end{scope}

}

\pgfkeys{
/network/atom/.cd,
square/.code={

\def\netshapepath{(-\tempsize,-\tempsize)rectangle (\tempsize,\tempsize)}
\def\netshapecoords{
\node[rectangle,wid=2*\tempsize,hei=2*\tempsize,inner sep=0,transform shape](\atomname)at(0,0){};
\coordinate(\atomname-c) at (0,0);
\coordinate(\atomname-r) at (\tempsize,0);
\coordinate(\atomname-l) at (-\tempsize,0);
\coordinate(\atomname-t) at (0,\tempsize);
\coordinate(\atomname-b) at (0,-\tempsize);
\coordinate(\atomname-br) at (\tempsize,-\tempsize);
\coordinate(\atomname-tr) at (\tempsize,\tempsize);
\coordinate(\atomname-bl) at (-\tempsize,-\tempsize);
\coordinate(\atomname-tl) at (-\tempsize,\tempsize);
}},
circ/.code={

\def\netshapepath{(0,0)circle(\tempsize)}
\def\netshapecoords{
\node[circle,wid=2*\tempsize,hei=2*\tempsize,inner sep=0,transform shape](\atomname)at(0,0){};
\coordinate(\atomname-c) at (0,0);
\coordinate(\atomname-r) at (\tempsize,0);
\coordinate(\atomname-l) at (-\tempsize,0);
\coordinate(\atomname-t) at (0,\tempsize);
\coordinate(\atomname-b) at (0,-\tempsize);
}},
triang/.code={

\def\netshapepath{(-30:\tempsize)--(90:\tempsize)--(-150:\tempsize)--cycle}
\def\netshapecoords{
\node[regular polygon,regular polygon sides=3,wid=2*\tempsize,inner sep=0,transform shape](\atomname)at(0,0){};
\coordinate(\atomname-c) at (0,0);
\coordinate(\atomname-cr) at (-30:\tempsize);
\coordinate(\atomname-cl) at (-150:\tempsize);
\coordinate(\atomname-ct) at (90:\tempsize);
\coordinate(\atomname-mb) at (-90:0.5*\tempsize);
\coordinate(\atomname-mr) at (30:0.5*\tempsize);
\coordinate(\atomname-ml) at (150:0.5*\tempsize);
}},
diamond/.code={

\def\netshapepath{(0,-\tempsize)--(\tempsize,0)--(0,\tempsize)--(-\tempsize,0)--cycle}
\def\netshapecoords{
\node[rotate=45,rectangle,wid=sqrt(2)*\tempsize,hei=sqrt(2)*\tempsize,inner sep=0,transform shape](\atomname)at(0,0){};
\coordinate(\atomname-c) at (0,0);
\coordinate(\atomname-r) at (\tempsize,0);
\coordinate(\atomname-l) at (-\tempsize,0);
\coordinate(\atomname-t) at (0,\tempsize);
\coordinate(\atomname-b) at (0,-\tempsize);
}},
pentagon/.code={

\def\netshapepath{(-126:\tempsize)--(-54:\tempsize)--(18:\tempsize)--(90:\tempsize)--(162:\tempsize)--cycle}
\def\netshapecoords{
\node[regular polygon,regular polygon sides=5,wid=2*\tempsize,inner sep=0,transform shape](\atomname)at(0,0){};
\coordinate(\atomname-c) at (0,0);
\coordinate (\atomname-mb)at(-90:{\tempsize*cos(36)});
\coordinate (\atomname-mbr)at(-18:{\tempsize*cos(36)});
\coordinate (\atomname-mtr)at(54:{\tempsize*cos(36)});
\coordinate (\atomname-mtl)at(126:{\tempsize*cos(36)});
\coordinate (\atomname-mbl)at(-162:{\tempsize*cos(36)});
\coordinate (\atomname-cbr)at(-54:\tempsize);
\coordinate (\atomname-cr)at(18:\tempsize);
\coordinate (\atomname-ct)at(90:\tempsize);
\coordinate (\atomname-cl)at(162:\tempsize);
\coordinate (\atomname-cbl)at(-126:\tempsize);
}},
halfcirc/.code={

\def\netshapepath{(\tempsize,0)arc(0:180:\tempsize)--++(0,-0.04)-|cycle}
\def\netshapecoords{
\node[circle,wid=2*\tempsize,hei=2*\tempsize,inner sep=0,transform shape](\atomname)at(0,0){};
\coordinate(\atomname-c) at (0,0);
\coordinate(\atomname-r) at (\tempsize,0);
\coordinate(\atomname-l) at (-\tempsize,0);
\coordinate(\atomname-t) at (0,\tempsize);
\coordinate(\atomname-b) at (0,0);
}},
void/.code={
\def\netshapepath{}
\def\netshapecoords{
\coordinate(\atomname) at (0,0);
\coordinate(\atomname-c) at (0,0);
}},
labbox/.code={
\def\netshapepath{(0,0)}
\def\netshapecoords{}
\tikzset{netpostdeco/.append style={netbox=#1}}
},
}

\pgfkeys{
/network/atom/.cd,
p/.code={\tikzset{netatompos/.append style={shift={(#1)}}}},
xscale/.code={\tikzset{nettrafo/.append style={xscale=#1}}},
yscale/.code={\tikzset{nettrafo/.append style={yscale=#1}}},
scale/.code={\tikzset{nettrafo/.append style={scale=#1}}},
rot/.code={\tikzset{nettrafo/.append style={rotate=#1}}},
hflip/.style={xscale=-1},
vflip/.style={yscale=-1},
big/.style={scale=3/2},
small/.style={scale=2/3},
huge/.style={scale=9/4},
tiny/.style={scale=4/9},
flat/.style={yscale=2/3},
fflat/.style={yscale=4/9},
}

\tikzset{
netbdstyle/.style={line width=0.15em}, % changed from pt(default)
netdecstyle/.style={},
netpostdecstyle/.style={},
netbackstyle/.style={white},
}
\pgfkeys{
/network/atom/.cd,
bdstyle/.code={\tikzset{netbdstyle/.style={#1}}},
bdastyle/.code={\tikzset{netbdstyle/.append style={#1}}},
decstyle/.code={\tikzset{netdecstyle/.style={#1}}},
decastyle/.code={\tikzset{netdecstyle/.append style={#1}}},
postdecstyle/.code={\tikzset{netpostdecstyle/.style={#1}}},
postdecastyle/.code={\tikzset{netpostdecstyle/.append style={#1}}},
backstyle/.code={\tikzset{netbackstyle/.style={#1}}},
backastyle/.code={\tikzset{netbackstyle/.append style={#1}}},
style/.style={bdstyle=#1, decstyle=#1, postdecstyle=#1},
astyle/.style={bdastyle=#1, decastyle=#1, postdecastyle=#1},
whiteblack/.style={decstyle=white,backstyle=black},
nobd/.style={bdstyle={line width=0}},
}

\tikzset{
netbscope/.code={\begin{scope}[#1]},
netescope/.code={\end{scope}},
}

\pgfkeys{
/network/atom/.cd,
bscope/.code={\tikzset{netdeco/.append style={netbscope={#1}}}},
escope/.code={\tikzset{netdeco/.append style={netescope}}},
dec/.style 2 args={bscope={#1},#2,escope}, % for applying transformation keys to certain decorations only
vbar/.style={dec={rotate=90}{hbar}},
dcross/.style={dec={rotate=45}{hbar},dec={rotate=-45}{hbar}},
cross/.style={hbar,vbar},
sect6/.style={sect=60},
sect8/.style={sect=45},
sect10/.style={sect=36},
}

\def\regdec#1{\pgfkeys{/network/atom/.cd,#1/.code={\tikzset{netdeco/.append style={net#1}}}}}
\regdec{all}\regdec{rhalf}\regdec{rquart}\regdec{brquart}\regdec{dot}\regdec{spiral}\regdec{swirl}\regdec{hstripe}\regdec{hbar}\regdec{rrey}
\pgfkeys{/network/atom/.cd,sect/.code={\tikzset{netdeco/.append style={netsect=#1}}}}

\tikzset{
netall/.code={\fill[netdecstyle] (-0.3,-0.3)rectangle (0.3,0.3);}, % fill all
netrhalf/.code={\fill[netdecstyle] (0,-0.3)rectangle (0.3,0.3);}, % right half
netrquart/.code={\fill[netdecstyle] (0.075,-0.3)rectangle (0.3,0.3);}, % right quarter
netbrquart/.code={\fill[netdecstyle] (0,0)rectangle (0.3,-0.3);}, % bottom right quarter
netsect/.code={\fill[netdecstyle] (0,0)--(0,-0.3)arc(-90:-90+#1:0.3)--cycle;}, % section of angle #1 starting from -90
netdot/.code={\fill[netdecstyle] (0,0)circle(0.07);}, % dot in the middle
netspiral/.code={\draw[netdecstyle] plot [variable=\t,domain=0:4] ({0.075*\t*cos(pi*(\t-0.5) r)},{0.075*\t*sin(pi*(\t-0.5) r)});}, % spiral
netswirl/.code={\fill[netdecstyle] plot [variable=\t,domain=0:2] ({0.15*\t*cos(pi*(\t-0.5) r)},{0.15*\t*sin(pi*(\t-0.5) r)}) arc(-90:-450:0.3)--cycle;}, % filled swirl
nethstripe/.code={\fill[netdecstyle] (-0.3,-0.05)rectangle(0.3,0.05);}, % horizontal stripe
nethbar/.code={\draw[netdecstyle] (-0.3,0)--(0.3,0);}, % horizontal line
netrrey/.code={\draw[netdecstyle] (0,0)--(0.3,0);} % line from the middle to the right
}

\pgfkeys{
/network/atom/.cd,
postbscope/.code={\tikzset{netpostdeco/.append style={netbscope={#1}}}},
postescope/.code={\tikzset{netpostdeco/.append style={netescope}}},
postdec/.style 2 args={postbscope={#1},#2,postescope}, % for applying transformation keys to certain decorations only
arc/.code={\tikzset{netpostdeco/.append style={netarc=#1}}},
lab/.code={\tikzset{netpostdeco/.append style={netlab={#1}}}},
shadecirc/.code={\tikzset{netpostdeco/.append style=netshadecirc}},
shaderect/.code={\tikzset{netpostdeco/.append style={netshaderect={#1}}}},
debug/.code={\tikzset{netpostdeco/.append style=netdebug}},
markline/.code 2 args={\tikzset{netpostdeco/.append style={netmarkline={#1}{#2}}}},
postcirc/.code={\tikzset{netpostdeco/.append style=netpostcirc}},
}

\tikzset{
netlab/.code={
\pgfkeys{/network/atom/lab/.cd,#1}
\node[netpostdecstyle] at (\ifdefined\netlabpos\netlabpos\else\netlabang:\netlabdist\fi) {\netlabwrap{\netlabtext}};
},
netarc/.code args={#1:#2:#3}{
\draw[netpostdecstyle] (#1:#3) arc (#1:#2:#3);
},
netshadecirc/.code= {
\fill[opacity=0.4,netpostdecstyle] (0,0)circle(0.4);
},
netpostcirc/.code= {
\draw[netpostdecstyle] (0,0)circle(0.15);
},
netshaderect/.code= {
\fill[rc,opacity=0.4,netpostdecstyle] ($-1*(#1)$) rectangle (#1);
},
netdebug/.code= {
\node[red] at (0,0){\atomname};
},
netmarkline/.code 2 args= {
\draw (\atomname)edge[mark={#2}]++(#1);
},
}

\def\netlabwrap#1{#1}
\pgfkeys{
/network/atom/lab/.cd,
t/.code={\def\netlabtext{#1}}, % label text
p/.code={\def\netlabpos{#1}}, % position of the label
ang/.code={\def\netlabang{#1}},
dist/.code={\def\netlabdist{#1}},
wrap/.code={\def\netlabwrap##1{#1}}, % store in doesn't work
}

\tikzset{
ind/.style={mark={lab=#1,a}}, % normal open index label
startind/.style={mark={lab=#1,b}}, % normal open index label
worldline/.style={red,line width=5,opacity=0.3,line join=round},
back/.style={opacity=0.5},
lightback/.style = {circle, fill=white, inner sep=0.5,path fading=fade out}, % lighten up background for label nodes
zmeasurement/.style = {line width=0.2cm},%,line cap=round},
xmeasurement/.style = {postaction = {draw,white,line width=0.15cm}, line width=0.2cm},%,line cap=round},
}

\setelements{
z2/.style={small,circ},
delta/.style={small,circ,all},
vertex/.style={tiny,circ,all},
charge/.style={postdecstyle=red,postdec={scale=0.5}{shadecirc}},
tgate/.style = {circ,small,all,astyle={tcol}},
tstargate/.style = {circ,small,all,astyle={tstarcol}},
plusinit/.style = {circ,small,cross},
singlexdestruct/.style = {circ,small,dot},
}

\newcommand\singlexmeas[1]{\draw[] (#1) circle(0.15);}

\begin{document}
\title{Planar fault-tolerant circuits for non-Clifford gates on the 2D color code}
\author{Andreas Bauer}
\email{andib@mit.edu}
\affiliation{\footnotesize Department of Mechanical Engineering, Massachusetts Institute of Technology, Cambridge, MA 02139, USA}
\author{Julio C. Magdalena de la Fuente}
\email{jm@juliomagdalena.de}
\affiliation{\footnotesize Dahlem Center for Complex Quantum Systems, Freie Universit\"at Berlin, 14195 Berlin, Germany}

\begin{abstract}
We introduce a family of scalable planar fault-tolerant circuits that implement logical non-Clifford operations on a 2D color code, such as a logical $T$ gate or a logical non-Pauli measurement that prepares a magic $\ket T$ state.
The circuits are relatively simple, consisting only of physical $T$ gates, $CX$ gates, and few-qubit measurements.
They can be implemented with an array of qubits on a 2D chip with nearest-neighbor couplings, and no wire crossings.
The construction is based on a spacetime path integral representation of a non-Abelian 2+1D topological phase, which is related to the 3D color code.
We turn the path integral into a circuit by expressing it as a spacetime $ZX$ tensor network, and then traversing it in some chosen time direction.
We describe in detail how fault tolerance is achieved using a ``just-in-time'' decoding strategy, for which we repurpose and extend state-of-the-art color-code matching decoders.
\end{abstract}

\maketitle
\tableofcontents

\vspace{-0.5cm}

\section{Introduction}
Topological quantum error correction is one of the most promising routes toward fault-tolerant quantum computation due to its intrinsic scalability with local, low-weight measurements.
The most prominent examples of 2D topological codes are the surface code~\cite{Kitaev1997,Dennis2001,Bravyi1998} and the color code~\cite{Bombin2006b}, small instances of which have been realized experimentally~\cite{Lacroix2024, Krinner2022, Acharya2024}.
One of their most attractive features is their ability to perform logical Clifford gates transversally and via lattice surgery using a 2D planar qubit connectivity~\cite{Horsman2011,Litinski2018,Landahl2014,Thomsen2022,Ryan2022,Bluvstein2024}.
Logical non-Clifford gates are more difficult:
One of the most-studied protocols is \emph{magic state distillation}~\cite{Bravyi2004, Bravyi2012a, Litinski2019}, which comes at the cost of significant additional overhead.

Recently, Ref.~\cite{Davydova2025} introduced a new approach to implement a variety of logical non-Clifford gates in a scalable, topological, fault-tolerant, purely 2D manner, using domain walls between (Abelian) stabilizer codes and non-Abelian codes.
Historically, this approach emerged through 3+0D dimension-jump protocols~\cite{Bombin2014,Beverland2021} using transversal non-Clifford gates of 3D codes~\cite{Bombin2006, Vasmer2018}, which can be turned into 2+1D protocols using \emph{just-in-time decoding}~\cite{Bombin2018, Brown2019, Scruby2020}.
From this perspective, the non-Abelian phases interpretation leads to a much greater flexibility in constructing both new global protocols for different logic gates, as well as new microscopic circuits.
Despite its promising features, there are still significant steps required to bring this proposal closer to a practical implementation.

\myparagraph{Contributions}
In this work, we make further progress in the design of low-overhead and practical protocols for logical non-Clifford gates:
We construct a family of fault-tolerant circuits, called \emph{twisted color circuits}, as microscopic implementation for the global logical non-Clifford gates proposed in Ref.~\cite{Davydova2025}.
The circuits are composed of simple physical operations such as $CX$ and $T$ gates, and few-qubit Pauli-$X$ or $Z$ measurements, and the initial and final state is encoded in a 2D color code.
Importantly, the required qubit connectivity is fully planar, so the protocols are implementable on a 2D chip with nearest-neighbor couplings and without wire crossings.
This makes our protocols particularly suitable for specific hardware architectures such as superconducting qubits~\cite{Lacroix2024, Krinner2022, Acharya2024}.

To derive twisted color circuits, we use the \emph{path-integral approach} to topological quantum error correction.
In this approach, a $+1$-post-selected syndrome-extraction circuit is viewed as a spacetime $ZX$ tensor network~\cite{Bombin2023,Teague2023} or a path-integral representation of the topological phase~\cite{path_integral_qec,twisted_double_code,xy_floquet}.
Vice versa, a path integral can be turned into a circuit by traversing it in some time direction, and the $+1$-post-selection can be removed by introducing defects.
The path-integral approach is well suited to construct fault-tolerant circuits for exotic topological phases, including non-Abelian ones, as was demonstrated in Ref.~\cite{twisted_double_code}.
The particular $ZX$ tensor-network path integral from which we derive twisted color circuits is called the \emph{color path integral}, and it is inspired from the 3D color code and its transversal $T$ gate.
There is a variety of different twisted color circuits, depending on a choice of (1) a 3-colex on which the underlying 3D color code is defined, (2) a time direction, (3) a way to place \emph{qubit worldlines}, and (4) a global spacetime topology that determines the overall logical action (see Ref.~\cite{Davydova2025}).

The path integral approach was already used in Refs.~\cite{twisted_double_code,Davydova2025} to construct low-overhead non-Abelian circuits, which are related to the transversal $CCZ$ gate on three copies of the 3D toric code rather than the transversal $T$ gate in the 3D color code.
Compared to these circuits, the two advantages of twisted color circuits are (1) that they use single-qubit physical $T$ gates instead of 3-qubit $CCZ$ gates, and (2) that their qubit connectivity is naturally planar.
Note that Refs.~\cite{Bombin2018,Davydova2025} also give protocols based on physical $T$ gates, but explicit circuits are not given and would have a larger qubit footprint and not be planar.

We also provide a direct physical interpretation of the color path integral as a lattice topological quantum field theory (TQFT), or a discrete gauge theory, for a non-Abelian topological phase.
We formulate the underlying cohomology theory, called \emph{color cohomology} (see also Ref.~\cite{Kubica2019}), establish the gauge invariance of the according action, show its equivalence to a known Dijkgraaf-Witten gauge theory, and study its flux and charge defects.
The latter are essential for the decoding of twisted color circuits.
The detailed interplay of defects that we derive can also be used to accurately and efficiently benchmark the logical performance of the protocol, despite the non-Clifford gates in the circuit.
As also noted in Ref.~\cite{Davydova2025}, it is not necessary to simulate the full circuit including the non-Clifford operations.

Like the protocols in Refs.~\cite{Bombin2018,Brown2019,Scruby2020,Davydova2025}, Pauli-$X$ errors in our twisted color circuits have to be decoded and corrected \emph{just-in-time} during the execution of the circuit, instead of globally at the end of the protocol.
We describe in detail how this decoding works microscopically, concretely implementing ideas of Ref.~\cite{Bombin2018}.
%Roughly, decoding works by reapplying a global matching-based 3D-color-code decoder at every time step, and using another global decoder to reconcile the result with corrections already committed at earlier time steps.
Our method of just-in-time decoding invokes an arbitrary 3D-color-code decoder twice at every time step:
Once to obtain a fresh global ``current best estimate'', and another time to join the previously committed corrections with the current best estimate.
This is to be contrasted with more heuristic RG-based decoders which are mainly discussed in Refs.~\cite{Brown2019,Scruby2020,Davydova2025}.
Our method is closer to the original ideas in Ref.~\cite{Bombin2018} and the matching-based algorithm mentioned in Ref.~\cite{Davydova2025}.
We revisit the restriction decoder~\cite{Kubica2019} which can be used as a matching-based global decoder for the $Z$ errors in our circuits, and propose a new global decoding algorithm for the $X$ errors.
Furthermore, we study how the knowledge of the Pauli-$X$ decoding can be used later to optimize the global decoding of the Pauli-$Z$ errors.

\myparagraph{Structure}
We give a complete self-contained pedagogical example of a twisted color circuit implementing a specific logical gate in Section~\ref{sec:simple_example}, which we hope to be accessible to a broad error-correction audience.
We provide deeper mathematical insights into the construction in the following Sections~\ref{sec:color_cohomology} and \ref{sec:path_integral}, where we introduce color cohomology and study the color path integral as a spacetime representation of a non-Abelian topological phase.
We give more details on decoding and fault tolerance of twisted color circuits using a just-in-time matching decoder in Section~\ref{sec:decoding}.
We provide further examples for twisted color circuits in Section~\ref{sec:circuits}.
An error-correction inclined reader might be particularly interested in Sections~\ref{sec:simple_example}, \ref{sec:decoding} and \ref{sec:circuits}, while a reader interested in the mathematical background and the relation to non-Abelian topological fixed-point path integrals might want to focus on Sections~\ref{sec:color_cohomology} and \ref{sec:path_integral}.

\section{Example: \texorpdfstring{$\ket T$}{T}-state measurement on hexagonal grid}
\label{sec:simple_example}
In this section, we provide a self-contained exposition of our construction, by focusing on one particular example of a twisted color circuit.
Some basic familiarity with (1) the 3-dimensional color code~\cite{Bombin2006a, Bombin2006b} and (2) the use of $ZX$ diagrams to represent and manipulate error-correcting protocols~\cite{Gidney2022, Kissinger2022, Bombin2023, Teague2023, xyzrubycode, path_integral_qec, xy_floquet} is helpful but not necessary.
We do not aim to provide the ``best'' example in terms of expected overhead and logical error rate, but instead the one that may be the most convenient for the reader to follow.
In Section~\ref{sec:circuits} we provide further examples of twisted color circuits.
We remark that depending on the hardware requirements further improvements of the local structure of the circuits can be made.

\subsection{The global protocol}
\label{sec:tmeasurement_protocol}
Twisted color circuits are geometrically local circuits of unitaries and measurements in 2+1D.
Moreover, they have a macroscopic fault distance which makes them \textit{topological}.
To create a protocol that does not only store but also process logical information, we define the circuits in a spacetime that is composed of different subregions.
This includes 3D \textit{bulk} regions that are interfaced along 2D \textit{boundary} and \textit{domain-wall} regions, that respectively are interfaced along one-dimensional \textit{corner} regions.
Note that all of these subregions and interfaces have to be defined carefully in order to give a topologically protected protocol.
Within each subregion, the circuit is uniform, that is, it is fully specified by a spacetime unit cell.
The \emph{global protocol} is specified by a spacetime diagram where we give a label to each region that describes the unit cell of the circuit.
Our example uses a global protocol from Ref.~\cite{Davydova2025} that performs a logical $\ket T$-state measurement:
\begin{equation}
\label{eq:global_protocol}
\begin{tikzpicture}[scale=0.8]
\draw[gray] (-0.3,2.3)edge[->]node[midway,rotate=90,anchor=south]{time} (-0.3,4.7);
\draw (0,0)rectangle++(3,2) (0,5)rectangle++(3,2) (0,2)--++(0,3) (3,2)--++(0,3) (0,7)--++(2,0.7)--(3,7);
\draw[dashed] (0,0)--++(2,0.7)--++(0,7) (2,0.7)--(3,0) (0,2)--++(2,0.7)--(3,2) (0,5)--++(2,0.7)--(3,5);
\draw[gray,align=left] (1.7,3.7)--++(-10:2.2)node[anchor=west]{twisted color circuit\\(3D color code\\+ transversal $T$)};
\draw[gray] (1.7,1.2)--++(-10:2.2)node[anchor=west]{2D color code};
\draw[gray] (1.7,6.2)--++(-10:2.2)node[anchor=west]{2D color code};
\node[colr] at (1.5,3.5){$r$};
\node[colb] at (1,1.35){$b$};
\node[colg] at (2.5,1.35){$g$};
\node[colr] at (1.5,1){$r$};
\node[colb] at (1,3.85){$b$};
\node[colg] at (2.5,3.85){$g$};
\node[colr] at (1.5,6){$r$};
\node[colb] at (1,6.35){$b$};
\node[colg] at (2.5,6.35){$g$};
\node[coly] at (1.8,2.3){$y$};
\node[coly] at (1.8,5.3){$y$};
\end{tikzpicture}
\end{equation}
That is, the circuit that we construct will act on qubits distributed over a triangle.
We start by performing the 2+1D syndrome-extraction circuit for the 2D color code, on the usual triangle with $r$ (red), $g$ (green), and $b$ (blue) color boundaries.
Through some constant-depth transitioning circuit, we switch to the twisted color circuit, related to 3D color code and its transversal $T$ gate.
After applying the twisted color circuit for some time that needs to be increased linearly with the desired fault distance of the protocol, we switch back to the 2D color code via another constant-depth transitioning circuit.
We label the interface between the 2D and 3D color-code regions with $y$ (yellow) since it is related to the $y$ color boundary of the 3D color code.
The spatial boundaries of the twisted color circuit are related to the $r$, $g$, and $b$ color boundaries of the 3D color code.
Overall (including error correction), the circuit performs a logical $\ket T$-state measurement of the qubit encoded in the 2D color code~\cite{Davydova2025}.
This is a measurement with eigenstates $\ket T\coloneqq \ket{0} + e^{2\pi i\frac18}\ket{1}$ and $\ket{0} + e^{2\pi i \frac58}\ket{1}$, which can be used to prepare the $\ket T$ magic state.

\myparagraph{Dimensional jumps}
We believe that it is most natural to describe our protocols directly in terms of a 2+1D non-Abelian topological phase as we do in Section~\ref{sec:path_integral}.
However, for this section, we motivate them from a measurement-based perspective, as in Refs.~\cite{Bombin2018, Brown2019}.
To this end, we reinterpret the spacetime diagram above, viewing the middle part as a 3+0D measurement-based protocol.
When we reach the temporal $y$ domain wall after performing some 2D-color-code syndrome-extraction circuit, we perform a dimension jump~\cite{Bombin2014}:
We prepare the 3D color code state in the middle triangle prism with the according $r$, $g$, $b$, and $y$ color boundaries, by preparing qubits in the $\ket+$ state and measuring all $Z$-type stabilizers.
In doing so, we inject the 2D color-code logic state at the $y$ boundary at the bottom.
Then we perform the transversal $T$ gate in the 3D volume, and collapse the logic state of the 3D color code onto a 2D color code at its top $y$ boundary.
Finally, we continue with the 2D color-code syndrome-extraction circuit.

\myparagraph{3+0=2+1}
Note that there is nothing 3+1D about the above protocol since the 3D color code in the middle only exists for constant time.
We can turn a 3+0D measurement-based protocol~\cite{Raussendorf2007, Nickerson2018} into a 2+1D circuit protocol, by keeping a few 2D slices of qubits ``active'' at a time, and sweeping through the 3D code by initializing a new slice on one side and measuring out the oldest slice on the other side.
As was first noticed in Ref.~\cite{Bombin2018}, this can also be done with the 3+0D color-code circuit including the transversal $T$ gate.
As we need to fix the Pauli frame before applying each physical $T$ gate, this requires continuous decoding during the execution of the circuit before the full syndrome history is available, a task called \emph{just-in-time decoding} in Ref.~\cite{Bombin2018}.
In this work, we provide a method to obtain such 2+1D circuits without keeping multiple 2D slices active.
As a consequence, the twisted color circuits we obtain from our method need a smaller number of qubits, and most importantly, allow for a fully planar connectivity.
In order to derive these circuits, we first turn the 3+0D protocol into a 3D $ZX$ tensor-network path integral in spacetime, which we refer to as the \emph{color path integral}.
Then we turn this $ZX$ tensor network into a circuit, using methods similar to those in Refs.~\cite{path_integral_qec, twisted_double_code, xy_floquet, Bombin2023, Teague2023}.

\subsection{The color path integral}
\label{sec:easy_path_integral}
In this subsection, we define the 3+0D color-code measurement-based protocol mentioned in the previous section more precisely, and then turn it into what we call the \textit{color path integral}, expressed as a $ZX$ tensor network.
We first focus on the bulk of the protocol, and later define appropriate boundary conditions.

\myparagraph{The 3D color code}
The 3D color code~\cite{Bombin2006a} can be defined on any \emph{3-colex}, which is a 3D cellulation that is Poincar\'e dual to a triangulation,%
\footnote{A 3D triangulation is a decomposition of a 3-manifold into tetrahedron volumes, which are glued at their triangle faces.
A 3D cellulation is defined similarly, without requiring the volumes and faces to be tetrahedra and triangles.}
and whose volumes are colored red ($r$), green ($g$), blue ($b$), and yellow ($y$) such that adjacent volumes always have different colors.

In every 3-colex, we denote
\begin{itemize}
\item the set of $i$-cells by $\Delta_i$,
\item the pair of colors (such as $rg$) of the volumes adjacent to a face $f\in \Delta_2$ by $\ccol_f$,
\item the triple of colors of the volumes adjacent to an edge $e\in \Delta_1$ by $\ccol_e$, where we denote the triple that does not contain a color $c$ by $\ovl c$ (such as $rgb\simeq \ovl y$),
\item the \textit{sign} of a vertex $p\in\Delta_0$ by $\sgn_p\in \{\pm 1\}$.
We have $\sgn_p=+1$ if for the $r$, $g$, and $b$ volumes around the $rgb=\ovl y$-colored edge adjacent to $p$, the clockwise ordering when looking away from $p$ coincides with the ordering $r-g-b$, and $\sgn_p=-1$ otherwise.
In particular, the vertices at the endpoints of an edge always have opposite sign.
\end{itemize}

\begin{figure}
\begin{tikzpicture}
\node[inner sep=0] (x0) at (4.7,0){
\includegraphics[width=5cm]{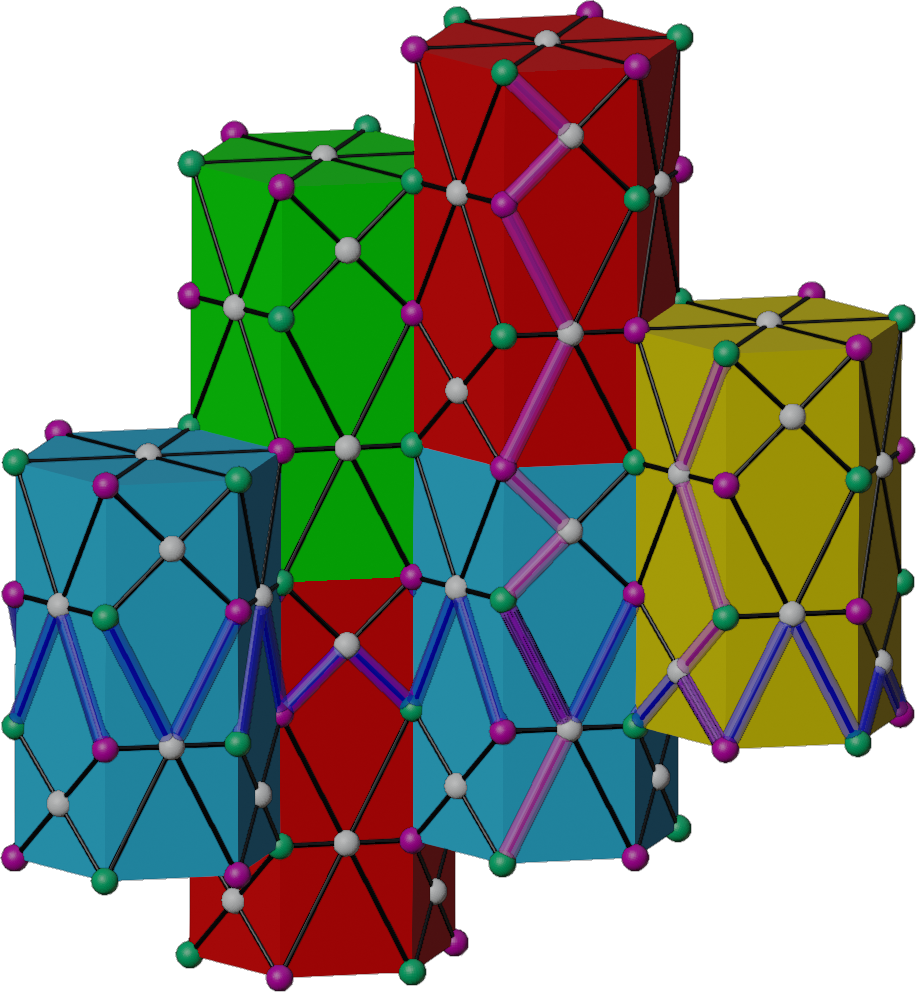}
};
\node[inner sep=0] (x1) at (0,1.6){
\includegraphics[width=1.6cm]{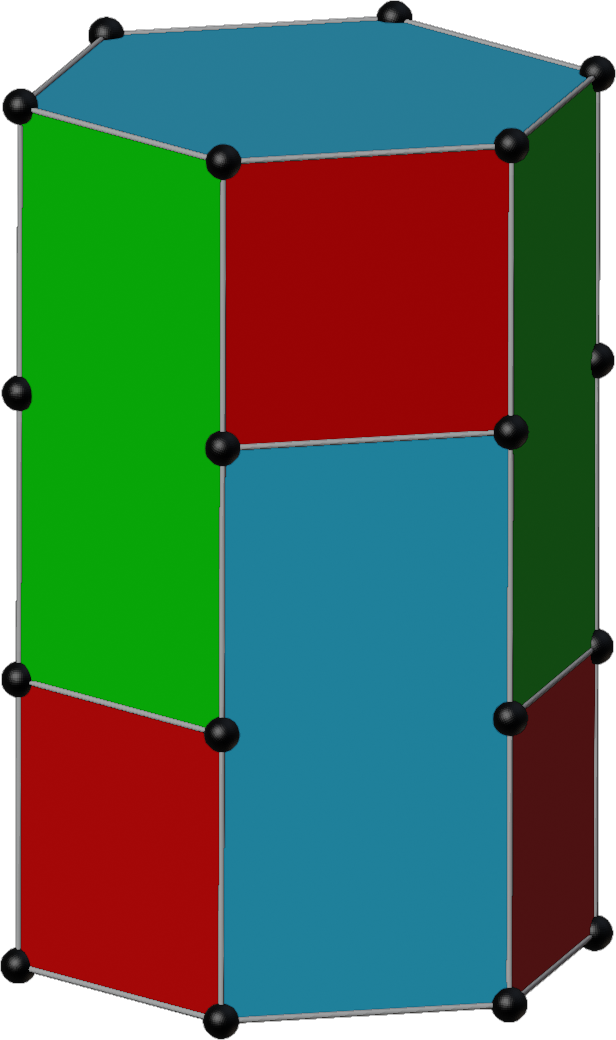}
};
\node[inner sep=0] (x2) at (0,-1.6){
\includegraphics[width=1.6cm]{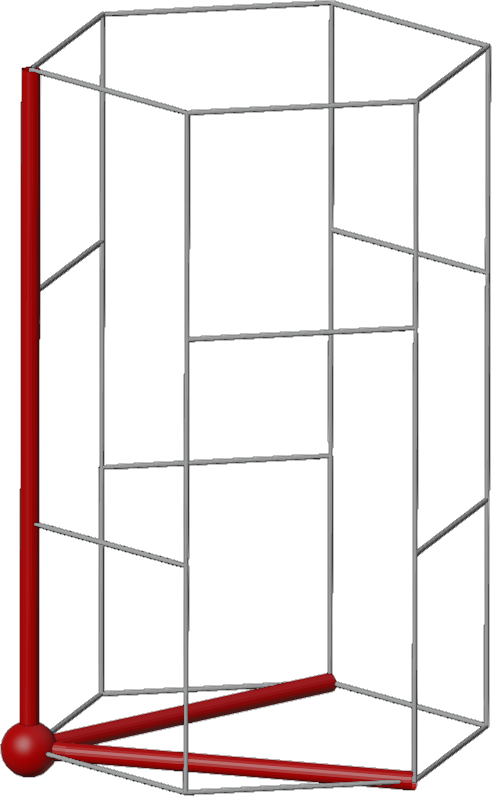}
};
\node[inner sep=0,anchor=east] at (x0.west){(a)};
\node[inner sep=0,anchor=east] at (x1.west){(b)};
\node[inner sep=0,anchor=east] at (x2.west){(c)};
\path (x2)++(-0.7,1.25)node {$t$};
\path (x2)++(0.7,-1.3)node {$x$};
\path (x2)++(0.4,-0.8)node {$y$};
\draw ($(x1)+(0,1.2)$)--++(90:0.4)node[anchor=south,inner sep=0]{hexagon};
\draw ($(x1)+(0.7,0.6)$)--++(30:0.8)node[anchor=south,inner sep=0,xshift=0.5cm]{brick-hexagon};
\draw ($(x1)+(0.7,-0.8)$)--++(30:0.8)node[anchor=south,inner sep=0]{square};
\end{tikzpicture}
\caption{
(a)
Snippet of the 3-colex and associated color path integral used to define the example twisted color circuit.
The volumes of the 3-colex are hexagon prisms colored accordingly.
The $X$-tensors of the $ZX$ diagram are white balls.
The $+\frac\pi4$ $Z$-tensors are purple balls, and the $-\frac\pi4$ $Z$-tensors are dark green balls.
Bonds are represented as black wires.
The two purple shaded paths of bonds are examples of qubit worldlines-- there are two such worldlines for each $t$-perpendicular edge.
The bonds shaded in dark blue are an example for a qubit timeslice.
(b)
Yellow volume of the 3-colex, whose boundary faces have been colored like the adjacent volume.
The boundary itself forms a 2-colex.
There are three types of faces in the 3-colex, which we will call ``hexagons'', ``squares'', and ``brick hexagons''.
(c)
The coordinate system on the left shows the time ($t$) direction, and defines the spatial $x$ and $y$ coordinates.
}
\label{fig:hexagon_zx}
\end{figure}
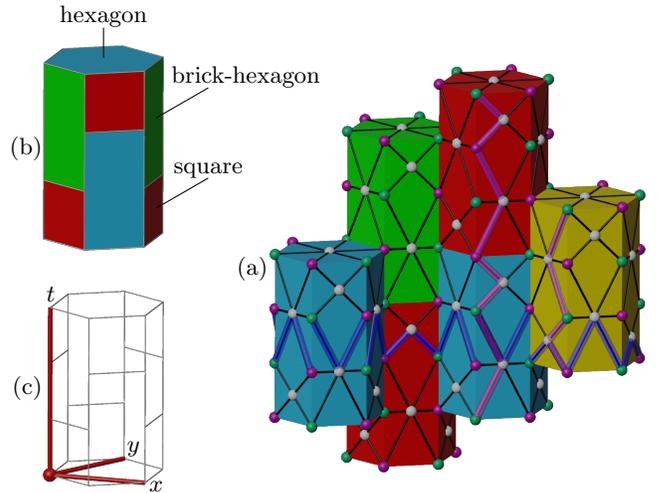

For our example twisted color circuit, we take the 3-colex shown in Fig.~\eqref{fig:hexagon_zx} consisting of hexagon-prism volumes, which is combinatorially equivalent to a \emph{bitruncated cubic honeycomb}~\cite{Wiki_bitrunc}.
The 3D color code is defined on one qubit for every vertex of the 3-colex.
There is one stabilizer for every face $f$, acting as a product of $Z$ at all the vertices $p$ of $f$,
\begin{equation}
S^Z_f = \prod_{p\in f} Z_p\;,
\end{equation}
and one stabilizer for every volume $v$, acting as a product of $X$ at all vertices of $v$,
\begin{equation}
S^X_v = \prod_{p\in v} X_p\;.
\end{equation}
The transversal $\ovl T$ gate~\cite{Bombin2006} acts as $T$ or $T^{-1}$ on every vertex:
\begin{equation}
\overline{T} = \prod_{p\in \Delta_0} T_p^{\sgn_p}\qcomma T = \mqty(\dmat[0]{1,e^{2\pi i\frac18 }})\;.
\end{equation}

\myparagraph{Color path integral from 3+0D protocol}
Based on the 3D color code, consider the following measurement-based protocol:
\begin{enumerate}
\item Prepare the qubit at every vertex in the $\ket+$ state.
\item Measure $S_f^Z$ at every face $f$, and apply $X$ corrections such that all measurement outcomes are flipped to $+1$.
This prepares a codestate of the 3D color code.
\item Apply the transversal $T$ gate.
\item Measure every qubit in the $X$ basis.
\end{enumerate}
Note that in order to make this protocol useful, we would have to add inputs and outputs at spatial boundaries as in the previous section, which we do not consider here.
The color path integral can be obtained from considering the $+1$-post-selected version of the above protocol.
That is, instead of applying corrections, we replace the measurements in the second and fourth step by the projector for the $+1$ outcome.
This post-selected circuit can be written as a $ZX$ tensor-network diagram.
A $ZX$ tensor network~\cite{Coecke2017,Wetering2020,Kissinger2022} is one consisting of \emph{$X$-tensors} and \emph{$Z$-tensors}.
Both $X$ and $Z$-tensors can be defined for an arbitrary number of indices and an arbitrary \emph{phase} $0\leq \alpha<2\pi$.
The $Z$-tensors force all their indices (valued in $\zz_2$) to take the same configuration,
\begin{equation}
\label{eq:delta_definition}
\begin{tikzpicture}
\atoms{delta}{0/lab={t=$\alpha$,p=45:0.25}}
\draw (0)edge[ind=$a$]++(0:0.5) (0)edge[ind=$b$]++(90:0.5) (0)edge[ind=$c$]++(180:0.5);
\node at (-90:0.5){$\ldots$};
\end{tikzpicture}
=
e^{i\alpha a}\cdot\delta_{a=b=c=\ldots}\;.
\end{equation}
Here, we are using a Kronecker-$\delta$-symbol which is $1$ if the condition in its subscript is fulfilled and $0$ otherwise.
The second family are \emph{$X$-tensors}, whose entries only depend on the total parity of its indices,
\begin{equation}
\begin{tikzpicture}
\atoms{z2}{0/lab={t=$\alpha$,p=45:0.25}}
\draw (0)edge[ind=$a$]++(0:0.5) (0)edge[ind=$b$]++(90:0.5) (0)edge[ind=$c$]++(180:0.5);
\node at (-90:0.5){$\ldots$};
\end{tikzpicture}
=
\sum_{x\in \zz_2} \frac12(1+e^{i(\alpha+\pi x)})\cdot \delta_{a+b+c+\ldots=x}\;.
\end{equation}
Note that the sum in the subscript is in $\zz_2$.
To represent the protocol that we just described we only need to some specific phases $\alpha\in\{0, \pm\pi/4,\pi\}$, and use colors instead of labels to indicate these phases:
\begin{equation}
\label{eq:zx_phases}
\begin{gathered}
\begin{tikzpicture}
\atoms{z2}{0/}
\draw (0)--++(0:0.4) (0)--++(90:0.4) (0)--++(180:0.4);
\node at (-90:0.2){$\ldots$};
\end{tikzpicture}
\coloneqq
\begin{tikzpicture}
\atoms{z2}{0/lab={t=$0$,p=45:0.25}}
\draw (0)--++(0:0.4) (0)--++(90:0.4) (0)--++(180:0.4);
\node at (-90:0.2){$\ldots$};
\end{tikzpicture}
\;,\quad
\begin{tikzpicture}
\atoms{delta}{0/}
\draw (0)--++(0:0.4) (0)--++(90:0.4) (0)--++(180:0.4);
\node at (-90:0.2){$\ldots$};
\end{tikzpicture}
\coloneqq
\begin{tikzpicture}
\atoms{delta}{0/lab={t=$0$,p=45:0.25}}
\draw (0)--++(0:0.4) (0)--++(90:0.4) (0)--++(180:0.4);
\node at (-90:0.2){$\ldots$};
\end{tikzpicture}\;,
\\
\begin{tikzpicture}
\atoms{delta,astyle=tcol}{0/}
\draw (0)--++(0:0.4) (0)--++(90:0.4) (0)--++(180:0.4);
\node at (-90:0.2){$\ldots$};
\end{tikzpicture}
\coloneqq
\begin{tikzpicture}
\atoms{delta}{0/lab={t=$\frac{\pi}{4}$,p=45:0.35}}
\draw (0)--++(0:0.4) (0)--++(90:0.4) (0)--++(180:0.4);
\node at (-90:0.2){$\ldots$};
\end{tikzpicture}
\;,\quad
\begin{tikzpicture}
\atoms{delta,astyle=tstarcol}{0/}
\draw (0)--++(0:0.4) (0)--++(90:0.4) (0)--++(180:0.4);
\node at (-90:0.2){$\ldots$};
\end{tikzpicture}
\coloneqq
\begin{tikzpicture}
\atoms{delta}{0/lab={t=$-\frac{\pi}{4}$,p=45:0.35}}
\draw (0)--++(0:0.4) (0)--++(90:0.4) (0)--++(180:0.4);
\node at (-90:0.2){$\ldots$};
\end{tikzpicture}\;,
\\
\begin{tikzpicture}
\atoms{z2,charge}{0/}
\draw (0)--++(0:0.4) (0)--++(90:0.4) (0)--++(180:0.4);
\node at (-90:0.2){$\ldots$};
\end{tikzpicture}
\coloneqq
\begin{tikzpicture}
\atoms{z2}{0/lab={t=$\pi$,p=45:0.25}}
\draw (0)--++(0:0.4) (0)--++(90:0.4) (0)--++(180:0.4);
\node at (-90:0.2){$\ldots$};
\end{tikzpicture}
\;,\quad
\begin{tikzpicture}
\atoms{delta,charge}{0/}
\draw (0)--++(0:0.4) (0)--++(90:0.4) (0)--++(180:0.4);
\node at (-90:0.2){$\ldots$};
\end{tikzpicture}
\coloneqq
\begin{tikzpicture}
\atoms{delta}{0/lab={t=$\pi$,p=45:0.25}}
\draw (0)--++(0:0.4) (0)--++(90:0.4) (0)--++(180:0.4);
\node at (-90:0.2){$\ldots$};
\end{tikzpicture}
\end{gathered}
\end{equation}
Note that the phase factor assigned to the purple and green $Z$-tensors is that of the $T$ and $T^{-1}$ gate.

The post-selected 3+0D protocol is a finite-depth circuit that creates qubits out of nothing, and in the end discards them by projecting them onto the $\ket+$ state.
The operators in this circuit can be represented as simple $ZX$ diagrams, which we interpret as linear operators from the bottom to the top in the following diagrams.
\begin{itemize}
\item The $\ket+$ state preparation is an operator mapping from the trivial vector space $\mathbb C$ to the vector space of a qubit $\mathbb C^2$, and can be diagrammatically represented using a 1-index $Z$-tensor:
\begin{equation}
\label{eq:plus_preparation}
\begin{tikzpicture}
\atoms{delta}{0/}
\draw (0)edge[ind=$a$]++(90:0.5);
\end{tikzpicture}
= 1 = \sqrt2 \braket{a}{+} \; \forall a
\;.
\end{equation}
\item The $+1$-post-selected $Z^{\otimes n}$-measurement ($M_{Z^{\otimes n}}$) acting on the $n$ vertices of a face has the familiar form~\cite{Kissinger2022} (here for $n=6$):
\begin{equation}
\label{eq:z6_measurement}
\begin{gathered}
\begin{tikzpicture}
\atoms{z2}{0/}
\atoms{delta}{1/p={-150:0.8}, 2/p={180:1.5}, 3/p={150:1.2},4/p={-30:0.8},5/p={0:1.5},6/p={30:1.2}}
\draw (0)--(1) (0)--(2) (0)--(3) (0)--(4) (0)--(5) (0)--(6);
\draw (1)edge[ind=$a$]++(-90:0.5) (1)edge[ind=$a'$]++(90:1) (2)edge[ind=$b$]++(-90:0.5) (2)edge[ind=$b'$]++(90:0.5) (3)edge[ind=$c$]++(-90:0.8) (3)edge[ind=$c'$]++(90:0.5) (4)edge[ind=$d$]++(-90:0.5) (4)edge[ind=$d'$]++(90:1) (5)edge[ind=$e$]++(-90:0.5) (5)edge[ind=$e'$]++(90:0.5) (6)edge[ind=$f$]++(-90:0.8) (6)edge[ind=$f'$]++(90:0.5);
\end{tikzpicture}\\
=
\delta_{a=a'} \delta_{b=b'} \delta_{c=c'} \delta_{d=d'} \delta_{e=e'} \delta_{f=f'} \delta_{a+b+c+d+e+f=0}\\
=
\bra{a',b',c',d',e',f'} (1+Z^{\otimes 6}) \ket{a,b,c,d,e,f}
\;.
\end{gathered}
\end{equation}
\item The $T$ gate at every vertex $p$ with $\sgn_p=+1$ is a 2-index $Z$-tensor with $\frac\pi4$ phase,
\begin{equation}
\label{eq:tgate}
\begin{tikzpicture}
\atoms{delta}{0/astyle=tcol}
\draw (0)edge[ind=$b$]++(90:0.5) (0)edge[ind=$a$]++(-90:0.5);
\end{tikzpicture}
=
\delta_{a=b} e^{\frac{2\pi i}{8}}
=
\bra{b}T\ket{a}\;,
\end{equation}
\item The $T^{-1}$ gate at every $\sgn_p=-1$ vertex is defined analogously,
\begin{equation}
\label{eq:tstargate}
\begin{tikzpicture}
\atoms{delta}{0/astyle=tstarcol}
\draw (0)edge[ind=$b$]++(90:0.5) (0)edge[ind=$a$]++(-90:0.5);
\end{tikzpicture}
=
\delta_{a=b} e^{-\frac{2\pi i}{8}}
=
\bra{b}T^{-1}\ket{a}
\;.
\end{equation}
\item The post-selected destructive single-qubit $X$ measurement is a projection onto the $\ket+$ state, and thus the diagram is the time-reversed version of Eq.~\eqref{eq:plus_preparation},
\begin{equation}
\label{eq:plus_projection}
\begin{tikzpicture}
\atoms{delta}{0/}
\draw (0)edge[ind=$a$]++(-90:0.5);
\end{tikzpicture}
= 1 = \sqrt2 \braket{+}{a}
\; \forall a \;.
\end{equation}
\end{itemize}
Composing the $ZX$ building blocks above according to the circuit yields a $ZX$ tensor-network diagram on the 3-colex.
At every vertex, we first have a $\ket+$ preparation, then the qubit is involved in 6 $Z^{\otimes n}$-measurements at the 6 adjacent faces, then we perform a $T$ or $T^{-1}$ gate, and, finally, a projection onto the $\ket+$ state.
Each of these operations involves one $Z$-tensor, and all these $Z$-tensors can be fused into a single one (using the ``spider-fusion'' rule of the $ZX$ calculus):
\begin{equation}
\label{eq:30d_path_integral_translation}
\begin{tikzpicture}
\atoms{delta}{0/, 1/p={0,0.4}, 2/p={0,0.8}, 3/p={0,1.2}, 4/p={0,1.6}, 5/p={0,2}, 6/p={0,2.4}, {7/p={0,2.8},astyle=tcol}, 8/p={0,3.2}}
\draw (0)--(1)--(2)--(3)--(4)--(5)--(6)--(7)--(8);
\draw (1)--++(0:1) (3)--++(0:1) (5)--++(0:1) (2)--++(180:1) (4)--++(180:1) (6)--++(180:1);
\end{tikzpicture}
=
\begin{tikzpicture}
\atoms{delta}{0/astyle=tcol}
\draw (0)--++(-30:1) (0)--++(0:1) (0)--++(30:1) (0)--++(-150:1) (0)--++(180:1) (0)--++(150:1);
\end{tikzpicture}
\;.
\end{equation}
After this fusion, the $ZX$ tensor-network diagram takes a very simple form, as shown in Fig.~\ref{fig:hexagon_zx}:
There is one $Z$-tensor at every vertex of the 3-colex, one $X$-tensor at every face, and one bond connecting the tensors at every vertex and adjacent face.
All $Z$-tensors have 6 indices, and the $X$-tensors have some even number of indices depending on the chosen 3-colex (either 4 or 6 in the 3-colex we consider for our example).
The $Z$-tensors have a phase of $\sgn_p\frac{\pi}{4}$.
We will to refer this $ZX$ tensor network as the \emph{color path integral}.

Note that we can also introduce an \emph{untwisted color path integral}, where all $Z$ tensors have trivial phases, corresponding to the same 3+0D protocol without $T$ gates.
This untwisted color path integral can also be turned into an \emph{untwisted color circuit}, which is interesting in its own right, but only yields Clifford gates.
We will use the untwisted path integral and circuit for pedagogical reasons at some places in Sections~\ref{sec:path_integral} and \ref{sec:decoding}.

\subsection{The twisted color circuit}
\label{sec:circuit_easy}
In this section we turn the $ZX$ color path integral back into a circuit, but one in 2+1D instead of 3+0D.
After choosing which of the directions in the 3-colex correspond to space and which to time, we apply some $ZX$ \emph{rewrite rules} to obtain a $+1$-post-selected circuit.
The real circuit is then obtained by simply replacing post-selected measurements by real measurements.
To retain fault tolerance in the non-post-selected case, we need to add decoding and corrections, as we discuss in Section~\ref{sec:decoding_easy} and in more detail in Section~\ref{sec:decoding}.

Our example twisted color circuit is obtained from the 3-colex in Fig.~\ref{fig:hexagon_zx}.
We start by choosing space and time directions in our 3-colex, indicated in the coordinate system in Fig.~\ref{fig:hexagon_zx}.
According to these chosen space and time coordinates we identify sets of bonds in the $ZX$-diagram that correspond to different qubits at a fixed time coordinate (called \emph{qubit timeslices}), as well as sets of bonds that are represented by a fixed qubit at different times (called \emph{qubit worldlines}).
Both are shown in Fig.~\ref{fig:hexagon_zx}.

\begin{figure}
\includegraphics{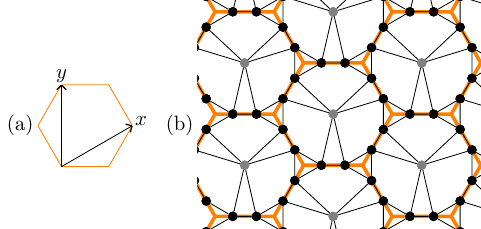}
\caption{
(a) Spatial coordinate system.
(b) Projection of the 3-colex of Fig.~\ref{fig:hexagon_zx} along time direction yields a hexagonal lattice shown in orange.
The qubits of the circuit are black dots.
The black lines show the planar connectivity required by the 2-qubit gates in the circuit.
In practice, the $Z^{\otimes 6}$ measurements require an ancilla qubit in the center of each face, which we indicate with gray dots.
}
\label{fig:hexagon_qubits}
\end{figure}

Projecting the 3-colex along the time direction yields a hexagonal lattice, as shown in Fig.~\ref{fig:hexagon_qubits}.
For each edge of the hexagonal lattice, there are two qubit worldlines, and accordingly two qubits.
The next step is to turn the $ZX$ diagram into a $+1$-post-selected circuit of operators with respect to the chosen qubit worldlines and timeslices, using the $ZX$ rewrite rules.
In particular, we use the spider fusion rules to split up some of the tensors into a group of ``smaller'' tensors, and then regroup the smaller tensors, such that each group can be viewed as an operator acting on the qubit worldlines.
The resulting post-selected circuit uses the following operators, expressed as $ZX$ diagrams:
\begin{itemize}
\item A unitary controlled-$X$ ($CX$) gate,
\begin{equation}
\begin{tikzpicture}
\atoms{delta}{0/}
\atoms{z2}{1/p={0.8,0}}
\draw (0)--(1) (0)edge[ind=$a$]++(-90:0.5) (1)edge[ind=$b$]++(-90:0.5) (0)edge[ind=$c$]++(90:0.5) (1)edge[ind=$d$]++(90:0.5);
\end{tikzpicture}
=
\delta_{a=c=b+d}
=
\bra{c,d} CX \ket{a,b}
\;.
\end{equation}
\item a $+1$-post-selected 2-qubit $XX$-measurement $M_{XX}$,
\begin{equation}
\begin{tikzpicture}
\atoms{z2}{0/}
\draw (0)edge[ind=$a$]++(-135:0.5) (0)edge[ind=$b$]++(-45:0.5) (0)edge[ind=$c$]++(135:0.5) (0)edge[ind=$d$]++(45:0.5);
\end{tikzpicture}
=
\delta_{a+b+c+d=0}
=
\bra{c,d} (1+XX) \ket{a,b}
\;.
\end{equation}
\item a $+1$-post-selected 2-qubit $ZZ$-measurement $M_{ZZ}$,
\begin{equation}
\label{eq:delta_tensor_operator}
\begin{tikzpicture}
\atoms{delta}{0/}
\draw (0)edge[ind=$a$]++(-135:0.5) (0)edge[ind=$b$]++(-45:0.5) (0)edge[ind=$c$]++(135:0.5) (0)edge[ind=$d$]++(45:0.5);
\end{tikzpicture}
=
\delta_{a=b=c=d}
=
\bra{c,d} \frac12(1+ZZ) \ket{a,b}
\;.
\end{equation}
\item the $+1$-post-selected measurement $M_{Z^{\otimes 6}}$ shown in Eq.~\eqref{eq:z6_measurement}, the $T$ gate in Eq.~\eqref{eq:tgate}, and the $T^{-1}$ gate in Eq.~\eqref{eq:tstargate}.
\end{itemize}
Concretely, we will have
\begin{itemize}
\item one $M_{ZZ}$ measurement for every vertex of the 3-colex,
\item one $CX$ gate for every vertex,
\item one $M_{XX}$ measurement for every time-parallel brick hexagon or square,
\item and one $M_{Z^{\otimes 6}}$ measurement for every time-perpendicular hexagon.
\end{itemize}
Let us now show how we regroup the tensors of the $ZX$ path integral to obtain these operators:
Every $Z$-tensor at a vertex $p\in \Delta_0$ has 6 indices connecting it to $X$-tensors at (1) a past time-parallel brick hexagon, (2) a past time-parallel square, (3) a future brick hexagon, (4) a future square, (5) a time-perpendicular hexagon, and (6) a brick hexagon whose center has the same time coordinate as $p$.
We split it up into one 4-index $Z$-tensor, two 3-index $Z$-tensors, and one 2-index $Z$ tensor carrying the $T^{\sgn_p}$ phase:
\begin{equation}
\label{eq:hexagon_ztensor_splitting}
\begin{tikzpicture}
\atoms{delta}{0/astyle=tcol}
\draw (0)edge[ind=$1$]++(-120:0.8)  (0)edge[ind=$2$]++(-60:0.8)  (0)edge[ind=$4$]++(120:0.8)  (0)edge[ind=$3$]++(60:0.8)  (0)edge[ind=$6$]++(0:0.8)  (0)edge[ind=$5$]++(180:0.8);
\end{tikzpicture}
=
\begin{tikzpicture}
\atoms{delta}{0/, 1/p={60:0.5}, 2/p={120:0.5}, {3/p={60:1.1},astyle=tcol}}
\draw (0)--(1) (0)--(2) (1)--(3) (0)--++(-120:0.8)  (0)--++(-60:0.8)  (2)--++(120:0.8)  (3)--++(60:0.4)  (1)--++(0:0.8)  (2)--++(180:0.8);
\begin{scope}
\clip (current bounding box.south west)rectangle(current bounding box.north east);
\draw[dashed,cyan] ($(60:0.5)+(0:0.8)$) ellipse (1cm and 0.3cm);
\draw[dashed,cyan] ($(120:0.5)+(180:0.8)$) ellipse (1cm and 0.3cm);
\end{scope}
\draw[cyan] node at ($(60:0.5)+(0:0.5)$) {$CX$};
\draw[cyan] node at ($(120:0.5)+(-0.65,-0.1)$) {$M_{Z^{\otimes 6}}$};
\draw[dashed,cyan] (0,0) node[shift={(-20:0.6)}] {$M_{ZZ}$} ellipse (0.35cm and 0.2cm);
\draw[cyan] node at ($(3)+(-0.35,0)$) {$T$};
\draw[dashed,cyan] (3)ellipse (0.2cm and 0.2cm);
\end{tikzpicture}
\;.
\end{equation}
The labels at the indices correspond to the numbers of the list above.
As shown, it gives rise to the $M_{ZZ}$ measurement at the vertex, contributes to half of the $CX$ gate at the vertex, and contributes to one of the 6 $Z$-tensors of the $M_{Z^{\otimes 6}}$ measurement at the adjacent time-perpendicular hexagon (5).
Next, the 6-index $X$-tensor at every time-parallel brick hexagon is split into one 4-index $X$-tensor and two 3-index $X$-tensors:
\begin{equation}
\label{eq:xtensor_splitting}
\begin{tikzpicture}
\atoms{z2}{0/}
\draw (0)--++(-120:0.8)  (0)--++(-60:0.8)  (0)--++(120:0.8)  (0)--++(60:0.8)  (0)--++(0:0.8)  (0)--++(180:0.8);
\end{tikzpicture}
=
\begin{tikzpicture}
\atoms{z2}{0/, 1/p={60:0.5}, 2/p={120:0.5}}
\draw (0)--(1) (0)--(2) (0)--++(-120:0.8)  (0)--++(-60:0.8)  (2)--++(120:0.8)  (1)--++(60:0.8)  (1)--++(0:0.8)  (2)--++(180:0.8);
\draw[cyan] node at ($(60:0.5)+(0:0.5)$) {$CX$};
\draw[cyan] node at ($(120:0.5)+(180:0.5)$) {$CX$};
\draw[dashed,cyan] (0,0) node[shift={(-20:0.6)}] {$M_{XX}$} ellipse (0.35cm and 0.2cm);
\clip (current bounding box.south west)rectangle(current bounding box.north east);
\draw[dashed,cyan] ($(60:0.5)+(0:0.5)$) ellipse (0.7cm and 0.3cm);
\draw[dashed,cyan] ($(120:0.5)+(180:0.5)$) ellipse (0.7cm and 0.3cm);
\end{tikzpicture}
\;.
\end{equation}
As shown, it gives rise to the $M_{XX}$ measurement of the hexagon, and contributes half of the $CX$ gates at both of its middle-time-coordinate vertices.
The 4-index $X$-tensor at every time-parallel square is not split up and simply gives rise to the $M_{XX}$ measurement at the square.
Finally, the 6-index $X$-tensor at every time-perpendicular hexagon forms the central $X$-tensor of the $Z^{\otimes 6}$ measurement.

\begin{figure}
\includegraphics[width=0.85\linewidth]{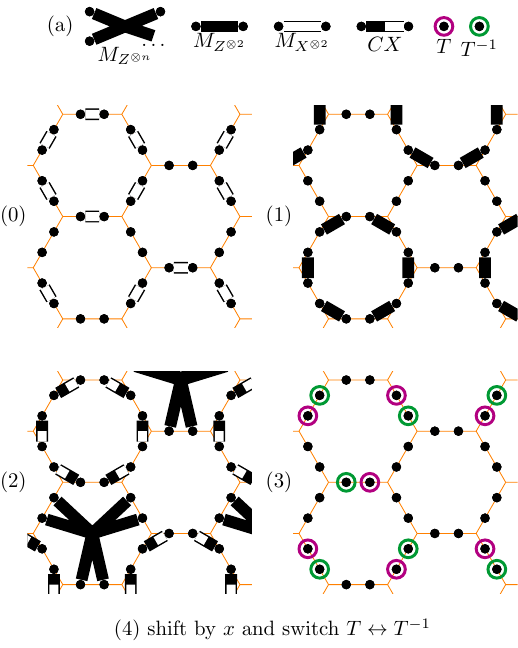}
\caption{
(a) Symbols used for $Z$-type measurements, $X$-type measurements, $CX$, $T$ and $T^{-1}$ gates.
(0)-(3) 4 time steps of the circuit corresponding to a time translation by $\frac13 t$ in the 3-colex in Fig.~\ref{fig:hexagon_zx}.
In step (0), we perform all the $M_{XX}$ measurements resulting from the $X$-tensors on squares and brick-hexagons.
In step (1), we perform all the $M_{ZZ}$ measurements resulting from the $Z$-tensors on every vertex.
In step (2), we perform all the $Z^{\otimes 6}$ measurements at the hexagons, as well as the $CX$ gates related to the time-perpendicular bonds between the $Z$-tensors and brick-hexagon $X$-tensors.
In step (3), we perform all the $T$ or $T^{-1}$ gates implementing the $e^{\pm\frac{2\pi i}{8}}$ phases of every $Z$-tensor.
After step (3), we continue with step (0), but with all operators shifted by $x$, and with $T$ and $T^{-1}$ in step (3) exchanged.
One full time period thus consists of 12 steps.
}
\label{fig:bulk_circuit}
\end{figure}

The last step in deriving the twisted color circuit is to observe on which qubits the operators above act in which order, which is best done by looking at Fig.~\ref{fig:hexagon_zx}.
The resulting circuit is shown in Fig.~\ref{fig:bulk_circuit}.
In practice, one might perform the $M_{Z^{\otimes 6}}$ measurement using an ancilla qubit in the middle of the hexagon, via 6 $CX$ gates and a single-qubit measurement, or more involved methods that use more ancillary qubits~\cite{Chamberland2018, Magdalena2024, rodatz2024}.
The resulting circuit then consists of 1 and 2-qubit operations and has fully planar connectivity as shown in Fig.~\ref{fig:hexagon_qubits}.

\myparagraph{Spacetime unit cell}
If we ignore the difference between $T$ and $T^{-1}$ in the circuit, and the 4-coloring and signs on the vertices of the 3-colex, then both have the same minimal unit cell.
Such a unit cell is spanned by translations under which the 3-colex in Fig.~\ref{fig:hexagon_zx} is invariant.
There are multiple ways of defining a minimal unit cell of the 3-colex.
We choose one spanned by $\frac13 t+x$, $2x-y$, and $2y-x$ since this leads to the most compact representation of the circuit in Fig.~\ref{fig:bulk_circuit}.%
\footnote{Alternatively, we could have chosen a unit cell spanned by $t$, $\frac13 t+x$, and $\frac13 t+y$.
In this case we would specify the circuit on a smaller patch of the spatial lattice but for 12 instead of 4 time steps.}
If we do distinguish between $T$ and $T^{-1}$, and accordingly include the vertex signs, the minimal unit cell is enlarged by a factor of 2:
The translation by $\frac13 t+x$ inverts the vertex signs, so we have to replace it by $2(\frac12 t+x)$.
In Fig.~\ref{fig:bulk_circuit} we incorporate this by swapping $T$ and $T^{-1}$ in step (4).

If we include the 4-coloring of the 3-colex, then the unit cell becomes larger.
For example, the 3-colex is no longer invariant under a displacement by $t$ but only by $4t$.
As a consequence, the 12 time steps of the circuit corresponding to a $t$ displacement implement some non-trivial logical Clifford unitary when put on a finite lattice with periodic boundary conditions, and only 48 time steps (or 16 time steps together with a shift by $x$) implement the logical identity.
Further, for certain open boundary conditions, such as these discussed in Section~\ref{sec:boundaries_easy}, the minimal unit cell is that of the 3-colex including the 4-coloring.
Accordingly, a full time period of the boundary circuit consists of 48 instead of 12 steps.
A third place where the 4-coloring matters is the decoding as we discuss in Section~\ref{sec:decoding_easy}.

\subsection{Boundaries of the twisted color circuit}
\label{sec:boundaries_easy}
Let us next describe how to define open boundary conditions for our circuit.
We cannot simply truncate the circuit in an arbitrary way, but we have to do it in a specific topologically protected way.
Further, there are multiple distinct boundary conditions implementing different boundary topological phases, and we have to choose the ones that result in the desired logical action.
For the $\ket T$-state measurement protocol shown in Eq.~\eqref{eq:global_protocol}, we need three different boundaries corresponding to three different boundary phases.
We name these boundaries the $r$, $g$, and $b$ \emph{color boundaries}, after the related boundaries of the 3D color code.
Note that the three boundaries are related by permuting the four 3-colex colors, so it suffices to focus on the $r$ boundary.%
\footnote{A fourth $y$ color boundary is completely analogous but not used in the protocol.}

Constructing the circuit on the boundary is straight-forward once we have defined suitable boundary conditions for the color path integral.
Such boundary conditions correspond to ways to terminate the $ZX$ diagram in a way that does not result in open indices at the boundary, and that yields a macroscopic fault distance.
Another way to view this is to cut the tensor network along some plane resulting in a \emph{state boundary} (see Section~\ref{sec:path_integral}) with open indices, and then close these open indices by contracting them with some auxiliary 2D boundary tensor network~\cite{xyzrubycode}.

\begin{figure}
\begin{tikzpicture}[scale=1]
\node[inner sep=0] (x0) at (4.3,0){
\includegraphics[width=4.75cm]{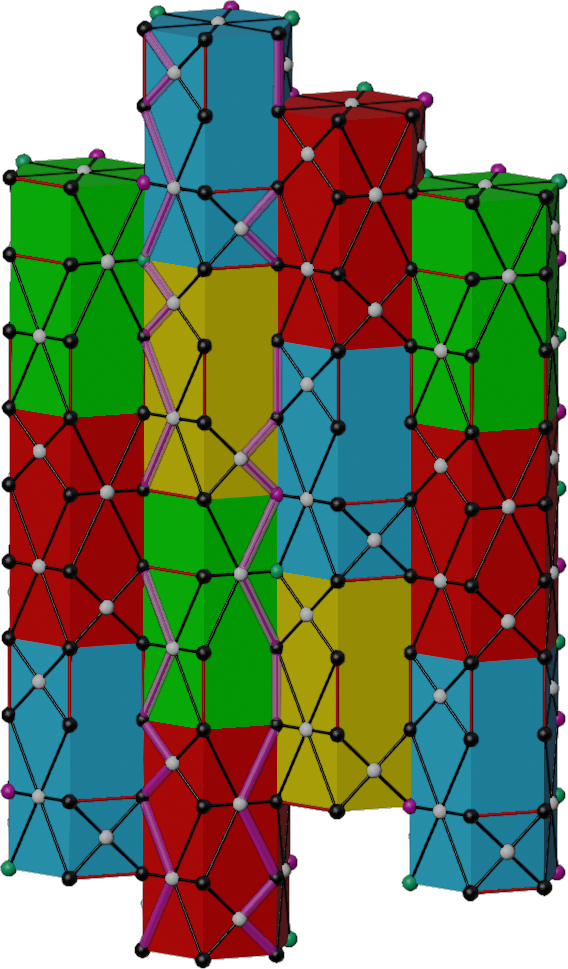}
};
\node[inner sep=0] (x1) at (0,-3){
\includegraphics[width=1.5cm]{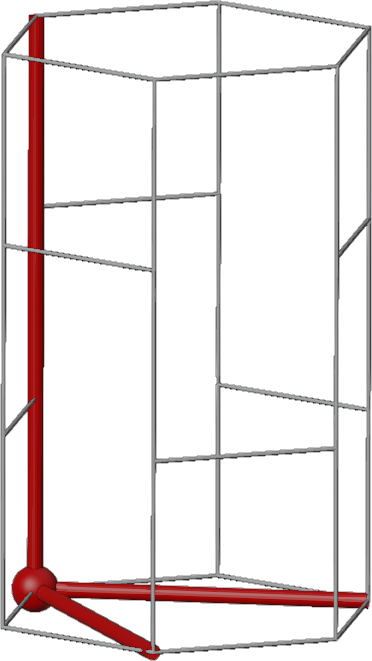}
};
\node[inner sep=0,anchor=east,yshift=1.5cm] at (x0.west){(a)};
\node[inner sep=0,anchor=east] at (x1.west){(b)};
\path (x1)++(-0.6,1.5)node {$t$};
\path (x1)++(0,-1.5)node {$x$};
\path (x1)++(0.9,-1.2)node {$y$};
\end{tikzpicture}
\caption{
(a)
Color path integral with $r$ color boundary on a snippet of 3-colex.
The boundary surface is in front, extending along the $(y,t)$ plane (in a zig-zag manner along $y$).
The $ZX$ diagram is a simple truncation of the bulk diagram (see Fig.~\ref{fig:hexagon_zx}) up to the following three considerations:
(1) For every $\ovl r$-colored boundary edge, we add a new bond (drawn in red) connecting the $Z$-tensors at its vertices.
(2) For every $gb$, $gy$, or $by$-colored boundary face, we may or may not include the $X$-tensor, which is the same up to rewrite rules.
We chose to keep the $X$-tensor at ``every second'' of these boundary faces, since this is convenient for turning the path integral into a circuit.
(3) We only include the $\sgn_p \frac\pi4$ phases at the boundary vertices that are adjacent to a $g$, $b$, and $y$ bulk volume, and the remaining boundary $Z$-tensors are phaseless.
Two representative qubit worldlines inside the boundary are shaded in purple.
(b)
Coordinate system.
}
\label{fig:r_boundary_zx}
\end{figure}
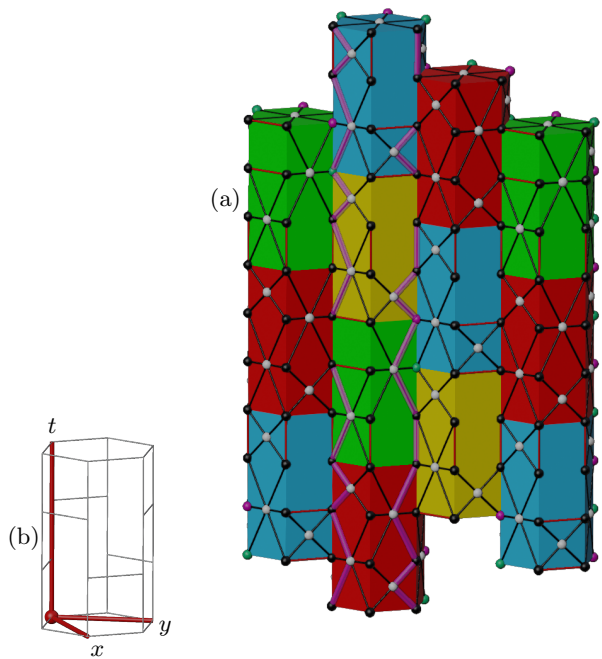

Geometrically, we define a color boundary along any surface consisting of faces of the 3-colex.
The specific time-parallel boundary surface that we choose in our hexagon-prism 3-colex extends in $t$ direction in a straight manner and in $y$ direction in a zig-zag manner, corresponding to the surface in front in Fig.~\eqref{fig:r_boundary_zx}.
The $ZX$ diagram for the $r$ color boundary on this surface is explained and depicted in Fig.~\ref{fig:r_boundary_zx}.
We will argue why these boundaries are topologically protected and give rise the correct logical action later in Section~\ref{sec:boundaries} and Appendix~\ref{sec:boundaries_derivation}.%
\footnote{One could derive these boundaries from analogous color boundaries of the 3D color code.
Unfortunately, $r$ color boundaries are commonly defined along a surface inside the triangulation dual to the 3-colex, which cuts only through $\ovl r$-colored edges of the primal 3-colex~\cite{Bombin2006a}.
Our color boundaries are defined on an arbitrary surface inside the primal 3-colex, and the additional bonds at the red boundary edges would correspond to additional $ZZ$ stabilizers at the boundary of the 3D color code.
Ref.~\cite{Song2024} defines similar looking boundaries with $ZZ$ stabilizers but still considers surfaces in the dual triangulation.}

\begin{figure}
\includegraphics[width=0.8\linewidth]{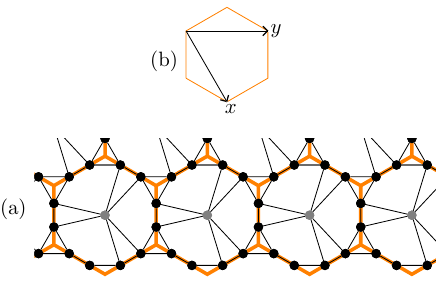}
\caption{
(a) Spatial projection of the bulk 3-colex with boundary from Fig.~\ref{fig:r_boundary_zx}, yielding a hexagonal lattice with ``zig-zag'' boundary at the bottom.
Left, right, and top are not boundaries but truncations of the drawing.
The qubits are represented by black dots, and there are two at every bulk and boundary edge.
The black lines represent the connectivity of 2-qubit operations required by the circuit.
(b) Coordinate system, with $x$ and $y$ coordinates consistent with Fig.~\ref{fig:r_boundary_zx} (b), such that the bottom edge of (a) coincides with the front surface of Fig.~\ref{fig:r_boundary_zx} (a).
In the coordinate system of Fig.~\ref{fig:hexagon_qubits}, the boundary is located on the right and extends in the vertical direction.
}
\label{fig:boundary_qubits}
\end{figure}

Next, we show how to turn the path-integral boundaries into a circuit, using the same method as in Section~\ref{sec:circuit_easy}.
Two representative qubit worldlines along the boundary are shown in Fig.~\ref{fig:r_boundary_zx}.
The projected spatial lattice is a hexagonal lattice with ``zig-zag'' boundary, as shown in Fig.~\ref{fig:boundary_qubits}.
Just like in the bulk, there are two qubits on every boundary edge.

\begin{figure*}
\includegraphics[height=0.9\textheight]{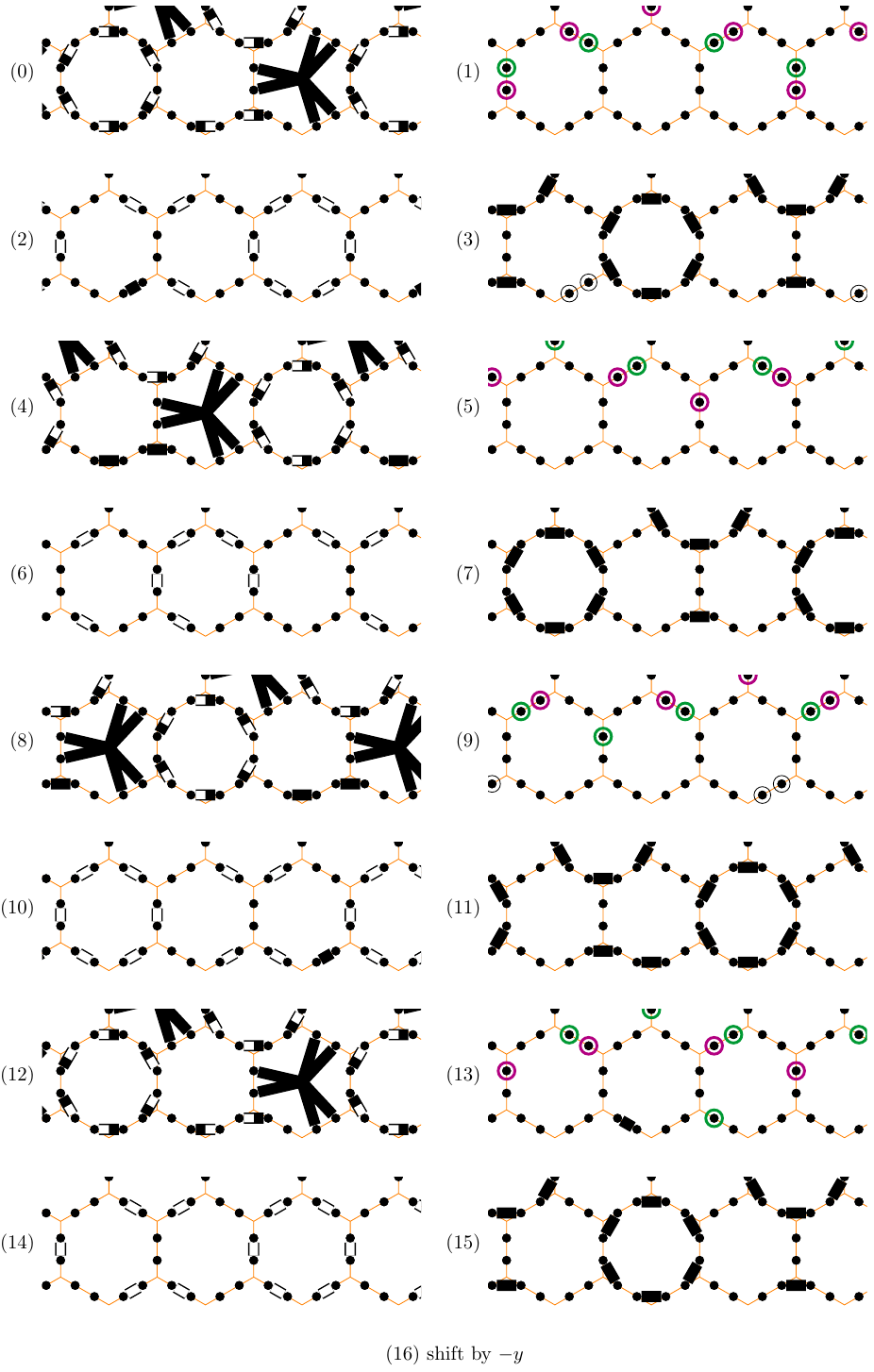}
\caption{
(0)-(15) Circuit implementing an $r$ color boundary on the qubits shown in Fig.~\ref{fig:boundary_qubits}.
The symbols for gates are the same as in Fig.~\ref{fig:bulk_circuit}, with
{\protect\tikz\protect\atoms{vertex}{0/}\protect\singlexmeas{0};}
representing a single-qubit $X$ measurement $M_x$.
}
\label{fig:hexagon_boundary_circuit}
\end{figure*}

After identifying qubit worldlines, we slightly reshape the $ZX$ diagram to turn it into a $+1$-post-selected circuit.
In addition to the operators introduced in Sections~\ref{sec:easy_path_integral} and \ref{sec:circuit_easy}, the circuit contains post-selected non-destructive single-qubit $X$ measurements which are expressed as a $ZX$ diagram as follows:
\begin{equation}
\begin{tikzpicture}
\atoms{delta}{0/, 1/p={0,0.5}}
\draw (0)edge[ind=$a$]++(-90:0.5) (1)edge[ind=$b$]++(90:0.5);
\end{tikzpicture}
=
1
=
\braket{a}{+}\braket{+}{b}\;\forall a,b\;.
\end{equation}
The square and brick hexagon boundary faces that still have $X$-tensors at their center give rise to operators in the same way as in the bulk.
The square boundary faces without $X$-tensor but two adjacent time-parallel red bonds correspond to doing nothing on the two qubits whose worldlines run through the bonds:
\begin{equation}
\begin{tikzpicture}
\atoms{delta}{0/lab={t=$\ldots$,p={-90:0.25}}, {1/p={0.8,0},lab={t=$\ldots$,p={-90:0.25}}}, {2/p={0,0.8},lab={t=$\ldots$,p={90:0.25}}}, {3/p={0.8,0.8},lab={t=$\ldots$,p={90:0.25}}}}
\draw[red] (0)--(2) (1)--(3);
\end{tikzpicture}
=
\begin{tikzpicture}
\atoms{delta}{0/lab={t=$\ldots$,p={-90:0.25}}, {1/p={0.8,0},lab={t=$\ldots$,p={-90:0.25}}}, {2/p={0,1},lab={t=$\ldots$,p={90:0.25}}}, {3/p={0.8,1},lab={t=$\ldots$,p={90:0.25}}}}
\draw[red] (0)--(2) (1)--(3);
\draw[cyan,dashed] (0.4,0.5)ellipse(0.6cm and 0.2cm) node{$\idop$};
\end{tikzpicture}
\;.
\end{equation}
The square boundary faces without $X$-tensor but two adjacent time-perpendicular red bonds correspond to a projection of the two qubits onto the Bell state, which can be written as a $+1$-post-selected $ZZ$ followed by a $XX$ measurement:
\begin{equation}
\begin{tikzpicture}
\atoms{delta}{0/lab={t=$\ldots$,p={-90:0.25}}, {1/p={0.8,0},lab={t=$\ldots$,p={-90:0.25}}}, {2/p={0,0.8},lab={t=$\ldots$,p={90:0.25}}}, {3/p={0.8,0.8},lab={t=$\ldots$,p={90:0.25}}}}
\draw[red] (0)--(1) (2)--(3);
\end{tikzpicture}
=
\begin{tikzpicture}
\atoms{delta}{0/lab={t=$\ldots$,p={-90:0.25}}, {1/p={0.8,0},lab={t=$\ldots$,p={-90:0.25}}}, {2/p={0,1.2},lab={t=$\ldots$,p={90:0.25}}}, {3/p={0.8,1.2},lab={t=$\ldots$,p={90:0.25}}}}
\atoms{delta}{a/p={0.4,0.3}}
\atoms{z2}{b/p={0.4,0.9}}
\draw (0)--(a) (1)--(a) (a)to[bend left=45](b) (a)to[bend right=45](b) (2)--(b) (3)--(b);
\draw[cyan,dashed] (a)circle(0.2) node[shift=(150:0.1),anchor=east]{$M_{ZZ}$};
\draw[cyan,dashed] (b)circle(0.2) node[shift=(-150:0.1),anchor=east]{$M_{XX}$};
\end{tikzpicture}
\;.
\end{equation}
The $ZX$ diagram around the boundary brick hexagons without $X$-tensor but a time-perpendicular red bond at its largest-time edge (in addition to two time-parallel red bonds) is rewritten such that it corresponds to two single-qubit $X$-measurements and one 2-qubit $M_{ZZ}$ measurement:
\begin{equation}
\begin{tikzpicture}
\atoms{delta}{0/lab={t=$\ldots$,p={-90:0.25}}, {1/p={0.8,0},lab={t=$\ldots$,p={-90:0.25}}}, {2/p={0,0.8},lab={t=$\ldots$,p={180:0.35}}}, {3/p={0.8,0.8},lab={t=$\ldots$,p={0:0.35}}}, {4/p={0,1.6},lab={t=$\ldots$,p={90:0.25}}}, {5/p={0.8,1.6},lab={t=$\ldots$,p={90:0.25}}}}
\draw[red] (0)--(2) (1)--(3) (4)--(5);
\end{tikzpicture}
=
\begin{tikzpicture}
\atoms{delta}{0/lab={t=$\ldots$,p={-90:0.25}}, {1/p={0.8,0},lab={t=$\ldots$,p={-90:0.25}}}, {2/p={0,0.6},lab={t=$\ldots$,p={180:0.35}}}, {3/p={0.8,0.6},lab={t=$\ldots$,p={0:0.35}}}, {4/p={0,2.2},lab={t=$\ldots$,p={90:0.25}}}, {5/p={0.8,2.2},lab={t=$\ldots$,p={90:0.25}}}}
\atoms{delta}{a/p={0,1}, b/p={0,1.4}, c/p={0.8,1}, d/p={0.8,1.4}, e/p={0.4,1.8}}
\draw (0)--(2) (1)--(3) (2)--(a) (b)--(e) (3)--(c) (d)--(e) (e)--(4) (e)--(5);
\draw[cyan,dashed] (e)circle(0.2) node[shift=(180:0.1),anchor=east]{$M_{ZZ}$};
\draw[cyan,dashed] (0,1.2)ellipse(0.2cm and 0.4cm) node[shift=(180:0.1),anchor=east]{$M_{X}$};
\draw[cyan,dashed] (0.8,1.2)ellipse(0.2cm and 0.4cm) node[shift=(0:0.1),anchor=west]{$M_{X}$};
\end{tikzpicture}
\;.
\end{equation}
The $ZX$ diagram around the brick hexagons without $X$-tensor but with a red bond at the smallest-time edge is rewritten similarly:
\begin{equation}
\begin{tikzpicture}
\atoms{delta}{0/lab={t=$\ldots$,p={-90:0.25}}, {1/p={0.8,0},lab={t=$\ldots$,p={-90:0.25}}}, {2/p={0,0.8},lab={t=$\ldots$,p={180:0.35}}}, {3/p={0.8,0.8},lab={t=$\ldots$,p={0:0.35}}}, {4/p={0,1.6},lab={t=$\ldots$,p={90:0.25}}}, {5/p={0.8,1.6},lab={t=$\ldots$,p={90:0.25}}}}
\draw[red] (0)--(1) (2)--(4) (3)--(5);
\end{tikzpicture}
=
\begin{tikzpicture}
\atoms{delta}{0/lab={t=$\ldots$,p={-90:0.25}}, {1/p={0.8,0},lab={t=$\ldots$,p={-90:0.25}}}, {2/p={0,1.6},lab={t=$\ldots$,p={180:0.35}}}, {3/p={0.8,1.6},lab={t=$\ldots$,p={0:0.35}}}, {4/p={0,2.2},lab={t=$\ldots$,p={90:0.25}}}, {5/p={0.8,2.2},lab={t=$\ldots$,p={90:0.25}}}}
\atoms{delta}{a/p={0,1.2}, b/p={0,0.8}, c/p={0.8,1.2}, d/p={0.8,0.8}, e/p={0.4,0.4}}
\draw (4)--(2) (5)--(3) (2)--(a) (b)--(e) (3)--(c) (d)--(e) (e)--(0) (e)--(1);
\draw[cyan,dashed] (e)circle(0.2) node[shift=(180:0.1),anchor=east]{$M_{ZZ}$};
\draw[cyan,dashed] (0,1)ellipse(0.2cm and 0.4cm) node[shift=(180:0.1),anchor=east]{$M_{X}$};
\draw[cyan,dashed] (0.8,1)ellipse(0.2cm and 0.4cm) node[shift=(0:0.1),anchor=west]{$M_{X}$};
\end{tikzpicture}
\;.
\end{equation}
After these modifications, we are ready to read off the boundary circuit from Fig.~\ref{fig:r_boundary_zx}.
Since the $ZX$ boundary path integral depends on the 4-coloring of the underlying 3-colex, the resulting circuit has an enlarged unit cell compared to the bulk.
A convenient choice of unit cell is given by translations by $\frac43 t-y$, and translations by $3y$.
So in order to specify a full unit cell of the circuit, we need to show how it acts on the boundary qubits of three spatial hexagons, and for a time period of $\frac43 t$, corresponding to four of the bulk cycles in Fig.~\ref{fig:bulk_circuit}.
The resulting circuit is shown in Fig.~\ref{fig:hexagon_boundary_circuit}.

\subsection{Input and output domain walls}
\label{sec:input_ouput_domain}
In this section we show how to transfer logical information from a 2D color code into the twisted color circuit and back, which will be the key to implementing a fault-tolerant non-Clifford operation.
In other words, we implement a 2+0D interface -- or \textit{domain wall} -- between a 2+1D syndrome-extraction circuit of the 2D color code and the twisted color circuit.
To construct the interfacing circuit, we will start with a 3+0D protocol performing a dimension jump from the 2D to the 3D color code.
Then we will turn the interface to the 3D color code into an interface to the color path integral from Section~\ref{sec:easy_path_integral}, and then finally turn this path integral back into the desired 2+0D interfacing section of the circuit.

\begin{figure}
\includegraphics{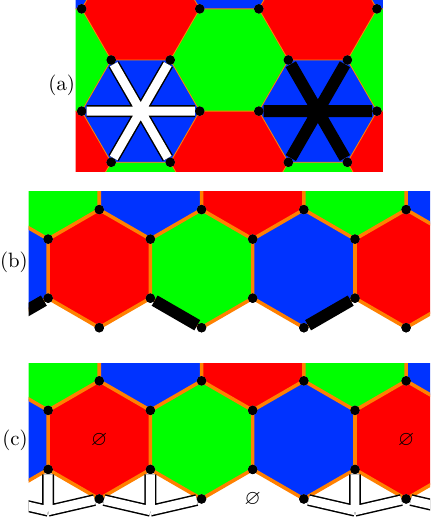}
\caption{
(a) 2D color code on hexagonal 2-colex, with one qubit at every vertex, and one $X$ and one $Z$-stabilizer at every face, using the notation from Fig.~\ref{fig:bulk_circuit}.
(b) Additional $ZZ$-stabilizers at the $\ovl r$-colored edges of the bottom boundary.
(c) Truncated $X$-type stabilizers.
As indicated by the $\varnothing$ symbol, the $X^{\otimes n}$ operators are not stabilizers as they anti-commute with the new $ZZ$ stabilizers.
}
\label{fig:color_code}
\end{figure}

\myparagraph{2D color code}
Let us recall the definition of a 2D color code on \emph{2-colex}, which is a 2-cellulation dual to a triangulation, whose faces are colored $r$, $g$, or $b$ such that no adjacent faces have the same color~\cite{Bombin2006b}.
There are two stabilizer generators for each face $f$, one acting as $Z$ and $X$ on all the vertices $p$ of $f$, respectively:
\begin{equation}
S^Z_f = \prod_{p\in f} Z_p
\;,\qquad
S^X_f = \prod_{p\in f} X_p\;.
\end{equation}
For our example, we use a 2D color code on a hexagonal lattice as depicted in Fig.~\ref{fig:color_code} (a).
In the global protocol of Eq.~\eqref{eq:global_protocol}, we define the 2D color code on a global triangle with \textit{color boundaries} labeled by $r$, $g$, and $b$.
We implement the color boundaries in a slightly unusual way, such that they can be interfaced easily with the 3-colex underlying the twisted color circuit.
In the following, we describe the $r$ color boundary.
The $g$ and $b$ boundaries are defined analogously.
Geometrically, the boundary can be placed along an arbitrary path of edges in the primal 2-colex.%
\footnote{This is different from the usual way of defining a color boundary along a path in the dual lattice that only cuts perpendicular through $\ovl r$-colored edges.
Such boundaries are usually placed along the ``arm-chair'' direction of the hexagonal lattice, which is rotated by 60 degrees relative to the zig-zag direction shown in Fig.~\ref{fig:color_code}.}
Specifically, we put the boundaries along a zig-zag direction of the hexagonal lattice as shown in Fig.~\ref{fig:color_code} (b) and (c).
As shown in Fig.~\ref{fig:color_code}, the stabilizers at the boundary are determined by (1) removing $Z$ stabilizers beyond the boundary, (2) truncating $X$ stabilizers, (3) adding a new $ZZ$ stabilizer at every $\ovl r$-colored boundary edge, and (4) removing the $X$ stabilizers that do not commute with the new $ZZ$ stabilizers.

\myparagraph{Dimension-jump protocol for non-Clifford gate}
Let us now discuss a 3+0D protocol that performs a non-Clifford gate by performing a dimension jump from the 2D to the 3D color code, then the transversal $T$ gate of the 3D color code, and then jumps back to 2D.
The logical information will be transferred into and from the 3D color code at a $y$ color boundary.
For simplicity, we will assume for the rest of this section that there are no spatial boundaries of the 2D color code and in the resulting 2+1D protocol.
To be concrete, let us assume that the 2D color code is supported on a torus $S_1\times S_1$, and the 3D color code is supported on $S_1\times S_1\times [0,1]$, with $y$-color boundaries at $S_1\times S_1\times 0$ and $S_1\times S_1\times 1$ where the logical information is transferred in and out.
Apart from the different overall topology, the protocol is the same as introduced in Ref.~\cite{Bombin2006}.%
\footnote{The logical action on $S_1\times S_1\times [0,1]$ will not be a $\ovl T$ gate, and the logical action of the protocol in Eq.~\eqref{eq:global_protocol} will be yet different, see Ref.~\cite{Davydova2025}.
We are not interested in the logical action of the 3+0D protocol here, but use it as tool to derive the interface of the color path integral.}

\begin{figure}
\includegraphics[width=0.75\linewidth]{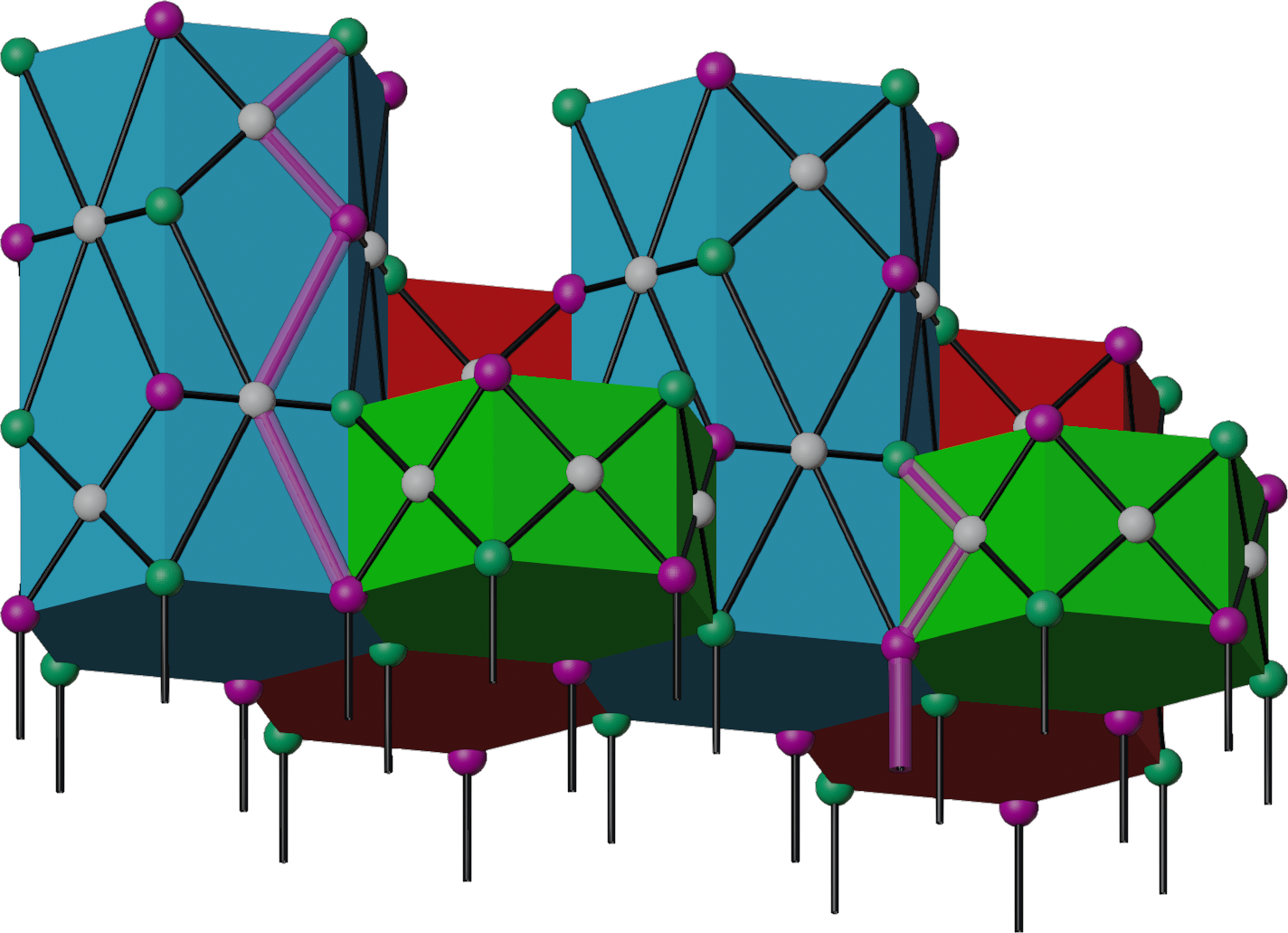}
\caption{
Snippet of the 3-colex with path-integral domain wall to the 2D color code at the bottom.
The open indices at the bottom support a 2D-color-code state on the bottom 2-colex.
The domain wall cuts the $rg$ brick hexagons down to squares, such that the according 6-index $X$-tensors become 4-index $X$-tensors.
One can add $X$-tensors to all hexagons inside the domain wall -- this is equivalent to the shown $ZX$ diagram up to rewrite rules.
Two representative worldlines are shaded in purple:
Only the worldlines contained in the $rg$-faces near the domain wall continue as 2D-color-code worldlines, the remaining ones begin or terminate on the domain wall.
}
\label{fig:zx_domain_wall}
\end{figure}

To start, let us recall the $y$ color boundaries of the 3D color code, which are analogous to the path-integral boundaries introduced in the previous section:
We add one $ZZ$ stabilizer for every $\ovl y$-colored boundary face, and remove all (truncated) $X$ stabilizers that do not commute with these new $ZZ$ stabilizers.
Geometrically, we define the boundary along a time-perpendicular plane through the smallest-time hexagons of one layer of $b$-colored volumes, as shown in Fig.~\eqref{fig:zx_domain_wall}.
Thereby, we introduce new hexagons along the boundary that truncate the $r$ and $g$ volumes.
All boundary faces have color $ry$, $by$, or $gy$, and form themselves an $rgb$-colored 2-colex.
There are no $\ovl y$-colored boundary edges, so the boundary is simply defined with $Z$-type stabilizers at all faces and $X$-type stabilizers at all volumes.

With this, the protocol goes as follows:
We start by identifying the qubits of the 2D color code with the qubits of the 3D color code at the 2-colex of the $S_1\times S_1\times 0$ boundary.
Then we initialize all remaining qubits of the 3D code in the $\ket+$ state, and perform all $Z$-type stabilizer measurements of the 3D code.
After performing corrections, we have switched from the 2D to the 3D color code.
We then apply the transversal $T^{\sgn_p}$ gate of the 3D color code, which acts on the boundary qubits in the same way as it does on the bulk qubits.
Finally, we perform a destructive single-qubit $X$ measurement on all the qubits except for those on the $S_1\times S_1\times 1$ boundary 2-colex.
After corrections, the resulting state on these 2-colex qubits will be a 2D color-code state.

\myparagraph{Path integral from dimension-jump protocol}
We now turn the above 3+0D protocol into a 2D domain wall between $ZX$ tensor-network path integrals.
Note that we do not work with an explicit 2D-color-code syndrome-extraction circuit or path integral.
Instead, we construct the domain wall as a $ZX$ tensor network with open indices, where we can attach the path integral for any 2D-color-code circuit.
If the tensor network is contracted, these open indices support a 2D color-code state.
The path integral is obtained by replacing all measurements with the according $+1$-post-selected projectors.
The resulting $ZX$ diagram is shown in Fig.~\ref{fig:zx_domain_wall}.
Every qubit becomes a $Z$-tensor, and every $Z$-type measurement becomes an $X$-tensor.
In the 3+0D bulk, the protocol and path integral look like those in Section~\ref{sec:easy_path_integral}.
Note that the path integrals for the input and output domain walls of the protocol at $S_1\times S_1\times 0$ and $S_1\times S_1\times 1$ are the same (apart from reflection).

\begin{figure}
\includegraphics[width=0.4\textwidth]{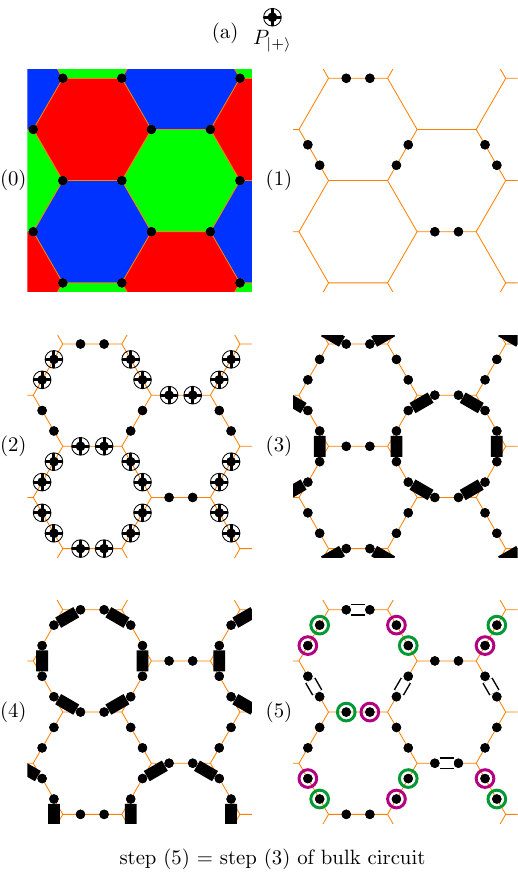}
\caption{
(a) Symbol for the $\ket+$ state preparation gate. The rest of the gate symbols are the same as in Fig.~\ref{fig:bulk_circuit}.
(0)-(5) Circuit transferring the logical information of the 2D color code to the twisted color circuit.
We start with the qubits of the 2D color code (0) and shift the qubits slightly onto the adjacent $\ovl b$-colored edges (1).
Then, we initialize the remaining qubits of the twisted color circuit in the $\ket+$ state (2) and perform the $ZZ$ measurements between the moved and the initialized qubits (3 and 4), which implements all the tensors on the right-hand side of Eq.~\eqref{eq:domain_wall_rewriting}.
In step (5), we perform $XX$ measurements on all $\ovl b$-colored edges of the 2-colex, which implement the $X$-tensors at the $rg$-squares of the 3-colex in Fig.~\ref{fig:zx_domain_wall}.
After this, we continue with step (3) of the bulk protocol in Fig.~\ref{fig:bulk_circuit}, whose $T$ and $T^{-1}$ gates we have already included in step (5) above.
}
\label{fig:hexagon_domainwall_circuit}
\end{figure}

\myparagraph{Circuit from path integral}
Next, we turn the above path integral into a circuit.
There are two circuits, one transferring logical information from the 2D color code into the twisted color circuit, and vice versa.
We start with the former.
We first choose how the qubit worldlines propagate through the domain wall.
There is one qubit per 2-colex vertex in the 2D color code, but there are three qubit worldlines per vertex in our twisted color circuit.
Thus, we need to choose one of the three worldlines to continue the 2D-color-code qubit worldline.
The other two twisted-color-circuit worldlines begin at the temporal domain wall, meaning that these qubits were unused in the 2D color code, and can be seen as auxiliary qubits needed to perform the non-Clifford logic operation.
Fig.~\eqref{fig:zx_domain_wall} shows one worldline that continues through, and one worldline that begins at the domain wall.

After choosing the qubit worldlines, we slightly rewrite the $ZX$ diagram above into a circuit, whose input is the color code state at the $rgb$ 2-colex corresponding to the bottom boundary.
Every $Z$-tensor at every domain-wall vertex has four indices:
(0) The open index, and the indices connecting it with the $X$-tensors at the adjacent time-parallel (1) $rb$ brick hexagon, (2) $rg$ square, and (3) $gb$ square.
We rewrite each such tensor as a little circuit consisting of two $\ket+$ initializations and two $+1$-post-selected $ZZ$ measurements:
\begin{equation}
\label{eq:domain_wall_rewriting}
\begin{tikzpicture}
\atoms{delta}{0/}
\draw (0)edge[ind=0]++(-90:0.5) (0)edge[ind=1]++(45:0.8) (0)edge[ind=2]++(90:0.8) (0)edge[ind=3]++(135:0.8);
\end{tikzpicture}
=
\begin{tikzpicture}
\atoms{delta}{f0/, f1/p={0,0.5}, i0/p={$(f0)+(-45:0.7)$}, i1/p={$(f1)+(-135:0.7)$}}
\draw (i0)--(f0) (i1)--(f1) (f0)--(f1) (f0)edge[ind=0]++(-90:0.5) (f0)edge[ind=1]++(45:0.8) (f1)edge[ind=2]++(90:0.5) (f1)edge[ind=3]++(135:0.6);
\draw[cyan,dashed] (i0)circle(0.2) node[shift=(0:0.1),anchor=west]{$\ket+$};
\draw[cyan,dashed] (i1)circle(0.2) node[shift=(180:0.1),anchor=east]{$\ket+$};
\draw[cyan,dashed] (f0)circle(0.2) node[shift=(0:0.1),anchor=west]{$M_{ZZ}$};
\draw[cyan,dashed] (f1)circle(0.2) node[shift=(180:0.1),anchor=east]{$M_{ZZ}$};
\end{tikzpicture}
\;.
\end{equation}
The resulting circuit implementing the domain wall is shown in Fig.~\ref{fig:hexagon_domainwall_circuit}.

\begin{figure}
\includegraphics[width=0.4\textwidth]{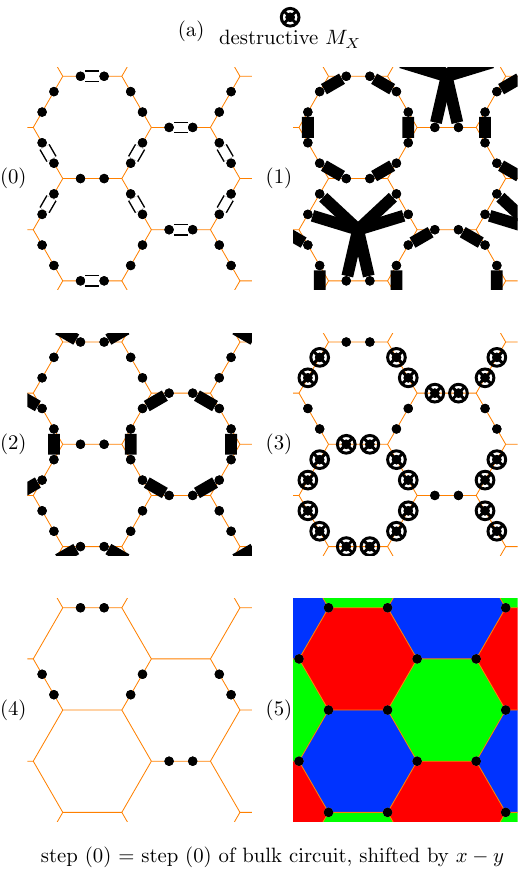}
\caption{
(a) Symbol for destructive single-qubit $M_X$ measurement, the remaining symbols are these in Fig.~\ref{fig:bulk_circuit}.
(0)-(5) Circuit transferring the logical information of the twisted color circuit back to the 2D color code.
Step (0) corresponds to step (0) of the twisted color circuit in Fig.~\ref{fig:bulk_circuit}.
}
\label{fig:hexagon_domainwall_circuitback}
\end{figure}

\myparagraph{Switching back}
We also need to construct the circuit to transition from the twisted color circuit to the 2D color code.
The $ZX$ diagram is the same as in Fig.~\ref{fig:zx_domain_wall}, just reflected vertically if time goes upwards.
The qubit worldlines are also the same as in Fig.~\ref{fig:zx_domain_wall}.
Similar to Eq.~\eqref{eq:domain_wall_rewriting}, the 4-index $Z$-tensors at the domain wall vertices are split up into two $+1$-post-selected $ZZ$ measurements, and two $+1$-post-selected destructive $X$ measurements:
\begin{equation}
\label{eq:domain_wall_rewriting2}
\begin{tikzpicture}
\atoms{delta}{0/}
\draw (0)edge[ind=0]++(90:0.5) (0)edge[ind=1]++(-45:0.8) (0)edge[ind=2]++(-90:0.8) (0)edge[ind=3]++(-135:0.8);
\end{tikzpicture}
=
\begin{tikzpicture}
\atoms{delta}{f0/, f1/p={0,0.5}, i0/p={$(f0)+(45:0.7)$}, i1/p={$(f1)+(135:0.7)$}}
\draw (i0)--(f0) (i1)--(f1) (f0)--(f1) (f0)edge[ind=0]++(-90:0.5) (f0)edge[ind=1]++(-45:0.6) (f1)edge[ind=2]++(90:0.5) (f1)edge[ind=3]++(-135:0.8);
\draw[cyan,dashed] (i0)circle(0.2) node[shift=(0:0.1),anchor=west]{$\bra+$};
\draw[cyan,dashed] (i1)circle(0.2) node[shift=(180:0.1),anchor=east]{$\bra+$};
\draw[cyan,dashed] (f0)circle(0.2) node[shift=(0:0.1),anchor=west]{$M_{ZZ}$};
\draw[cyan,dashed] (f1)circle(0.2) node[shift=(180:0.1),anchor=east]{$M_{ZZ}$};
\end{tikzpicture}
\;.
\end{equation}
The resulting circuit is shown in Fig.~\ref{fig:hexagon_domainwall_circuitback}.

\subsection{Corners}
\label{sec:corners}
\begin{figure*}
\includegraphics[width=0.7\linewidth]{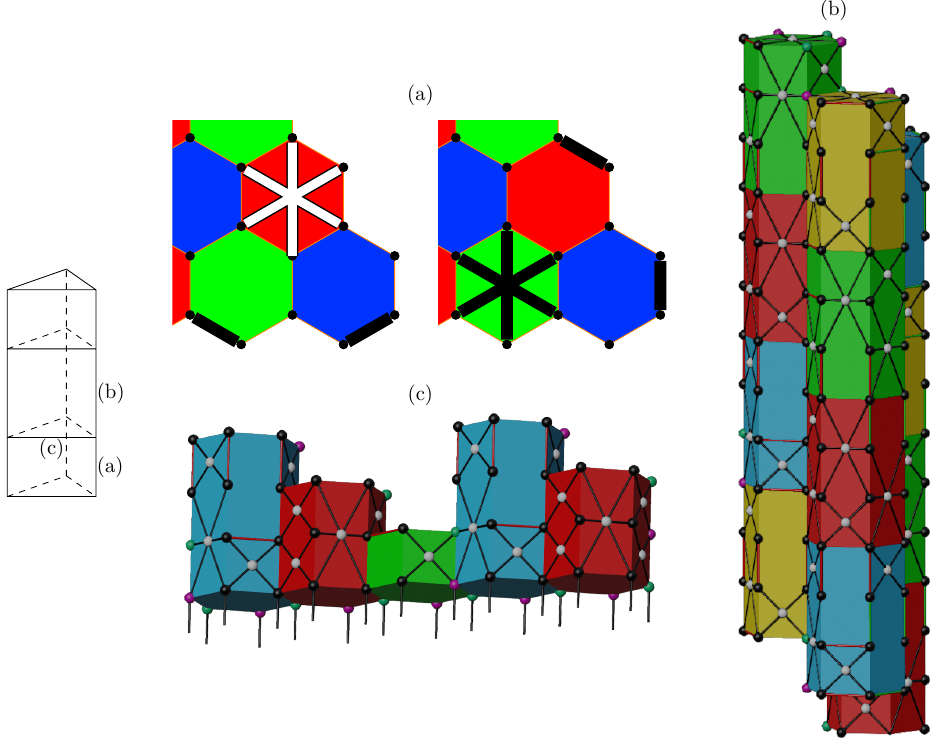}
\caption{
(Left) Three different types of corners in the protocol, labeled (a), (b), and (c).
(a) Corner of the 2D color code between a $r$ (bottom) and a $g$ (top right) zig-zag color boundary.
Each of the two pictures shows some of the stabilizer generators near the corner, which is located at the bottom-right vertex of the bottom-right blue face.
The vertex on the corner is involved in two $ZZ$ stabilizers, since it is adjacent to a $\ovl r$ edge of the $r$ boundary, and a $\ovl g$ edge of the $g$ boundary.
(b) Corner between a $r$ (front left) and a $g$ (front right) color boundary of the color path integral.
(c) Corner joining the $r$ boundary of the 2D color code and the $r$ boundary of the color path integral (front surface), and the domain wall between 2D color code and color path integral (bottom surface).
}
\label{fig:hexagon_corner}
\end{figure*}

So far, we have constructed the microscopic circuits for the 3D bulk volumes of the global protocol in Eq.~\eqref{eq:global_protocol} as well as for the 2D boundaries and domain walls.
The next step is to say what happens at the 1D lines where different boundary conditions or domain walls meet.
Such interfaces between boundary conditions are lines in spacetime but points in space, and are therefore known as \emph{corners} in quantum error correction.
There are three kinds of corners in our protocol: (a) the corners of the 2D color code (b) time-parallel corners separating two color boundaries of the twisted color circuit, and (c) time-perpendicular corners between the 2D color-code boundary and the according twisted-color-circuit boundary, where also the domain wall terminates.
We will not describe these corners in detail here, but the according circuits can be constructed straight-forwardly from $ZX$ tensor networks as done for the bulk, boundaries and domain-wall in the preceding subsections.
Fig.~\ref{fig:hexagon_corner} illustrates the stabilizers and $ZX$ tensor networks for the different kinds of corners, from which we can read off the circuits.

\subsection{Decoding}
\label{sec:decoding_easy}

\begin{figure*}
\includegraphics[width=0.8\linewidth]{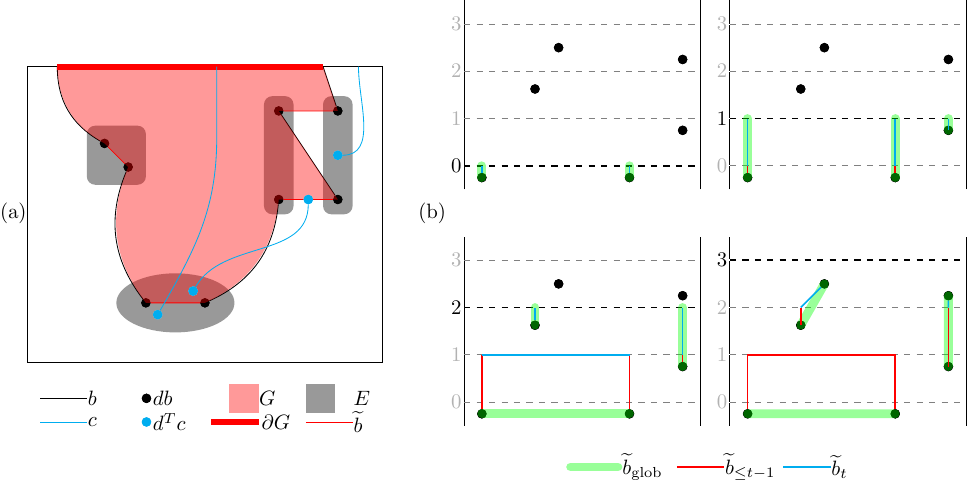}
\caption{
(a) 2D toy picture illustrating the decoding and error correction in the 3+0D protocol, with output state at the top boundary.
$b$ is the flux configuration corresponding to the face $Z$ measurements.
Its endpoints $db$ form the flux syndrome which can only be supported in the vicinity of the errors $E$.
$\widetilde b$ is the closing flux configuration, which is a low-weight string configuration chosen by the decoding algorithm, with the same endpoints $db$ as $b$.
$G$ is the gauge-fixing configuration, an arbitrary membrane configuration whose boundary is given by $b+\widetilde b$.
$c$ is the charge configuration corresponding to the vertex $X$ measurements.
Its endpoints $d^Tc$ form the charge syndrome which can be supported in the vicinity of both the errors $E$ and $\widetilde b$.
(b) Four time steps of just-in-time decoding algorithm.
Time is going upwards, and the spatial directions extend horizontally.
The time slices are the horizontal gray dashed lines, with the current time slice drawn in thick and black.
$\widetilde b_{\leq t-1}$ is the portion of $\widetilde b$ chosen before time step $t-1$.
$\widetilde b_t$ is the portion of $\widetilde b$ between $t-1$ and $t$, which is chosen at the current time step.
As shown, if an error creates a pair of flux syndrome points that are far apart spatially, then $\widetilde b$ as chosen by the just-in-time decoder will not connect them immediately, but with some temporal delay proportional to their spatial separation.
}
\label{fig:decoding_easy}
\end{figure*}

In this section, we will sketch how to decode twisted color circuits.
The aim here is to address most important aspects of the decoding procedure, without going into the combinatorial details.
These details will be discussed later in Section~\ref{sec:decoding}, after introducing the appropriate vocabulary in Sections~\ref{sec:color_cohomology} and \ref{sec:path_integral}.

\myparagraph{3+0D protocol}
We first recall how decoding works for the 3+0D measurement-based dimension-jump protocol~\cite{Bombin2006} that we have used to derive the 2+1D twisted color circuit.
Suppose that every $Z$-type face measurement and every single-qubit $X$ measurement is faulty with a small probability.
We denote the subset of faces that experience a $Z$-measurement error by $b_E$, and the subset of faces yielding a $-1$ measurement outcome by $b$.
Both $b_E$ and $b$ can be pictured as a configuration of strings on the 3D spacetime lattice, and we refer to them as \emph{flux configurations}.%
\footnote{More precisely, the 3D color code is equivalent to three copies of the 3D toric code, so $b$ maps to three separate string configurations as we show in Section~\ref{sec:color_cohomology}.
We will pretend that there is only one string configuration to simplify the following discussion.}
Taken together, the symmetric difference $b+b_E$ must fulfill certain local parity or ``Gauss law'' constraints, which ensure that the strings form closed loops.
That is, by measuring $b$, we measure the endpoints of the error configuration $b_E$.
The endpoints (denoted $db$) of $b$ (and $b_E$) are often called \emph{detection events} and form the \emph{flux syndrome}.
The task of the decoder is to find an estimate $\widetilde b$ for $b_E$, which must have the same endpoints $db$, and which we thus refer to as the \emph{closing flux configuration}.
Since larger errors have a small probability, a good strategy is to choose $\widetilde b$ to be a low-weight configuration.
Note that this task is equal to the syndrome-fixing step in the single-shot decoding of $X$ errors in the 3D color code~\cite{Bombin2014a}.

After performing all the $Z$ face measurements and decoding, we need to apply corrections that map the state onto a 3D color-code state with all $Z$-type measurement outcomes equal to $+1$.
The corrections consist of Pauli-$X$ operators acting on a subset $G$ of vertices, which we call the \emph{gauge fixing configuration}, and which we can picture as a membrane configuration.
Since the effective configuration of $-1$ outcomes is given by $b+b_E$, choosing a membrane configuration $G$ whose boundary is the closed-loop configuration is $b+b_E$ yields an exact code state.
As we do not have access to $b_E$, we replace it with its estimate $\widetilde b$, and choose an arbitrary $G$ that bounds $b+\widetilde b$ instead.
The state after corrections is then one with a \emph{residual flux configuration} $b_E+\widetilde b$.

Let us next describe the decoding for measurement errors in the single-qubit $X$ measurements at the end of the protocol.
We denote the set of vertices that experience a measurement error by $c_E$, and the set of vertices with a $-1$ outcome by $c$.
Again, both $c$ and $c_E$ can be pictured as string patterns, and we refer to them as \emph{charge configurations}.
If we omit the transversal $T$ gate in the protocol, then the overall charge configuration $c+c_E$ must form closed loops.
So a decoder can estimate $c_E$ by finding a low-weight string pattern $\widetilde c$, whose endpoints are given by the endpoints $d^Tc=d^Tc_E$ of $c$, forming the charge syndrome.
This decoding problem is equal to the decoding of on-site $Z$ errors in the 3D color code without measurement errors.

If we do apply the transversal $T$ gate before the single-qubit $X$ measurements, the overall charge configuration $c+c_E$ no longer needs to be closed~\cite{Bombin2018a}.
Instead, $c+c_E$ forms a string pattern that may terminate anywhere on the residual flux configuration $b_E+\widetilde b$.%
\footnote{More precisely, there are three sorts of flux and charge strings each, and the charge strings of one sort can terminate on the residual flux strings of the other two sorts~\cite{Bombin2018a}.}
The reason for this is that after applying the transversal $T$ gate to a state with flux configuration $b_E+\widetilde b$, it is no longer an eigenstate of the $X$-stabilizers of the 3D color code that detect the charge syndrome.
Thus, the residual flux configuration $b_E+\widetilde b$ behaves similar to a heralded erasure, where a random charge error happens with high probability.
In other words, the residual flux configuration reduces the fault distance for $Z$ errors (or $X$ measurement errors) in the protocol.
If the error rate for $Z$ measurement errors is low, then the $X$ measurement errors are still decodable.
Also note that it is possible to use the knowledge of $\widetilde b$ to improve the performance of the $X$-measurement decoding.
Further, we would like to point out that when decoding the protocol without $T$ gates, only the cohomology class of $\widetilde b$ matters.
In contrast, with the $T$ gate, $\widetilde b$ itself matters and should have low weight.

Finally, we note that even though it is convenient to explain decoding for Pauli-$X$ and $Z$ errors, the path-integral perspective allows us to consider arbitrary errors~\cite{twisted_double_code}.
Let $E$ denote the subset of locations in the 3-colex where some arbitrary errors happen.
It is still true that the endpoints of $b$ must be located inside $E$, and the endpoints of $c$ must be located either inside $E$ or on $\widetilde b$, as shown in Figure~\ref{fig:decoding_easy}(a).
This is sufficient to show fault tolerance, see Ref.~\cite{Davydova2025}, and we elaborate on further on this in App.~\ref{app:local_fault_tolerance}.

\myparagraph{2+1D protocol}
Let us now describe how decoding changes if we turn the 3+0D protocol into a 2+1D protocol using the method of Refs.~\cite{Bombin2018,Brown2019}, by keeping only a few 2D slices of qubits active at a time.
We need to apply the Pauli-$X$ corrections along the gauge fixing configuration $G$ while we execute the 2+1D circuit.
Since $G$ is chosen to bound $b+\widetilde b$, we must also choose $\widetilde b$ while executing the circuit.
That is, our choice of $\widetilde b_t$ at a time $t$ can only depend on $b$ (or better, its syndrome $db$) at times earlier than $t$.
In particular, we cannot use a ``global'' decoder operating on the 3D spacetime at the end.
Instead, we have to use \emph{just-in-time} decoding as proposed in Ref.~\cite{Bombin2018}:
Let $\widetilde b_t$ denote of the portion of $\widetilde b$ located in between time steps $t-1$ and $t$, and let $\widetilde b_{\leq t}$ denote the portion of $\widetilde b$ before time $t$.
Then at time step $t$, the decoder must finalize its choice of $\widetilde b_t$, depending only on $(db)_{\leq t}$.
This choice is made in the following two subroutines:
(1) The first step is to use a global decoder on $(db)_{\leq t}$, which yields a closing flux configuration $b_{\text{glob}}$, such that $b_{\leq t}+b_{\text{glob}}$ forms strings that may terminate at time $t$ but are otherwise closed.
In general, this global choice of $b_{\text{glob}}$ is not directly compatible with the choices $\widetilde b_{\leq t-1}$ made by the just-in-time decoder so far.
(2) The next step is to reconcile $b_{\text{glob}}$ with $\widetilde b_{\leq t-1}$.
To this end, we note that the endpoints of $b_{\text{glob}}$ and $\widetilde b_{\leq t-1}$ differ only between $t-1$ and $t$.
We can use our global decoding algorithm again to match up these endpoints.
The result is $\widetilde b_t$.
The procedure is illustrated in Fig.~\ref{fig:decoding_easy} (b).
After decoding the flux syndrome in a just-in-time way, the charge syndrome is decoded as in the 3+0D protocol with a global decoder.

\myparagraph{Twisted color circuits}
For our twisted color circuits, there are some small additional differences to the circuits in Refs.~\cite{Bombin2018,Brown2019}.
First, the configuration of $Z$-type measurement results is no longer the same as the flux configuration $b$ (which we define as a subset of 3-colex faces).
Instead, some of the $Z$-measurements contribute multiple 3-colex faces to $b$.
The same is true for the $X$-measurements and the charge configuration $c$.
Second, the ``gauge fixing'' operation is not given by applying a Pauli-$X$ operator along the gauge fixing configuration $G$.
Instead, in our circuit, it is given by conjugating the $T^{\sgn_p}$ gate by $X$, or equivalently, exchanging $T$ and $T^{-1}$ along $G$.
We elaborate more on the just-in-time decoding procedure for twisted color circuits in Sec.~\ref{sec:decoding}.

\section{Color cohomology}
\label{sec:color_cohomology}
In this section, we introduce \emph{color cohomology}, which is a cohomology theory defined on $3$-colexes.
As we show in Section~\ref{sec:color_cohomology_equivalence}, color cohomology is equivalent to $\zz_2^3$-valued \emph{cellular cohomology}, the type of cohomology theory that describes toric-code protocols.
Color cohomology is essential for studying the topological nature of the path integral and its defects in Section~\ref{sec:path_integral}, as well as for formulating the decoding algorithms in Section~\ref{sec:decoding}.
A length-2 subcomplex of this cohomology theory can also be found in Ref.~\cite{Kubica2019}, and the 2D case was first introduced in Ref.~\cite{Delfosse2013}.
\subsection{\texorpdfstring{$\zz_2$}{Z2}-valued cohomology}
\label{sec:general_cohomology}
Let us start by introducing general $\zz_2$-valued cohomology theories.
A cohomology theory is defined by a \emph{chain complex}, which is determined by
\begin{itemize}
\item a \emph{length} $n\in\zz_+$,
\item a sequence of finite \emph{generator sets},
\begin{equation}
\{X_i\}_{0\leq i\leq n}\;,
\end{equation}
\item a sequence of \emph{boundary matrices}
\begin{equation}
\{d_i\in \zz_2^{X_{i+1}\times X_{i}}\}_{0\leq i<n}\;,
\end{equation}
\item which satisfy
\begin{equation}
\label{eq:cohomology_axiom}
\sum_{y\in X_{i+1}} (d_{i+1})_{zy} (d_{i})_{yx}=0 \quad\forall i\leq n-2, x\in X_i,z\in X_{i+2}\;.
\end{equation}
\end{itemize}
Usually, the boundary matrices are interpreted as group homomorphisms (or, similarly, as $\zz_2$-linear maps between finite-field vector spaces),
\begin{equation}
d_i: \zz_2^{X_i}\rightarrow \zz_2^{X_{i+1}}\;,\quad (d_i v)_x = \sum_y (d_i)_{xy} v_y\;.
\end{equation}
Then, Eq.~\eqref{eq:cohomology_axiom} simply becomes
\begin{equation}
d_{i-1} d_{i}=0\;.
\end{equation}
Sometimes, when no confusion will arise, we will drop the subscript $i$ of $d_i$ and we just write $d$.
Also, note that we will sometimes let the sequence $X_i$ start from $i=-1$ rather than $i=0$.
It is custom to denote a chain complex as a sequence of groups with the homomorphisms as arrows,
\begin{equation}
\zz_2^{X_0}\xrightarrow{d_0}\zz_2^{X_1}\xrightarrow{d_1}\zz_2^{X_2}\xrightarrow{d_2}\ldots\;.
\end{equation}
Will just write $X_i$ instead of $\zz_2^{X_i}$ in such sequences for brevity.

Elements of the group $\zz_2^{X_i}$ are called \emph{$i$-(co-)chains}.%
\footnote{Usually, the cochains are defined with coefficients in the Pontryagin dual group of the chains.
For us, the coefficient group is $\zz_2$ which can be identified with its Pontryagin dual via the pairing (or ``Fourier transform'' or ``Hadamard gate'') $(a,b)\rightarrow (-1)^{ab}$, so we identify the cochains with chains.
If we sometimes call a chain ``cochain'', this is only to signal that we intend to apply the coboundary operator $d_i$ to it instead of the boundary operator $d_{i-1}^T$.}
Note that, for some $x\in X_i$, we will sometimes use $x$ to also denote the $i$-chain $c$ that is non-zero only on $x$, $c_y=\delta_{y=x}$.
In this sense, the elements $x\in X_i$ can be understood as generating $i$-chains.
An $i$-chain $A$ is called an \emph{$i$-cocycle} if $d_iA=0$, and an \emph{$i$-cycle} if $d_{i-1}^TA=0$.
Further, $A$ is called \emph{$i$-coboundary} if there exists a $i-1$-chain $B$ such that $d_{i-1}B=A$, and \emph{$i$-boundary} if there is an $i+1$-chain $B$ such that $d_i^TB = A$.
Due to Eq.~\eqref{eq:cohomology_axiom}, every $i$-(co)boundary is a $i$-(co)cycle.
Just like the (co-)chains, the (co-)cycles and (co-)boundaries form groups under element-wise $\zz_2$ addition.
The quotient of the group of cocycles by the group of coboundaries,
\begin{equation}
H^i\simeq \operatorname{Kernel}(d_{i}) / \operatorname{Image}(d_{i-1})
\end{equation}
is called the \emph{$i$th cohomology} group, and its elements are called \emph{$i$th cohomology classes}.
Dually, the quotient
\begin{equation}
H_i\simeq \operatorname{Kernel}(d_{i-1}^T) / \operatorname{Image}(d_i^T)
\end{equation}
is the \emph{$i$th homology group} and its elements are the \emph{$i$th homology classes}.
Homology and cohomology groups can be identified.%
\footnote{We could define homology to be the cohomology of the Pontryagin dual chain complex consisting of $(\zz_2^{X_i})^*$ and $d_i^*$, then we simply have $H_i=(H^i)^*$.
Since we define it via the transpose matrices, there is a canonical pairing between $H_i$ and $H^i$.}

\myparagraph{Chain maps}
Finally, there also is a way to map between different cohomology theories, whose generator sets and boundary maps we distinguish by superscripts $(a)$ and $(b)$:
A \emph{chain map} $M$ from $(a)$ to $(b)$ consists of a collection of matrices
\begin{equation}
\{M_i\in \zz_2^{X^{(b)}_i\times X^{(a)}_i}\}_{0\leq i\leq n}\;,
\end{equation}
which we can interpret as homomorphisms from the group of $(a)$ $i$-(co-)chains to the group of $(b)$ $i$-(co-)chains, such that
\begin{equation}
\label{eq:chain_map_definition}
d^{(b)}_i M_i=M_{i+1}d^{(a)}_i\qquad \forall 0\leq i<n\;.
\end{equation}
A chain map can also be denoted as a commutative diagram involving the two chain complexes,
\begin{equation}
\begin{tikzpicture}
\node(x0a) at (0,0){$X_{0}^{(b)}$};
\node(x0b) at (0,-1.3){$X_{0}^{(a)}$};
\node(x1a) at (2,0){$X_{1}^{(b)}$};
\node(x1b) at (2,-1.3){$X_{1}^{(a)}$};
\node(x2a) at (4,0){$X_{2}^{(b)}$};
\node(x2b) at (4,-1.3){$X_{2}^{(a)}$};
\node(x3a) at (6,0){$\ldots$};
\node(x3b) at (6,-1.3){$\ldots$};
\draw (x0a)edge[->]node[midway,above]{$\scriptstyle{d_0^{(b)}}$}(x1a);
\draw (x1a)edge[->]node[midway,above]{$\scriptstyle{d_1^{(b)}}$}(x2a);
\draw (x2a)edge[->]node[midway,above]{$\scriptstyle{d_2^{(b)}}$}(x3a);
\draw (x0b)edge[->]node[midway,above]{$\scriptstyle{d_0^{(a)}}$}(x1b);
\draw (x1b)edge[->]node[midway,above]{$\scriptstyle{d_1^{(a)}}$}(x2b);
\draw (x2b)edge[->]node[midway,above]{$\scriptstyle{d_2^{(a)}}$}(x3b);
\draw (x0b)edge[->]node[midway,right]{$\scriptstyle{M_0}$}(x0a);
\draw (x1b)edge[->]node[midway,right]{$\scriptstyle{M_1}$}(x1a);
\draw (x2b)edge[->]node[midway,right]{$\scriptstyle{M_2}$}(x2a);
\end{tikzpicture}
\;.
\end{equation}
$M_i$ maps $(a)$ $i$-cocycles $(b)$ $i$-cocycles and $(a)$ $i$-coboundaries to $(b)$ $i$-coboundaries, so it yields a homomorphism from $(H^i)^{(a)}$ to $(H^i)^{(b)}$.
Similarly, $M_i^T$ maps $(b)$ $i$-cycles $(a)$ $i$-cycles and $(b)$ $i$-boundaries to $(a)$ $i$-boundaries, and yields a homomorphism from $(H_i)^{(b)}$ to $(H_i)^{(a)}$.

\myparagraph{Example: cellular cohomology}
After introducing $\zz_2$-valued cohomology theories in general, let us briefly consider \emph{cellular cohomology}, which is a length-$n$ cohomology theory defined for any $n$-cellulation.We denote by $\Delta_i$ the set of $i$-cells, and define (for $0\leq i,j\leq n$) the $(i,j)$-adjacency matrix $\partial_{ij}$ as%
\footnote{More precisely, $(\partial_{ij})_{xy}$ is the $\mmod 2$ number of times $y$ is contained in $x$, but this is only relevant for very irregular lattices or very small lattices with periodic boundary conditions.}
\begin{align}
\begin{split}
\partial_{ij}\in& \zz_2^{\Delta_i\times \Delta_j}\\
(\partial_{ij})_{xy} =&
\begin{cases}
1 & \text{if } x\subseteq y \text{ or }y\subseteq x\\
0 & \text{otherwise}
\end{cases}
\;.
\end{split}
\end{align}
Here, $y\subseteq x$ means that the $i$-cell $y$ is contained in the boundary of the $j$-cell $x$, or that $x=y$.
Note that we will usually just write $\partial_{xy}$ instead of $(\partial_{ij})_{xy}$ if it is understood that $x\in \Delta_i$ and $y\in \Delta_j$.
To help with this, we will usually denote vertices by $p$ (``point''), edges by $e$, faces by $f$, and volumes by $v$.
With this, the generator sets and boundary matrices for cellular cohomology are given by
\begin{equation}
\label{eq:cellular_cohomology}
X_i\coloneqq \Delta_i\;\forall 0\leq i\leq n\;,\quad d_i\coloneqq \partial_{i+1,i}\;\forall 0\leq i<n\;.
\end{equation}
Eq.~\eqref{eq:cohomology_axiom} becomes
\begin{equation}
\sum_{y\in \Delta_{i+1}} \partial_{zy}\partial_{yx} = 0\;\quad \forall x\in \Delta_i,z\in \Delta_{i+2}\;.
\end{equation}
The left-hand side is the $\mmod 2$ number of $i+1$-cells $y$ adjacent to both the $i$-cell $x$ and the $i+2$-cell $z$.
If $x\subset z$, then there are two such $i+1$-cells $y$, otherwise there are none, so the equation holds.

All considered cohomology theories have a property that we call \emph{local exactness}:%
\footnote{Note that a chain complex is called \emph{exact} if every (co)cycle is a (co)boundary, that is, if the (co)homology is trivial.
Here we demand that this is true only for locally supported (co)cycles, such that the (co)homology can be non-trivial, but can only depend on the global properties (usually, the topology) of the lattice.}
First, the generators are associated to places on some sort of lattice or graph with a notion of distance, and the coboundary maps are local, that is, zero beyond some constant distance.
Second, any (co)cycle that is locally supported on the lattice or graph is a (co)boundary.
This holds in particular for cellular cohomology.

Let us briefly give some intuition for cellular cohomology.
We note that for an $i$-cell $c\in \Delta_i$, $d^Tc$ consists of the boundary $i-1$-cells of $c$.
Thus, $i$-cocycles are subsets of $i$-cells with no boundary.
In particular, 1-cycles are superpositions of closed loops consisting of lattice edges, 2-cycles are superpositions of closed membranes consisting of faces, and so on.
Homology classes are equivalence classes of closed loops or membranes under local deformations.
For example, the homology group on a torus is $\zz_2^2$, and each element of $\zz_2^2$ corresponds to an equivalence class of closed-loop patterns on the torus, which can be characterized by their vertical and horizontal $\mmod 2$ winding number.
Homological codes encode quantum information non-locally into the homology class~\cite{Kitaev1997}.
For example, in the toric code, there is a basis of codestates labeled by cohomology classes:
Each computational basis state is the equal-weight superposition of all 1-cycles in the same cohomology class.

We can extend cellular cohomology to a length-$n+2$ chain complex by adding two coboundary maps at the beginning and end, corresponding to
\begin{equation}
\label{eq:cellular_cohomology_extension}
\begin{gathered}
X_{-1}=X_{n+1}\coloneqq \{0\}\;,\\
(d_{-1})_{p0}=1\;\forall p\in\Delta_0\;,\quad (d_n)_{0v}=1\;\forall v\in \Delta_n\;.
\end{gathered}
\end{equation}
Roughly, $d_{-1}$ and $d_n$ can be viewed as adding ``global conservation laws'' for 0-cycles and $n$-cocycles.

\subsection{Color cohomology: Definition}
\label{eq:color_cohomology_definition}
After introducing general $\zz_2$-valued cohomology, and cellular cohomology as a paradigmatic example, let us now define color cohomology.
We start by revisiting and extending our notation related to the 4-coloring of a 3-colex given at the beginning of Section~\ref{sec:easy_path_integral}.
Let $\col_i$ for $0\leq i\leq 4$ denote the set of subsets of $\col\coloneqq \{r,g,b,y\}$ with $i$ elements.
We define the \emph{color} $\ccol_x\in\col_{4-i}$ of a 3-colex $i$-cell $x\in \Delta_i$ as the subset of colors of its $4-i$ adjacent volumes, and we denote the set of $i$-cells $x\in \Delta_i$ with $\ccol_x=c$ by $\Delta_i^c$.
We will denote the union of two disjoint subsets $c_i\in \col_i$ and $c_j\in \col_j$ by $c_ic_j\in \col_{i+j}$ or by $c_jc_i$, and the complement of $c_i\in \col_i$ by $\ovl c\in \col_{4-i}$, and we identify $\col_1$ with $\col$, e.g., we write $\{r\}$ simply as $r$.
Taken together, we denote $c\in\col_i$ by listing its $\col$-elements in arbitrary order, for example $rgy\in\col_3$.
We use analogous notation for 2-colexes.

\begin{figure}
\includegraphics[width=0.8\linewidth]{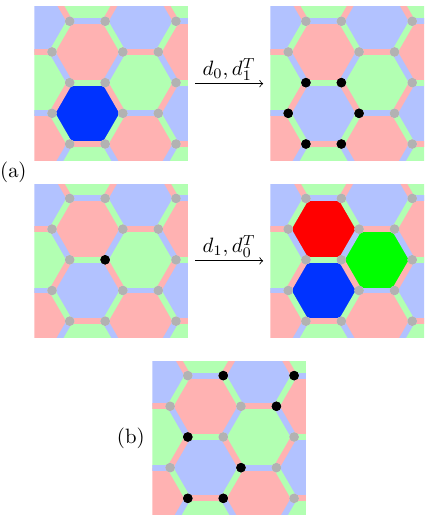}
\caption{
Illustration of the (co)boundary maps of 2D color cohomology.
(a) Boundary maps (as well as their transpose) acting on individual generators, that is, faces and vertices.
In other words, each picture on the right represents a row or a column of the boundary matrix.
(b) Example for a 1-cocycle.
Note that 1-cycles are the same as 1-cocycles since $d_0=d_1^T$.
}
\label{fig:2d_color_cohomology}
\end{figure}

\myparagraph{2D}
To prepare for the 3D case, let us start by discussing the analogous \emph{2D color cohomology}, which is a locally exact length-2 chain complex defined for any 2-colex.
It is instructive to define a larger structure of generator sets and homomorphisms between them:
\begin{equation}
\begin{tikzpicture}
\node(x-1p) at (-1.5,-1){$X_{-1}''$};
\node(x-1) at (-1.5,0){$X_{-1}$};
\node(x0) at (0,0){$X_{0}$};
\node(x1) at (1.5,0){$X_{1}$};
\node(x2) at (3,0){$X_{2}$};
\node(x3) at (4.5,0){$X_{3}$};
\node(x3p) at (4.5,1){$X_{3}'$};
\draw (x-1p)edge[->]node[midway,right,inner sep=0.05cm]{$\scriptstyle{m_{-1}'}$}(x-1);
\draw (x-1)edge[->]node[midway,above,inner sep=0.05cm]{$\scriptstyle{d_{-1}}$}(x0);
\draw (x0)edge[->]node[midway,above,inner sep=0.05cm]{$\scriptstyle{d_{0}}$}(x1);
\draw (x1)edge[->]node[midway,above,inner sep=0.05cm]{$\scriptstyle{d_{1}}$}(x2);
\draw (x2)edge[->]node[midway,above,inner sep=0.05cm]{$\scriptstyle{d_{2}}$}(x3);
\draw (x3)edge[->]node[midway,right,inner sep=0.05cm]{$\scriptstyle{m_3}$}(x3p);
\end{tikzpicture}
\;.
\end{equation}
The generator sets are given by
\begin{equation}
\label{eq:2d_color_generators}
\begin{aligned}
X_{-1} &\coloneqq\col_2\;,&
X_0 &\coloneqq \Delta_2\;,\\
X_1 &\coloneqq \Delta_0\;,&
X_2 &\coloneqq \Delta_2\;,\\
X_3&\coloneqq \col_2\;,&&\\
X_{-1}''&\coloneqq \{0\}\;,&
X_{3}'&\coloneqq \{0\}\;.
\end{aligned}
\end{equation}
The maps are defined by
\begin{equation}
\label{eq:2d_color_boundaries}
\begin{aligned}
(d_{-1})_{f\kappa}&\coloneqq \delta_{\ccol_f\in\kappa}\;,&
d_0&\coloneqq \partial_{02}\;,\\
d_1&\coloneqq \partial_{20}\;,&
(d_2)_{\kappa f}&\coloneqq \delta_{\ccol_f\in\kappa}\;,\\
(m_{-1}')_{\kappa 0}&\coloneqq 1\forall\kappa\;,&
(m_3)_{0\kappa}&\coloneqq 1\forall\kappa\;.\\
\end{aligned}
\end{equation}
The full sequence of maps above define a length-6 chain complex.
However, the middle length-2 chain complex consisting of $d_0$ and $d_1$ is most important, and only this part is locally exact.
The boundary maps $d_0$ and $d_1$ are illustrated in Fig.~\ref{fig:2d_color_cohomology}.

The maps $m_{-1}'$ and $m_3$ are somewhat trivial, and we could incorporate them by reducing the 3-generator sets $X_{-1}$ and $X_3$ to 2-generator sets.
Then, schematically, the chain complex would look like
\begin{equation}
2\rightarrow \Delta_2 \rightarrow \Delta_0\rightarrow \Delta_2\rightarrow 2\;,
\end{equation}
where ``2'' denotes a 2-element generator set.
The only reason why we do not do this is that it would require us to break the permutation symmetry between the colors.
$d_{-1}$ and $d_2$ can be understood as global conservation laws, similar to the extension of cellular cohomology in Eq.~\eqref{eq:cellular_cohomology_extension}.

Eq.~\eqref{eq:cohomology_axiom} for $i=0$ becomes
\begin{equation}
\sum_{p\in \Delta_0} \partial_{f_1p} \partial_{pf_0}=0\;\quad \forall f_0,f_1\in \Delta_2\;.
\end{equation}
The left-hand side is the $\mmod 2$ number of vertices $p$ that are adjacent to both faces $f_0$ and $f_1$.
Since the 2-cellulation is trivalent, any two distinct faces $f_0$ and $f_1$ either share $0$ or $2=0\mmod2$ common vertices, so the equation holds.
If $f_0=f_1$, the equation still holds since 3-colorability implies that all faces of a 2-colex have an even number of vertices.
So we note that even though the 3-coloring does not show up in the definition of $d_0$ and $d_1$ in Eqs.~\ref{eq:2d_color_generators} and \ref{eq:2d_color_boundaries}, it is important to ensure that this definition forms a valid cohomology theory.

Note that 2D color cohomology is equal to the length-2 chain complex underlying the 2D color code:
$X_0=\Delta_2$, $X_1=\Delta_0$, and $X_2=\Delta_2$ correspond to $X$-stabilizer generators, qubits, and $Z$-stabilizer generators, respectively.
The $X$ check matrix is given by $d_0$, and the $Z$ check matrix by $d_1^T$.
The ground states are superpositions of color 1-cocycles with equal amplitude within each cohomology class.
$X_{-1}$ and $X_3$ are global redundancies between the $X$ and $Z$ stabilizer generators, respectively, and $X_{-1}''$ and $X_3'$ are redundancies among these redundancies.

\myparagraph{3D}
After introducing 2D color cohomology, we are now ready to define 3D color cohomology, which is a locally exact length-3 chain complex defined on any 3-colex.
Again, it is instructive to define a larger structure of generator sets and homomorphisms between them as follows:
\begin{equation}
\label{eq:3d_color_cohomology_diagram}
\begin{tikzpicture}
\node(x-3) at (-1.3,-2){$X_{-1}'''$};
\node(x-2) at (-1.3,-1){$X_{-1}''$};
\node(x-1) at (-1.3,0){$X_{-1}$};
\node(x0) at (0,0){$X_{0}$};
\node(x1) at (1.3,0){$X_{1}$};
\node(x2) at (2.6,0){$X_{2}$};
\node(x3) at (3.9,0){$X_{3}$};
\node(x4) at (5.2,0){$X_{4}$};
\node(x3p) at (3.9,1){$X_{3}'$};
\node(x4p) at (5.2,1){$X_{4}'$};
\draw (x-3)edge[->]node[midway,right,inner sep=0.05cm]{$\scriptstyle{m_{-1}''}$}(x-2);
\draw (x-2)edge[->]node[midway,right,inner sep=0.05cm]{$\scriptstyle{m_{-1}'}$}(x-1);
\draw (x-1)edge[->]node[midway,above,inner sep=0.05cm]{$\scriptstyle{d_{-1}}$}(x0);
\draw (x0)edge[->]node[midway,above,inner sep=0.05cm]{$\scriptstyle{d_{0}}$}(x1);
\draw (x1)edge[->]node[midway,above,inner sep=0.05cm]{$\scriptstyle{d_{1}}$}(x2);
\draw (x2)edge[->]node[midway,above,inner sep=0.05cm]{$\scriptstyle{d_{2}}$}(x3);
\draw (x3)edge[->]node[midway,above,inner sep=0.05cm]{$\scriptstyle{d_{3}}$}(x4);
\draw (x3p)edge[->]node[midway,above,inner sep=0.05cm]{$\scriptstyle{d_{3}'}$}(x4p);
\draw (x3)edge[->]node[midway,right,inner sep=0.05cm]{$\scriptstyle{m_3}$}(x3p);
\draw (x4)edge[->]node[midway,right,inner sep=0.05cm]{$\scriptstyle{m_4}$}(x4p);
\end{tikzpicture}
\;.
\end{equation}
The generator sets are given by
\begin{equation}
\begin{aligned}
X_{-1} &\coloneqq \col_2\;,&
X_0 &\coloneqq \Delta_3\;,\\
X_1 &\coloneqq \Delta_0\;,&
X_2 &\coloneqq \Delta_2\;,\\
X_3 &\coloneqq \bigsqcup_{c\in\col_1} \Delta_3^c\times \ovl c\;,&
X_4&\coloneqq \col_1\\
X_{-1}''' &\coloneqq \{0\}\;,&
X_{-1}'' &\coloneqq \col_1\;,\\
X_3'&\coloneqq \Delta_3\;,&
X_4'&\coloneqq \{0\}\;.
\end{aligned}
\end{equation}
Here, $\sqcup$ denotes disjoint union, and we recall that $\ovl c$ denotes the 3-element set of colors that are not $c$.
That is, a 3-chain consists of three $\zz_2$-variables at every $c$-colored volume, each labeled by one of the three other colors.
The homomorphisms are given as follows:
\begin{equation}
\label{eq:color_cohomology_boundaries}
\begin{aligned}
(d_{-1})_{v\kappa}&\coloneqq \delta_{\ccol_v\in\kappa}\;,&
d_0 &\coloneqq \partial_{03}\;,\\
d_1 &\coloneqq \partial_{20}\;,&
(d_2)_{(v,c),f} &\coloneqq \partial_{vf} \delta_{c\not\in \mathbf c_f}\;,\\
(d_3)_{c',(v,c)} &\coloneqq \partial_{c=c'}\;,&
(d_3')_{0v}&\coloneqq 1\forall v\;,\\
(m_{-1}'')_{c 0}&\coloneqq 1\forall c\;,&
(m_{-1}')_{\kappa c}&\coloneqq \delta_{c\notin\kappa}\;,\\
(m_3)_{v',(v,c)} &\coloneqq \delta_{v'=v}\;,&
(m_4)_{0c}&\coloneqq 1\forall c\;.
\end{aligned}
\end{equation}
That is, $d_{-1}$ maps a color pair $\kappa$ to all volumes whose color is in $\kappa$.
$d_0$ maps a volume to all of its vertices, and $d_1$ maps a vertex to all adjacent faces.
$d_2^T$ maps a volume $v$ and a color $c\neq \ccol_v$ to all faces of $v$ whose color does not contain $c$.
For example, if $v$ has color $\ccol_v=r$, then $d_2(v,g)$ consists of all $rb$ and $ry$ faces of $v$.
$d_3^T$ maps a color $c'$ to all pairs $(v,c')$ (one for each volume $v$ with $\ccol_v\neq c'$).
$(d_3')^T$ maps the unique generator $0$ to all volumes.
$m_3^T$ maps a volume $v'$ to all three pairs $(v',c)$ (one for each color $c$ with $c\neq \ccol_{v'}$).
$m_4^T$ maps the unique generator $0$ to all four colors $c\in \col_1$.
$m_{-1}'$ maps a color $c$ to the three color pairs $\kappa$ that do not contain $c$.
$m_{-1}''$ maps the unique generator $0$ to all four colors.

\begin{figure*}
\begin{tikzpicture}[scale=0.8]
\node(x0cl) at (0,0){
\includegraphics[width=2cm]{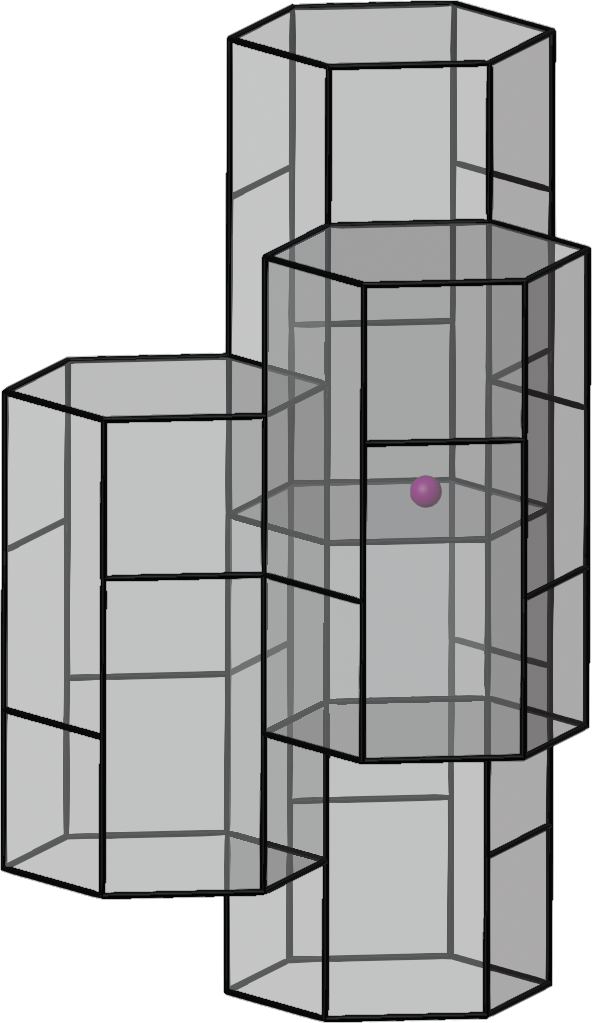}
};
\node(x0cr) at (4,0){
\includegraphics[width=2cm]{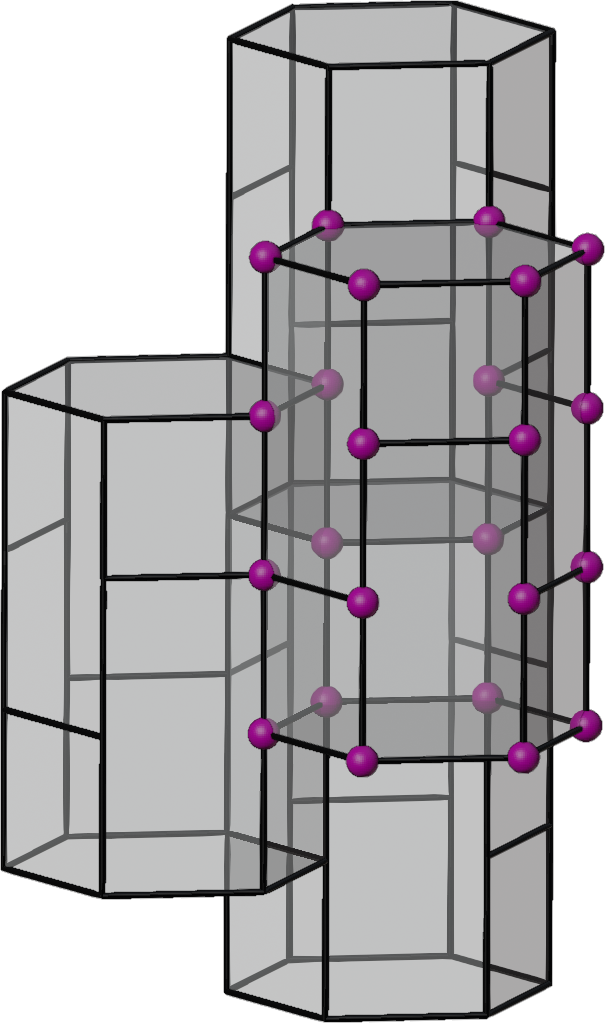}
};
\node(x0l) at (8,0){
\includegraphics[width=2cm]{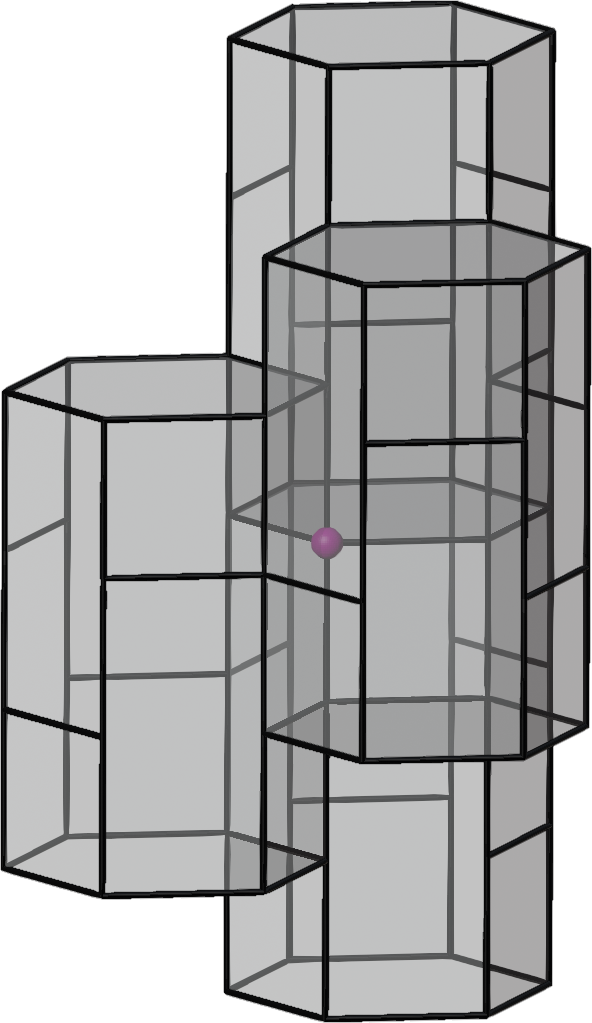}
};
\node(x0r) at (12,0){
\includegraphics[width=2cm]{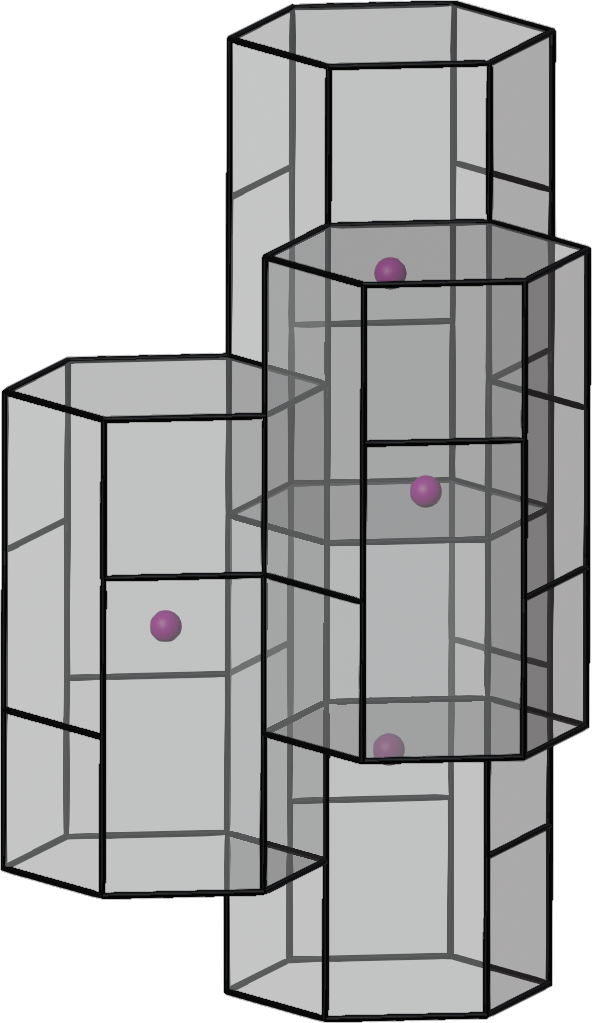}
};
\node(x1cl) at (0,-5){
\includegraphics[width=2cm]{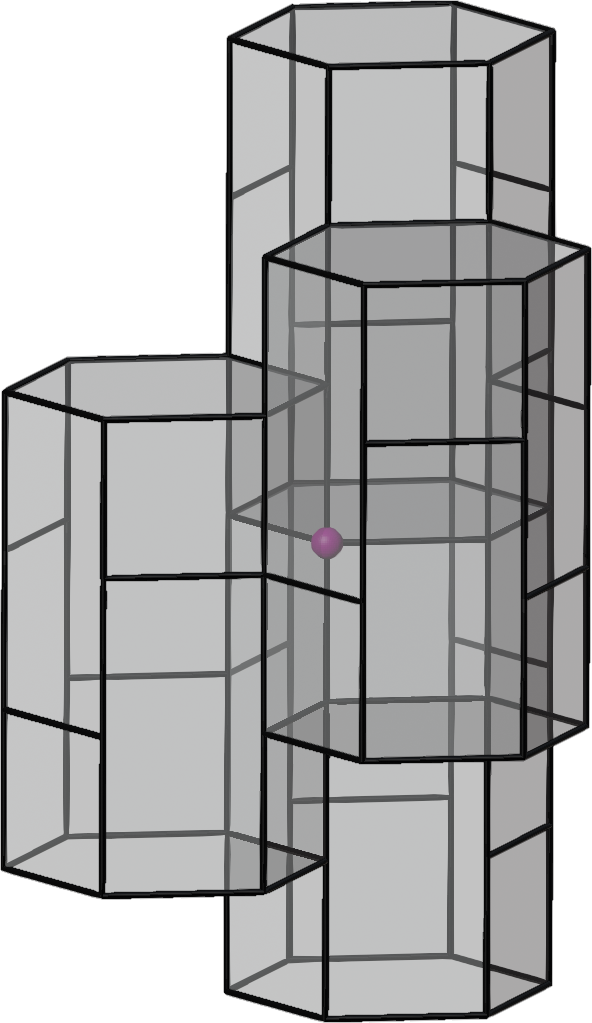}
};
\node(x1cr) at (4,-5){
\includegraphics[width=2cm]{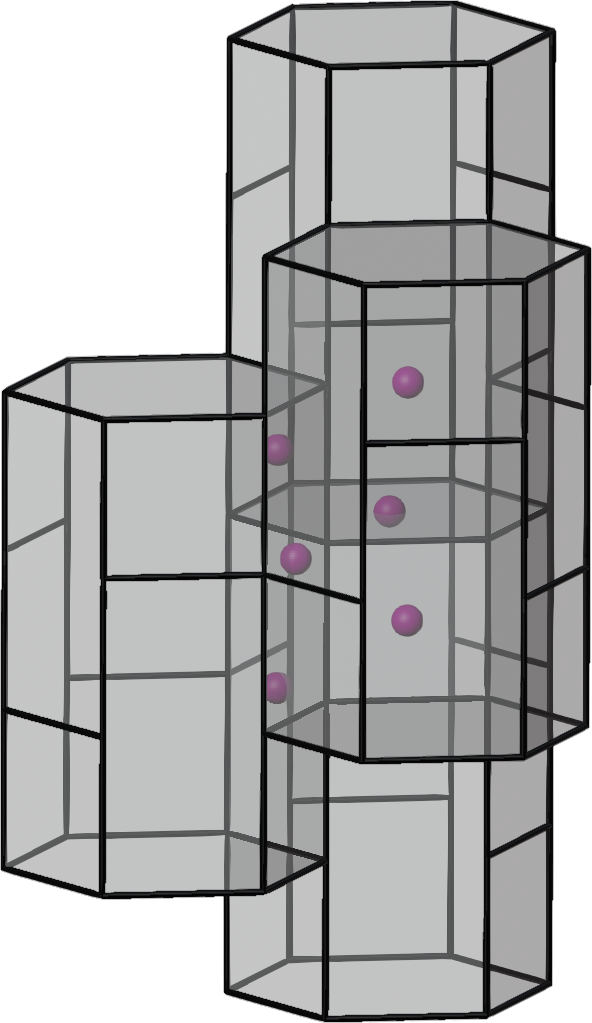}
};
\node(x1l) at (8,-5){
\includegraphics[width=2cm]{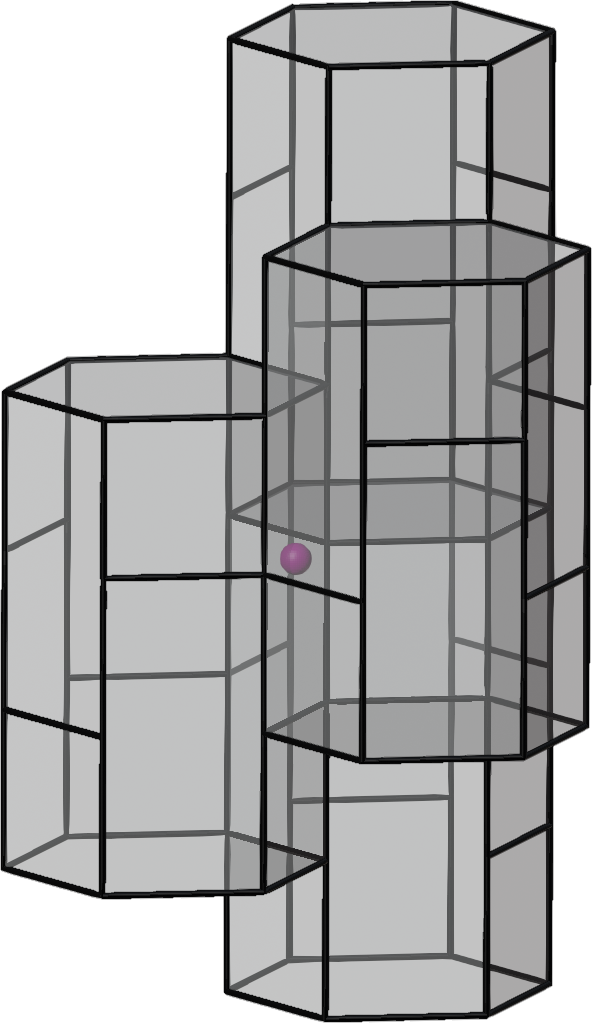}
};
\node(x1r) at (12,-5){
\includegraphics[width=2cm]{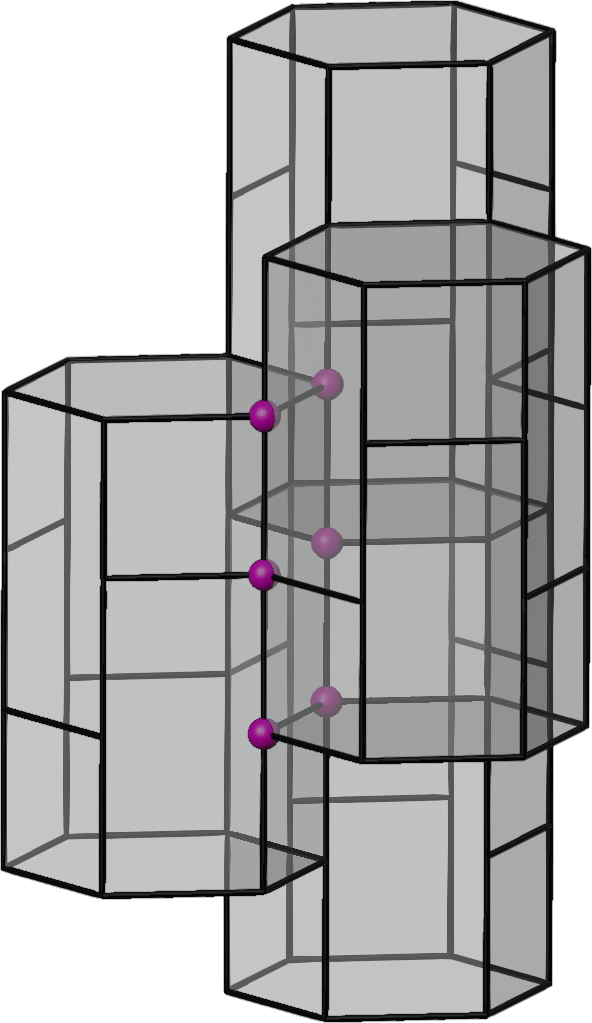}
};
\node(x2cl) at (0,-10){
\includegraphics[width=2cm]{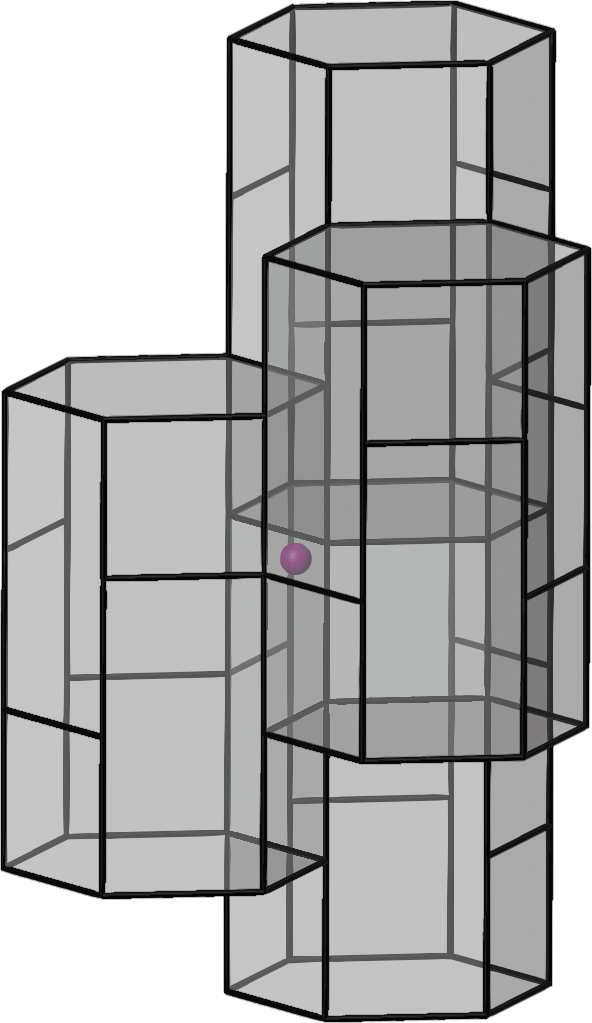}
};
\node(x2cr) at (4,-10){
\includegraphics[width=2cm]{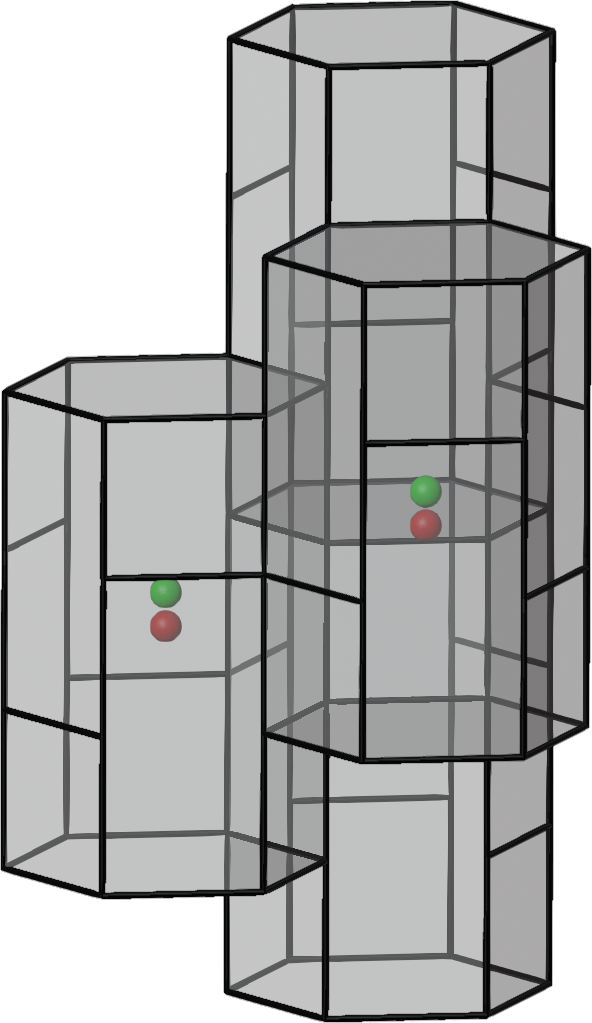}
};
\node(x2l) at (8,-10){
\includegraphics[width=2cm]{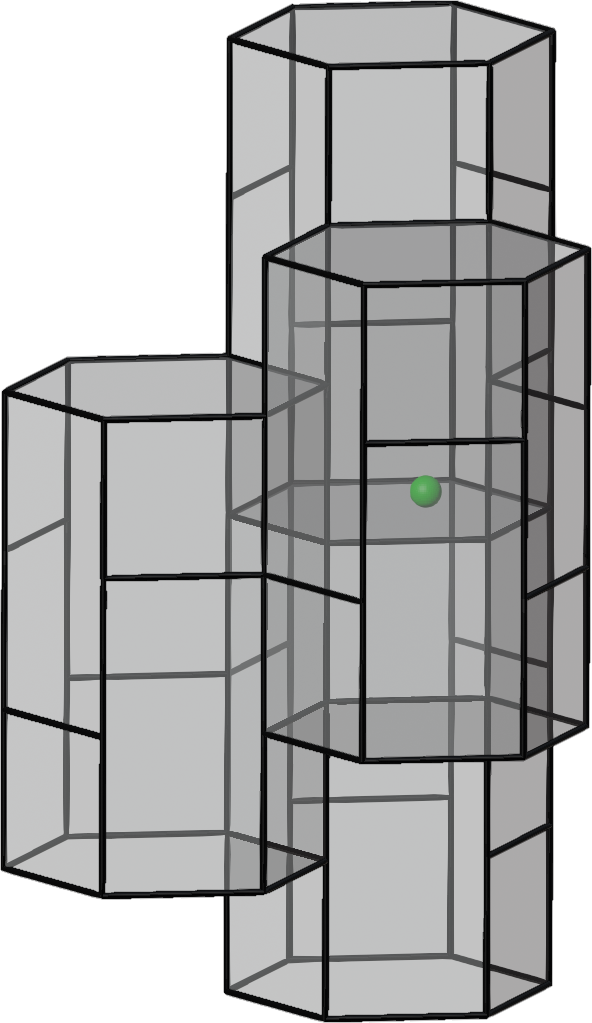}
};
\node(x2r) at (12,-10){
\includegraphics[width=2cm]{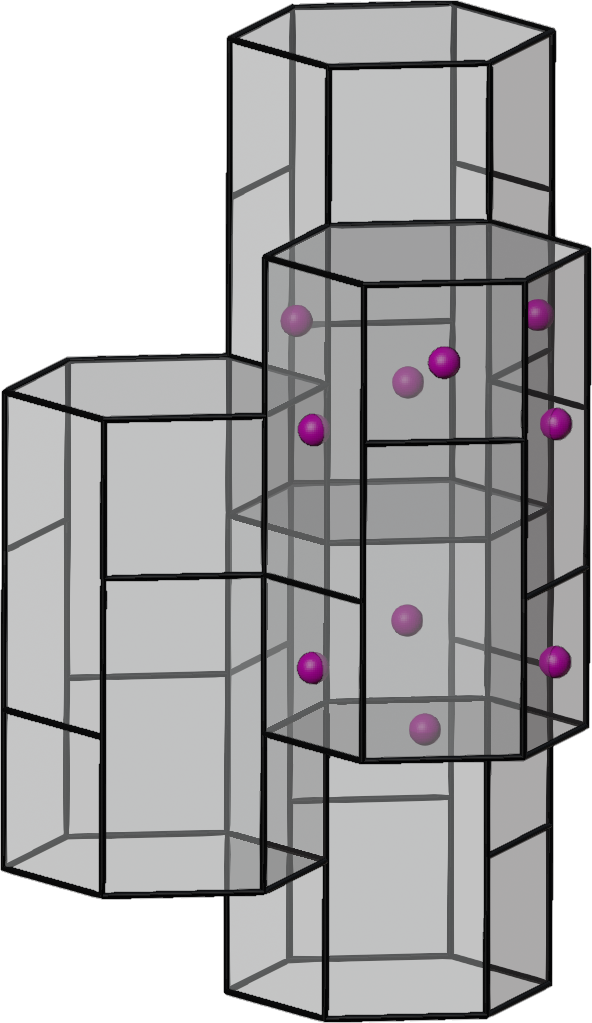}
};
\draw (x0cl)edge[->]node[midway,above]{$d_0$}(x0cr) (x0l)edge[->]node[midway,above]{$d_0^T$}(x0r);
\draw (x1cl)edge[->]node[midway,above]{$d_1$}(x1cr) (x1l)edge[->]node[midway,above]{$d_1^T$}(x1r);
\draw (x2cl)edge[->]node[midway,above]{$d_2$}(x2cr) (x2l)edge[->]node[midway,above]{$d_2^T$}(x2r);
\end{tikzpicture}
\caption{
Boundary maps $d_0$, $d_1$, and $d_2$, as well as their transposes, acting on a single generator.
Each generator is indicated by placing a little purple ball at the center of the vertex, volume, or face.
There are three $X_3$ generators inside each volume $v$ corresponding the three colors $c\neq \ccol_v$, which we distinguish by coloring the ball with color $c$.
}
\label{fig:3d_color_cohomology}
\end{figure*}
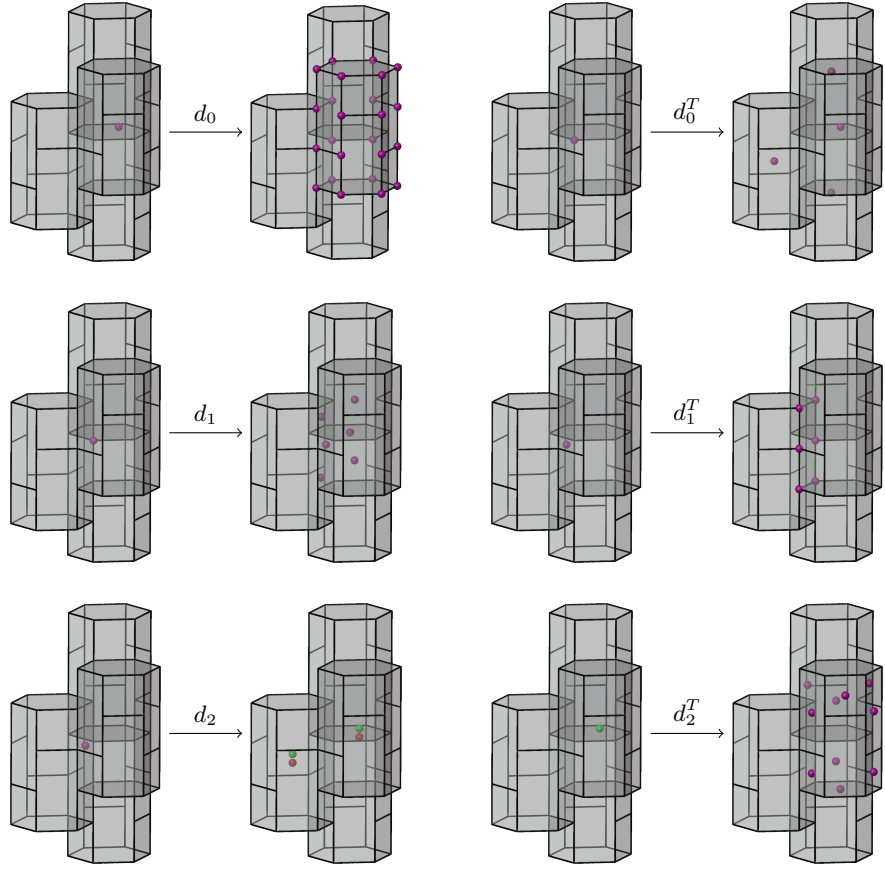

Each horizontal and each vertical line in Eq.~\eqref{eq:3d_color_cohomology_diagram} defines a chain complex, and each square commutes.
This holds also for squares obtained after filling gaps with trivial 1-elements sets $\{0\}$ and zero homomorphisms.
Explicitly, we have the following relations:
\begin{equation}
\begin{aligned}
m_{-1}'m_{-1}''&=0\;,& d_{-1}m_{-1}'&=0\;,\\m_3d_2&=0\;,& m_4d_3-d_3'm_3&=0\;.
\end{aligned}
\end{equation}
The middle length-3 chain complex consisting of $d_0$, $d_1$, and $d_2$ is the most important.
Only this part fulfills the local exactness property that every locally supported (co-)cycle is a (co-)boundary.
We illustrate the coboundary maps $d_0$, $d_1$, and $d_2$ in Fig.~\ref{fig:3d_color_cohomology}.

Note that $m_3$ acts within each volume separately and thus does not couple different unit cells of the 3-colex.
The same holds trivially for $m_{-1}'$, $m_{-1}''$, and $m_4$, since they not related to the 3-colex at all.
So, we can use these maps to cancel some of the generators in $X_{-1}$, $X_3$, and $X_4$, while still maintaining a cell complex that is local on the lattice.
In particular, using $m_{-1}'$ and $m_{-1}''$, we can reduce $X_{-1}$ from 6 color pairs to 3 generators, using $m_3$, we can reduce $X_3$ from 3 to 2 generators per volume, and using $m_4$, we can reduce $X_4$ from 4 colors to 3 generators.
After this, we obtain a single length-5 chain complex, which schematically looks like
\begin{equation}
3\rightarrow \Delta_3 \rightarrow \Delta_0 \rightarrow \Delta_2 \rightarrow 2\times\Delta_3\rightarrow 3\;,
\end{equation}
where ``2'' or ``3'' denotes a 2-element or 3-element set.
The only reason why we have not performed this reduction is that it would require us to break the symmetry under color permutations.
Note that while the middle three boundary maps define circuit and path integral of our protocol as well as the decoding graphs for $X$ and $Z$ errors, the outer two boundary maps define global parity constraints of the syndrome and are thus important for constructing decoding algorithms based on matching decoders.

Let us briefly verify the chain complex axiom in Eq.~\eqref{eq:cohomology_axiom} for some cases.
For $i=0$, we get
\begin{equation}
\sum_{p\in \Delta_0} \partial_{fp} \partial_{pv}=0\quad\forall f\in \Delta_2,v\in \Delta_3\;.
\end{equation}
The left-hand side is the $\mmod 2$ number of vertices $p$ adjacent to both the face $f$ and the volume $v$.
Due to the 4-coloring, $v$ and $f$ are either disjoint, or they share an edge, or $f\subset t$, and in all of these cases the number of vertices $p$ they share is divisible by 2.
For $i=1$, we have
\begin{equation}
\sum_{f\in \Delta_2} \delta_{c\not\in \ccol_f} \partial_{vf}\partial_{fp}=0\quad\forall v\in \Delta_3,c\in \ovl{\ccol_v},p\in \Delta_0\;.
\end{equation}
The left-hand side is the $\mmod 2$ number of faces $f$ that are adjacent to both the vertex $p$ and the volume $v$, and whose color $\ccol_f$ does not contain $c$.
If $p\not\subset v$, then there are no such faces, and the equation holds.
If $p\subset v$, they share three adjacent faces but only $2=0\mmod 2$ have the right color (namely $\ccol_v$ together with either of the two colors that are not $c$), so the equation holds.
For $i=2$, we have
\begin{align}
\begin{split}
\sum_{v\in \Delta_3,c\in \ovl{\ccol_v}} \partial_{v'v} \delta_{c\not\in \ccol_f} \partial_{vf}
=&\sum_{c\in \ovl{\ccol_v}: c\not\in \ccol_f} \partial_{v'f}
=0\\
&\qq{} \forall f\in \Delta_2,v'\in \Delta_3\;.
\end{split}
\end{align}
The expression in the middle is $0$ immediately if $f\not\subset v'$.
If $f\subset v'$, it is the number of colors which are not $\ccol_{v'}$ and not in $\ccol_f$, and there are always $2=0\mmod 2$ such colors.

Note that 3D color cohomology is the cohomology theory underlying the 3D color code.
That is, $X_0$, $X_1$, and $X_2$ correspond to the $X$ stabilizer generators, qubits, and $Z$ stabilizer generators, respectively.
$d_0$ and $d_1^T$ are the $X$ and $Z$ check matrices respectively.
$X_3$ corresponds to the \emph{meta-checks} of the $Z$-stabilizer (or $X$-error) syndrome, which makes the 3D color code single-shot decodable, and $d_2^T$ is the associated meta-check matrix.

Similarly, color cohomology is a convenient language to describe twisted color circuits, such as the one in Section~\ref{sec:simple_example}.
For example, the configuration of $Z$ measurement results can be mapped to a color 2-cochain, and the configuration of $X$ results to a color 1-cycle.
Therefore, color cohomology turns out to be the natural language to formulate the decoding of twisted color circuits, as we do in Section~\ref{sec:decoding}.

\subsection{Color cohomology: Equivalence to cellular cohomology}
\label{sec:color_cohomology_equivalence}
In this section, we show that color cohomology is equivalent to three copies of cellular cohomology (or in other words, to cellular cohomology with coefficient group $\zz_2^3$).
Technically, we show that there is a chain map $M$ from color to cellular cohomology, as defined around Eq.~\eqref{eq:chain_map_definition}.
Importantly, all $M_i$ are \emph{local} meaning that the matrix entries are only non-zero for neighboring pairs of cells, and the chain map corresponds to an isomorphism (not just a homomorphism) of cohomology groups.
A chain map very similar to $M$ was used in Ref.~\cite{Kubica2019} to show the equivalence between the 3+1D color code and three copies of the 3+1D toric code.

\begin{figure}
\includegraphics{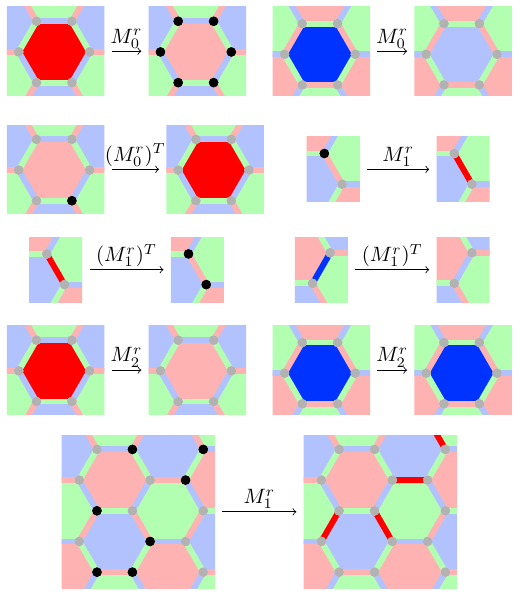}
\caption{
Action of $M_i^r$ or $(M_i^r)^T$ on color chains formed by a single generator.
Note that $(M_2^r)^T=(M_2^r)$.
Permuting colors yields analogous illustrations for $M_i^g$ and $M_i^b$.
}
\label{fig:2d_cohomology_map}
\end{figure}

\myparagraph{2D}
As usual, we will start with the analogous 2D case, and give a chain map $M_i$ from 2D color cohomology to $\zz_2^2$-valued (not $\zz_2^3$-valued) cellular cohomology.
The 2-cellulation on which the cellular $i$-chains are defined is just the 2-colex with the coloring removed.
More precisely, we define one map $M_i^c$ for every $c\in \{r,g,b\}$ from color $i$-chains to $\zz_2$-valued cellular $i$-chains, such that $M_i$ can be built from taking any two of the three $M_i^c$, while the third $M_i^c$ is redundant.
The $M_i^c$ are given by
\begin{equation}
\begin{aligned}
(M_{-1}^c)_{0\kappa}&\coloneqq\delta_{c\in\kappa}\;,&
(M_0^c)_{pf}&\coloneqq\delta_{\ccol_f=c}\partial_{pf}\;,\\
(M_1^c)_{ep}&\coloneqq\delta_{\ccol_e=\ovl c} \partial_{ep}\;,&
(M_2^c)_{f'f}&\coloneqq\delta_{\ccol_{f}\neq c} \partial_{f'f}\;,\\
(M_3^c)_{0\kappa}&\coloneqq\delta_{c\in\kappa}\;,
\end{aligned}
\end{equation}
and illustrated in Fig.~\ref{fig:2d_cohomology_map}.

Let us now check that Eq.~\eqref{eq:chain_map_definition} indeed holds for some cases.
For $i=0$, we have
\begin{equation}
\sum_{p\in \Delta_0} \partial_{ep} \delta_{\ccol_f=c} \partial_{pf}
=
\sum_{p\in \Delta_0} \delta_{\ccol_e=\ovl c} \partial_{ep} \partial_{pf}
\;.
\end{equation}
We confirm this by going through the different cases.
If $\ccol_f\neq c$ and $\ccol_e\neq \ovl c$, both sides are immediately zero.
If $\ccol_f\neq c$ and $\ccol_e=\ovl c$, the left-hand side is zero immediately.
The right-hand side is zero as well, since any $\ovl c$-colored edge $e$ either shares no vertex with a non-$c$-colored face $f$, or $e\subset f$ and they share two vertices.
If $\ccol_f= c$ and $\ccol_e\neq \ovl c$, the right-hand is zero immediately.
The left-hand side is zero as well, since any non-$\ovl c$-colored edge $e$ either shares no vertex with a $c$-colored face, or $e\subset f$ and they share two vertices.
Finally, if $\ccol_f= c$ and $\ccol_e= \ovl c$, both sides are immediately identical.
For $i=1$, we have
\begin{align}
\begin{split}
    \sum_{e\in \Delta_1} \partial_{fe} \delta_{\ccol_e=\ovl c} \partial_{ep}
=&
\sum_{f'\in \Delta_2} \delta_{\ccol_{f}\neq c} \partial_{ff'} \partial_{f'p}\\
=&\delta_{\ccol_{f}\neq c} \partial_{fp}
\;.
\end{split}
\end{align}
We confirm this equation again by going through the different cases.
If $\ccol_f=c$, then the right-hand side is immediately zero.
The left-hand side is zero as well since there are no edges of color $\ovl c$ in the boundary of the face $f$.
If $\ccol_f\neq c$, then both sides are $1$ if $p\in f$, and both sides are zero if $p\not\in f$.

Finally, we note that $\ovl M\coloneqq M^T$ defines a chain map from transposed cellular cohomology to color cohomology.
This is because color cohomology is symmetric under transposition, $d_i=d_{1-i}^T$.

\myparagraph{3D}
Having discussed the equivalence between color and cellular cohomology in 2D, the generalization to 3D is straight-forward:
We start by defining one chain map $M^c$ for every $c\in \col$ from color to $\zz_2$-valued cellular cohomology, such that $M$ can be constructed from any triple, while the fourth $M^c$ is redundant.
The $M_i^c$ are given by
\begin{equation}
\label{eq:color_to_cell_chain_map}
\begin{aligned}
(M_{-1}^c)_{0\kappa}&\coloneqq \delta_{c\in \kappa}\;,&
(M_0^c)_{pv} &\coloneqq \delta_{\ccol_v=c} \partial_{pv}\;,\\
(M_1^c)_{ep} &\coloneqq \delta_{\ccol_e=\ovl c} \partial_{ep}\;,&
(M_2^c)_{f'f} &\coloneqq \delta_{c\not\in\ccol_f} \partial_{f'f}\;,\\
(M_3^c)_{v',(v,c')} &\coloneqq \delta_{c= c'} \partial_{v'v}\;,&
(M_4^c)_{0c'} &\coloneqq \delta_{c= c'}\;.
\end{aligned}
\end{equation}
Let us verify Eq.~\eqref{eq:chain_map_definition}.
For $i=-1$, we have
\begin{equation}
\delta_{c\in \kappa}
=
\sum_{v\in \Delta_3} \delta_{\ccol_v=c} \partial_{pv} \delta_{\ccol_v\in\kappa}
=
\delta_{c\in \kappa} \sum_{v\in \Delta_3} \delta_{\ccol_v=c} \partial_{pv}\;.
\end{equation}
The equation holds because the sum on the right is the number of $c$-colored volumes adjacent to the vertex $p$, which is always $1$.
For $i=0$, we have
\begin{equation}
\sum_{p\in \Delta_0} \partial_{ep} \delta_{\ccol_v=c} \partial_{pv}
=
\sum_{p\in \Delta_0} \delta_{\ccol_e=\ovl c} \partial_{ep} \partial_{pv}\;.
\end{equation}
If $\ccol_v\neq c$ and $\ccol_e\neq \ovl c$, both sides are zero immediately.
If $\ccol_v= c$ and $\ccol_e\neq \ovl c$, the left-hand side is zero immediately.
The right-hand side is zero as well because $e$ shares either no vertex with the $c$-colored volume $v$, or $e\subset v$ and they share 2 vertices.
The case $\ccol_v\neq c$ and $\ccol_e= \ovl c$ is the same up to color permutation.
If $\ccol_v= c$ and $\ccol_e= \ovl c$, both sides are identical immediately.
For $i=1$, we have
\begin{equation}
\sum_{e\in \Delta_1} \partial_{fe} \delta_{\ccol_e=\ovl c} \partial_{ep}
=
\sum_{f'\in \Delta_2} \delta_{c\not\in\ccol_f} \partial_{ff'} \partial_{f'p}
= \delta_{c\not\in\ccol_f} \partial_{fp}
\;.
\end{equation}
If $c\in \ccol_f$, the right-hand side is zero immediately.
The left-hand side is zero as well since the $\ccol_f$-colored face $f$ has no $\ovl c$-colored edges.
If $c\not\in \ccol_f$, then both sides are $1$ if $p\in f$, and both sides are zero if $p\not\in f$.
For $i=2$, we have
\begin{align}
\begin{split}
\partial_{vf} \delta_{c\not\in \ccol_f}
=&
\sum_{f'\in \Delta_2} \partial_{vf'} \delta_{c\not\in \ccol_f} \partial_{f'f}\\
=&
\sum_{v'\in \Delta_3,c'\in\col} \delta_{c=c'} \partial_{vv'} \delta_{c'\not\in \ccol_f} \partial_{v'f}
=
\delta_{c\not\in \ccol_f} \partial_{vf}\;,
\end{split}
\end{align}
and both sides are identical immediately.
For $i=3$, we have
\begin{equation}
\sum_{v'\in \Delta_3} \delta_{c=c'}\partial_{vv'}
=
\sum_{c''\in \col} \delta_{c=c''} \delta_{c'=c''}\;,
\end{equation}
which holds immediately.

We can also define a reverse chain map $\ovl M$ from ($\zz_2^3$-valued) cellular cohomology to color cohomology.
More precisely, we can define one chain map $\ovl M^\kappa$ for each color pair $\kappa\in \col_2$.
Any triple of color pairs $\kappa$ that cover all four colors can be combined into $\ovl M$, while the other three chain maps are redundant.
The 3-cellulation is obtained from the 3-colex as follows:
The cellulation vertices are at the centers of the 3-colex volumes $v$ whose color is in $\kappa$, $\ccol_v\in \kappa$.
The cellulation edges are dual to the 3-colex faces $f$ of color $\ccol_f=\kappa$, connecting the vertices at the centers of the volumes adjacent to $f$.
The cellulation faces are the 3-colex faces of opposite color $\ovl\kappa$, which are enlarged such that their vertices are at the centers of the adjacent $\ccol_v\in\kappa$ volumes.
The cellulation volumes are the 3-colex volumes $v$ of color $\ccol_v\notin\kappa$, again enlarged.
So for each cellulation cell $c$, there is an associated 3-colex cell, which we denote by $x(c)$.

Now, $\ovl M^\kappa$ is defined as follows:
\begin{equation}
\label{eq:color_inverse_chain_map}
\begin{aligned}
(\ovl M_{-1}^\kappa)_{\kappa' 0}&\coloneqq \delta_{\kappa= \kappa'}\;,&
(\ovl M_0^\kappa)_{vp} &\coloneqq \delta_{v=x(p)}\;,\\
(\ovl M_1^\kappa)_{pe} &\coloneqq \partial_{p,x(e)}\;,&
(\ovl M_2^\kappa)_{f'f} &\coloneqq \delta_{f'=x(f)}\;,\\
(\ovl M_3^\kappa)_{(v',c)v} &\coloneqq \delta_{v'=x(v)} \delta_{c\in\kappa}\;,&
(\ovl M_4^\kappa)_{c0} &\coloneqq \delta_{c\in\kappa}\;.
\end{aligned}
\end{equation}

Physically, the equivalence between the two cohomology theories implies that the codes or circuits built from both can be understood as different microscopic realizations of the same topological phase.
For example, mappings $M$ and $\ovl M$ define local isometries between the 3D color code (which is based on 3D color cohomology) and three copies of the toric code (each based on 3D cellular cohomology), which show that they are in the same topological phase.
We will later use the cohomology equivalence in Section~\ref{sec:path_integral_equivalence} to show that the color path integral represents a particular non-Abelian topological phase.

\section{The color path integral as a non-Abelian phase}
\label{sec:path_integral}
In Section~\ref{sec:simple_example}, we motivated the color path integral in terms of a 3+0D protocol based on the 3D color code.
In this section, we will give a more natural and direct physical interpretation of the path integral, namely as a spacetime fixed-point model for a 2+1D non-Abelian topological phase.
To this end, we introduce a state-sum language for general path integrals in Section~\ref{sec:path_integrals_general}, after which the color path integral can be conveniently described in terms of color cohomology in Section~\ref{sec:color_path_integral_cohomology}.
Using the cohomology equivalence from Section~\ref{sec:color_cohomology_equivalence}, we then show in Section~\ref{sec:path_integral_equivalence} that the color path integral is equivalent to a known model for a non-Abelian phase, namely the Dijkgraaf-Witten state sum for the type-$III$ group 3-cocycle on $\zz_2^3$.
In Section~\ref{sec:flux_and_charge}, we equip the path integral with flux and charge defect configurations, which is necessary to turn its tensor network representation into a fault-tolerant circuit.
Finally, in Section~\ref{sec:boundaries}, we discuss how to equip the path integral with boundaries.

\subsection{Discrete path integrals}
\label{sec:path_integrals_general}
Before we get to the color path integral, we introduce path integrals more generally, and explain how they describe topological phases.
We introduce their \emph{fixed-point} property, which implies that isolated errors do not affect the logical action of the circuit.
One way to define a path integral is as a tensor network with tensors assigned to certain places of a spacetime lattice, and index contractions between nearby tensors.
We have defined the color path integral in this way in Section~\ref{sec:easy_path_integral}.
For this section it will be more convenient to use an alternative, but equivalent, language to describe path integrals, which in the context of topological phases is known as \emph{state sums}~\cite{Turaev1992,Barrett1993,Fukuma1992,Dijkgraaf1990,Crane1993,liquid_intro,thesis}.

\myparagraph{State-sum path integrals}
A \emph{state-sum path integral} associates variables to certain places of the spacetime lattice, such as all the vertices, all the edges, or all the corners of volumes.
The variables take values in some finite set, such as $\{0,1\}$.
Evaluating the path ``integral'' means summing over all spacetime configurations $\vec c$ of the variables.
The summand is a product of weights $\omega_x$, which are also associated to certain places $x$ of the spacetime lattice.
These weights are complex numbers $\omega$ that depend on the configuration of the variables within some constant-size neighborhood.
So, schematically, a state-sum path integral can be written as
\begin{equation}
\label{eq:general_path_integral}
Z=\sum_{\substack{\text{confs. }\vec c\ }}
\prod_{\text{places } x} \omega_x(\vec c|_{\text{near } x})\;.
\end{equation}
The path integrals we consider are uniform, that is, the distribution of variables and weights only depends on what the lattice looks like locally.
In particular, on a translation-invariant spacetime lattice the path integral is translation invariant as well, and specified by a single unit cell.
In physics, state sums are commonly used in statistical mechanics, where they are referred to as ``partition functions''.
For example, the classical 2D Ising model at inverse temperature $\beta\in\rr^+$ is a state sum on a 2D square lattice, with one $\{\pm 1\}$-valued variable at each vertex, and one ``Boltzmann'' weight $\omega_e(c) = e^{-\beta J c(e_0) c(e_1)}$ at each edge $e$ with endpoints $e_0$ and $e_1$.

Let us briefly point out how state sums are equivalent to tensor networks.
A tensor network is evaluated by summing over all configurations of contracted index pairs.
The summand is the product over all tensor entries for the corresponding index configuration.
So the contracted index pairs in the tensor network are equivalent to the variables in the state sum, and the tensors themselves are equivalent to the weights.
State sums may seem more general, as an index contraction can only involve two indices, whereas a variable can enter into an arbitrary number of weights.
However, we can define the contraction over a larger number of indices by connecting them all to one common \emph{Kronecker $\delta$-tensor}, which is the generalization of the $Z$-tensor in Eq.~\eqref{eq:delta_definition} to arbitrary bond dimensions.

Evaluating the path integral on a closed spacetime lattice with periodic boundary conditions yields a number $Z$ as in Eq.~\eqref{eq:general_path_integral}.
This number is not particularly relevant for our purposes.
What is relevant instead is the evaluation of the path integral on a spacetime lattice with \emph{state boundaries}.
In this case, we freeze a configuration $\vec c_{\text{boundary}}$ of state-sum variables within a constant distance from the boundary, and sum over all other variables in the interior.
This way, we get a vector of numbers $Z(\vec c_{\text{boundary}})$, instead of a single number, which can be interpreted as the state vector of a state defined on degrees of freedom living on the boundary.%
\footnote{If we divide the boundary into an \emph{input} and an \emph{output} component, the state can also be interpreted as a vectorized linear map from the input to the output vector space.}
For a tensor-network path integral, this corresponds to cutting the indices near the boundary and leaving them open.

\myparagraph{Physical interpretation of path integrals}
The path integrals that we consider are defined in a 3D spacetime and describe the zero-temperature physics of a 2D quantum model, in contrast to thermal physics for the Ising-model path integral illustrated above.
So our path integrals can be viewed as an alternative language to 2D lattice Hamiltonians.%
\footnote{Physically, our path integrals have the same qualitative properties as the imaginary-time evolution of a 2D Hamiltonian~\cite{liquid_intro}.
It should be noted though, that our path integrals are defined in discrete time, whereas the imaginary-time evolution is defined in continuous time.
One could in principle obtain a path integral from an imaginary-time evolution by Trotterization, but our path integrals do not directly arise in this way.}
The state $Z(\vec c_{\text{boundary}})$ then plays a role analogous to the ground state of a Hamiltonian model.

In particular, we consider path integrals that represent non-trivial 2D topological phases.%
\footnote{Note that topological phases can be defined directly on the level of path integrals:
Two path integrals are in the same phase if they are related by a continuous deformation that does maintain a constant spectral gap of the transfer operator (corresponding to a 2D layer of the 3D path integral).
We believe that the resulting classification of topological phases is closely related to that for lattice Hamiltonians.}
While lattice Hamiltonians are more commonly used to describe topological phases in condensed-matter physics, the path-integral language makes it easier to construct syndrome-extraction circuits for these topological phases.
More precisely, our path integrals are directly equal to a syndrome-extraction circuit that is post-selected onto a particular ($+1$) outcome.
In this context, $Z(\vec c_{\text{boundary}})$ corresponds to a code state of an underlying error correcting code.
This code will, in general, be microscopically different depending on the chosen state boundary but is guaranteed to inherit necessary properties for error correction from the path integral.

\myparagraph{Fixed-point path integrals}
We use a strongly restricted subclass of path integrals, which are particularly useful for constructing error-correcting circuits.
Namely, we use ones that represent the desired topological phase in a particularly pure manner, at zero correlation length.
We call them \emph{fixed-point} path integrals~\cite{liquid_intro,thesis}, and define them in the following.
Consider evaluating the path integral on a 3-ball with a smaller 3-ball removed from its center.
We consider both interior and exterior boundary 2-spheres as state boundaries, and assume that they are separated by at least some constant lattice distance $d_0$.%
\footnote{For usual examples, $d_0$ is very small, for example it might suffice if the interior and exterior are separated by at least one lattice edge.}
Evaluating the path integral on this lattice yields a state $\ket T$ on the boundary, which can be considered a bipartite state where one part corresponds to the interior boundary and the other to the exterior boundary.
The defining property of a fixed-point path integral is that $\ket T$ is a product state, $\ket T = \ket{T_{in}}\otimes \ket{T_{out}}$, as illustrated by following 2D toy picture:
\begin{equation}
\label{eq:fixed_point_property}
\begin{tikzpicture}
\fill[gray] (0,0)circle(1.2);
\fill[white] (0,0)circle(0.6);
\draw (0.6,0)edge[mark={arr,s},mark={arr,e},mark={slab=$d_0$}](1.2,0);
\node[anchor=west] at(-0.6,0){$\vec a$};
\node[anchor=east] at(-1.2,0){$\vec b$};
\end{tikzpicture}
\quad\rightarrow\quad
\langle\vec a,\vec b|T\rangle = \langle\vec a|T_{in}\rangle \langle\vec b|T_{out}\rangle\;.
\end{equation}
As a consequence, these path integrals have \emph{zero correlation length} in spacetime:
If we insert a observable $O$ (or alter the path integral in any other way) inside some simply-connected region, this only changes the path integral by a global prefactor $\alpha_O$:
\begin{equation}
\begin{tikzpicture}
\fill[gray] (0,0)circle(1);
\fill[green] (0,0)circle(0.4);
\draw (0.4,0)edge[mark={arr,s},mark={arr,e},mark={slab=$d_0$}](1,0);
\node[anchor=east] at(-1,0){$\vec a$};
\node at(0,0){$O$};
\end{tikzpicture}
=
\alpha_O\cdot
\begin{tikzpicture}
\fill[gray] (0,0)circle(1);
\node[anchor=east] at(-1,0){$\vec a$};
\end{tikzpicture}
\;.
\end{equation}
As a consequence, all correlators%
\footnote{Here, we mean the connected correlator, $\langle O_0O_1\rangle - \langle O_0\rangle \langle O_1\rangle$.}
between any two observables $O_0$ and $O_1$ disappear if they are separated by at least $d_0$:
\begin{equation}
\begin{tikzpicture}
\fill[gray] (0,0)rectangle(3,2);
\fill[green] (0.7,0.7)circle(0.4) (2.3,1.3)circle(0.4);
\node at (0.7,0.7){$O_0$};
\node at (2.3,1.3){$O_1$};
\draw ($(0.7,0.7)+(20:0.4)$)edge[mark={arr,s},mark={arr,e},mark={slab=$d_0$}]($(2.3,1.3)+(-160:0.4)$);
\end{tikzpicture}
=
\alpha_{O_1}\cdot\alpha_{O_2}\cdot
\begin{tikzpicture}
\fill[gray] (0,0)rectangle(3,2);
\end{tikzpicture}
\;.
\end{equation}
The fixed-point property in Eq.~\eqref{eq:fixed_point_property} is the underlying reason why the circuits that we construct from the path integral have a fault-tolerant threshold:
If we view the inserted operator $O$ as an error, then this means that isolated errors are irrelevant.

Note that typical examples of fixed-point path integrals have nice algebraic properties, such as an exact notion of topological invariance.
That is, we can freely deform the underlying spacetime lattice by locally applying \emph{rewrite rules}, which are equations between small tensor networks, such as the $ZX$ rules~\cite{liquid_intro,thesis,path_integral_qec}.
Similarly, if two fixed-point path integrals are in the same topological phase, then they are typically related via local rewrite rules as well.

\myparagraph{Boundaries and other defects}
The state boundaries discussed above are to be contrasted with \emph{physical boundaries}, which we have discussed in Section~\ref{sec:boundaries_easy} and will revisit in Section~\ref{sec:boundaries}.
The main difference is that for physical boundaries we do sum over all boundary variables instead of keeping them fixed.
To define a physical boundary, we do not only remove the variables outside the boundary and the weights depending on them, but we may also introduce new variables and weights on the boundary.%
\footnote{Since we sum over all boundary variables, evaluating the path integral with physical boundary yields again a number $Z$.
We can also evaluate the path integral on a lattice whose boundary is divided into a physical-boundary part and a state-boundary part, yielding a state supported only on the state-boundary part.
In this case, the state itself has a physical boundary, which is the region separating state and physical boundary in spacetime.}
In the tensor-network language, a physical boundary is a way to close off the open indices obtained from merely cutting the tensor network along some surface (corresponding to a state boundary).
Similarly, we can consider \emph{domain walls}, which are 2D surfaces separating two different 3D regions where the path integral is defined differently, one on each side.

Boundaries and domain walls for the path integral are needed to construct a protocol with a globally planar layout, which implements a non-Clifford logic gate.
Eq.~\eqref{eq:global_protocol} shows an example of a configuration of boundaries and domain walls that corresponds to a particular logic operation.
Like the topological phases for the bulk, the different physical boundaries (or domain walls) can be classified into different \emph{boundary phases}.
The logical action of the resulting circuit only depends on the topological phases of the bulk, boundaries, and domain walls in the spacetime diagram.
The boundaries and domain walls we consider are such that they obey the fixed-point property in Eq.~\eqref{eq:fixed_point_property} also on the boundary or domain wall.

In a conceptually similar way, we can equip a path integral with \emph{cohomological defects} as in Section~\ref{sec:flux_and_charge}.
That is, in addition to the variables, we let the weights depend on some cohomological chain (see Section~\ref{sec:general_cohomology}), which we do not sum over.
Such cohomological chains represent the configurations of measurement outcomes of the circuit.

\subsection{The color path integral via color cohomology}
\label{sec:color_path_integral_cohomology}
In this section, we reintroduce the color path integral from Section~\ref{sec:easy_path_integral}, formulated as a state sum using color cohomology.
We first introduce the 2D analog of the 3D color path integral, both for pedagogical reasons and because we will use it later to show an important property (``gauge invariance'') of the 3D color path integral.

\myparagraph{2D}
The \emph{2D untwisted color path integral} is defined on any 2-colex as
\begin{equation}
\label{eq:2d_untwisted_color_path_integral}
Z_{\text{untw}}=\sum_{A\in \zz_2^{\Delta_0}} \prod_{f\in \Delta_2} \delta_{(d A)_f=0} 
=\sum_{A\in \zz_2^{\Delta_0}: d A=0} 1 \;.
\end{equation}
This is a state sum with one $\zz_2$-valued variable at every vertex, and one weight at every face $f$ enforcing that $(d A)_f=0$, where $d=d_1$ is the coboundary of 2D color cohomology, see Section~\ref{eq:color_cohomology_definition}.
As shown, this weight can be suppressed by summing only over 2D color 1-cocycles $A$.
The \emph{2D (twisted) color path integral} includes an additional weight at every vertex $p\in \Delta_0$:
\begin{equation}
\label{eq:2d_color_path_integral}
Z=\sum_{A\in \zz_2^{\Delta_0}: d A=0} \prod_{p\in \Delta_0} e^{\sgn_p \frac{2\pi i}{4}\ovl{A_p}}\;.
\end{equation}
Here, $\sgn_p=+1$ if cycling through the adjacent faces in the ordering $r-g-b$ is clockwise, and $\sgn_p=-1$ otherwise.
Also, $\ovl{A_t}$ means that we consider $A_t$ as an element of $\zz$ instead of $\zz_2$.
The overall weight is the exponential of the following \emph{action}:
\begin{equation}
\label{eq:2d_color_action}
S[A]=\frac{1}{4} \sum_{p\in \Delta_0} \sgn_p \ovl{A_p}\mod 1\in \rr/\zz\;.
\end{equation}
The central property of the action is its \emph{gauge invariance},
\begin{equation}
\label{eq:gauge_invariance_definition}
A'\coloneqq A+d\alpha
\quad\Rightarrow\quad
S[A']=S[A]\;,
\end{equation}
for any 2D color 1-cocycle $A$ and color 0-cochain $\alpha$.
Gauge invariance directly implies that $S$ only depends on the cohomology class of the color 1-cocycle $A$.
It suffices to show gauge invariance for the generating 0-chains $\alpha=f$ formed by a single face $f$.
We indeed find:
\begin{align}
\begin{split}
S[A']-S[A] =& \frac14 \sum_{p\in \Delta_0} \sgn_p (\ovl{A_p+(df)_p}-\ovl{A_p})\\
\overset{\eqref{eq:integer_lift_expansion}}=& \frac14 \sum_{p\in \Delta_0}\sgn_p \big(\ovl{(df)_p}+2 \ovl{A_p} \ovl{(df)_p})\big)\\
\overset{*}{=}& \frac14 \sum_{p\in f}\sgn_p+\frac12 \sum_{p\in f} \ovl{A_p} \overset{**}{=} \frac12 \ovl{(dA)_f}\\
=& 0\;.
\end{split}
\end{align}
For the middle equality, we have used that
\begin{equation}
\label{eq:integer_lift_expansion}
\ovl{a+b}-\ovl a=\ovl b -2 \ovl a\ovl b
\end{equation}
for $a,b\in \zz_2$.
For ($*$) we have used that the summand is only supported on the vertices $p\in f$.
For ($**$) we have used that the first summand is zero, since the 3-colorability implies that $f$ has an equal number of vertices $p\in f$ with $\sgn_p=1$ and $\sgn_p=-1$.
We have also used the definition of $d_1$ and that $A$ is a color 1-cocycle.
Finally, we note that the action in Eq.~\eqref{eq:2d_color_action} (or better, the according weight in Eq.~\eqref{eq:2d_color_path_integral}) is analogous to the transversal $S$ gate in the 2D color code, and the gauge invariance is analogous to the fact that the transversal gate preserves the code space.

\myparagraph{3D}
After describing the 2D color path integral, the 3D generalization is rather straight-forward.
The \emph{(3D) untwisted color path integral} is defined on any 3-colex as
\begin{equation}
\label{eq:untwisted_color_path_integral}
Z_{\text{untw}}
=\sum_{A\in \zz_2^{\Delta_0}} \prod_{f\in \Delta_2} \delta_{(d A)_f=0}
=\sum_{A\in \zz_2^{\Delta_0}: d A=0} 1
\;.
\end{equation}
The \emph{(3D twisted) color path integral} includes an additional weight:
\begin{equation}
\label{eq:color_path_integral}
Z=\sum_{A\in \zz_2^{\Delta_0}: d A=0} \prod_{p\in \Delta_0} e^{\sgn_p \frac{2\pi i}{8}\ovl{A_p}}\;,
\end{equation}
recalling the sign $\sgn_p$ of a vertex $p$ from the beginning of Section~\ref{sec:easy_path_integral}.

Let us briefly confirm that this is indeed the same as the color path integral introduced in Section~\ref{sec:easy_path_integral} as a $ZX$ tensor network.
Indeed, the weight $\delta_{(dA)_f=0}$ at a face $f$ is $1$ if the variables at all the vertices of $p\in f$ sum to $0\mmod 2$, and $0$ otherwise.
So these weights correspond to the $X$ tensors at every face in Fig.~\ref{fig:hexagon_zx}.
The summation over the variable at every vertex corresponds to a ``multi-index contraction'', which is implemented by a $Z$-tensor with indices connected to all adjacent face weights.
The state-sum weight $e^{\sgn_p \frac{2\pi i}{8}\ovl{A_p}}$ at every vertex $p$ is implemented by adding a phase of $\sgn_p \frac{2\pi}8$ to the according $Z$-tensor in the tensor network.

The overall weight in the path integral is the exponential of the following action:
\begin{equation}
\label{eq:twisted_action}
S[A]=\frac{1}{8} \sum_{p\in \Delta_0} \sgn_p \ovl{A_p}\;.
\end{equation}
The action is again gauge invariant as defined in Eq.~\eqref{eq:gauge_invariance_definition}, and thus only depends on the cohomology class of $A$.
In the 3D case, gauge invariance ensures that the resulting path integral is a fixed-point path integral and obeys the zero-correlation length condition in Eq.~\eqref{eq:fixed_point_property}.
It suffices to show gauge invariance for the generating 0-cochains $\alpha=v$ consisting of a single volume $v\in \Delta_3$.
We indeed find:
\begin{align}
\label{eq:3d_gauge_invariance}
\begin{split}
S[A']-S[A] =& \frac18 \sum_{p\in \Delta_0} \sgn_p (\ovl{A_p+(dv)_p}-\ovl{A_p})\\
\overset{\eqref{eq:integer_lift_expansion}}{=}& \frac18 \sum_{p\in \Delta_0} \sgn_p \big(\ovl{(dv)_p}+2 \ovl{A_p} \ovl{(dv)_p})\big)\\
=& \frac18 \sum_{p\in v}\sgn_p + \frac14 \sum_{p\in v} \sgn_p \ovl{A_p}\\
\overset{*}{=}& S_{\text{2D},v}[A|_v]
=0\;.
\end{split}
\end{align}
For ($*$) we have used that due to the 4-colorability the number of $\sgn_p=+1$ and $\sgn_p=-1$ vertices of $v$ is equal, so the first summand vanishes.
We have also used that the second term is the action $S_{\text{2D},v}$ of the 2D color path integral restricted to the boundary of $v$, which is a 2-colex when colored according to the adjacent volumes.
Since (1) the color 1-cohomology group on any 2-sphere is trivial and (2) the 2D action is gauge invariant and only depends on the 1-cohomology class, the 2D action is the same as for the $A=0$ color 1-cocycle, namely $0$.
Finally, we note that the action in Eq.~\eqref{eq:twisted_action} (or better, the corresponding phase factor in Eq.~\eqref{eq:color_path_integral}) is the same as occurs in the transversal $T$ gate of the 3D color code.

\subsection{Equivalence to type-III twisted path integral}
\label{sec:path_integral_equivalence}
In this section, we establish that the color path integral represents a non-Abelian topological phase.
We will do this by showing its equivalence to a known path integral, namely the \emph{Dijkgraaf-Witten} state sum~\cite{Dijkgraaf1990} for the group $\zz_2^3$ and the type-III group 3-cocycle.
The latter is known to host a non-Abelian anyon theory~\cite{Propitius1995}.
This equivalence is an extension of the equivalence between color and cellular cohomology discussed in Section~\ref{sec:color_cohomology_equivalence}.

\myparagraph{2D cellular path integral}
As usual, we start with a simpler 2D analogy of the 3D equivalence.
We show the equivalence of the 2D twisted color path integral with the following \emph{2D twisted cellular path integral}, defined on any 2-cellulation:
\begin{equation}
\label{eq:2d_cellular_path_integral}
\begin{multlined}
Z_{\text{cell}}=
\sum_{A_r,A_g\in \zz_2^{\Delta_1}} \prod_{f\in \Delta_2} \delta_{(d^{\text{cell}} A_r)_f=0}\; \delta_{(d^{\text{cell}} A_g)_f=0}\\ \times \prod_{f\in \Delta_2} e^{2\pi i\frac12 (A_r\cup A_g)(f)}\;.
\end{multlined}
\end{equation}
So the path integral is the sum over all pairs of (cellular) 1-cocycles $A_r$ and $A_g$, with an action
\begin{equation}
S[A_r,A_g] = \frac12 \sum_{f\in \Delta_2} (A_r\cup A_g)(f)\;.
\end{equation}
Here $\cup$ denotes the \emph{cup product}, whose value on a face $f$ only depends on the values of $A_r$ and $A_g$ on the edges of $f$.
There are explicit formulas for the cup product on triangulations~\cite{Steenrod1947}, cubulations~\cite{Chen2021} or general cellulations~\cite{twisted_double_code}.
Intuitively, the cup product counts the number of intersections of the closed-loop patterns $A_r$ and $A_g$.
To turn this intuition into a concrete formula, we divide the edges $e$ of the face $f$ with $(A_r)_e=1$ into pairs in an arbitrary way.
Then, we connect the two edge centers of each edge pair with a red line.
Analogously, we connect the $(A_g)_e=1$ edge centers with green lines.
$(A_r\cup A_g)(f)$ is then given by the $\mmod 2$ number of red-green line intersections inside the face.
Note that this number is independent on how we pair up the $(A_r)_e=1$ ($(A_g)_e=1$) edges.

For example, in the 8-gon face shown below, we have 4 $(A_r)_e=1$ edges (marked in thick red) and 2 $(A_g)_e=1$ edges (marked in thick green).
The connecting lines are shown as thin red and green lines, and there is one intersection:
\begin{equation}
(A_r\cup A_g)
\Big(\;
\begin{tikzpicture}
\draw (0,0)coordinate(0)--++(0:0.5)coordinate(1)--++(45:0.5)coordinate(2)--++(90:0.5)coordinate(3)--++(135:0.5)coordinate(4)--++(180:0.5)coordinate(5)--++(-135:0.5)coordinate(6)--++(-90:0.5)coordinate(7)--cycle;
\draw[line width=0.05cm,colr] (0)--(1) (1)--(2) (3)--(4) (5)--(6);
\draw[line width=0.05cm,colg] (4)--(5) (7)--(0);
\draw[colr] ($(0)!0.5!(1)$)to[out=90,in=135]($(1)!0.5!(2)$) ($(3)!0.5!(4)$)to[out=-135,in=-45]($(5)!0.5!(6)$);
\draw[colg] ($(4)!0.5!(5)$)to[out=-90,in=45]($(7)!0.5!(0)$);
\end{tikzpicture}
\;\Big)
=1\mmod 2
\;.
\end{equation}
Thus, the cup product evaluates to $1\mmod 2$ and the according path-integral weight is $e^{2\pi i\frac12 1}=-1$.%
\footnote{If there is an edge $e$ with both $(A_r)_e=1$ and $(A_g)_e=1$, it becomes ambiguous whether there is a red-green intersection at the edge center or not, which can be resolved by shifting the red and green edge centers with respect to another.
However, this case does not appear when showing the equivalence of the color and twisted cellular path integrals.}

\myparagraph{2D equivalence}
Let us now consider the 2D color path integral, which is a sum over color 1-cocycles $A$.
Using the chain map defined in Section~\ref{sec:color_cohomology_equivalence}, we can turn $A$ into a pair of cellular 1-cocycles $A_r\coloneqq M_1^r A$ and $A_g\coloneqq M_1^g A$ on the same 2-colex without the coloring.
We will now show that the actions are equivalent, i.e.,
\begin{equation}
\label{eq:2dcolor_action_equivalence}
\frac14 \sum_{p\in \Delta_0} \sgn_p \ovl{A_p} = \frac12 \sum_{f\in \Delta_2} (A_r\cup A_g)(f)\;.
\end{equation}
In fact, we will show the stronger statement that this equation holds locally:
It holds if we restrict the right-hand side to a single $b$-colored face $f$, and the left-hand side to the vertices $p\in f$:
\begin{equation}
\label{eq:2dcolor_action_equivalence_helper}
\frac14 \sum_{p\in f} \sgn_p \ovl{A_p} = \frac12 (M^r A\cup M^g A)(f)\;.
\end{equation}
Note that (1) each vertex $p$ is part of exactly one $b$-colored face, and (2) $M^rA$ is only supported on the $\ovl r$-colored edges, and $M^gA$ only on the $\ovl g$-colored edges, so the right-hand side is only non-zero on the $b$-colored faces where $rg$ intersections can happen.
Thus summing the local equalities Eq.~\eqref{eq:2dcolor_action_equivalence_helper} over all $b$-colored faces yields Eq.~\eqref{eq:2dcolor_action_equivalence}.
To show Eq.~\eqref{eq:2dcolor_action_equivalence_helper}, we notice that both sides are $0$ if the number of vertices $p\in f$ with $A_p=1$ and $\sgn_p=+1$ is equal to that with $A_p=1$ and $\sgn_p=-1$, and $\frac12$ otherwise.
For the left-hand side, this is rather obvious.
For the right-hand side, we notice that every $A_p=1$ vertex gives rise to one $(A_r)_e=1$ edge and one $(A_g)_e=1$ edge.
If $\sgn_p=+1$, then the $(A_r)_e=1$ edge is next to the $(A_g)_e=1$ edge in clockwise direction, otherwise their ordering is counter-clockwise.
Note that we start a red line from every $(A_r)_e=1$ edge center and a green line from every $(A_g)_e=1$ edge center.
If there is one clockwise-ordered pair of $(A_r)_e=1,(A_g)_e=1$ edges and another counter-clockwise pair, then the green and red lines connecting them do not intersect.
More generally, the number of $rg$ intersections is $0\mmod 2$ if the number of clockwise pairs is equal to the number of counter-clockwise pairs, and $1\mmod 2$ otherwise.

\myparagraph{3D cellular path integral}
Let us now go back to the 3D color path integral and show that it is equivalent to the following \emph{twisted cellular path integral}:
\begin{align}
\label{eq:3d_cellular_path_integral}
\begin{split}
Z^{\text{cell}}=&
\sum_{A_r,A_g,A_b\in \zz_2^{\Delta_1}} \prod_{f\in \Delta_2} \delta_{(d^{\text{cell}} A_r)_f=0}\; \delta_{(d^{\text{cell}} A_g)_f=0}\\ &\times \delta_{(d^{\text{cell}} A_b)_f=0}
\prod_{v\in \Delta_3} e^{2\pi i\frac12 (A_r\cup A_g\cup A_b)(v)}\;.
\end{split}
\end{align}
So the path integral is the sum over all triples of cellular 1-cocycles $A_r$, $A_g$, and $A_b$, with an action
\begin{equation}
\label{eq:3d_cellular_action}
S[A_r,A_g,A_b] = \frac12 \sum_{v\in \Delta_3} (A_r\cup A_g\cup A_b)(v)\;.
\end{equation}
Again, $\cup$ denotes is the cup product, and the value of the action on a volume $v$ only depends on the values of $A_r$, $A_g$, and $A_b$ on the edges of $v$.
On a tetrahedron with vertices labeled $0$, $1$, $2$, and $3$, and edges labeled accordingly by $01$, $02$, $03$, $12$, $13$, and $23$, the path-integral weight is given by the type-III group 3-cocycle~\cite{Propitius1995} $\omega\in Z^3(B\zz_2^3,U(1))$:
\begin{equation}
\begin{aligned}
&\begin{multlined}
\omega\Big(\big((A_r)_{01},(A_g)_{01},(A_b)_{01}\big),\big((A_r)_{12},(A_g)_{12},(A_b)_{12}\big),\\\big((A_r)_{23},(A_g)_{23},(A_b)_{23}\big)\Big)
\end{multlined}\\
&= (-1)^{(A_r)_{01} (A_g)_{12} (A_b)_{23}}\;.
\end{aligned}
\end{equation}
Intuitively, $A_r$, $A_g$, and $A_b$ can be pictured as closed-membrane patterns, $A_r\cup A_g$ is a closed-loop pattern formed by the intersection of $A_r$ and $A_g$, and $A_r\cup A_g\cup A_b$ is the set of points where this closed-loop pattern intersects with the closed-membrane pattern $A_b$.
In other words, the action counts the number of triple-intersections between the three closed-membrane patterns $A_r$, $A_g$, and $A_b$.
To make this intuition into a concrete formula, we need to make some local choices.
(1) We need to choose three distinct points on every edge, labeled $r$, $g$, and $b$.
(2) For every configuration of $A_r$ on the edges of each face, choose a way to connect the $r$-points on the $A_r=1$-edges with ``$r$'' lines.
We make the analogous choices for $A_g$ and $A_b$ such that the $r-g$, $r-b$, and $g-b$ intersection points are always distinct.
With these choices, we can compute $(A_r\cup A_g\cup A_b)(v)$ as follows:
The $r$ lines inside every face $f\subset v$ (for the corresponding $A_r$-configuration) together form a closed-loop pattern inside the boundary of $v$, and analogously for $g$ and $b$.
Choose a ``$r$'' closed-membrane pattern inside the volume $v$ whose boundary is the $r$ closed-loop pattern, and analogously for $g$ and $b$.
Now, $(A_r\cup A_g\cup A_b)(v)$ is the $\mmod 2$ number of triple-intersections of the $r$, $g$, and $b$ membranes inside the volume $v$.
Note that the number of triple-intersections is independent of the chosen membranes, since $v$ is a 3-ball with trivial homology.

\myparagraph{3D equivalence}
Having defined the twisted cellular path integral, let us now show that it is equivalent to the color path integral.
We translate the color 1-cocycle $A$ summed in the color path integral into three $\zz_2$-valued cellular 1-cocycles $A^r\coloneqq M^rA$, $A^g\coloneqq M^gA$, and $A^b\coloneqq M^bA$.
Then, we show that the actions are equivalent, i.e.
\begin{equation}
\label{eq:color_action_equivalence}
\frac18 \sum_{p\in \Delta_0} \sgn_p \ovl{A_p} = \frac12 \sum_{v\in \Delta_3} (A^r\cup A^g\cup A^b)(v)\;.
\end{equation}
In fact, we show the stronger condition that this equation holds locally:
The equivalence also holds if we restrict to a single $y$-colored volume $v$ on the right-hand side, and to all vertices of $v$ on the left-hand side:
\begin{equation}
\label{eq:color_action_equivalence_helper}
\frac18 \sum_{p\in v} \sgn_p \ovl{A_p} = \frac12 (M^r A\cup M^g A\cup M^bA)(v)\;.
\end{equation}
Note that (1) every vertex is adjacent to exactly one $y$-colored volume, and (2) the right-hand side is zero on every non-$y$-colored volume, since one of $M^rA$, $M^gA$, or $M^bA$ is not supported there, and so there are no triple-$rgb$ intersections inside these volumes.
Thus, summing Eq.~\eqref{eq:color_action_equivalence_helper} over all $y$-colored volumes yields Eq.~\eqref{eq:color_action_equivalence}.
To show Eq.~\eqref{eq:color_action_equivalence_helper}, we start by noting that it holds if $A_p=0$ for all $p\in v$, since both sides are zero.
Next, we note that the boundary of $v$ is a 2-colex, and $A$ defines a color 1-cocycle on this 2-colex.
Since the 2D color cohomology of a 2-sphere is trivial, any color 1-cocycle $A$ can be obtained from $A=0$ by applying gauge transformations $A'= A+d_{\text{2D,color}}f$ for different faces $f\subset v$.
Now, consider such a gauge transformation for a $by$-colored face $f\subset v$.
Then, we study how both sides of Eq.~\eqref{eq:color_action_equivalence_helper} change if we change $A$ to $A'$:
\begin{align}
\label{eq:color_action_equivalence_proof}
\begin{split}
\frac18 \sum_{p\in v} &\sgn_p \ovl{A_p'} - \frac18 \sum_{p\in v} \sgn_p \ovl{A_p}\\
\overset{\eqref{eq:integer_lift_expansion}}{=}& \frac18 \sum_{p\in v} \sgn_p (\ovl{(df)_p} + 2\ovl{(df)_p} \ovl{A_p})\\
=& \frac18 \sum_{p\in f} \sgn_p + \frac14 \sum_{p\in f} \ovl{A_p}
\overset{\eqref{eq:2dcolor_action_equivalence_helper}}{=}\frac12 (A^r\cup_{\text{2D}} A^g)(f)\\
\overset{*}{=}& \frac12 ((A')^r\cup (A')^g\cup (A')^b)(v) - \frac12 (A^r\cup A^g\cup A^b)(v)\;.
\end{split}
\end{align}
We find that they change by the same amount.
For the equation labeled (*), we have used that $(A')^r=A^r$ and $(A')^g=A^g$, since each $\ovl r$-colored or $\ovl g$-colored edge shares two or no vertices with the $by$-colored face $f$.
$(A')^b$ differs from $A^b$ by all edges that share one vertex with $f$.
Thus, the closed-loop configurations on $v$ for $(A')^b$ and $A^b$ differ by a closed loop surrounding $f$.
Adding a closed loop around $f$ to $A^b$ leads to an additional triple-membrane crossing inside $v$ precisely if $A^r$ and $A^g$ have an odd number of line crossings inside $f$, which is measured by $(A^r\cup_{\text{2D}} A^g)(f)$.
Eq.~\eqref{eq:color_action_equivalence_proof} holds analogously if $f$ has color $yr$ or $yg$.
Since any color 1-cocycle $A$ on the boundary of $v$ can be obtained by a sequence of gauge transformations starting with $A=0$, Eq.~\eqref{eq:color_action_equivalence_helper} holds for all $A$.

\subsection{Flux and charge defects}
\label{sec:flux_and_charge}
In this section, we will equip the color path integral with \emph{flux and charge defects}.
While the color path integral is equal to the twisted color circuit from Section~\ref{sec:simple_example} where all measurement outcomes are post-selected to $+1$, the color path integral with flux and charge defects correspond to the circuit post-selected onto some non-trivial configuration of outcomes.
Adding flux and charge defects is therefore essential for turning the path integral into a circuit of non-post-selected measurements, as well as for discussing decoding as in Section~\ref{sec:decoding}.
Inserting flux and charge defects means modifying the path integral such that it depends on a \emph{flux configuration} $b$, and a \emph{charge configuration} $c$.
$b$ is a color 2-cochain on the 3-colex, and $c$ is a color 1-chain.

\myparagraph{Untwisted case: Definition}
We start by introducing flux and charge defects in the untwisted color path integral in Eq.~\eqref{eq:untwisted_color_path_integral}:
\begin{equation}
\label{eq:untwisted_path_integral_1form}
Z_{\text{untw}}[b,c]=\sum_{A\in \zz_2^{\Delta_0}: d A=b} \prod_{p\in \Delta_0} (-1)^{\frac12A_p c_p}\;.
\end{equation}
The flux and charge configurations modify the path integral in two ways:
First, instead of summing over all color 1-co\emph{cycles} $A$, we sum over all color 1-co\emph{chains} whose boundary is the flux configuration, $dA=b$.
Second, there is an additional path-integral weight of $(-1)^{A_pc_p}$ at every vertex $p\in \Delta_0$, which depends on the charge configuration.%
\footnote{Alternatively, we can interpret $c$ as taking values in the Pontryagin dual $\zz_2^*$.
Then this weight is equal to the evaluation of $c$ on $A$.}
Note that we recover the original path integral for $b=0$ and $c=0$, $Z_{\text{untw}}[0,0]=Z_{\text{untw}}$.

\myparagraph{Untwisted case: Properties}
The flux and charge defects each obey two central properties, which are important for decoding and fault tolerance of the untwisted color circuits.
The first property is that the flux configuration must be a color 2-cocycle, otherwise the untwisted color path integral evaluates to zero:
\begin{equation}
\label{eq:flux_gauss_law}
db\neq 0\Rightarrow Z_{\text{untw}}[b,c]=0\;.
\end{equation}
The simple reason for this is that if $db\neq 0$, then there exists no color 1-cochain $A$ such that $dA=b$ due to Eq.~\eqref{eq:cohomology_axiom}, so the summation in Eq.~\eqref{eq:untwisted_path_integral_1form} is empty.
In fact, this equation holds locally, meaning that if $db\neq 0$ at some location in the 3-colex, then the path integral evaluated on a small neighborhood of this location is zero.
Specifically, if $(db)_{(v,c)}\neq 0$ for some volume $v\in\Delta_3$ and color $c\neq \ccol_v$, then the path integral only on $v$ is zero, where all $A_p$ with $p\in v$ are state-boundary variables.
The analogous property for the charges is that the charge configuration must be a color 1-cycle,
\begin{equation}
\label{eq:charge_gauss_law}
d^T c\neq 0\Rightarrow Z_{\text{untw}}[b,c]=0\;.
\end{equation}
Again, this equation holds locally:
Assume that $(d^Tc)_v\neq 0$ on some volume $v$, and evaluate the path integral on any neighborhood such that all vertices $p\in v$ are not boundary variables, and thus summed over.
Then, the summation over all color 1-cocycles $A$ cancels:
\begin{align}
\begin{split}
Z_{\text{untw}}[b,c] =& \sum_A \prod_{p\in \Delta_0} (-1)^{A_pc_p}\\
=& \frac12 \sum_A \big( \prod_{p\in \Delta_0} (-1)^{A_pc_p} +\prod_{p\in \Delta_0} (-1)^{(A+dv)_pc_p}\big)\\
=& \frac12 \sum_A\prod_{p\in \Delta_0} (-1)^{A_pc_p} \big(1+\prod_{p\in \Delta_0} (-1)^{(dv)_pc_p}\big)\\
=& \frac12 \sum_A\prod_{p\in \Delta_0} (-1)^{A_pc_p} \big(1+\prod_{p\in v} (-1)^{c_p}\big)\\
=& \frac12 \sum_A\prod_{p\in \Delta_0} (-1)^{A_pc_p} \big(1+ (-1)^{(d^Tc)_v}\big)
= 0\;.
\end{split}
\end{align}
The second property is that the path integral is invariant under adding the coboundary of a 1-chain $\beta$ to $b$:
\begin{equation}
\label{eq:flux_invariance}
Z_{\text{untw}}[b+d\beta,c] \propto Z_{\text{untw}}[b,c]\;.
\end{equation}
As usual, this holds locally:
If $\beta$ consists of a single vertex $p'$, then it holds for the path integral on a small neighborhood of $p'$:
\begin{align}
\label{eq:flux_invariance_derivation}
\begin{split}
Z_{\text{untw}}[b+dp',c]
=& \sum_{A:dA=b+dp'} \prod_{p\in \Delta_0} (-1)^{A_pc_p}\\
\overset{*}{=}& \sum_{A:dA=b} \prod_{p\in \Delta_0} (-1)^{(A+p')_pc_p}\\
=& Z_{\text{untw}}[b,c] \prod_{p\in \Delta_0} (-1)^{(p')_pc_p}\\
=& Z_{\text{untw}}[b,c] (-1)^{c_{p'}}\;.
\end{split}
\end{align}
In the equation labeled $*$, we have performed a change of summation variable from $A$ to $A+p'$.
An analogous property holds for the charges:
The path integral is invariant under adding the boundary of a 2-chain $\gamma$ to $c$:
\begin{equation}
\label{eq:charge_invariance}
Z_{\text{untw}}[b,c+d^T\gamma] \propto Z_{\text{untw}}[b,c]\;.
\end{equation}
Again, this equation holds locally:
If $\gamma$ consists of a single face $f$, then the equation holds for $Z$ on a constant-size neighborhood of $f$:
\begin{align}
\label{eq:charge_invariance_derivation}
\begin{split}
&Z_{\text{untw}}[b,c+d^Tf]\\
=& \sum_{A:dA=b} \prod_{p\in \Delta_0} (-1)^{A_p(c+d^Tf)_p}\\
=& \sum_{A:dA=b} \prod_{p\in \Delta_0} (-1)^{A_p c_p} \prod_{p\in \Delta_0} (-1)^{A_p (d^Tf)_p}\\
=& Z_{\text{untw}}[b,c] (-1)^{(dA)_{f}}\\
=& Z_{\text{untw}}[b,c] (-1)^{b_{f}}\;.
\end{split}
\end{align}

\myparagraph{Relation to anyons and 1-form symmetries}
The flux and charge configurations $b$ and $c$ can be viewed as \emph{anomalous 1-form symmetry defects}~\cite{twisted_double_code,Roberts2020,Bulmash2020} in the untwisted color path integral.
Eqs.~\eqref{eq:flux_invariance} and \eqref{eq:charge_invariance} make the defects ``symmetric''.
The ``anomaly'' corresponds to the fact that the equations only hold up to a global prefactor, as shown in Eqs.~\eqref{eq:flux_invariance_derivation} and \eqref{eq:charge_invariance_derivation}.
The symmetries are ``1-form'' because color 2-cocycles and 1-cycles can be mapped to (three copies of) cellular 2-cocycles and 1-cycles, as shown in Section~\ref{sec:color_cohomology_equivalence}.
These can be pictured as a closed-loop pattern inside the 3D spacetime.
In contrast, an ordinary global ``0-form'' symmetry would correspond to domain walls in spacetime, that is, closed-membrane pattern.
Physically, the closed loops in spacetime can be interpreted as anyon worldlines, and the $(-1)^{c_\beta}$ and $(-1)^{b_\gamma}$ factors in Eqs.~\eqref{eq:flux_invariance_derivation} and \eqref{eq:charge_invariance_derivation} correspond to their braiding.

\myparagraph{Defects and Pauli flows}
Let us briefly discuss what charge and flux defects look like if we describe the untwisted color path integral as a $ZX$ tensor network as in Section~\ref{sec:simple_example}, and relate them to the formalism of \emph{Pauli flows} (also called ``Pauli webs'' or ``stabilizer flows'')~\cite{McEven2023, Bombin2023, xyzrubycode}.
If a face $f$ carries a flux defect, $b_f=1$, then this means that the path integral weight $\delta_{(dA)_f=0}$ is replaced by $\delta_{(dA)_f=1}$.
Thus, this modified weight is represented by an $X$-tensor with a $\pi$-phase.
If a vertex $p$ carries a charge defect, $c_p=1$, this corresponds to an additional weight $(-1)^{A_p}$, which can be implemented in the $ZX$ diagram by adding a $\pi$-phase to the according $Z$-tensor.

In the context of $ZX$ tensor networks, flux (charge) defects are related to \emph{Pauli-$X$ (Pauli-$Z$) insertions}.
A Pauli-$X$ (Pauli-$Z$) insertion corresponds to a set of bonds in the tensor network, where we insert a Pauli-$X$ (Pauli-$Z$) operator.
Note that Pauli-$Z$ and $X$ operators are 2-index $Z$ and $X$-tensors with a $\pi$-phase,
\begin{equation}
\label{eq:pauli_operators}
\begin{gathered}
\begin{tikzpicture}
\atoms{delta,charge}{0/}
\draw (0)edge[ind=$a$]++(-90:0.5) (0)edge[ind=$b$]++(90:0.5);
\end{tikzpicture}
=
\delta_{a=b} (-1)^a
=
\bra{b} Z \ket{a}
\;,\\
\begin{tikzpicture}
\atoms{z2,charge}{0/}
\draw (0)edge[ind=$a$]++(-90:0.5) (0)edge[ind=$b$]++(90:0.5);
\end{tikzpicture}
=
\delta_{a+b=1}
=
\bra{b} X \ket{a}
\;,
\end{gathered}
\end{equation}
using the notation in Eq.~\eqref{eq:zx_phases}.
Each Pauli-$X$ (Pauli-$Z$) operator can be combined with the adjacent $X$-tensor ($Z$-tensor),
\begin{equation}
\begin{tikzpicture}
\atoms{delta}{0/lab={t=$\ldots$,p=-90:0.3}}
\atoms{z2}{{1/p={0,1},lab={t=$\ldots$,p=90:0.3}}}
\atoms{z2,charge}{c/p={0,0.5}}
\draw (0)--(c)--(1);
\end{tikzpicture}
=
\begin{tikzpicture}
\atoms{delta}{0/lab={t=$\ldots$,p=-90:0.3}}
\atoms{z2}{{1/p={0,1},lab={t=$\ldots$,p=90:0.3}}}
\atoms{void,charge}{c/p={0,1}}
\draw (0)--(1);
\end{tikzpicture}
\;,\qquad
\begin{tikzpicture}
\atoms{delta}{0/lab={t=$\ldots$,p=-90:0.3}}
\atoms{z2}{{1/p={0,1},lab={t=$\ldots$,p=90:0.3}}}
\atoms{delta,charge}{c/p={0,0.5}}
\draw (0)--(c)--(1);
\end{tikzpicture}
=
\begin{tikzpicture}
\atoms{delta}{0/lab={t=$\ldots$,p=-90:0.3}}
\atoms{z2}{{1/p={0,1},lab={t=$\ldots$,p=90:0.3}}}
\atoms{void,charge}{c/p={0,0}}
\draw (0)--(1);
\end{tikzpicture}
\;.
\end{equation}
This way, each Pauli-$X$ (Pauli-$Z$) insertion is mapped to a flux configuration $b$ (charge configuration $c$).

Pauli insertions can be deformed using the $ZX$ rewrite rules:
We can add a $\pi$-phase to the $X$-tensors at all 6 faces adjacent to a vertex $p$, by
\begin{equation}
\begin{multlined}
\begin{tikzpicture}
\atoms{delta}{0/}
\atoms{z2}{{1/p={0:1},lab={t=$\ldots$,p={0:0.4}}}, {2/p={60:1},lab={t=$\ldots$,p={90:0.3}}}, {3/p={120:1},lab={t=$\ldots$,p={90:0.3}}}, {4/p={-60:1},lab={t=$\ldots$,p={-90:0.4}}}, {5/p={-120:1},lab={t=$\ldots$,p={-90:0.4}}}, {6/p={180:1},lab={t=$\ldots$,p={180:0.4}}}}
\draw (0)--(1) (0)--(2) (0)--(3) (0)--(4) (0)--(5) (0)--(6);
\end{tikzpicture}
=
\begin{tikzpicture}
\atoms{delta}{0/}
\atoms{z2}{{1/p={0:1},lab={t=$\ldots$,p={0:0.4}}}, {2/p={60:1},lab={t=$\ldots$,p={90:0.3}}}, {3/p={120:1},lab={t=$\ldots$,p={90:0.3}}}, {4/p={-60:1},lab={t=$\ldots$,p={-90:0.4}}}, {5/p={-120:1},lab={t=$\ldots$,p={-90:0.4}}}, {6/p={180:1},lab={t=$\ldots$,p={180:0.4}}}}
\atoms{z2,charge}{x1/p={0:0.5}, x2/p={60:0.5}, x3/p={120:0.5}, x4/p={-60:0.5}, x5/p={-120:0.5}, x6/p={180:0.5}}
\draw (0)--(x1)--(1) (0)--(x2)--(2) (0)--(x3)--(3) (0)--(x4)--(4) (0)--(x5)--(5) (0)--(x6)--(6);
\end{tikzpicture}\\
=
\begin{tikzpicture}
\atoms{delta}{0/}
\atoms{z2}{{1/p={0:1},lab={t=$\ldots$,p={0:0.4}}}, {2/p={60:1},lab={t=$\ldots$,p={90:0.3}}}, {3/p={120:1},lab={t=$\ldots$,p={90:0.3}}}, {4/p={-60:1},lab={t=$\ldots$,p={-90:0.4}}}, {5/p={-120:1},lab={t=$\ldots$,p={-90:0.4}}}, {6/p={180:1},lab={t=$\ldots$,p={180:0.4}}}}
\atoms{void,charge}{x1/p={0:1}, x2/p={60:1}, x3/p={120:1}, x4/p={-60:1}, x5/p={-120:1}, x6/p={180:1}}
\draw (0)--(1) (0)--(2) (0)--(3) (0)--(4) (0)--(5) (0)--(6);
\end{tikzpicture}
\;.
\end{multlined}
\end{equation}
This is the tensor-network formulation of the invariance in Eq.~\eqref{eq:flux_invariance} for $\beta=p$.
Analogously, we can add a $\pi$-phase to all the $X$-tensors at all the vertices around a face $f$, corresponding to the invariance in Eq.~\eqref{eq:flux_invariance} for $\gamma=f$.

Now, we can also consider applying a set of invariances that leave $b$ ($c$) unchanged, namely a set of vertices $\beta$ (faces $\gamma$) which is a 1-cocycle with $d\beta=0$ (or a 2-cycle with $d^T\gamma=0$).
Such sets of invariances are an example for \emph{Pauli flows}~\cite{McEven2023, Bombin2023, xyzrubycode}, which have been used extensively for the construction of fault-tolerant circuits.
Pauli flows can be defined for an arbitrary \emph{stabilizer tensor network}, whose tensors are stabilizer states~\cite{xyzrubycode}, and which is the kind of tensor network corresponding to circuits of Clifford gates and post-selected Pauli measurements.
Pauli flows are usually drawn by coloring each bond in the tensor network by the Pauli operator ($\{\idop, X,Y,Z\}$ in general) that is inserted (and removed) at that bond during the application of the invariances.
In our $ZX$ tensor network, the Pauli flows consist either purely of $X$, or of $Z$ operators, which we color in red and blue, respectively.
For every volume $v$, there is one $X$ Pauli flow given by $\beta=dv$.
That is, $\beta$ consists of all vertices of $v$, and the Pauli flow is given by marking all bonds adjacent to the according $Z$-tensors in red.
Similarly, for each volume $v$ and color $c\neq \ccol_v$, there is one $Z$ Pauli flow with $\gamma=d^T(v,c)$.
That is, $\gamma$ consists of all faces of $v$ whose color does not contain $c$, and the Pauli flow is given by marking all bonds adjacent to the according $X$-tensors in blue.
Such local Pauli flows are called \emph{detector flows}.
The following shows an example of the $X$ detector flow, and one of the $Z$ detector flows at a yellow volume in the 3-colex:
\begin{equation}
\includegraphics[width=3cm,valign=c]{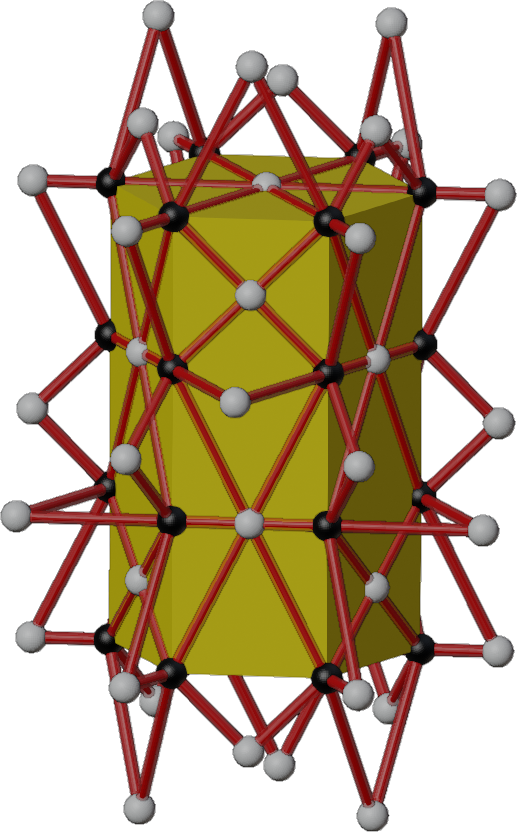}
\quad,
\qquad
\includegraphics[width=2cm,valign=c]{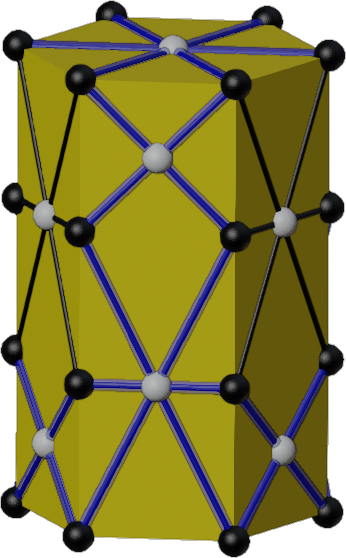}
\;.
\end{equation}
If we insert a Pauli-$Z$ operator at a red edge of a (Pauli-$X$) detector flow, applying the according invariance yields a global factor of $-1$ due to the anti-commutativity of $X$ and $Z$.
The analogous holds when inserting a Pauli-$X$ operator at a blue edge of a (Pauli-$Z$) detector flow.
In this sense, $X$ detector flows (red) detect Pauli-$Z$ (and $Y$) insertions and $Z$ detector flows (blue) detect Pauli-$X$ (and $Y$) insertions.
We find that any local Pauli insertion that is not equivalent to the trivial insertion via $ZX$ rewrite rules, is detectable, and hence, the protocols have a macroscopic fault distance.
Note that this also is guaranteed more generally by the fixed-point property of the path-integral, which manifests in terms of linear constraints on the flux and charge configurations as discussed above.

Pauli flows also describe the logical action of a Clifford protocol~\cite{Bombin2023, xyzrubycode}.
We have established that the local equivalences among Pauli insertions in the $ZX$ diagram, that combine to Pauli flows, agree with 1-cocycle and 2-cycle condition in the color path integral.
This means that the $X$ ($Z$) logical flows are given by non-trivial color 1-cocycles (2-cycles).

\myparagraph{Twisted case: Definition}
Let us now return to the original, twisted, color path integral used to construct twisted color circuits.
For each non-trivial flux configuration $b$ and charge configurations $c$ we define a modified (twisted) color path integral, similar to the untwisted case,
\begin{equation}
\label{eq:path_integral_1form}
Z[b,c]=\sum_{A\in \zz_2^{\Delta_0}: d A=b} \prod_{p\in \Delta_0} e^{2\pi i (\frac18\sgn_p\ovl{A_p}+\frac12A_p c_p)}\;.
\end{equation}
Note that we recover the color path integral in Eq.~\eqref{eq:color_path_integral} for $b,c=0$.
Compared to the untwisted case, the properties in Eqs.~\eqref{eq:flux_gauss_law}, \eqref{eq:charge_gauss_law}, \eqref{eq:flux_invariance} and \eqref{eq:charge_invariance} are modified due to the non-trivial action entering $Z[0,0]$.
Eq.~\eqref{eq:flux_gauss_law} holds unchanged, since it only relies on the fact that $db=ddA=0$.
Eq.~\eqref{eq:flux_invariance} does not hold anymore, and it is unclear whether there is a meaningful way to fix it.
Luckily, we do not need to rely on this property for fault tolerance.
The constraint and invariance in Eqs.~\eqref{eq:charge_gauss_law} and \eqref{eq:charge_invariance} are modified in the twisted case in the presence of non-trivial flux, $b\neq 0$:
While the constraint in Eq.~\eqref{eq:charge_gauss_law} gets replaced by a weaker constraint, the invariance in Eq.~\eqref{eq:charge_invariance} can be extended to a stronger invariance.
Note that $b$ and $c$ correspond to configurations of measurement outcomes in the twisted color circuits, and the modified $c$ constraints correspond to modified detectors generating a modified syndrome.
Using this modified syndrome may be important for effective decoding, as we discuss in Section~\ref{sec:decoding}.
In the following we study these modified constraints, which we refer to as \emph{twisted charge constraints}.

\myparagraph{Twisted charge constraints}
As we saw earlier, the (generating) charge constraints correspond to detectors in the context of quantum error correction.
We will therefore refer to the generating twisted charge constraints as \emph{twisted charge detectors}, and denote their set by $X$.
The twisted charge constraint is of the form
\begin{equation}
\label{eq:weakened_charge_constraint}
K^b d^Tc=k^b\;,
\end{equation}
where the matrix $K^b\in \zz_2^{X\times \Delta_3}$ and the vector $k^b\in \zz_2^X$ are defined as follows.
First, consider the symmetric matrix $M^b\in \zz_2^{\Delta_3\times \Delta_3}$, defined by
\begin{equation}
\label{eq:gauge_variance_2form}
M^b_{vv'} = \sum_{f\in \Delta_2, f\in v, v'} b_f\;.
\end{equation}
That is, if $v$ and $v'$ do not share a face, then $M^b_{vv'}=0$.
If they share a face $f$, then $M^b_{vv'}$ equals the value of $b$ on $f$.
For $v=v'$, $M^b_{vv}$ is equal to the sum of $b$ over all faces of $v$.

Given $M^b$, $K^b$ can be chosen to be any maximum-rank matrix such that
\begin{equation}
\label{eq:kmatrix_definition}
(K^bM^b)_{xv} = \sum_{v'\in \Delta_3} K^b_{xv'} M^b_{v'v} = 0\quad\forall x\in X,v\in \Delta_3\;.
\end{equation}
In other words, we choose $K^b$ such that the vectors $\{K^b_x\}_{x\in X}$ form a complete basis for the kernel of $M^b$.
Finally, in order to define $k^b$, we need to choose a fixed $A_0$ such that $dA_0=b$.
Then, we set
\begin{align}
\label{eq:modified_detector_charge}
\begin{split}
k^b_x \coloneqq& \frac12 \sum_{v\in \Delta_3} \sum_{p\in \Delta_0, p\in v} (A_0)_p K^b_{xv}\\
&+ \sum_{v<v'} M^b_{vv'} K^b_{xv'} K^b_{xv''}
\;.
\end{split}
\end{align}
The expression above takes integer values, as we show below, and is interpreted $\mmod 2$.

Before we verify that the twisted charge constraint in Eq.~\eqref{eq:weakened_charge_constraint} indeed holds, let us discuss how it differs from the charge constraint in the untwisted color path integral.
First, we find that for every volume $v$ such that $b_f=0$ on all faces $f\in v$, we have $M^b_{vv'}=0\forall v'$.
Thus, each such volume $v$ also gives rise to a twisted detector $x$ with $K^b_{xv}=\delta_{x=v}$.
More generally, consider a connected component $V\subset \Delta_3$, where two volumes are considered connected if they share a face $f$ with $b_f=1$.
Then, we have $M^b_{vv'}=0\;\forall v\in V,v'\neq V$.
Thus, we can choose a basis of twisted detectors, such that twisted detector is supported only on one connected component.
On a high level, the twisted charge constraint can be summarized as follows:
The volumes adjacent to a non-trivial flux face do not yield twisted charge detectors individually as in the untwisted case, but instead each connected component of such volumes together yields a small number of twisted charge detectors.

After stating the twisted charge constraints, let us now show that they hold.
We begin by defining the \emph{gauge variance} of the action $S$,
\begin{equation}
\label{eq:gauge_variance_definition}
\Delta S_\alpha[A] = S[A+d\alpha]-S[A]\;,
\end{equation}
for any color 1-cocycle $A$ and 0-cochain $\alpha$.
We find for any pair of volumes $v$ and $v'$,
\begin{widetext}
\begin{align}
\label{eq:quadratic_refinement}
\begin{split}
\Delta S_{v+v'}[A]-\Delta S_{v}[A]-\Delta S_{v'}[A]
\overset{\eqref{eq:gauge_variance_definition}}{=}&S[A+dv+dv']- S[A+dv] - S[A+dv'] + S[A]\\
=& \frac18 \sum_p \sgn_p \big(\ovl{(A+dv+dv')_p} - \ovl{(A+dv)_p} - \ovl{(A+dv')_p} + \ovl{A_p} \big)\\
\overset{\eqref{eq:integer_lift_expansion}}{=}& \frac18 \sum_p \sgn_p \big(\ovl{(dv')_p} - 2\ovl{(A+dv)_p}\ovl{dv'}_p - \ovl{(dv')_p} + 2 \ovl{A_p}\ovl{(dv')_p} \big)\\
\overset{\eqref{eq:integer_lift_expansion}}{=}& \frac14 \sum_p \sgn_p \big(-(\ovl{A_p}+ \ovl{(dv)_p}-2\ovl{A_p}\ovl{(dv)_p}) \ovl{(dv')_p} +\ovl{A_p}\ovl{(dv')_p} \big)\\
=& \sum_p \sgn_p \big(-\frac14\ovl{(dv)_p}\ovl{(dv')_p} + \frac12 \ovl{A_p}\ovl{(dv)_p}\ovl{(dv')_p}\big)\\
\overset{*}{=}& \frac12 \sum_p \ovl{A_p}\ovl{(dv)_p}\ovl{(dv')_p}
= \frac12 \sum_{f: f\in v,f\in v'} \sum_{p\in f} A_p\\
=& \frac12 \sum_{f: f\in v,f\in v'} b_f = \frac12 M^b_{vv'}\;.
\end{split}
\end{align}
\end{widetext}
For the equality labeled ($*$), we have used that two volumes $v$ and $v'$ share either no vertices $p$, or they share a face and all adjacent vertices.
Since each face has equal number of vertices with $\sgn_p=1$ and $\sgn_p=-1$, the first term vanishes.
In words, the equation says that $\Delta S_\alpha[A]$, as a function of $\alpha$ for a fixed $A$, is a \emph{quadratic refinement}~\cite{Quadratic_refinement_nlab} of the $\rr/\zz$-valued bi-linear form defined by the matrix $M^b$.
Using this, we find
\begin{align}
\label{eq:twisted_constraint_offset_gaugevariance}
\begin{split}
\Delta S_{K^b_x}[A_0]
\overset{\eqref{eq:quadratic_refinement}}{=}& \sum_v K^b_{xv}\Delta S_{v}[A_0]
+ \frac12 \sum_{v<v'} M^b_{vv'} K^b_{xv} K^b_{xv'}\\
\overset{\eqref{eq:modified_detector_charge}}{=}& \frac12 k^b_x\;,
\end{split}
\end{align}
using the notation $K^b_x\coloneqq \sum_v K^b_{xv}\cdot v$ for the $x$th row of the matrix $K^b$.
We can now use Eq.~\eqref{eq:quadratic_refinement} to show that (1) $k^b$ is integral, and (2) it can only depend on the cohomology class of $A_0$.
For (1), we observe that
\begin{align}
\begin{split}
-2\Delta S_{K^b_x}[A_0]
=& \Delta S_0[A_0]-\Delta S_{K^b_x}[A_0]-\Delta S_{K^b_x}[A_0]\\
\overset{\eqref{eq:quadratic_refinement}}{=}&\frac12 (K^b_x)^T M^b K^b_x = 0\;,
\end{split}
\end{align}
which implies that $k^b_x=-2\frac12 k^b_x=0\mmod 1$, recalling that both equations above are in $\rr/\zz$.
For (2), we calculate
\begin{align}
\begin{split}
\Delta S_{K^b_x}[A_0+d\alpha]
\overset{\eqref{eq:gauge_variance_definition}}{=}& \Delta S_{K^b_x+\alpha}[A_0] - \Delta S_{K^b_x}[A_0] \\ &- \Delta S_{\alpha}[A_0]
+ \Delta S_{K^b_x}[A_0]\\
\overset{\eqref{eq:quadratic_refinement}}{=}& \frac12 \alpha^T M^b K^b_x + \Delta S_{K^b_x}[A_0]\\
\overset{\eqref{eq:kmatrix_definition}}{=}&\Delta S_{K^b_x}[A_0]\;.
\end{split}
\end{align}
To show the twisted charge constraint in Eq.~\eqref{eq:weakened_charge_constraint}, consider a patch of the spacetime 3-colex such that $K^b_x$ is supported on its interior.
If $(K^bd^Tc)_x\neq k^b_x$, then we find
\begin{align}
\begin{split}
Z[b,c] =& \sum_A e^{2\pi i(S[A]+\frac12 \sum_p A_p c_p)}\\
=&\frac12 \big(\sum_A e^{2\pi i(S[A]+\frac12 \sum_p A_p c_p)}\\ &+ e^{2\pi i(S[A+dK^b_x]+\frac12 \sum_p (A+dK^b_x)_p c_p)}\big)\\
\overset{\eqref{eq:gauge_variance_definition}}{=}&\frac12 \sum_A e^{2\pi iS[A]} (1+ e^{2\pi i(\Delta S_{K^b_x} + \frac12 \sum_p (dK^b_x)_p c_p)})\\
\overset{\eqref{eq:twisted_constraint_offset_gaugevariance}}{=}&\frac12 \sum_A e^{2\pi iS[A]} (1+ (-1)^{k^b_x+(K^bd^T c)_x})\\
=&\frac12 \sum_A e^{2\pi iS[A]} (1-1)=0\;.
\end{split}
\end{align}
In addition to the weakened (twisted) constraints, the invariance in Eq.~\eqref{eq:charge_invariance} can be extended:
For any locally supported 1-chain $\Gamma$ such that $K^bd^T\Gamma=0$, we have
\begin{equation}
Z[b,c+\Gamma] \propto Z[b,c]\;.
\end{equation}
This includes 1-chains of the form $\Gamma=d^T\gamma$, which form the invariance in the untwisted case.

\myparagraph{Physical interpretation of twisted constraints}
Let us briefly discuss the physics interpretation of our findings.
The twisted constraint for the charge configuration in the (twisted) color path integral mirrors the fusion rules of the anyon theory of the underlying non-Abelian phase.
Recall that in Section~\ref{sec:path_integral_equivalence}, we showed the phase equivalence between the color path integral and the type-III twisted $\zz_2^{3}$ Dijkgraaf-Witten state sum, whose anyon model is known~\cite{Propitius1995}.
Using the cohomology mapping in Section~\ref{sec:color_cohomology_equivalence}, we can generate special charge or flux configurations that look like isolated strings, which we can interpret as \emph{anyon worldlines}.
Each of the three generating flux strings is associated with two non-Abelian anyons of quantum dimension $2$.%
\footnote{In fact, a flux string corresponds to the direct sum of both non-Abelian flux anyons.}
The fact that the flux constraint in Eq.~\eqref{eq:flux_gauss_law} is unchanged mirrors the fact that the fusion of two non-Abelian flux anyons of the same type always yields an anyon with trivial flux.
However, due to their non-Abelian nature, there is only a topological invariance for isolated flux worldlines, but no cohomological invariance for some (possibly dense) network of fluxes as for the untwisted case in Eq.~\eqref{eq:flux_invariance}.%
\footnote{If the fluxes in the model would correspond to Abelian worldlines, then the invariance in Eq.~\eqref{eq:flux_invariance} can be rescued by adding further terms to the action as it was done in Ref.~\cite{twisted_double_code}.}
The charge configuration still corresponds to a pattern of Abelian anyon worldlines in the path integral.
The fact that the charge constraint $d^Tc=0$ is weakened and does not need to hold on volumes with non-trivial flux faces mirrors the fact that the fusion of two non-Abelian fluxes can result in Abelian charges.
Or, in other words, the Abelian charge worldlines can terminate on the flux worldlines.

\myparagraph{Charges and fluxes in 3D color code}
The color path integral is closely related to the 3D color code and its transversal $T$ gate~\cite{Bombin2006a}, see Section~\ref{sec:simple_example}.
Let us briefly discuss to what kind of objects the charge and flux configurations and their constraints correspond in the 3D color code.
The flux configuration $b$ corresponds to a $\pm 1$ configuration of measurement outcomes of the $Z$ stabilizers $S_f^Z$, or in other words, a $Z$ syndrome or $Z$ excitation of the 3D color code.
The charge configuration $c$, on the other hand, does not correspond to the syndrome for the $X$ stabilizers $S_v^X$.
Rather, the operator acting as $Z^{c_p}$ at every 3-colex vertex $p$ maps the code state of the 3D color code to one with $X$ syndrome given by $d^Tc$.

Let us next study what happens if we apply the transversal $T$ gate $\overline{T} = \prod_{p\in\Delta_0} T_p^{\sgn_p}$ to a color-code state with $Z$ syndrome $b$ and $X$ syndrome $d^Tc$~\cite{Bombin2018a}.
The action of the transversal $T$ gate on the $X$ syndrome is trivial, since the $Z^{c_p}$ operator that creates it commutes with $T$.
It thus suffices to study the action on a state $\ket b$ with a definite flux syndrome $b$ but trivial charge syndrome.
After applying the $\ovl T$ gate, $\ovl T\ket b$ is an eigenstate of the conjugated $S^X$ stabilizers
\begin{align}
\widetilde{S}_v^X\coloneqq \ovl T S_v^X \ovl T^\dagger = \prod_{p\in v} S_p^{\sgn_p} X_p,
\end{align}
where $S = i^{\ketbra{1}}$ denotes the single-qubit Clifford phase gate.
In order to determine the $X$ syndrome of $\ovl T\ket b$, we measure the original stabilizers $S_v^X$.
If $S_v^X$ and $\widetilde S_v^X$ commute, then the measurement outcome will be deterministic, and if they anti-commute it will be fully random.
Computing the group commutator $\comm{A}{B}_G \coloneqq ABA^{-1}B^{-1}$ yields
\begin{align}
\label{eq:group_comm_Cliffordstab}
\comm{S_v^X}{\widetilde{S}_{v'}^X}_G = \prod_{f: f\in v, f\in v'}S_f^Z.
\end{align}
Accordingly, we find
\begin{align}
\begin{split}
    \bra{b}\prod_{f: f\in v, f\in v'}S_f^Z\ket{b} =& (-1)^{\sum_{f: f\in v, f\in v'}b_f}\\
    =& (-1)^{M_{vv'}^b}\;,
\end{split}
\end{align}
recovering the quadratic form from Eq.~\eqref{eq:quadratic_refinement} that describes the gauge variance.
$S_v^X$ will yield a deterministic outcome on $\ovl T\ket b$ exactly if it commutes with all $\widetilde S_{v'}^X$, which is the case if $M^b_{vv'}=0$ for all $v'$.
More generally, a product
\begin{equation}
\prod_{v\in \Delta_3} (S_v^X)^{\alpha_v}
\end{equation}
of generating $S_v^X$ stabilizers yields a deterministic outcome if the vector $\alpha\in \zz^{\Delta_3}$ is in the kernel of $M^b$.
So the subgroup of the original $X$-stabilizer group that yields deterministic outcomes is generated by
\begin{equation}
\left\{\prod_{v\in \Delta_3} K^b_{xv} S_v^X \right\}_{x\in X}\;,
\end{equation}
using the matrix $K^b$ from Eq.~\eqref{eq:kmatrix_definition}.
The deterministic outcome of the measurement does not have to be $+1$.
Instead, it is given by $k_x^b$ from Eq.~\eqref{eq:modified_detector_charge}:
\begin{equation}
\prod_{v\in \Delta_3} K^b_{xv} S_v^X \ovl T\ket b = (-1)^{k^b_x} \ovl T\ket b\;.
\end{equation}
Concretely, such deterministic $-1$ outcomes can occur for configurations $b$ consisting of linked loops, a phenomenon which was called \emph{linking charge} in Ref.~\cite{Bombin2018a}.

\subsection{Boundaries}
\label{sec:boundaries}
In this section, we revisit the boundaries of the path integral in Section~\ref{sec:boundaries_easy} in more detail, and describe other constructions for topological boundaries.
In order to define a boundary condition for the color path integral, we need to (1) extend color cohomology to 3-colexes with boundaries, and (2) extend the action in Eq.~\eqref{eq:twisted_action} to the boundaries.
(1) and (2) need to satisfy two crucial conditions:
(1) the extension of color cohomology to the boundary needs to be locally exact, and (2) the extension of the action to the boundary needs to be gauge invariant.
We will here only state the different boundaries, and derive them explicitly in Appendix~\ref{sec:boundaries_derivation}.
There, we also discuss examples of boundaries in 2D and boundaries that do not fulfill local exactness or gauge invariance, and we show that we have found all possible boundaries using their equivalence to boundaries of a cellular twisted path integral.

Geometrically, we will define the boundaries on an arbitrary 2D surface formed by faces of the 3-colex.
For any such surface, we introduce the following notation:
\begin{itemize}
\item $\Delta_i^{\mbd,c}$ for $i=0,1,2$ denotes the set of $i$-cells of the 3-colex that are part of the boundary and have coloring $c$.
\item $\Delta_3^{\bdin,c}\subset \Delta_3$ denotes the set of volumes that are adjacent to a boundary face.
\item $\Delta_3^{\bdout,c}\not\subset\Delta_3$ denotes the set of ``phantom'' volumes that are not part of the 3-colex, but adjacent to a boundary face on the ``outside''.
\item $\Delta_2^{\bdout}\not\subset\Delta_2$ denotes the set of ``phantom'' faces that are adjacent to one or more boundary edges in $\Delta_1^{\mbd}$ but outside the boundary.
\item $\Delta_0^{\bdout,c}\subset \Delta_0^{\mbd}$ ($\Delta_0^{\bdin,c}\subset \Delta_0^{\mbd}$) denotes the set of boundary vertices that are adjacent to a single volume in $\Delta_2^{\bdout,c}$ ($\Delta_2^{\bdin,c}$), which is of color $c$.
\end{itemize}

\myparagraph{All-rough boundary}
The first boundary type is the \emph{all-rough boundary}.
The generator sets of the boundary cohomology are
\begin{equation}
\label{eq:colorcohomology_allrough_boundary}
\begin{aligned}
X_0 &\coloneqq \Delta_3\;,&
X_1&\coloneqq\Delta_0\;,\\
X_2&\coloneqq\Delta_2+\Delta_2^{\bdout}\;,&
X_3&\coloneqq\bigsqcup_{c\in\col_1} (\Delta_3^c+\Delta_3^{\bdout,c} )\times \ovl c\;,\\
\end{aligned}
\end{equation}
and the boundary maps $d_0$, $d_1$, $d_2$ are simple restrictions of these in the bulk.
The boundary action looks the same as the bulk one in Eq.~\eqref{eq:twisted_action}, where the summation runs over all vertices $p\in \Delta_0$ including those on the boundary.
The $ZX$ tensor network for this boundary is discussed later in Section~\ref{sec:circuits}, where it is depicted in Fig.~\ref{fig:other_boundaries}.

\myparagraph{Color boundary}
The second family of boundaries are \emph{color boundaries}.
There are four such boundaries, one for every color $c$, where we set $c=r$ (red) without loss of generality.
The cohomology theory is
\begin{equation}
\label{eq:colorcohomology_color_boundary}
\begin{aligned}
X_0&\coloneqq \Delta_3-\Delta_3^{\bdin,r} + \Delta_3^{\bdout,g} + \Delta_3^{\bdout,b} + \Delta_3^{\bdout,y}\;,\\
X_1&\coloneqq\Delta_0\;,\\
X_2&\coloneqq\Delta_2-\Delta_2^{\mbd,gb}-\Delta_2^{\mbd,gy}-\Delta_2^{\mbd,by}+\Delta_1^{\mbd,\ovl r}\;,\\
X_3&\coloneqq\bigsqcup_{c\in\col_1} \Delta_3^c\times \ovl c% + \Delta_3^{\bdout,g} + \Delta_3^{\bdout,b} + \Delta_3^{\bdout,y}
\;.
\end{aligned}
\end{equation}
$d_1^T$ maps an edge $e\in \Delta_1^{\mbd,\ovl r}$ to both of its endpoints.
$d_2$ acts on an edge $e\in \Delta_1^{\mbd,\ovl r}$ as follows:
First, we map $e$ to all adjacent faces in $\Delta_2^{\mbd,gb}$, $\Delta_2^{\mbd,gy}$, and $\Delta_2^{\mbd,by}$.
Then, we act on these faces with the bulk coboundary map $d_2$ from Eq.~\eqref{eq:color_cohomology_boundaries}.
Otherwise, the coboundary maps are restrictions of the bulk ones in Eq.~\eqref{eq:color_cohomology_boundaries}.
The boundary action extends from Eq.~\eqref{eq:twisted_action} to the boundary as follows:
\begin{equation}
\label{eq:color_boundary_action}
S^{\mbd,r}[A] = \frac18 \sum_{p\in \Delta_0-\Delta_0^{\bdin,r}} \sgn_p \ovl{A_p}\;.
\end{equation}
This boundary is the one used in Section~\ref{sec:simple_example}.

\myparagraph{Double-color boundary}
The third family of boundaries are \emph{double-color boundaries}.
There are 6 such boundaries, one for each color pair $\kappa$.
Without loss of generality, we choose $\kappa=rg$.
The cohomology theory is
\begin{align}
\label{eq:colorcohomology_doublecolor_boundary}
\begin{split}
X_0\coloneqq& \Delta_3-\Delta_3^{\bdin,r} - \Delta_3^{\bdin,g} + \Delta_3^{\bdout,b} + \Delta_3^{\bdout,y}\;,\\
X_1\coloneqq&\Delta_0\;,\\
X_2\coloneqq&\Delta_2-\Delta_2^{\mbd}+\Delta_2^{\mbd,rg}+\Delta_1^{\mbd,\ovl r}+\Delta_1^{\mbd,\ovl g}\;,\\
X_3\coloneqq&\bigsqcup_{c\in\col_1} \Delta_3^c\times \ovl c + \Delta_2^{\mbd,by}% \Delta_3^{\bdout,r}\times \{b,g,y\}\\&+ \Delta_3^{\bdout,g}\times \{r,b,y\}
\;.
\end{split}
\end{align}
$d_2^T$ maps a boundary face $f\in\Delta_2^{\mbd,by}$ to all of its edges.
Apart from this, the coboundary maps are defined similarly to Eq.~\eqref{eq:colorcohomology_color_boundary}.
The boundary action looks the same as in Eq.~\eqref{eq:color_boundary_action}.
The according $ZX$ tensor network is shown in Section~\ref{sec:circuits} in Fig.~\ref{fig:other_boundaries}.

In total, there are 11 distinct boundaries of the 3D twisted color path integral corresponding to three families, namely 1 all-rough, 4 color, and 6 double-color boundaries.

\section{Decoding and fault tolerance}
\label{sec:decoding}
In this section, we discuss in more detail how to decode the twisted color circuit from Section~\ref{sec:simple_example}.
We describe decoding in five different steps:
In Section~\ref{sec:decoding_without_t} we explain how to decode the untwisted color circuit without $T$ gates.
In Section~\ref{sec:x_decoding}, we discuss how the presence of $T$ gates in the twisted color circuit makes it necessary decode and apply $X$ corrections while running the circuit, and how this can be achieved using a \emph{just-in-time} decoder.
While $Z$ errors could be corrected in the same way as in the untwisted circuit, we show in Section~\ref{charge_decoding} how the output of just-in-time $X$ decoding can be used to improve the $Z$ decoding of the twisted circuits.
In Section~\ref{sec:circuit_decoding}, we discuss global aspects of the decoding procedure for a logical circuit using our non-Clifford logical elements.
We phrase the subroutines of the decoding problem in all these sections as hypergraph matching problems, for which we propose concrete (efficient) algorithms in Section~\ref{sec:hypergraph_matching}.

\subsection{Decoding of the untwisted color circuits}
\label{sec:decoding_without_t}
Let us warm up by discussing how to perform decoding and corrections in the \emph{untwisted} color circuits, where we omit the physical $T$ gates.
Similarly to other circuits for Abelian topological phases, 2D toric code~\cite{Dennis2001}, color code~\cite{Sahay2021, Gidney2023, Kubica2019}, or honeycomb Floquet code~\cite{Hastings2021,Paetznick2022}, it suffices to record all measurement outcomes while running the circuit, and decode at a later time.
This is the main difference to the \emph{twisted} color circuits which we discuss in the following sections, where we need to perform decoding and corrections while executing the circuit.

\myparagraph{Decoding in 3+0D}
Let us recall how decoding works for the 3+0D measurement-based protocol discussed in Section~\ref{sec:easy_path_integral}, without the $T$ gates.
For concreteness, we assume a 3-colex of the topology $S_1\times S_1\times [0,1]$, where the task is to reliably teleport the logical information from the input boundary $S_1\times S_1\times 0$ to the output boundary $S_1\times S_1\times 1$.
After running the circuit and recording all measurement results, the task is to apply corrections to the output state such that it represents the logical state of the input.

\myparagraph{Measurements in 3+0D}
We start by translating the measurement outputs into a flux configuration $b$ and charge configuration $c$ in the color path integral, as discussed in Section~\ref{sec:flux_and_charge}:
We recall that, if all $X$ and $Z$ measurement outcomes are $+1$, the post-selected circuit is equal to the untwisted color path integral as discussed in Section~\ref{sec:easy_path_integral}.
If a $Z$ measurement outcome at a face is $-1$, then this corresponds to adding a $\pi$ phase to the $X$-tensor at that face, for example, for a 6-gon face:
\begin{equation}
\label{eq:z6_measurement_minus}
\begin{gathered}
\begin{tikzpicture}
\atoms{z2}{0/charge}
\atoms{delta}{1/p={-150:0.8}, 2/p={180:1.5}, 3/p={150:1.2},4/p={-30:0.8},5/p={0:1.5},6/p={30:1.2}}
\draw (0)--(1) (0)--(2) (0)--(3) (0)--(4) (0)--(5) (0)--(6);
\draw (1)edge[ind=$a$]++(-90:0.5) (1)edge[ind=$a'$]++(90:1) (2)edge[ind=$b$]++(-90:0.5) (2)edge[ind=$b'$]++(90:0.5) (3)edge[ind=$c$]++(-90:0.8) (3)edge[ind=$c'$]++(90:0.5) (4)edge[ind=$d$]++(-90:0.5) (4)edge[ind=$d'$]++(90:1) (5)edge[ind=$e$]++(-90:0.5) (5)edge[ind=$e'$]++(90:0.5) (6)edge[ind=$f$]++(-90:0.8) (6)edge[ind=$f'$]++(90:0.5);
\end{tikzpicture}\\
=
\delta_{a=a'} \delta_{b=b'} \delta_{c=c'} \delta_{d=d'} \delta_{e=e'} \delta_{f=f'} \delta_{a+b+c+d+e+b=1}\\
=
\bra{a',b',c',d',e',f'} \frac12(1-Z^{\otimes 6}) \ket{a,b,c,d,e,f}
\;,
\end{gathered}
\end{equation}
using notation from Eq.~\eqref{eq:zx_phases}.
As discussed in Section~\ref{sec:flux_and_charge}, this $\pi$ phase corresponds to a non-trivial flux ($b_f=1$) on the face $f$.
Similarly, a $-1$ outcome of a destructive $X$ measurement corresponds to adding a $\pi$ phase to the according $Z$-tensor, which corresponds to a non-trivial charge ($c_p=1$) on the vertex $p$:
\begin{equation}
\label{eq:minus_projection}
\begin{tikzpicture}
\atoms{delta}{0/charge}
\draw (0)edge[ind=$a$]++(-90:0.5);
\end{tikzpicture}
= (-1)^a = \sqrt2 \braket{-}{a}
\;.
\end{equation}
All in all, the $Z$ measurement outcomes (relabeled as $\{1,-1\}\rightarrow\{0,1\}\simeq \zz_2$) form a flux configuration $b$ which is a color 2-chain, and the $X$ measurement outcomes form a charge configuration $c$ which is a color 1-chain.

\myparagraph{Errors in 3+0D}
Let us next look at Pauli-$X$ and $Z$ errors in the circuit.
Pauli-$X$ and $Z$ operators correspond to 2-index $X$ or $Z$-tensors with a $\pi$ phase as shown in Eq.~\eqref{eq:pauli_operators}.
The circuit with Pauli-$X$ and $Z$ errors corresponds to the $ZX$ tensor network flux and charge configurations $b_E$ and $c_E$, similar to the Pauli insertions discussed in Section~\ref{sec:flux_and_charge}.
Concretely, using the relation between the protocol and the path integral in Eq.~\eqref{eq:30d_path_integral_translation}, a $Z$ error at a qubit at any point in the protocol adds to the value $c_E$ on the corresponding vertex,
\begin{equation}
\begin{tikzpicture}
\atoms{delta}{0/, 1/p={0,0.4}, 2/p={0,0.8}, 3/p={0,1.2}, 4/p={0,1.6}, {5/p={0,2},charge}, 6/p={0,2.4}, 7/p={0,2.8}, 8/p={0,3.2}}
\draw (0)--(1)--(2)--(3)--(4)--(5)--(6)--(7)--(8);
\draw (1)--++(0:1) (3)--++(0:1) (6)--++(0:1) (2)--++(180:1) (4)--++(180:1) (7)--++(180:1);
\end{tikzpicture}
=
\begin{tikzpicture}
\atoms{delta}{0/charge}
\draw (0)--++(-30:1) (0)--++(0:1) (0)--++(30:1) (0)--++(-150:1) (0)--++(180:1) (0)--++(150:1);
\end{tikzpicture}
\;.
\end{equation}
As another example, an $X$ error occurring after the qubit has been involved in 4 of the 6 face measurements adds to the value of $b_E$ at the two remaining faces,
\begin{equation}
\begin{tikzpicture}
\atoms{delta}{0/, 1/p={0,0.4}, 2/p={0,0.8}, 3/p={0,1.2}, 4/p={0,1.6}, 6/p={0,2.4}, 7/p={0,2.8}, 8/p={0,3.2}}
\atoms{z2,lab={t=$\ldots$,p={-90:0.25}}}{1x/p={-1,0.4}, 2x/p={1,0.8}, 3x/p={-1,1.2}, 4x/p={1,1.6}, 6x/p={-1,2.4}, 7x/p={1,2.8}}
\atoms{z2,charge}{{5x/p={0,2}}}
\draw (0)--(1)--(2)--(3)--(4)--(5x)--(6)--(7)--(8);
\draw (1)--(1x) (2)--(2x) (3)--(3x) (4)--(4x) (6)--(6x) (7)--(7x);
\end{tikzpicture}
=
\begin{tikzpicture}
\atoms{delta}{0/}
\atoms{z2,lab={t=$\ldots$,p={-90:0.25}}}{1/p={-30:1}, 2/p={0:1}, 3/p={30:1}, 4/p={-150:1}, 5/p={180:1}, 6/p={150:1}}
\atoms{z2,charge}{3/p={30:1}, 6/p={150:1}}
\draw (0)--(1) (0)--(2) (0)--(3) (0)--(4) (0)--(5) (0)--(6);
\end{tikzpicture}
\;.
\end{equation}
Equivalently, the $X$ error contributes to the value of $b_E$ on the four other faces.
All in all, an $X$ ($Z$) error configuration can be represented as a $\zz_2$-valued vector $b_{\text{err}}$ ($c_{\text{err}}$) with one entry for every location in the circuit where an error can occur.
The redistribution of $\pi$-phases above defines a local $\zz_2$-linear map which takes $b_{\text{err}}$ ($c_{\text{err}}$) to a color 2-cochain $b_E$ (1-chain $c_E$).
This map itself is not unique, as we can locally change $b_E'=b_E+d\beta$ ($c_E'=c_E+d^T\gamma$) via the invariance in Eq.~\eqref{eq:flux_invariance} (Eq.~\eqref{eq:charge_invariance}).
However, the syndrome $db_E$ ($d^Tc_E$) is uniquely determined by $b_{\text{err}}$ ($c_{\text{err}}$), due to $dd=0$.

\myparagraph{Decoding algorithm in 3+0D}
The output state of the protocol at $S_1\times S_1\times 1$ is not a color code state, but a state with $X$ ($Z$) syndrome given by the termination of $b+b_E$ ($c+c_E$) at the state boundary $S_1\times S_1\times 1$.
So we need to apply a correction that maps us back into the code space.
This correction is equal to inserting another flux (charge) configuration $b_c$ ($c_c$) which is only supported near $S_1\times S_1\times 1$.
The whole circuit with measurements, errors, and corrections, forms a path integral with a total flux (charge) configuration given by $b+b_E+b_c$ ($c+c_E+c_c$).
Due to the constraints in Eq.~\eqref{eq:flux_gauss_law} (Eq.~\eqref{eq:charge_gauss_law}), we know that this effective flux configuration $b+b_E+b_c$ (charge configuration $c+c_E+c_c$) is a color 2-cocycle (1-cycle), otherwise measuring $b$ ($c$) has zero probability.
Due to the invariance in Eq.~\eqref{eq:flux_invariance} (Eq.~\eqref{eq:charge_invariance}), a logical error occurs only if the effective flux (charge) configuration is (co)homologically non-trivial.
In other words, no logical fault occurs if we choose $b_c$ ($c_c$) to be cohomologically equivalent to $b+b_E$ ($c+c_E$).
Unfortunately, we do not know $b_E$ ($c_E$), but only its syndrome $db_E=db$ ($d^Tc_E=d^Tc$).
So the best we can do is to replace it with an estimate $\widetilde b$ ($\widetilde c$) that also satisfies
\begin{equation}
\label{eq:fluxcharge_fix}
d\widetilde b = db\;,\qquad (d^T\widetilde c=d^Tc)\;.
\end{equation}
It is the task of the decoder to find this estimate, and it succeeds if $\widetilde b$ ($\widetilde c$) is cohomologically equal to $b_E$ ($c_E$).
A strategy that generally works is to pick the $\widetilde b$ ($\widetilde c$) fulfilling Eq.~\eqref{eq:fluxcharge_fix} with the smallest weight, or at least as small as we can find within runtime limitations.
Here, the weight is the number of faces $f$ (vertices $v$) with $\widetilde b_f=1$ ($\widetilde c_p=1$).
We will refer to the problem of low-weight finding solutions to Eq.~\eqref{eq:fluxcharge_fix} as \emph{hypergraph matching}, and provide concrete algorithms in Section~\ref{sec:hypergraph_matching}.
Very roughly, the reason why this works is that for low enough error rate, $b_E$ ($c_E$) has low weight with high probability, thus if we choose $\widetilde b$ with low weight it will have the same (co)homology class than $b_E$ ($c_E$) with high probability.
Finally, we note that the fault-tolerance argument can also be made for any arbitrary $p$-bounded local circuit-level noise model, without expressing it in terms of Pauli errors, which we do in Appendix~\ref{app:local_fault_tolerance}.

\myparagraph{String-pattern picture}
After applying the mapping in Section~\ref{sec:color_cohomology_equivalence}, coarse-graining and abstracting away the lattice, both $b$ and $c$ can be imagined as string patterns.
More precisely, each of $b$ and $c$ corresponds to 3 independent string patterns of three different colors.
$db$ and $d^Tc$ are the endpoints of these string patterns.
$b_E$ and $c_E$, as well as $\widetilde b$ and $\widetilde c$ are low-weight string patterns with the same endpoints.
$b+\widetilde b$ and $c+\widetilde c$ are string patterns that are closed everywhere except for the points where they terminate on the output state boundary $S_1\times S_1\times 1$.
$b_c$ and $c_c$ close off these termination points inside the output state boundary.

\myparagraph{Decoding in 2+1D}
Decoding of a 2+1D untwisted color circuit (still without physical $T$ gates) works essentially in the same way as described above for the 3+0D protocol.
There are a few small differences which we list below.
First, we do not get all the measurement outcomes at the same time, but we record them over time to then perform decoding at a later time step.
This may be either when we pass the logical information into some other gadget that requires a clean corrected state, or when determining the classical outcome of a logical measurement, which is encoded in the charge configuration.

\myparagraph{Measurement outcomes in 2+1D}
The second difference is that there is no one-to-one correspondence between individual $Z$-type ($X$-type) measurement outcomes in the circuit and the flux (charge) value at an individual face (vertex) in the 3-colex.
Instead, the history of all $Z$ measurement outcomes ($X$ measurement outcomes) form a $\zz_2$-valued vector $b_{\text{meas}}$ ($c_{\text{meas}}$), which is related to the flux configuration $b$ (charge configuration $c$) by a local $\zz_2$-linear map, just like the relation between $b_{\text{err}}$ and $b_E$ ($c_{\text{err}}$ and $c_E$) above.
Again, $b$ ($c$) is only defined up to local 2-coboundaries (1-boundaries), but its syndrome $db$ ($d^Tc$) is uniquely determined by $b_{\text{meas}}$ ($c_{\text{meas}}$).

To get a map between $b_{\text{meas}}$ and $b$ ($c_{\text{meas}}$ and $c$), we express the $-1$-post-selected projectors of each measurement in the circuit as a $ZX$ diagram, and move the corresponding $\pi$-phases to the nearby tensors.
Let us give a few examples of this map for the circuit in Fig.~\ref{fig:bulk_circuit} discussed in Section~\ref{sec:simple_example}.
The $-1$ projector of a 2-qubit $XX$ measurement can be written with two additional $\pi$-phase $Z$-tensors:
\begin{align}
\begin{split}
\bra{c,d} (1-XX) \ket{a,b}
=&
(-1)^{d+b}\cdot \delta_{a+b+c+d=0}\\
=&
\begin{tikzpicture}
\atoms{z2}{0/}
\atoms{delta,charge}{a/p={45:0.4}, b/p={-45:0.4}}
\draw (0)edge[ind=$a$]++(-135:0.5) (b)edge[ind=$b$]++(-45:0.3) (0)--(a) (0)--(b) (0)edge[ind=$c$]++(135:0.5) (a)edge[ind=$d$]++(45:0.3);
\end{tikzpicture}
\;.
\end{split}
\end{align}
We can choose the $\pi$ $Z$-tensors to be on either side.
The $\pi$-phase $Z$-tensors can be fused with the nearby $Z$-tensors in the $ZX$ path integral.
With this, a $-1$ outcome of the $XX$ measurement in Fig.~\ref{fig:bulk_circuit} corresponding to a time-parallel square in Fig.~\ref{fig:hexagon_zx} contributes a charge to two of its vertices:
\begin{equation}
\label{eq:b_to_bmeas_example}
\begin{tikzpicture}
\atoms{z2}{0/}
\atoms{delta}{{x0/p={45:0.8},lab={t=$\ldots$,p=90:0.3}}, {x1/p={135:0.8},lab={t=$\ldots$,p=90:0.3}}, {x2/p={-45:0.8},lab={t=$\ldots$,p=-90:0.3}}, {x3/p={-135:0.8},lab={t=$\ldots$,p=-90:0.3}}}
\atoms{delta,charge}{a/p={45:0.4}, b/p={-45:0.4}}
\draw (0)--(a)--(x0) (0)--(b)--(x2) (0)--(x1) (0)--(x3);
\end{tikzpicture}
=
\begin{tikzpicture}
\atoms{z2}{0/}
\atoms{delta}{{x0/p={45:0.8},lab={t=$\ldots$,p=90:0.3}}, {x1/p={135:0.8},lab={t=$\ldots$,p=90:0.3}}, {x2/p={-45:0.8},lab={t=$\ldots$,p=-90:0.3}}, {x3/p={-135:0.8},lab={t=$\ldots$,p=-90:0.3}}}
\atoms{delta,charge}{a/p={45:0.8}, b/p={-45:0.8}}
\draw (0)--(x0) (0)--(x2) (0)--(x1) (0)--(x3);
\end{tikzpicture}
\;.
\end{equation}
Note that we could also have chosen the two vertices on the left rather than those on the right.
Similarly, a $-1$ outcome of the $XX$ measurement at a time-parallel brick hexagon contributes a $\pi$-phase to three of its vertices, since pushing the $\pi$-phase through part of the $CX$ gate yields a third $\pi$-phase:
\begin{equation}
\begin{tikzpicture}
\atoms{z2}{0/, 3/p={0.4,0.8}}
\atoms{delta,charge}{a/p={0.2,0.4}, b/p={-45:0.4}}
\atoms{delta}{{1/p={0.6,1.2},lab={t=$\ldots$,p=90:0.25}}, {2/p={1.2,0.8},lab={t=$\ldots$,p=90:0.25}}, {4/p={-45:0.8},lab={t=$\ldots$,p=-90:0.25}}}
\draw (0)--(b)--(4) (0)--++(-135:0.7) (0)--(a) (a)--(3) (0)--++(135:0.7) (3)--(1) (3)--(2);
\end{tikzpicture}
=
\begin{tikzpicture}
\atoms{z2}{0/, 3/p={0.2,0.4}}
\atoms{delta}{{1/p={0.45,0.9},lab={t=$\ldots$,p=90:0.25}}, {2/p={0.8,0.4},lab={t=$\ldots$,p=90:0.25}}, {4/p={-45:0.6},lab={t=$\ldots$,p=-90:0.25}}}
\atoms{void,charge}{x1/p={0.45,0.9}, x2/p={0.8,0.4}, x3/p=4}
\draw (0)--++(-135:0.7) (0)--(4) (0)--(3) (0)--++(135:0.7) (3)--(1) (3)--(2);
\end{tikzpicture}
\;.
\end{equation}
A $-1$ outcome of the $Z^{\otimes 6}$ measurement at a time-perpendicular hexagon still yields a single $\pi$-phase at the corresponding $Z$-tensor, as shown in Eq.~\eqref{eq:z6_measurement_minus}.
On the other hand, the $-1$ projector of a 2-qubit $ZZ$ measurement can be written with two additional $\pi$-phase $X$-tensors:
\begin{equation}
\bra{c,d} \frac12(1-ZZ) \ket{a,b}
=
(-1)^{a}\cdot \delta_{a=b=c=d}
=
\begin{tikzpicture}
\atoms{delta}{0/}
\atoms{z2,charge}{a/p={45:0.4}, b/p={-45:0.4}}
\draw (0)edge[ind=$a$]++(-135:0.5) (b)edge[ind=$b$]++(-45:0.3) (0)--(a) (0)--(b) (0)edge[ind=$c$]++(135:0.5) (a)edge[ind=$d$]++(45:0.3);
\end{tikzpicture}
\;.
\end{equation}
Again, we can move the $\pi$-phases of these $X$-tensors onto the nearby $X$-tensors.
For example, the $ZZ$-measurement in Fig.~\ref{fig:bulk_circuit} corresponding to a 3-colex vertex gives rise to a $\pi$-phase at three adjacent faces, namely one past square, one future brick hexagon, and the time-perpendicular hexagon.

\myparagraph{Pauli errors in 2+1D}
As for $b_{\text{meas}}$ ($c_{\text{meas}}$), also the local $\zz_2$-linear map from $b_{\text{err}}$ to $b_E$ ($c_{\text{err}}$ to $c_E$) is different for the 2+1D untwisted color circuit.
For example, consider a $Z$ error acting on a qubit after an $XX$ measurement and before the following $CX$ gate in the untwisted color circuit (shown in Fig.~\ref{fig:bulk_circuit} if we ignore the $T$ gates).
Recall that these $XX$ measurements and $CX$ gates originate from splitting up the $X$-tensor associated to brick hexagon face as shown in Eq.~\eqref{eq:xtensor_splitting}.
Such a $Z$ error yields $\pi$-phases at two of the vertices of the brick hexagon:
\begin{equation}
\begin{tikzpicture}
\atoms{z2}{0/, 3/p={0.4,0.8}}
\atoms{delta,charge}{a/p={0.2,0.4}}
\atoms{delta}{{1/p={0.6,1.2},lab={t=$\ldots$,p=90:0.25}}, {2/p={1.2,0.8},lab={t=$\ldots$,p=90:0.25}}}
\draw (0)--++(-135:0.5) (0)--++(-45:0.5) (0)--(a) (a)--(3) (0)--++(135:0.5) (3)--(1) (3)--(2);
\end{tikzpicture}
=
\begin{tikzpicture}
\atoms{z2}{0/, 3/p={0.2,0.4}}
\atoms{delta}{{1/p={0.45,0.9},lab={t=$\ldots$,p=90:0.25}}, {2/p={0.8,0.4},lab={t=$\ldots$,p=90:0.25}}}
\atoms{void,charge}{1/p={0.45,0.9}, 2/p={0.8,0.4}}
\draw (0)--++(-135:0.5) (0)--++(-45:0.5) (0)--(3) (0)--++(135:0.5) (3)--(1) (3)--(2);
\end{tikzpicture}
\;.
\end{equation}
As another example, apply an $X$ error in Fig.~\ref{fig:bulk_circuit} to a qubit, after it is involved in the $ZZ$ measurement (originating from splitting the $Z$-tensor at a vertex) and before a $Z^{\otimes 6}$ measurement.
Such an $X$ error gives rise to two $\pi$-phases at two of the 3-colex faces adjacent to the vertex, namely the future square as well as the time-perpendicular hexagon face,
\begin{equation}
\begin{tikzpicture}
\atoms{delta}{0/, 3/p={0.4,0.8}}
\atoms{z2,charge}{a/p={0.2,0.4}}
\atoms{z2}{{1/p={0.6,1.2},lab={t=$\ldots$,p=90:0.25}}, {2/p={1.2,0.8},lab={t=$\ldots$,p=90:0.25}}}
\draw (0)--++(-135:0.5) (0)--++(-45:0.5) (0)--(a) (a)--(3) (0)--++(135:0.5) (3)--(1) (3)--(2);
\end{tikzpicture}
=
\begin{tikzpicture}
\atoms{delta}{0/, 3/p={0.2,0.4}}
\atoms{z2}{{1/p={0.45,0.9},lab={t=$\ldots$,p=90:0.25}}, {2/p={0.8,0.4},lab={t=$\ldots$,p=90:0.25}}}
\atoms{void,charge}{1x/p={0.45,0.9}, 2x/p={0.8,0.4}}
\draw (0)--++(-135:0.5) (0)--++(-45:0.5) (0)--(3) (0)--++(135:0.5) (3)--(1) (3)--(2);
\end{tikzpicture}
\;.
\end{equation}
At most other places in the circuit, a single-qubit $X$ or $Z$ error just corresponds to a $\pi$-phase at a single neighboring $X$ or $Z$ tensor.

Apart from the different maps from $b_{\text{err}}$ to $b_E$ ($c_{\text{err}}$ to $c_E$) and from $b_{\text{meas}}$ to $b$ ($c_{\text{meas}}$ to $c$), decoding works as in the 3+0D case.

\subsection{Just-in-time decoding of flux configuration}
\label{sec:x_decoding}
So far, we have described the effect of (Pauli) errors and their decoding in an untwisted color circuit.
Now, we consider how decoding of $X$ errors changes for the \emph{twisted} color circuit including the $T$ gates.
Roughly speaking, applying the $T$ gates creates an additional random charge syndrome in $d^Tc$ supported near the flux configuration $b+b_E$.
In order to contain the damage caused by this additional charge syndrome, it is necessary to choose $\widetilde b$ and apply corrections while running the circuit, in contrast to the untwisted case described in the previous section.

\myparagraph{Flux corrections in 3+0D}
In the 3+0D protocol, we need to apply corrections before applying the transversal $T^{\sgn_p}$ gate.
The reason for this is that otherwise the protocol corresponds to the color path integral with an overall flux configuration $b+b_E$.
As we have seen around Eq.~\eqref{eq:weakened_charge_constraint}, the flux configuration modifies the constraint for the charge configuration, effectively decreasing the fault distance for $Z$ errors.
While $b_E$ is a low-weight configuration for low error rates, $b$ is completely random apart from the constraint $db=db_E$.
Thus, a typical flux configuration of $b+b_E$ is dense everywhere, and effectively decreases the $Z$ fault distance to a constant, spoiling fault tolerance completely.

The correction operator we need to apply to trivialize the flux configuration is of the form $\prod_p X_p^{G_p}$, that is, we need to apply a Pauli-$X$ operator at every vertex with $G_p=1$, where $G$ is a color 1-cochain which we call the \emph{gauge fixing configuration}.
The correction correctly prepares a 3D color code state with trivial flux configuration for any choice of $G$ such that $dG=b+b_E$.
Since we do not know $b_E$, we have to replace it with its estimate $\widetilde b$.
In that case, applying a correction with $G$ such that $dG=b+\widetilde b$ does not yield an ideal code state, but one with a \emph{residual flux configuration} of $\widetilde b+b_E$.
To see this precisely, we realize that the whole circuit including $X$ errors and corrections, with measurement outcomes post-selected onto $b$ and $c$, is equal to the path integral
\begin{align}
\begin{split}
Z =& \sum_{A:dA=b+b_E} e^{2\pi i S[A+G] + \frac12 c\cdot A}\\ \propto& \sum_{A:dA=\widetilde b +b_E} e^{2\pi i S[A] + \frac12 c\cdot A}\;,
\end{split}
\end{align}
where $c\cdot A$ is the generator-wise product of $c$ and $A$.
In the second line, we have performed a change of summation variable $A\rightarrow A+G$, and ignored a global prefactor of $(-1)^{c\cdot G}$.
All in all, the flux decoding does effectively reduce the code distance for $Z$ errors, but only by a tolerable amount since $\widetilde b$ and $b_E$ are both low weight.

Note that without the transversal $T^{\sgn_p}$ gate, the $X$ corrections are unnecessary since they have no effect when combined with the subsequent $X$ measurements.
Also note that instead of applying the $X^{G_p}$ correction, we can replace the transversal $T^{\sgn_p}$ gate with $X^{G_p}T^{\sgn_p} X^{G_p}\propto T^{\sgn_p \cdot (-1)^{G_p}}$.
That is, we swap between $T$ and $T^{-1}$ at all the vertices $p$ with $G_p=1$.
If we do not apply the $X$ corrections before performing the transversal $T^{\sgn_p}$ gate, this leads to an uncorrectable propagation of $Z$ errors as discussed in Section~\ref{sec:flux_and_charge}.
Finally, note the following important difference between the untwisted (without $T$) and twisted protocols, which is especially important in the 2+1D case discussed in the following paragraphs:
In the untwisted protocol, the map from the measurement outcomes $b_{\text{meas}}$ to the flux configuration $b$ is ambiguous as discussed in Section~\ref{sec:decoding_without_t}.
For example, as discussed after Eq.~\eqref{eq:b_to_bmeas_example}, a $-1$ outcome of a $ZZ$ measurement can contribute to $b$ at two different pairs of faces.
In general, only the coboundary $db$ is relevant for decoding and corrections, and we can make arbitrary changes of the form $b\rightarrow b'=b+d\beta$.
In the twisted protocol, there is a unique configuration $b$ associated to $b_{\text{meas}}$, as the non-trivial action in the path integral spoils invariance under changes of the form $b'=b+d\beta$.
For example, if we add the $e^{2\pi i/8}$-phase $Z$-tensor corresponding to the $T$ gate in Eq.~\eqref{eq:b_to_bmeas_example}, we can only move the $e^{2\pi i/2}$ $X$-tensor phases to one particular side.
Accordingly, the corrections do depend not only on $db$, but on $b$ itself as $dG=b+\widetilde b$.

\myparagraph{Flux corrections in 2+1D}
In the 2+1D circuit from Section~\ref{sec:simple_example}, we have to apply the analogous corrections.
Namely, we need to choose the gauge fixing configuration $G$ with $dG=b+\widetilde b$, and apply $T^{\sgn_p\cdot (-1)^{G_p}}$ instead of $T^{\sgn_p}$.
However, these corrections now need to be applied while we run the circuit:
When we apply the $T^{\sgn_p\cdot (-1)^{G_p}}$ gate in the circuit, we need to choose the value of $G_p$ at the according vertex $p$.
At this time we do not have access to the value of $b$ on all faces of the 3-colex, but only on a subset.
Roughly, the subset of faces with a known $b$ value are those whose time coordinate is smaller or similar to that of $p$.
We need to show that it is still possible to choose $G_p$ given these limitations.

In the following, we show that this can be done for our example twisted color circuit from Section~\ref{sec:simple_example}.
A similar strategy also works for the other examples of twisted color circuits in Section~\ref{sec:circuits}.
Let $P_t$ be the set of 3-colex vertices with the same time coordinate $t$, located inside a single common time-parallel plane.
Let $F_t$ be the set of faces $f$ such that the vertices of $f$ with the largest time coordinate are in $P_t$.
Write $P_{<t}\coloneqq \bigcup_{t'<t} P_t$, $F_{\leq t}\coloneqq \bigcup_{t'\leq t} F_t$, and so on.
Also, write $G_{<t}$ for $G$ restricted to $P_{<t}$ and zero otherwise, $b_t$ for $b$ restricted to $F_t$, and so on.
We need to choose $G_t$ such that $dG=b+\widetilde b$ holds on all faces $f\in F_t$.
This is because $dG$ does not depend on $G_p$ for any vertex $p\in P_{>t}$.
Luckily, at the moment where we apply the $T^{\sgn_p\cdot (-1)^{G_p}}$ gate corresponding to the vertices $p\in P_t$, we do already know the value of $b$ on all faces $f\in F_{\leq t}$.
This is because $(b+\widetilde b)_f$ for a face $f$ only depends on the $ZZ$ measurement outcomes performed at the vertices $p\in f$, and the $Z^{\otimes 6}$ measurement outcome of $f$ itself if $f$ is a time-perpendicular hexagon.
Let us now concretely describe how we choose $G_p$ on all vertices $p\in P_t$ in our 3-colex:
Each vertex $p\in P_t$ is part of one unique time-perpendicular hexagon $h$.
We label the 6 vertices that are part of $h$ by $p_h^{(0)},\ldots,p_h^{(5)}$ in clockwise order.
Each face $f\in F_t$ is either equal to some time-perpendicular hexagon $h$, or shares an edge with $h$.
We label the 6 faces $f\in F_t$ that share an edge with $h$ by $f_h^{(0)},\ldots,f_h^{(5)}$ in clockwise order, such that both $f_h^{(i)}$ and $f_h^{(i-1\mmod 6)}$ contain the vertex $p_h^{(i)}$.
The following picture shows the vertices $p_h^{(i)}$ and faces $h$ and $f_h^{(i)}$:
\begin{equation}
\label{eq:gaugefix_choice_labeling}
\begin{tikzpicture}
\node at (0,0){
\includegraphics[valign=c,width=2.5cm]{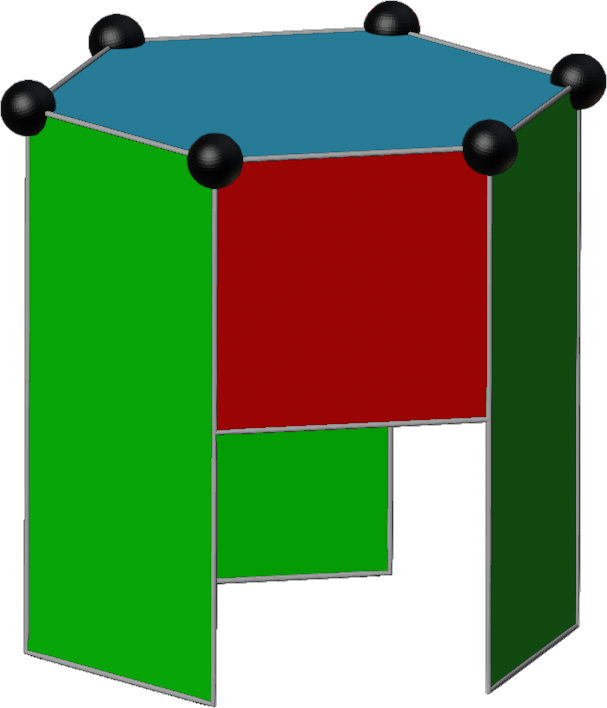}
};
\draw (0.35,1.5)edge[ind=$p_h^{(3)}$]++(80:0.4);
\draw (-0.8,1.45)edge[ind=$p_h^{(2)}$]++(90:0.4);
\draw (-1.2,1.15)edge[ind=$p_h^{(1)}$]++(120:0.4);
\draw (1.25,1.25)edge[ind=$p_h^{(4)}$]++(60:0.4);
\draw (0.8,1)edge[ind=$p_h^{(5)}$]++(70:0.7);
\draw (-0.35,0.95)edge[ind=$p_h^{(0)}$]++(90:0.4);
\draw (0,1.1)edge[ind=$h$]++(90:1);
\draw (0.2,0.25)edge[ind=$f_h^{(5)}$]++(-80:1.3);
\draw (-0.8,-0.3)edge[ind=$f_h^{(0)}$]++(180:0.6);
\draw (0.95,-0.3)edge[ind=$f_h^{(4)}$]++(0:0.5);
\draw (-0.05,-0.6)edge[ind=$f_h^{(2)}$]++(-90:0.5);
\draw (-1.2,0.6)edge[ind=$f_h^{(1)}$]++(160:0.5);
\draw (1.15,0.7)edge[ind=$f_h^{(3)}$]++(20:0.5);
\end{tikzpicture}
\;.
\end{equation}
In the example above, all faces belong to the boundary of a $y$-colored volume, and are colored according to the adjacent volumes.
We now choose
\begin{equation}
\label{eq:gauge_fix_choice}
G_{p_h^{(i)}} = \sum_{0 \leq x < i} (b+\widetilde b+dG_{<t})_{f_h^{(x)}}\;.
\end{equation}
We indeed find that
\begin{align}
\begin{split}
&(b+\widetilde b+ dG_{\leq t})_{f_h^{(i)}}\\
 =&(b+\widetilde b)_{f_h^{(i)}} + (dG_{<t})_{f_h^{(i)}}+ (dG_{t})_{f_h^{(i)}}\\
 =& (b+\widetilde b)_{f_h^{(i)}} + (dG_{<t})_{f_h^{(i)}} + G_{p_h^{(i+1\mmod 6)}} + G_{p_h^{(i)}}\\
\overset{\eqref{eq:gauge_fix_choice}}{=}& (b+\widetilde b)_{f_h^{(i)}} + (dG_{<t})_{f_h^{(i)}} + (b+\widetilde b+dG_{<t})_{f_h^{(i)}}=0\;,
\end{split}
\end{align}
and
\begin{align}
\begin{split}
&(b+\widetilde b+dG_{\leq t})_h\\
=& (b+\widetilde b)_h + \sum_i G_{p_h^{(i)}}\\
\overset{\eqref{eq:gauge_fix_choice}}{=}& (b+\widetilde b)_h + \sum_{i\in \{1,3,5\}} (b+\widetilde b+dG_{<t})_{f_h^{(i)}}\\
=&\sum_{f\in \{h,f_h^{(1)},f_h^{(3)},f_h^{(5)}\}} (b+\widetilde b+dG_{<t})_f\\
\overset{*}{=}& \sum_{f\in \Delta_2: \ccol_f\in \{yb,yg\}, \partial_{v_h,f}=1} (b+\widetilde b+dG_{<t})_f\\
=& d(b+\widetilde b+dG_{<t})_{v_h,r}\overset{**}{=}0\;.\\
\end{split}
\end{align}
For the equality labeled $*$, we have assumed that finding $G$ was successful at all previous steps, and we thus have $(b+\widetilde b+dG_{<t})_f=0$ for all faces $f\in F_{<t}$.
$v_h\in \Delta_3$ denotes the volume whose top face is $h$.
For the equation labeled $**$, we have used that $d(b+\widetilde b)=0$ by construction.
Summarizing, we have shown that it is possible to choose the gauge fixing configuration $G$ while running the circuit.

\myparagraph{2+1D: Just-in-time decoding}
Since we have to choose $G$ as we go, we also need to choose $\widetilde b$ as we go.
In particular, our choice of $G_t$ at time step $t$ according to Eq.~\eqref{eq:gauge_fix_choice} depends on $\widetilde b_{\leq t}$.
At the same time step $t$, we do not know the full flux configuration $b$ but only the portion that is determined by the measurement outcomes collected so far.
This portion is given by $b_{\leq t}$, since the $b_f$ on some face $f$ only depends on the $ZZ$ measurements associated with the vertices of $f$, as well as the $Z^{\otimes 6}$ measurement associated with $f$ itself if $f$ is a time-perpendicular hexagon.
So the challenge for the decoder is to find a low-weight estimate for $\widetilde b$ under the ``causality'' constraint that each portion $\widetilde b_t$ can only depend on $b_{\leq t}$.
A solution to this challenge was first proposed in Ref.~\cite{Bombin2018}, namely so-called \emph{just-in-time decoding}.
In Section~\ref{sec:decoding_easy}, we have given an intuitive explanation of just-in-time decoding when picturing flux configurations as string patterns.
Here, we describe a detailed decoding algorithm using the language introduced since then.

\myparagraph{Just-in-time algorithm}
The just-in-time decoding algorithm is executed once at every time step $t$.
It takes the measured flux configuration $b_{\leq t}$ as input, and computes $\widetilde b_t$ as output.
For convenience and performance, it also remembers the collective outputs $\widetilde b_{<t}$ from all previous time steps, even though these could in principle be re-computed.

The algorithm proceeds in two steps:
The first step is to find a 2-cochain $b_{\text{glob}}$ which is defined on the overall 3-colex, such that $(b_{\text{glob}})_{\leq t}$ has low weight, and $d b_{\text{glob}}=db_{\leq t}$.
In other words, we decode the flux configuration as if we were doing global decoding at time $t$ like in Section~\ref{sec:decoding_without_t}, ignoring all choices of $\widetilde b$ from previous time steps.
The second step is to find a low-weight 2-cochain $b_{\text{join}}$ that is only supported on $F_{\geq t}$ such that $db_{\text{join}}=d(\widetilde b_{<t} + (b_{\text{glob}})_{\leq t})$.
In other words, we consider the choices $\widetilde b_{<t}$ committed in previous time steps, and the current global choice $b_{\text{glob}}$.
$\widetilde b_{<t}$ terminates at time $t-1$, and $b_{\text{glob}}$ terminates at time $t$.
We extend $\widetilde b_{<t}$ by one time step, such that it terminates at time $t$ in the same way as $b_{\text{glob}}$.
After this, the output is given by $\widetilde b_t \coloneqq (b_{\text{join}})_t$.
The procedure is summarized in Algorithm.~\ref{alg:just_in_time}.
We defer the actual algorithm for the two hypergraph matching problems to Section~\ref{sec:hypergraph_matching}.

We should confirm that a solution for $b_{\text{join}}$ in the second step exists even if we restrict ourselves to $F_{\geq t}$.
This is the case because $db_{\text{glob}}$ coincides with $db$ at times earlier than $t$, and $d(\widetilde b_{<t} + (b_{\text{glob}})_{\leq t})$ is only supported on volumes that contain faces in $F_{\geq t}$.
More formally, we observe
\begin{align}
\begin{split}
d(\widetilde b_{<t} + (b_{\text{glob}})_{\leq t})
=& db_{<t} + db_{\leq t} + d(b_{\text{glob}})_{\geq t}\\
=& d(b_t+(b_{\text{glob}})_{\geq t})
\;,
\end{split}
\end{align}
so $b_{\text{join}}\coloneqq b_t+(b_{\text{glob}})_{\geq t}$ is a valid solution supported on $F_{\geq t}$, though not necessarily a low-weight one.

\begin{algorithm}[H]
\caption{Just-in-time flux decoding}
\label{alg:just_in_time}
At each time step $t \geq 0$:\\
Input: $(db)_{\leq t}$.\\
Output: $G_t$.\\
Memorize from previous steps: $\widetilde b_{<t}$.
\begin{algorithmic}[1]
\item Using Algorithm~\ref{alg:flux_matching} with open/smooth boundary conditions at time $t$, find a 2-chain $b_{\text{glob}}$ which is defined on the overall 3-colex, such that $(b_{\text{glob}})_{\leq t}$ has low weight, and $d b_{\text{glob}}=db_{\leq t}$.
\item Using Algorithm~\ref{alg:flux_matching} with closed/rough boundary conditions at time $t$, find a low-weight 2-chain $b_{\text{join}}$ that is only supported on $F_{\geq t}$ such that $db_{\text{join}}=d(\widetilde b_{<t} + (b_{\text{glob}})_{\leq t})$.
\item Set $\widetilde b_t \coloneqq (b_{\text{join}})_t$.
Set $\widetilde b_{\leq t}\coloneqq \widetilde b_{<t}+\widetilde b_t$, and memorize for time $t+1$.
\item The output $G_t$ is given by Eq.~\eqref{eq:gauge_fix_choice}, using the notation from Eq.~\eqref{eq:gaugefix_choice_labeling}.
\end{algorithmic}
\end{algorithm}

Finally, let us briefly give some intuition for how this algorithm works when imagining $b$ and $\widetilde b$ as string configurations.
Consider a configuration of $db$ consisting of two points at the same time $t_0$ but spatial separation $x$, like the bottom two points in Fig.~\ref{fig:decoding_easy}.
At each time step $t$ with $t_0\leq t< t_0+\frac12 x$, $b_{\text{glob}}$ matches each of the two points to the time-$t$ state boundary separately, since this has less weight than matching the two points to each other, $2(t-t_0)<x$.
Thus, $\widetilde b_t$ consists of two strings propagating from the two points in time direction.
At the first time step with $t>t_0+\frac12 x$, $b_{\text{glob}}$ instead consists of a string matching the two points in $db$.
So $\widetilde b_{<t} + (b_{\text{glob}})_{\leq t}$ now has a form shaped like $\sqcup$ (if time flows upward as in Fig.~\ref{fig:decoding_easy}), and $b_{\text{join}}$ connects the two end points at time $t$.
After this, $\widetilde b_{\leq t}$ will be shaped like $\sqcap$.

\subsection{Global decoding of charge configuration}
\label{charge_decoding}
Let us now describe how to decode the charge syndrome, created by $Z$ errors, in our twisted color circuits.
This decoding can be done globally, like for the untwisted color circuit in Section~\ref{sec:decoding_without_t}.
In fact, we could still decode by solving the same hypergraph matching problem $d^T\widetilde c=d^Tc$.
Even though the residual flux configuration $\widetilde b+b_E$ effectively decreases the fault distance for $Z$ errors, it is of low weight, so we still expect this method to have a fault-tolerance threshold.
We can however, mitigate the decrease of fault distance to some extent by using the information we get from the just-in-time flux decoding.

Let us assume for a moment that we are told the actual $X$ error configuration $b_E$, after just-in-time decoding with $\widetilde b$.
Then, the residual flux configuration $\widetilde b+b_E$, which leads to missing charge detectors around its support, can be treated similar to a heralded erasure.
In order to use the heralded information to optimize charge decoding, we simply replace the charge constraint $d^Tc=0$ with the twisted charge constraint in Eq.~\eqref{eq:weakened_charge_constraint}.
That is, instead of $d^T\widetilde c=dc$, we solve the twisted low-weight hypergraph matching problem
\begin{equation}
K^{b_E+\widetilde b} d^T\widetilde c = K^{b_E+\widetilde b} d^Tc+k^{b_E+\widetilde b}\;,
\end{equation}
where $K$ and $k$ are defined in Eqs.~\eqref{eq:kmatrix_definition} and \eqref{eq:modified_detector_charge}.

In reality, we only know the estimated error $\widetilde b$, but not the actual error $b_E$.
With this partial information, it is not so clear how to best use it to optimize charge decoding, and we have to resort to some heuristics.
First of all, we note that our interpretation of $M^b$ in Eq.~\eqref{eq:gauge_variance_2form} in terms of gauge variances of the color path integral only makes sense when $db=0$, since the path integral evaluates to zero otherwise.
Hence, we need to choose a heuristic to define $M^{\widetilde b}$ if $d\widetilde b\neq 0$.
A choice that seems sensible is $(M^{\widetilde b})_{vv'}=0$ if $(d\widetilde b)_v\neq 0$, or if $(d\widetilde b)_{v'}\neq 0$.
Given such a heuristic, we would proceed in the same way as in the $db=0$ case:
We define the matrix $K^{\widetilde b}$ via the kernel of $M^{\widetilde b}$.
When $\widetilde b$ terminates at some volume $v$ ($(d\widetilde b)_v=1$), then the choice above implies that any twisted charge detector can be truncated at $v$.

Also in our definition of $k^b$ in Eq.~\eqref{eq:modified_detector_charge}, the assumption $db=0$ was essential, since it required a choice of $A_0$ such that $dA_0=b$.
Again, we propose a heuristic, namely, to use a global decoder to obtain a low-weight $\widetilde b_{\text{glob}}$ with $d\widetilde b_{\text{glob}}=db$, and then use $A_0$ such that $dA_0=d(\widetilde b + \widetilde b_{\text{glob}})$ in order to define $k^{\widetilde b}$.
With these two choices, we then solve the hypergraph matching problem
\begin{equation}
\label{eq:twisted_hypergraph_matching}
K^{\widetilde b} d^T\widetilde c = K^{\widetilde b} d^Tc+k^{\widetilde b}\;.
\end{equation}
We do not provide a detailed algorithm for how to efficiently solve this hypergraph matching problem here, but we do provide some ideas in Section~\ref{sec:hypergraph_matching}.

\subsection{Decoding in a logical circuit}
\label{sec:circuit_decoding}

In this section we describe how decoding works when we use twisted color circuits as logical blocks that realize non-Clifford gates in a logical circuit.
Concretely, we assume an end-to-end logical circuit with only classical inputs and outputs, consisting of two kinds of logical blocks:
First, we use logical Clifford gates and Pauli measurements that are implemented by 2D CSS-type Abelian topological protocols, for example, by transversal Clifford gates or lattice surgery in the 2D color or surface code.
Second, we use logical non-Clifford gates or non-Pauli measurements which are implemented by twisted color circuits, such as the $\ket T$-state measurement from Section~\ref{sec:simple_example}.
The full physical circuit can be pictured as one overall 3D spacetime diagram whose volumes are labeled by either Abelian syndrome-extraction or twisted color circuits, which can have physical boundaries and are interfaced at domain walls.

The $X$ and $Z$ measurement outcomes of both the Abelian parts of the protocol and the twisted color circuits yield a flux configuration $b$ and a charge configuration $c$.
$b$ ($c$) is a configuration of $\zz_2$-variables on the overall spacetime lattice, and can be viewed as a 2-cochain (1-chain) of some cohomology theory defined on this spacetime lattice.%
\footnote{Concretely, this cohomology looks like 3D color cohomology on spacetime volumes with twisted color circuits, and like 3D cellular cohomology on volumes with surface-code syndrome-extraction circuits.
For volumes with 2D-color-code syndrome-extraction circuits, the cohomology theory corresponds to some sort of ``skeletal color cohomology'', defined on a ``skeletal 3-colex'' which is dual to a cellulation with only triangle faces and 3-colorable vertices but arbitrary volumes.
Transversal $H$ or $S$ gates on Abelian codes correspond to domain walls with an action in addition to the cohomology theory.
In contrast to the action of a $T$ or $CCZ$ gate, this action is quadratic and leads to a deterministic coupling between the constraints for $b$ and $c$, which can be decoded globally using methods similar to these in Ref.~\cite{twisted_double_code}.}
The different 3D cohomology theories have different boundaries, and are coupled at their domain walls.

\myparagraph{Just-in-time decoding}
We need to apply just-in-time decoding to choose the gauge fixing membrane $G$ inside all volumes of twisted color circuit.
A strategy that generally works is to apply just-in-time decoding to the full flux configuration $b$ during all of the protocol, including both the Abelian and the twisted-color-circuit volumes.
The resulting $G$ is then a 1-cocycle in the overall cohomology theory defined on all of spacetime.
However, $G$ will only have an effect inside the volumes with twisted color circuit, as only these contain the $T^{\sgn_p\cdot (-1)^{G_p}}$ gates.
As we will discuss later, this allows us to restrict just-in-time decoding to the vicinity of the spacetime volumes with twisted color circuits, but in general we have to include a large-enough buffer zone of abelian circuits.

\myparagraph{Global decoding for logical measurements}
In the end-to-end logical circuit, the only reason why we need to perform global decoding is to fault-tolerantly deduce logical measurement outcomes from $b$ and $c$.
After choosing global low-weight $X$ and $Z$ error estimates $\widetilde b$ and $\widetilde c$, the configuration of logical measurement outcomes is encoded in the cohomology class of $b+\widetilde b$ and $c+\widetilde c$.
In other words, each logical outcome corresponds to one generating cohomology class, which is usually supported locally on one logical block.
Note that we could also use the $\widetilde b$ resulting from just-in-time decoding instead of a global $\widetilde b$.
However, it is better to use a $\widetilde b$ from global decoding, because the latter leads to a lower logical error rate as it does not suffer from the causality constraints of just-in-time decoding.
Note that if the just-in-time decoded $\widetilde b$ and the global $\widetilde b$ disagree on a twisted-color-circuit volume, this indicates that we probably chose the wrong $G$ and already made a logical error.
If just-in-time and global decoding disagree on the Abelian volumes, this does not imply a likely logical error, and we get a lower logical error rate if we take the global decoding.
We further note that for the optimized decoding of the charge syndrome $c$ in the twisted-color-circuit volumes as described in Section~\ref{charge_decoding}, we need to use the $\widetilde b$ from just-in-time decoding, not the one from the global decoding.

We need to perform global decoding at two kinds of times $t$:
\begin{enumerate}
\item At the very end of the circuit where the logical measurement outcome is a final classical output, or
\item at an intermediate time where a logical measurement outcome is used as a classical control, for an adaptive logical circuit.
\end{enumerate}
In the second case, the value of the according generating (co)homology class must be deducible from $b_{\leq t}$ or $c_{\leq t}$.
In this case, we perform the global decoding with a state boundary condition at time $t$, that is, $b+\widetilde b$ and $c+\widetilde c$ are allowed to freely terminate at time $t$.

\myparagraph{Computational cost of decoding}
While the described decoding procedure has a runtime that is polynomial in the size of the overall circuit, there are two aspects that require a large amount of classical computation resources:
First, for a large logical circuit, global decoding in all of spacetime is slow, in particular if we consider decoding algorithms based on minimum-weight perfect matching which scale much worse than linear in the overall lattice size.
This can be solved by decomposing the overall spacetime into overlapping blocks and decoding each block separately.
As long as the blocks are large enough and have sufficient overlap, the decomposition will not increase the logical error rate by much~\cite{Dennis2001, Bombin2023a}.

The second aspect is that just-in-time decoding requires running a global decoder (in fact, two) at every single time step.
It may be possible to recycle some information of one global decoding to speed up the global decoding in the next time step, but this is highly non-trivial and depends on the implementation details of the decoder.
However, if we decompose spacetime into blocks, then we only need to apply just-in-time decoding inside blocks that contain twisted color circuits.
Note that in general, it is not possible to restrict just-in-time decoding to precisely the twisted-color-circuit volumes:
If a twisted color circuit has a domain wall with an Abelian circuit that is oriented parallel or diagonal to the time direction in spacetime, then we need to use just-in-time decoding also for the Abelian circuit, at least inside some buffer zone that scales with the fault distance.
Note that we do not describe any protocols with such domain walls in this paper, but examples can be found in Refs.~\cite{Bombin2018, Brown2019, Davydova2025}.

\myparagraph{Fault distance}
In order to suppress the logical error rate by scaling up a family of protocols, we need to make sure that we proportionally scale up the fault distance of the overall physical circuit.
We can define a notion of fault distance as follows:
The $X$ or flux ($Z$ or charge) fault distance is given by the lowest weight of a 2-cocycle (1-cycle) of non-trivial (co)homology in the overall spacetime cohomology theory.
So the scaling of the fault distance is a purely topological property, depending only on the underlying cohomology theory or path integral.

\myparagraph{Closure condition}
There are some further requirements for a scalable logical error suppression, which come from the just-in-time decoder rather than from the path integral.
Accordingly, these requirements are of a geometric rather than topological nature.
Consider a domain wall or boundary in the overall circuit that is positioned time-perpendicular.
Then such a boundary must not have a cohomology that allows for cocycles in the future volume to freely terminate.
This is similar to a condition called \emph{closure condition} in Ref.~\cite{Bombin2018}.
More precisely, any Abelian circuit is in the same phase as some number of toric codes.
Each domain wall can be classified by a subgroup $H\subset \zz_2^i\times \zz_2^o$, where $i$ ($o$) is the number of toric codes in the Abelian circuit before (after) the domain wall.
We also set $i=3$ or $o=3$ for a twisted color circuit (as if it was untwisted), and boundaries are domain walls with $i=0$ or $o=0$ on one side.
With this, we must not have a time-perpendicular domain wall for which $H$ contains an element of the form $(0, x)$ with $0\in \zz_2^i$ and $x\in \zz_2^o$.

The reason for this is that such a domain wall creates a random initial configuration of $b$, which leads to a high-weight $\widetilde b$, which spoils fault tolerance inside a twisted color circuit.
In fact, such domain walls do maintain the macroscopic fault distance if $\zz_2^o$ does not correspond to a twisted color circuit, but only to an Abelian circuit.
In this case the domain wall needs to have a macroscopic distance to any volume with twisted color circuit.
Concretely, we can turn our $\ket T$-state measurement protocol from Section~\ref{sec:tmeasurement_protocol} into a $\ket T$-state preparation protocol by initializing the 2D color code in a $\ket+$ state.%
\footnote{This has the benefit that such magic states can be prepared offline, in parallel to the remaining logical circuit.}
This initialization corresponds to a time-perpendicular smooth boundary for the 2D-color-code volume at the bottom of Eq.~\eqref{eq:global_protocol}.
Thus, as a consequence of the above considerations, we need to keep a macroscopic distance between this smooth boundary and the twisted-color-circuit volume.
In other words, after the $\ket+$ state initialization, we have to perform 2D-color-code syndrome extraction for a time that scales linear in the fault distance before introducing the domain wall with the twisted color circuit discussed in Section~\ref{sec:input_ouput_domain}.

\subsection{Hypergraph matching algorithms}
\label{sec:hypergraph_matching}
In order to implement the decoding algorithms from the previous sections in practice, we need efficient subroutines that find approximate-minimum-weight solutions to hypergraph matching problems related to color cohomology.
In this section, we propose such subroutines.
At the end of the section we comment on related decoding problems in the literature, and on the expected thresholds and possible improvements.

On a high level our hypergraph-matching algorithms works as follows:
We first apply the mappings from Section~\ref{sec:color_cohomology_equivalence} to map the syndrome color (co-)chains $db$ or $d^Tc$ to a triple $\vec y$ of cellular chains.
Then we solve three decoding problems of the form $d^{\text{cell}}\vec x=\vec y$, where $d^{\text{cell}}$ is the cellular (co-)boundary map.
Since the cellular (co-)boundary is an ordinary graph (instead of a hypergraph), we can use minimum-weight graph matching algorithms~\cite{Edmonds1965,Dennis2001,Higgott2023} to find the exact minimum-weight solution for $\vec x$.
Finally, we map the cellular triple $\vec x$ back to obtain a color chain $\widetilde b$ or $\widetilde c$.

\myparagraph{Hypergraph matching for fluxes}
We start with the algorithm that finds an approximate-minimum-weight $\widetilde b$ with $d\widetilde b=db$, on a 3-colex without any boundaries.
Up to dealing with the boundaries, this algorithm can be used as a global decoder for the flux syndrome in the untwisted color circuit, but also as a subroutine in every just-in-time decoding step for the twisted color circuit.
The algorithm depends on some choices:
First, we need to choose a ``special'' color, which we take to be $y$.
Second, for every $y$-colored volume $v$, we need to choose a special $ry$ face $f_r(v)\subset v$, a special $gy$ face $f_g(v)\subset v$, and a special $by$ face $f_b(v)\subset v$.
The algorithm starts by defining the color 2-cochain
\begin{equation}
\label{eq:flux_syndrome_shift}
\begin{aligned}
\ovl b &\coloneqq\sum_v\Big( (db)_{(v,r)}(db)_{(v,g)}f_b(v)\\
& + (db)_{(v,r)}(db)_{(v,b)}f_g(v)
+ (db)_{(v,g)}(db)_{(v,b)}f_r(v)\Big)\;.
\end{aligned}
\end{equation}
That is, each $y$ volume can have either no flux syndrome points, or a pair $(r,g)$, $(r,b)$, or $(g,b)$ of syndrome points.
If it has a $(r,g)$ pair, then we add the special $f_b(v)$ face to $\ovl b$, and analogously for other color pairs.
$\ovl b$ is constructed such that the syndrome of $b+\ovl b$ is supported only on the non-$y$ colored volumes,
\begin{equation}
\label{eq:bbar_redshift}
(d(b+\ovl b))_{(v,c)}=0\quad\forall c\in \{r,g,b\}\;,\; \forall v\in \Delta_3:\ccol_v=y\;.
\end{equation}
Next, for each color $c\in \{r,g,b\}\subset \col_1$, we apply the chain map $M^c$ from Eq.~\eqref{eq:color_to_cell_chain_map} to obtain a cellular 3-chain $s^c\coloneqq M^c d(b+\ovl b)$, or
\begin{equation}
\label{eq:flux_matching_syndrome}
(s^c)_v \coloneqq
\begin{cases}
(d(b+\ovl b))_{(v,c)}&\text{if } \ccol_v\neq c\\
0&\text{if } \ccol_v= c
\end{cases}
\;.
\end{equation}
Due to Eq.~\eqref{eq:bbar_redshift}, $s^c$ can only be non-zero on the volumes $v$ with $\ccol_v\notin \{c,y\}$.
Then, for each $c$, we find a minimum-weight cellular 2-chain $\widetilde b^c$ supported on the $\ovl{cy}$-colored faces, such that $d^{\text{cell}}\widetilde b^c=s^c$.
This can be done using algorithms as in Refs.~\cite{Edmonds1965,Dennis2001,Higgott2023} on the graph whose vertices are the volumes $v$ with $\ccol_v\notin \{c,y\}$ and whose edges are the $\ovl{cy}$ faces.
Finally, we set
\begin{equation}
\label{eq:fluxmatching_output}
\widetilde b = \ovl b + \sum_{c\in \{r,g,b\}} \widetilde b^c\;.
\end{equation}
The procedure is summarized in Algorithm~\ref{alg:flux_matching}.

\begin{algorithm}[H]
\caption{Hypergraph matching for fluxes}
\label{alg:flux_matching}
Input: Syndrome color 3-cochain $db$.\\
Output: Low-weight color 2-cochain $\widetilde b$ such that $d\widetilde b=db$.
\begin{algorithmic}[1]
\State Determine $\ovl b$ according to Eq.~\eqref{eq:flux_syndrome_shift}.
\State For every $c\in\{r,g,b\}$, set $s^c\coloneqq M^c d(b+\ovl b)$ (c.f.~Eq.~\eqref{eq:flux_matching_syndrome}).
\State For every $c\in\{r,g,b\}$, use matching algorithms~\cite{Edmonds1965,Dennis2001,Higgott2023} to compute a cellular 2-cochain $\widetilde b^c$ supported on $\ovl{cy}$-colored faces such that $d^{\text{cell}}\widetilde b^c=s^c$.
\State The output is given by $\widetilde b = \ovl b + \sum_{c\in \{r,g,b\}} \widetilde b^c$.
\end{algorithmic}
\end{algorithm}
Let us briefly argue why Eq.~\eqref{eq:fluxmatching_output} indeed satisfies $d\widetilde b=db$.
For a generator $(v,b)$ consisting of a $g$-colored volume $v$, and the color $b$ (blue), we indeed find
\begin{equation}
\begin{aligned}
(d\widetilde b)_{(v,b)} &= (d\ovl b)_{(v,b)} + \sum_{f\subset v, \ccol_f\in \{gr,gy\}} \sum_{c\in \{r,g,b\}} (\widetilde b^c)_f\\
&= (d\ovl b)_{(v,b)} + \sum_{f\subset v, \ccol_f=gr} (\widetilde b^b)_f\\
&= (d\ovl b)_{(v,b)} + (d^{\text{cell}}\widetilde b^b)_v\
= (d\ovl b)_{(v,b)} + (s^b)_v\\
&= (d\ovl b)_{(v,b)} + (d(b+\ovl b))_{(v,b)}
=(db)_{(v,b)}\;.
\end{aligned}
\end{equation}
For all other generators, the equation holds analogously, except for $\ccol_v=y$ where it holds trivially.

\myparagraph{Hypergraph matching for charges}
Let us now describe the algorithm that finds an approximate-minimum-weight $\widetilde c$ with $d^T\widetilde c = d^Tc$, on a 3-colex without any boundaries.
Up to dealing with boundaries, this algorithm can be used to perform the charge decoding in the untwisted color circuit, but also for a naive sub-optimal charge decoding in the twisted color circuit.
The algorithm is equivalent to the 3D restriction decoder proposed in Ref.~\cite{Kubica2019}.
After choosing a special color to be $y$, the algorithm proceeds as follows.
For each of color pair $\kappa\in \{ry, gy, by\}\subset\col_2$, we use the chain map $\ovl M^\kappa$ from Eq.~\eqref{eq:color_inverse_chain_map} to obtain a cellular 3-chain $s^{\kappa}\coloneqq \ovl M^\kappa d^Tc$, or
\begin{equation}
\label{eq:charge_matching_syndrome}
(s^\kappa)_v
= \begin{cases}
(d^Tc)_v&\text{if } \ccol_v\in \kappa\\
0&\text{if } \ccol_v\notin \kappa
\end{cases}
\;.
\end{equation}
Then, for each $\kappa$, we find a cellular 2-chain $\widetilde c^\kappa$ whose support is restricted to the $\kappa$-colored faces, such that $d_{\text{cell}} \widetilde c^\kappa=s^\kappa$.
We do this using algorithms as in Refs.~\cite{Edmonds1965,Dennis2001,Higgott2023} on the graph whose vertices are volumes $v$ with $\ccol_v\in\kappa$ and whose edges are the $\kappa$-colored faces.
We proceed by considering the sum
\begin{equation}
\widetilde c^{\Sigma}\coloneqq \widetilde c^{ry}+\widetilde c^{gy}+\widetilde c^{by}\;.
\end{equation}
Then, on each $y$-colored volume $v$, let $\widetilde c^{\Sigma}|_v$ denote the 2D color 2-chain obtained from restricting $\widetilde c^{\Sigma}$ to the boundary 2-colex of $v$.
We find a color 1-chain $\widetilde c_v$ on the boundary 2-colex of $v$ such that $d_{\text{2D}}\widetilde c_v=\widetilde c^{\Sigma}|_v$.
The choice of $\widetilde c_v$ does not affect the cohomology class of the final output, and we can find one by Gaussian elimination.
The output is then given by the sum of all the $\widetilde c_v$:
\begin{equation}
\label{eq:charge_matching_final}
\widetilde c\coloneqq \sum_{v\in \Delta_3: \ccol_v=y} \widetilde c_v\;.
\end{equation}
The procedure is summarized in Algorithm~\ref{alg:charge_matching}.

\begin{algorithm}[H]
\caption{Hypergraph matching for charges.}
\label{alg:charge_matching}
Input: Syndrome color 0-chain $d^Tc$.\\
Output: Low-weight color 1-chain $\widetilde c$ such that $d^T\widetilde c=d^Tc$.
\begin{algorithmic}[1]
\State For every $\kappa\in\{ry,gy,by\}$, compute the cellular 3-cocycle $(s^\kappa)_v\coloneqq (d^Tc)_v \delta_{\ccol_v\in\kappa}$.
\State For every $\kappa\in\{ry,gy,by\}$, use minimum-weight matching~\cite{Edmonds1965,Dennis2001,Higgott2023} to compute a cellular 2-cochain $\widetilde c^\kappa$ supported on the $\kappa$-colored faces, such that $d_{\text{cell}}\widetilde c^\kappa=s^\kappa$.
\State For each $y$-colored volume $v$, find a 2D color 1-chain $\widetilde c_v$ on the boundary 2-colex of $v$, such that $d_{\text{2D}}\widetilde c_v=(\widetilde c^{ry}+\widetilde c^{gy}+\widetilde c^{by})|_v$.
\State The output is given by $\widetilde c\coloneqq \sum_{v\in \Delta_3: \ccol_v=y} \widetilde c_v$.
\end{algorithmic}
\end{algorithm}
Let us verify that the above indeed satisfies $d^T\widetilde c=d^Tc$.
For a $y$-colored volume $v$, we find
\begin{equation}
\begin{aligned}
(d^T\widetilde c)_v
&= (d^T\widetilde c_v)_v
= (d_{\text{cell}}(d_{\text{2D}}\widetilde c_v))_v
= (d_{\text{cell}}\widetilde c^\Sigma)_v\\
&= (d_{\text{cell}}(\widetilde c^{ry}+\widetilde c^{gy}+\widetilde c^{by}))_v
= (s^{ry}+s^{gy}+s^{by})_v\\
&= 3(d^Tc)_v
= (d^Tc)_v\;.
\end{aligned}
\end{equation}
For a $g$-colored volume $v$, we have
\begin{equation}
\begin{aligned}
(d^T\widetilde c)_v
&= \sum_{p\in v}\; \sum_{v'\in \Delta_3: \ccol_{v'}=y, p\in v'} (\widetilde c_{v'})_p\\
&= \sum_{f\subset v: \ccol_f=gy} \sum_{p\in f} (\widetilde c_{v'(f)})_p
= \sum_{f\subset v: \ccol_f=gy} (d_{\text{2D}} \widetilde c_{v'(f)})_f\\
&= \sum_{f\subset v: \ccol_f=gy} (\widetilde c^{gy})_f
= (s^{gy})_v
=(d^Tc)_v\;,
\end{aligned}
\end{equation}
where $v'(f)$ denotes the $y$-colored volume containing the $gy$-colored face $f$.
For $r$ and $b$-colored volumes, the equation holds analogously.

\myparagraph{Matching with boundaries}
After describing decoding algorithms in the bulk, let us sketch how they can be extended to the various types of boundaries that may occur in our protocols, including both physical and state boundaries.
As we have discussed in Section~\ref{sec:boundaries}, each boundary type corresponds to a boundary condition for color cohomology.
So the approximate-minimum-weight hypergraph matching problems that we need to solve look the same as before, $d\widetilde b=db$ and $d^T\widetilde c=d^Tc$, only the (co)boundary operators are that of a different cohomology theory.
Let us now consider some examples for boundary conditions from Eq.~\eqref{sec:boundaries}, and discuss how the algorithms above can be adapted to incorporate these boundaries.

As a first example, let us consider flux decoding in the presence of an $r$ color boundary.
The hypergraph matching problem is given by the boundary map $d_2$ in Eq.~\eqref{eq:colorcohomology_color_boundary}.
When choosing $f_r(v)$, $f_g(v)$, and $f_b(v)$ for Eq.~\eqref{eq:flux_syndrome_shift} for a volume $v\in \Delta_3^{\bdin}$,
% or $v\in \Delta_2^{\bdout}$, 
we need to choose a face in $X_2$.
%When we compute the cellular 3-cochain $s^r$ as in Eq.~\eqref{eq:flux_matching_syndrome}, we add the generators $\Delta_3^{\bdout,g} + \Delta_3^{\bdout,b} + \Delta_3^{\bdout,y}\subset X_3$.
We solve the three cellular-cohomology matching problems $d_{\text{cell}} (\widetilde b^r,\widetilde b^g,\widetilde b^b)=(s^r,s^g,s^b)$.
Looking at Fig.~\ref{fig:boundary_equivalence}, the corresponding boundary of the cellular cohomology is determined by the subgroup $H=\langle (0,1,0),(0,0,1)\rangle$.
Thus, matching of $s^r$ is with respect to a rough boundary condition, whereas matching of $s^g$ and $s^b$ is with respect to a smooth boundary condition.
$d_{\text{cell}}$ with rough boundary still defines a graph, so we can directly use algorithms like in Refs.~\cite{Edmonds1965,Dennis2001,Higgott2023}.
$d_{\text{cell}}$ with smooth boundary defines a hypergraph, where some edges near the boundary are hyperedges with only a single vertex.
Concretely, the hyperedges are the $rb$-colored ($rg$-colored) boundary faces, and the missing vertex is the missing $X_3$ generator $(v,g)$ for the adjacent volume $v\in\Delta_3^{\bdout}$ ($(v,b)$ for the adjacent volume $v\in\Delta_3^{\bdout}$).
In this case, we have to add one global vertex to the hypergraph connected to all the single-vertex hyperedges, which turns the hypergraph into a graph.
After solving the three minimum-weight graph matching problems, the final result is obtained via Eq.~\eqref{eq:fluxmatching_output} as before.

As a second example, consider a the charge decoding for the same $r$ color boundary.
The hypergraph matching problem is given by the boundary map $d_0^T$ in Eq.~\eqref{eq:colorcohomology_color_boundary}.
We compute the cellular 3-cochains $s^{ry}$, $s^{gy}$, and $s^{by}$ as in Eq.~\eqref{eq:charge_matching_syndrome}, restricting ourselves to the generators in $X_0$.
Then we solve the three cellular-cohomology matching problems $d_{\text{cell}} (\widetilde c^{ry}, \widetilde c^{gy}, \widetilde c^{by})=(s^{ry}, s^{gy}, s^{by})$.
The boundary condition for this cellular cohomology is described by the subgroup $H^\perp=\langle (1,0,0)\rangle$, orthogonal to the subgroup $H$ in the flux decoding.
So $s^{ry}$ is matched with respect to a smooth boundary condition, whereas $s^{gy}$ and $s^{by}$ are matched with respect to a rough boundary condition.
Concretely, the missing vertices in the hypergraph for the smooth boundary condition are the $r$-volumes missing from $X_0$ in Eq.~\eqref{eq:colorcohomology_color_boundary}.
Finally, when computing $\widetilde c$ via Eq.~\eqref{eq:charge_matching_final}, $\widetilde c_v$ is defined on all $y$-colored volumes of $X_0$ including these in $\Delta_3^{\bdin,y}$ and $\Delta_3^{\bdout,y}$.
The boundary of the latter two kinds of volumes forms itself a 2-colex with boundary.
So in the equation $d_{\text{2D}}\widetilde c_v=\widetilde c^{\Sigma}|_v$, $d_{\text{2D}}$ is the 2D color coboundary map on a 2-colex with a $r$ color boundary as shown in Eq.~\eqref{eq:2d_color_boundary}.

As a third example, consider flux decoding for the $y$ color boundary.
This case is different from the $r$ boundary, because $y$ coincides with the ``special'' color in our flux decoding algorithm.
Of course, we could just pick a different special color that is not $y$.
However, if we consider a protocol that includes all 4 different color boundaries such as the logical $\ovl T$ gate in Eq.~\eqref{eq:logical_tgate}, then one of the color boundaries must coincide with the special color.
The hypergraph matching problem is given by the boundary map $d_2$ in Eq.~\eqref{eq:colorcohomology_color_boundary}, with $c=r$ exchanged by $c=y$.
According to Fig.~\ref{fig:boundary_equivalence}, the cellular-cohomology matching problem has a boundary condition corresponding to the subgroup $H=\langle (1,1,0), (0,1,1)\rangle$.
So instead of solving three uncoupled cellular-cohomology matching problems, the three problems are coupled at the $y$ color boundary.
The three coupled cellular boundary maps still define a graph (as opposed to a general hypergraph), so we can still apply algorithms like in Refs.~\cite{Edmonds1965,Dennis2001,Higgott2023}.
%Concretely, the vertices that are attached to edges in different graphs are the volumes in $\Delta_3^{\bdout,r}$,  $\Delta_3^{\bdout,g}$, and $\Delta_3^{\bdout,b}$.

As a fourth example, consider the subroutine of the just-in-time decoder in Algorithm~\ref{alg:just_in_time} for finding $b_{\text{glob}}$ at every time step.
The hypergraph matching problem is modified by (1) restricting $\widetilde b$ to $F_{\leq t}$, and (2) removing all (triples of) flux detectors at volumes containing faces in $F_{>t}$.
That is, it is given by the coboundary map $d_2$ of the all-smooth boundary in Eq.~\eqref{eq:colorcohomology_free_boundary}.
As the name suggests, the associated cellular cohomology corresponds to the full subgroup $H=\zz_2^3$.
So the matching of each cellular 3-cochain $s^c$ is with respect to a smooth boundary condition.
For each $s^c$, the matching can be turned into a graph-matching problem by adding one global vertex connected to all the $cy$-colored faces in $F_{>t}$.

As a fifth example, consider the subroutine of the just-in-time decoder for finding $b_{\text{join}}$.
The hypergraph matching problem is modified by restricting $\widetilde b$ to $F_{\geq t}$.
That is, the hypergraph is the coboundary map $d_2$ in Eq.~\eqref{eq:colorcohomology_allrough_boundary}.
The according cellular cohomology on which we match $s^c$ has a rough boundary condition for every $c\in \{r,g,b\}$, and thus corresponds to three separate graphs.
Concretely, the graph can be obtained from the boundary-free graph by removing all faces in $F_{<t}$.

For a general boundary, the hypergraph matching problem is defined by one of the (co-)boundary operators $d_2$ or $d_0^T$ in Eqs.~\eqref{eq:colorcohomology_free_boundary}, \eqref{eq:colorcohomology_color_boundary}, \eqref{eq:colorcohomology_doublecolor_boundary}, or \eqref{eq:colorcohomology_allrough_boundary}.
The corresponding cellular cohomology in which we have to match $s^c$ or $s^\kappa$ is determined by some subgroup $H$ listed in Fig.~\ref{fig:boundary_equivalence}, or its orthogonal complement $H^\perp$.
If $H$ or $H^\perp$ is not generated by elements of the form $(1,0,0)$, $(0,1,0)$, or $(0,0,1)$, the three cellular matching problems for $s^c$ or $s^\kappa$ are coupled at the boundary.
With one exception, the cellular matching problem can be turned into a graph by adding one global vertex for every smooth component, that is, for every $(1,0,0)$, $(0,1,0)$, or $(0,0,1)$ generator.
The exception is the charge decoding for the $y$ color boundary, whose cellular boundary cohomology is determined by $H^\perp=\langle (1,1,1)\rangle$.
In this case, the cellular boundary faces form hyperedges adjacent to three vertices, and there is no easy way to turn the hypergraph into a graph.
A possible but probably suboptimal solution for this case is to first match as if the boundary condition was $H^\perp=\langle (1,1,0)\rangle$, and then add match the third component consistently with this choice.

\myparagraph{Twisted charge matching}
After discussing boundaries, let us now make some comments on solving the optimized twisted hypergraph matching problem for the charges in Eq.~\eqref{eq:twisted_hypergraph_matching}, which takes into account the output $\widetilde b$ of the just-in-time flux decoding.
Note that the charge hypergraph matching decoder proposed above works in principle, and the following discussion is only to further improve its performance.

First of all, we note that the twisted charge constraint in Eq.~\eqref{eq:twisted_hypergraph_matching} is independent of a choice of generating constraints, or twisted charge detectors.
However, for designing an efficient hypergraph matching algorithm, it is essential to understand the structure of the twisted constraints, and to choose some ``canonical'' set of generators.
Let us now study the structure of the twisted constraints at hand of some examples.
First of all, we recall that every volume without $b_f=1$ faces defines a twisted constraint, and it is natural to take all these volumes as generators, like in the untwisted case.
The remaining volumes $b_{\text{vol}}\subset \Delta_3$ with at least one $b_f=1$ face are no longer individual generators, but combine to larger generators.
Now, consider a simple flux configuration that consists of an isolated loop of $rg$ faces, schematically depicted as
\begin{align}
\raisebox{-0.5\height}{\includegraphics[width=0.5\linewidth]{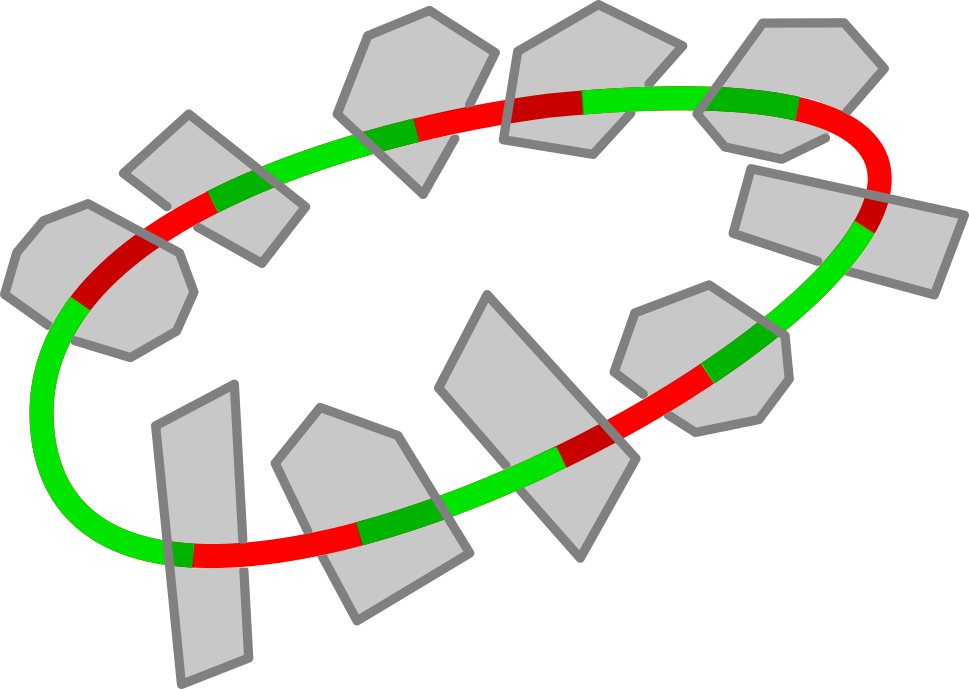}}.
\end{align}
In that case, $M^{b}_{v,v'} = 1$ if $v$ and $v'$ are neighboring volumes along the loop and $0$ for other pairs of volumes.
The volumes with flux faces together define two combined generating twisted constraints which we can label by $0$ and $1$:
\begin{align}
K^b_{0v} := (b_{\text{vol}})_v \delta_{\ccol_v=r}\;,\quad
K^b_{1v} := (b_{\text{vol}})_v \delta_{\ccol_v=g}\;.
\end{align}
In other words, the generator $0$ consists of all red volumes on the flux loop, and the generator $1$ consists of all green volumes.
We further find
\begin{equation}
k^{b}_0 = k^{b}_1 = 0\;.
\end{equation}
For this simple flux configuration $b$, the twisted hypergraph matching still works in the same way as for the untwisted case, apart from the following modification:
In each of the three graph-matching problems, we simply unify all volumes of $K^b_{0v}$ into a single vertex.
We do the same for all volumes in $K^b_{1v}$.
When unifying vertices, the graph remains a graph.
The method works for any flux configuration $b$ that consists of loops of faces, each for a different color pair, which are sufficiently isolated.
Note that if two loops are isolated but linked, we can get $k^b_x=1$ for some generators $x$ due to the linking charge phenomenon~\cite{Bombin2018a}.

In contrast, a generic flux configuration $b$ does not consist of isolated color-pair loops, but looks instead like a network of strings.
We do not know how to choose a natural set of twisted charge detectors for a generic flux configuration, and how to incorporate these twisted charge detectors into the hypergraph matching algorithm.
Note that for every connected component of $b$, there is one combined twisted charge constraint for every color pair $\kappa$, consisting of all volumes $v$ with $\ccol_v\in\kappa$.
Up to three of these constraints are independent.
However, a connected component can give rise to more than three linearly independent charge constraints, including ones involving volumes of more than two colors.
We leave a more detailed analytical study of the twisted charge constraints for a generic flux configuration and their influence on the $Z$-error threshold to future work.

Finally, we remark that we may be able to improve the threshold we get from decoding with the untwisted charge constraint, without using the full precise twisted charge constraints.
For example, we could simply remove all untwisted detectors of the volumes in $b_{\text{vol}}$, without adding any new combined detectors.
Or, we could only add the combined detector for every color pair $\kappa$ at every connected component, as discussed above.

\myparagraph{Expected and ideal performance}
Let us finally speculate what thresholds we could expect for the untwisted color circuit using the decoders above, and how they might compare to the threshold for an ideal maximum-likelihood decoder.%
\footnote{We expect the twisted color circuits to have a somewhat lower threshold since the just-in-time decoded $X$ errors effectively lead to further $Z$ errors.}
Our (untwisted) charge decoding problem is equivalent to that for the 3+1D color code for $Z$ errors in the absence of measurement errors.
Ref.~\cite{Kubica2017} uses a statistical-mechanics mapping and Monte-Carlo simulation to estimate the threshold for a maximum-likelihood decoder under iid on-site $Z$ noise to be $\sim 1.9\%$, which is similar to that of the 3D toric code at $\sim 3.3\%$.
This coincides with the maximum-likelihood threshold of our untwisted color circuit under a phenomenological noise model where every vertex in the underlying 3-colex experiences one potential $Z$ error.
Unfortunately, there are no known efficient decoding algorithms that come close to this ideal threshold.
Ref.~\cite{Beverland2021} benchmarks the 3D restriction decoder from Ref.~\cite{Kubica2019} on a 3D color code with boundaries and on-site $Z$ noise, and finds a threshold of $0.8\%$ which is only half of the maximum-likelihood threshold of $1.9\%$.
For a more realistic circuit-level noise model, one would expect the maximum-likelihood threshold to be a factor of the order of 2 to 5 smaller.
One reason for this is that in the 3+1D on-site or 2+1D phenomenological noise model, each $Z$ error contributes to $c$ at a single vertex, but for circuit-level noise in our untwisted color circuit multiple error locations may contribute to $c$ at a vertex.

Our flux decoding problem is equivalent to the $X$ syndrome fixing in the single-shot decoding of the 3D color code.
We are not aware of any results on the maximum-likelihood threshold for this decoding problem in the literature, but we expect that it will be of a similar order of magnitude to that of the charge decoding at $\sim 1.9\%$.%
\footnote{Note that this decoding has nothing to do with the flux decoding in the measurement-error-free case for the 3+1D color code, whose maximum-likelihood threshold is much higher.}
Again, efficiently implementable decoders for our flux-decoding problem achieve a much lower threshold than the (expected) ideal one.
Ref.~\cite{Brown2015} benchmarks a clustering-based decoder for on-site $X$ noise, and finds a threshold of $0.3\%$.
Another decoding strategy for the same problem was sketched in appendix F.5 of Ref.~\cite{Sahay2021}, but not benchmarked.
We hope that our proposed flux decoder has a somewhat higher threshold under this noise model, but we do not expect it to come close to the (unknown) maximum-likelihood threshold.

More generally, designing efficient decoders for decoding problems related to the 2D or 3D color code with thresholds close to the maximum-likelihood threshold is an ongoing topic of research.
For the measurement-error-free 2D color code, this goal has been achieved using the restriction decoder in Ref.~\cite{Kubica2019}:
Under on-site $X$ or $Z$ noise, the restriction-decoder threshold of $10.2\%$ is very close to the maximum-likelihood threshold of $11\%$~\cite{Katzgraber2009}.
Other decoders for other colorful decoding problems (including higher dimensions) have been proposed~\cite{Delfosse2013, Beverland2021, Gidney2023, Kubica2019, Sahay2021} and partially benchmarked but perform significantly worse than a maximum-likelihood decoder.

\section{More examples of twisted color circuits}
\label{sec:circuits}
In this section we briefly provide some further examples of twisted color circuits.

\subsection{\texorpdfstring{$x+y$}{x+y}-direction in bitruncated cubic lattice}
\label{sec:untwisted_circuit}
\begin{figure}
\centering
\begin{tikzpicture}
\node (x0) at (0,0){
\includegraphics[width=6cm]{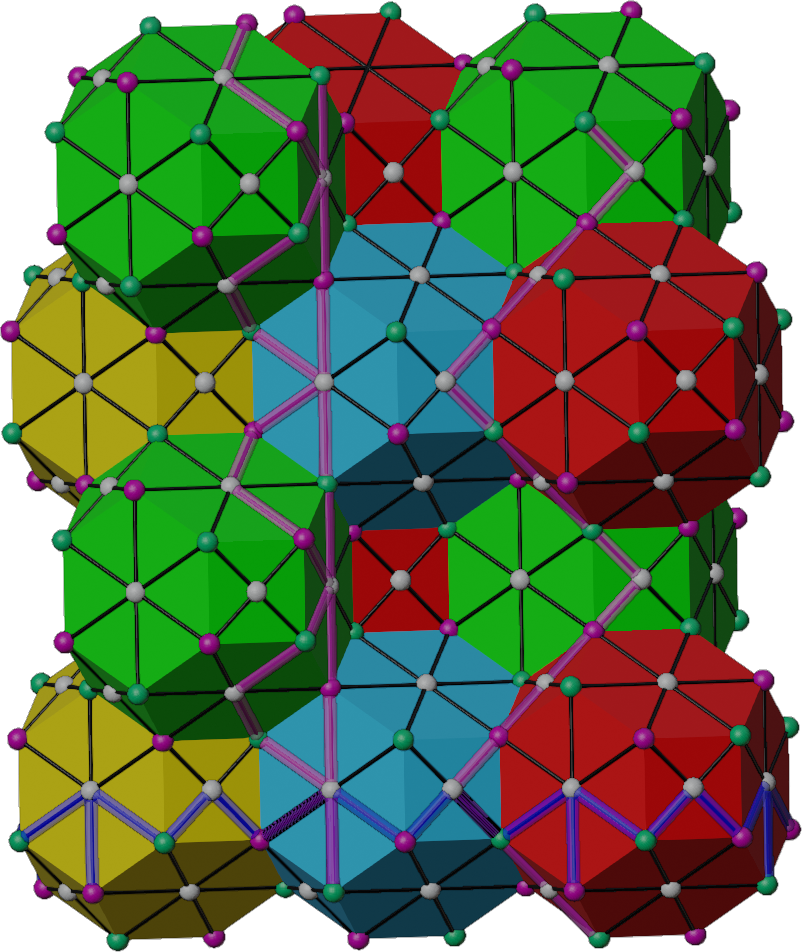}
};
\node (x1) at (-1,5){
\begin{tikzpicture}
\node (xx0) at (0,0){
\includegraphics[width=1.2cm]{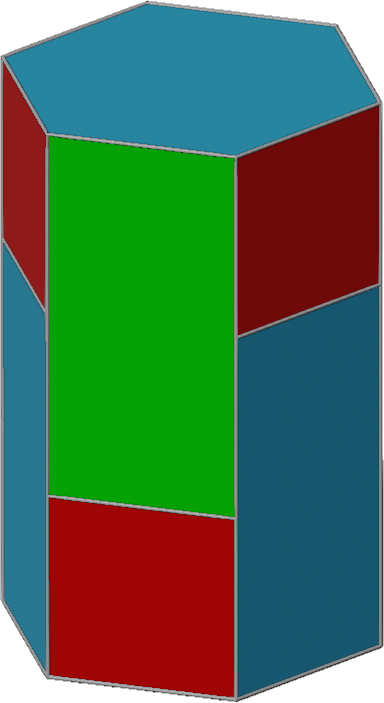}
};
\node (xx1) at (2.2,0){
\includegraphics[width=1.8cm]{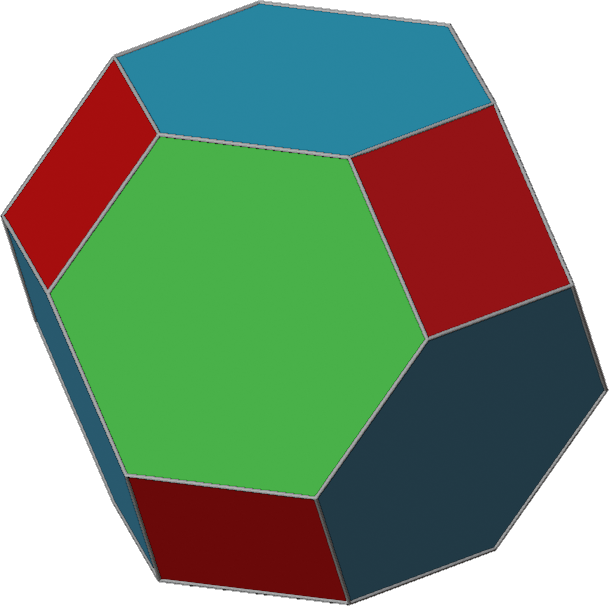}
};
\draw (xx0.east)edge[->](xx1.west);
\end{tikzpicture}
};
\node (x2) at (2.5,5){
\includegraphics[width=1.7cm]{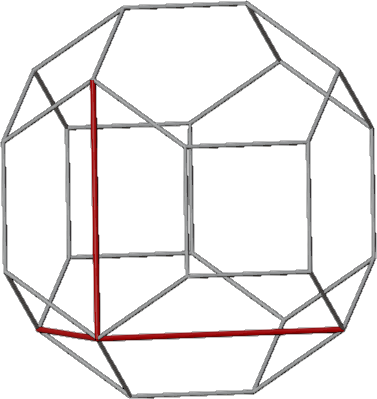}
};
\node[inner sep=0,anchor=east] at (x0.west){(b)};
\node[inner sep=0,anchor=east] at (x1.west){(a)};
\node[inner sep=0,anchor=east] at (x2.west){(c)};
\node at ($(x2)+(-0.3,0.6)$) {$t$};
\node at ($(x2)+(0.9,-0.6)$) {$\ovl x$};
\node at ($(x2)+(-0.8,-0.7)$) {$\ovl y$};
\end{tikzpicture}
\caption{
(a) The hexagon-prism lattice from Section~\ref{sec:simple_example} is related to a bitruncated cubic lattice by deforming the hexagon prisms into a more ``sphere-like'' shape.
(b) Snippet of the 3-colex in the bitruncated cubic geometry, together with the color path integral.
Three representative qubit worldlines are marked in purple, and one qubit timeslice is shown in blue.
(c) Coordinate system with space and time coordinates.
}
\label{fig:xy_direction_spacetime}
\end{figure}

For the first example, it turns out to be more convenient to consider a different geometric realization of the 3-colex in Section~\ref{sec:simple_example}.
That is, the 3-colex is combinatorially equivalent to the \emph{bitruncated cubic honeycomb}~\cite{Wiki_bitrunc} as shown in Fig.~\ref{fig:xy_direction_spacetime}.
Note that the Poincar\'e dual 3-triangulation to this lattice is a body-centered cubic lattice, consisting of two superimposed cubic lattices (one shifted by $(\frac12,\frac12,\frac12)$), with additional edges between nearest-neighbor vertices from the two different lattices.
We choose to work with the basis vectors $x$, $y$, and $z$ of this dual cubic lattice as a coordinate system.
The first example circuit we consider is obtained from traversing the path integral in the $t=x+y$ direction in this coordinate system.
Fig.~\ref{fig:xy_direction_spacetime} shows a snippet of the corresponding $ZX$ diagram, as well as three representative qubit worldlines.

\begin{figure}
\includegraphics{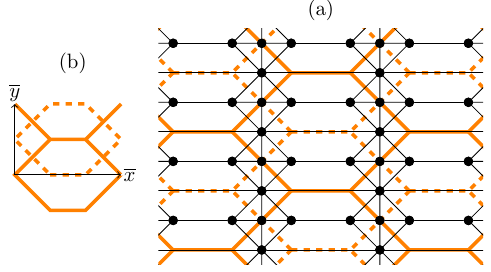}
\caption{
(a) Qubits (shown as dots) distributed over the 2D spatial lattice (shown as orange lines).
The 2D lattice is obtained by projecting the 3-colex along the $t$ direction onto the perpendicular $\ovl x$ and $\ovl y$ directions, which yields two superimposed stretched hexagonal lattices.
(b) Coordinate system, obtained from projecting the one in Fig.~\ref{fig:xy_direction_spacetime}
}
\label{fig:xplusy_qubits}
\end{figure}

From the qubit timeslices and qubit worldlines shown in Fig.~\ref{fig:xy_direction_spacetime}, we can read off how the qubits are distributed over a 2-dimensional spatial lattice as shown in Fig.~\ref{fig:xplusy_qubits}.

\begin{figure}
\includegraphics[width=0.85\linewidth]{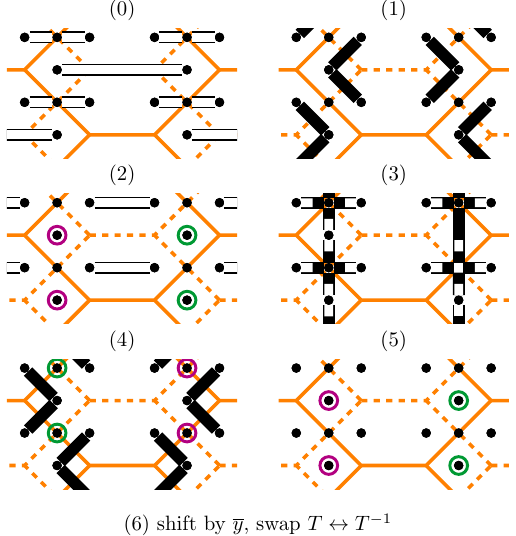}
\caption{
(0)--(5):
Twisted color circuit derived from the $ZX$ diagram, time direction, and qubit worldlines in Fig.~\ref{fig:xy_direction_spacetime}.
The symbols for Pauli measurements, $CX$ gates, and $T$ gates are the same as in Fig.~\ref{fig:bulk_circuit}.
Some of the time steps above contain multiple gates or measurements that act on the same qubits.
We can draw these overlapping gates because they commute, and thus the order in which we apply them does not matter.
However, in practice, it may not be possible to apply these gates simultaneously, and we will have to distribute them over multiple steps.
}
\label{fig:xplusy_circuit}
\end{figure}

Next, we discuss how to locally apply rewrite rules to split some of the $ZX$ tensors, and regroup them, in order to turn the path integral into a circuit.
There is one 6-index $X$-tensor at every time-parallel hexagon, which we split into two 4-index $X$ tensors:
\begin{equation}
\begin{tikzpicture}
\atoms{z2}{0/}
\draw (0)--++(-135:0.8) (0)--++(-90:0.8) (0)--++(-45:0.8) (0)--++(135:0.8) (0)--++(90:0.8) (0)--++(45:0.8);
\end{tikzpicture}
=
\begin{tikzpicture}
\atoms{z2}{0/, 1/p={0.5,0.5}}
\draw (0)--(1) (0)--++(-135:0.8) (0)--++(-70:0.8) (1)--++(-45:1) (0)--++(135:1) (1)--++(110:0.8) (1)--++(45:0.8);
\draw[dashed,cyan] (0) node[shift={(180:0.75)}] {$M_{XX}$} ellipse (0.35cm and 0.2cm);
\draw[dashed,cyan] (1) node[shift={(0:0.75)}] {$M_{XX}$} ellipse (0.35cm and 0.2cm);
\end{tikzpicture}
\;.
\end{equation}
As shown, this yields two $M_{XX}$-measurements.
There is one 6-index $X$-tensor at every time-diagonal hexagon, which we split into one 4-index $X$ tensor and two 3-index $X$-tensors:
\begin{equation}
\label{eq:time_diagonal_hexagon_splitting}
\begin{tikzpicture}
\atoms{z2}{0/}
\draw (0)--++(-120:0.8)  (0)--++(-60:0.8)  (0)--++(120:0.8)  (0)--++(60:0.8)  (0)--++(0:0.8)  (0)--++(180:0.8);
\end{tikzpicture}
=
\begin{tikzpicture}
\atoms{z2}{0/, 1/p={60:0.5}, 2/p={120:0.5}}
\draw (0)--(1) (0)--(2) (0)--++(-120:0.8)  (0)--++(-60:0.8)  (2)--++(120:0.8)  (1)--++(60:0.8)  (1)--++(0:0.8)  (2)--++(180:0.8);
\draw[cyan] node at ($(60:0.5)+(0:0.5)$) {$CX$};
\draw[cyan] node at ($(120:0.5)+(180:0.5)$) {$CX$};
\draw[dashed,cyan] (0,0) node[shift={(-20:0.6)}] {$M_{XX}$} ellipse (0.35cm and 0.2cm);
\clip (current bounding box.south west)rectangle(current bounding box.north east);
\draw[dashed,cyan] ($(60:0.5)+(0:0.5)$) ellipse (0.7cm and 0.3cm);
\draw[dashed,cyan] ($(120:0.5)+(180:0.5)$) ellipse (0.7cm and 0.3cm);
\end{tikzpicture}
\;.
\end{equation}
As depicted, it gives rise to a $M_{XX}$ measurement, and contributes half of two $CX$ gates.
The $4$-index $X$-tensor at each time-parallel square simply becomes a $M_{XX}$ measurement.
The $4$-index $X$-tensor at each time-diagonal square is split into two 3-index $X$-tensors:
\begin{equation}
\label{eq:time_diagonal_square_splitting}
\begin{tikzpicture}
\atoms{z2}{0/}
\draw (0)--++(180:0.8) (0)--++(-90:0.8) (0)--++(0:0.8) (0)--++(90:0.8);
\end{tikzpicture}
=
\begin{tikzpicture}
\atoms{z2}{0/, 1/p={0,0.5}}
\draw (0)--(1) (0)--++(-90:0.5) (0)--++(180:0.8) (1)--++(0:0.8) (1)--++(90:0.5);
\draw[cyan] node at ($(0)+(180:0.5)$) {$CX$};
\draw[cyan] node at ($(1)+(0:0.5)$) {$CX$};
\clip (current bounding box.south west)rectangle(current bounding box.north east);
\draw[dashed,cyan] ($(0)+(180:1)$) ellipse (1.3cm and 0.3cm);
\draw[dashed,cyan] ($(1)+(0:1)$) ellipse (1.3cm and 0.3cm);
\end{tikzpicture}
\;.
\end{equation}
As depicted, each of these 3-index $X$-tensors contributes half of one $CX$ gate in the circuit.
Next, there are 3 types of vertices in the 3-colex:
Type-1 vertices are adjacent to two time-parallel hexagons (one in the past (0) and one in the future (1)), two time-diagonal squares (2 and 3), as well as two time-diagonal hexagons (4 and 5).
The 6-index $Z$-tensor at each type-1 vertex $p$ is split into four phaseless 3-index $Z$ tensors, and one 2-index $Z$ tensor with a $\sgn_p \frac{2\pi}{8}$ phase:
\begin{equation}
\begin{tikzpicture}
\atoms{delta,astyle=tcol}{0/}
\draw (0)edge[ind=2]++(-150:0.8) (0)edge[ind=0]++(-90:0.8) (0)edge[ind=4]++(-30:0.8) (0)edge[ind=3]++(150:0.8) (0)edge[ind=1]++(90:0.8) (0)edge[ind=5]++(30:0.8);
\end{tikzpicture}
=
\begin{tikzpicture}
\atoms{delta}{0/, 1/p={0,0.5}, 2/p={0,1}, 3/p={0,1.5}, {4/p={0,2},astyle=tcol}}
\draw (0)--(1)--(2)--(3)--(4) (0)edge[ind=2]++(180:0.8) (0)edge[ind=0]++(-90:0.5) (1)edge[ind=4]++(0:0.8) (2)edge[ind=3]++(180:0.8) (4)edge[ind=1]++(90:0.5) (3)edge[ind=5]++(0:0.8);
\draw[cyan] node at ($(0)+(180:0.5)$) {$CX$};
\draw[cyan] node at ($(1)+(0:0.5)$) {$CX$};
\draw[cyan] node at ($(2)+(180:0.5)$) {$CX$};
\draw[cyan] node at ($(3)+(0:0.5)$) {$CX$};
\clip (current bounding box.south west)rectangle(current bounding box.north east);
\draw[dashed,cyan] ($(0)+(180:1)$) ellipse (1.3cm and 0.3cm);
\draw[dashed,cyan] ($(1)+(0:1)$) ellipse (1.3cm and 0.3cm);
\draw[dashed,cyan] ($(2)+(180:1)$) ellipse (1.3cm and 0.3cm);
\draw[dashed,cyan] ($(3)+(0:1)$) ellipse (1.3cm and 0.3cm);
\draw[cyan] node at ($(4)+(-0.35,0)$) {$T$};
\draw[dashed,cyan] (4)ellipse (0.2cm and 0.2cm);
\end{tikzpicture}
\;.
\end{equation}
As depicted, each of these 3-index $Z$-tensors contributes the control part of one $CX$ operation in the circuit.
The other halves of the $CX$ gates are these at the according time-diagonal hexagons and time-diagonal squares in Eqs.~\eqref{eq:time_diagonal_hexagon_splitting} and \eqref{eq:time_diagonal_square_splitting}.
Type-2 vertices are adjacent to one past time-parallel square (0), two past time-parallel hexagons (1 and 2), one future time-diagonal square (3) and two future time-diagonal hexagons (4 and 5).
The according 6-index tensor is split into two 4-index tensors, and one phase-$\sgn_v\frac{2\pi}{8}$ 2-index $Z$-tensor:
\begin{equation}
\label{eq:type2_vertex_splitting}
\begin{tikzpicture}
\atoms{delta}{0/astyle=tcol}
\draw (0)edge[ind=1]++(-135:0.8) (0)edge[ind=0]++(-90:0.8) (0)edge[ind=2]++(-45:0.8) (0)edge[ind=4]++(135:0.8) (0)edge[ind=3]++(90:0.8) (0)edge[ind=5]++(45:0.8);
\end{tikzpicture}
=
\begin{tikzpicture}
\atoms{delta}{0/, 1/p={0.5,0.5}, {2/p={0.25,1},astyle=tcol}}
\draw (0)--(1)--(2) (0)--++(-135:0.8) (0)--++(-70:0.8) (1)--++(-45:1) (0)--++(135:1) (2)--++(90:0.5) (1)--++(45:0.8);
\draw[dashed,cyan] (0) node[shift={(180:0.75)}] {$M_{ZZ}$} ellipse (0.35cm and 0.2cm);
\draw[dashed,cyan] (1) node[shift={(0:0.75)}] {$M_{ZZ}$} ellipse (0.35cm and 0.2cm);
\draw[cyan] node at ($(2)+(-0.35,0)$) {$T$};
\draw[dashed,cyan] (2)ellipse (0.2cm and 0.2cm);
\end{tikzpicture}
\;.
\end{equation}
As shown, it gives rise to two $M_{ZZ}$ measurements and one $T^{\sgn_v}$-gate.
Type-3 vertices are the same as type-2 vertices after reflecting along the $t$ direction.
They are split up in the same way as type-2 vertices in Eq.~\eqref{eq:type2_vertex_splitting}.
All in all, the circuit consists of
\begin{itemize}
\item two $M_{XX}$-measurements at every time-parallel hexagon,
\item one $M_{XX}$-measurement at every time-parallel square,
\item two $M_{ZZ}$-measurements at every type-2 and type-3 vertex,
\item one $CX$ gate for each pair of a type-1 vertex and adjacent time-diagonal hexagon or time-diagonal square.
\end{itemize}
In order to construct the circuit, we need to read off from Fig.~\ref{fig:xy_direction_spacetime} in which order these operators act on the according qubit worldlines.
The resulting circuit is shown in Fig.~\ref{fig:xplusy_circuit}, and has the qubit connectivity shown in Fig.~\ref{fig:xplusy_qubits}.
Notably, the circuit already naturally involves only 2-qubit measurements and $CX$ gates.
The spacetime unit cell of the circuit is that of the uncolored 3-colex, and can be chosen as spanned by $t+\ovl y$, $2\ovl x$, and $2\ovl y$.
This unit cell is the one used in Fig.~\ref{fig:xplusy_circuit}.
A shift by $t+\ovl y$ swaps the vertex signs, which is why we have to exchange $T$ and $T^{-1}$ in addition to shifting by $\ovl y$.
Note that a shift by $2t$ also leaves the 4-coloring of the 3-colex invariant, so the circuit already implements the logical identity already after two of the rounds shown in Fig.~\ref{fig:xplusy_circuit}.

\subsection{\texorpdfstring{$x$}{x} direction in bitruncated cubic lattice}
\begin{figure}
\begin{tikzpicture}
\node (x0) at (0,0){
\includegraphics[width=6cm]{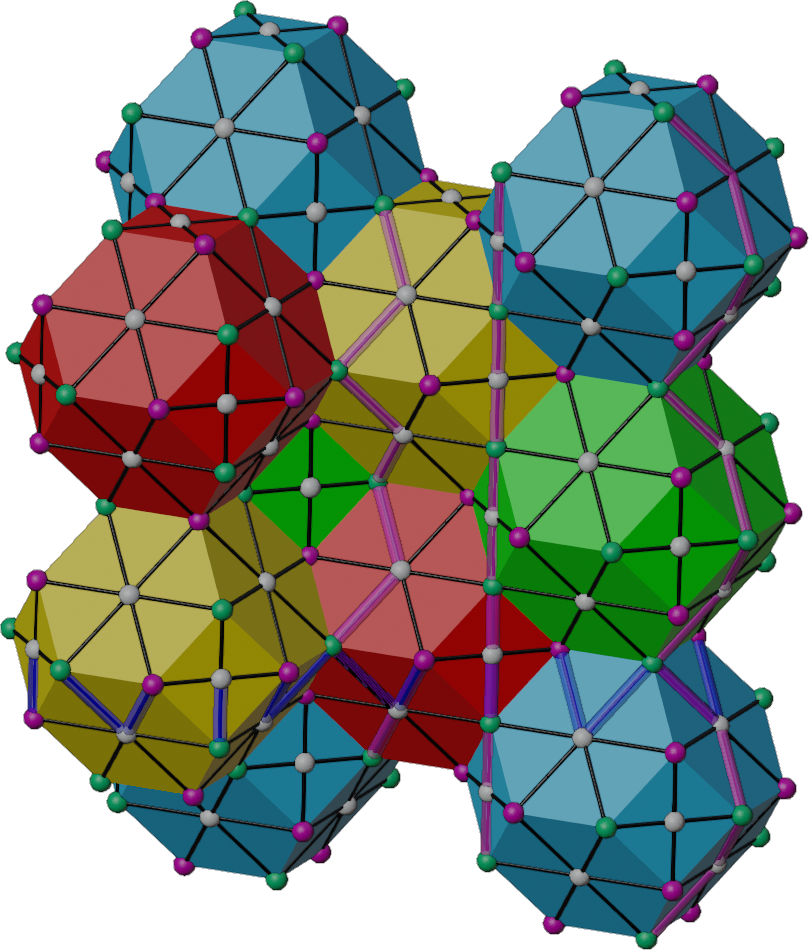}
};
\node (x1) at (-4.4,-2){
\includegraphics[width=2cm]{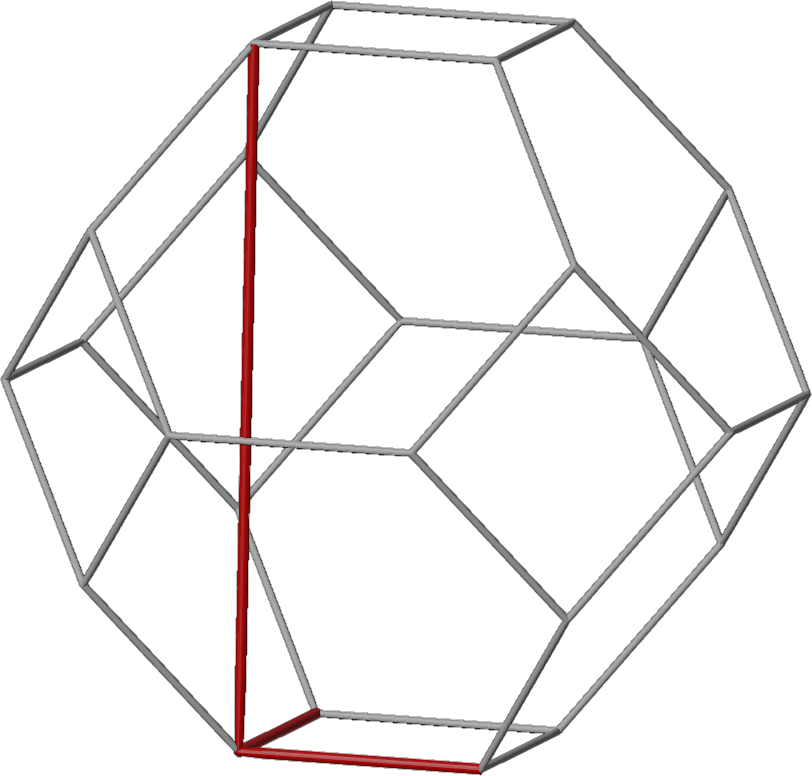}
};
\node at ($(x1)+(-0.4,1)$){$t$};
\node at ($(x1)+(-0.1,-0.6)$){$\ovl y$};
\node at ($(x1)+(0.4,-1)$){$\ovl x$};
\node[inner sep=0,anchor=south] at (x0.north){(a)};
\node[inner sep=0,anchor=south] at (x1.north){(b)};
\end{tikzpicture}
\caption{
(a) Color path integral on a snippet of the bitruncated cubic lattice.
Three representative qubit worldlines for the chosen $t$ direction are marked in purple.
One representative qubit timeslice is marked in dark blue.
(b) Coordinate system including the time coordinate $t=x$, and two orthogonal spatial coordinates $\ovl x$ and $\ovl y$.
}
\label{fig:bcc_100_bulk}
\end{figure}

As a next example, let us take the same bitruncated cubic lattice from Fig.~\ref{fig:xy_direction_spacetime} and traverse it in the $t=x$ direction.
The color path integral on this lattice with representative qubit worldlines and qubit timeslices is shown in Fig.~\ref{fig:bcc_100_bulk}.

There are three types of faces:
(1) Hexagons, all of which are neither perpendicular nor parallel to $t$, (2) time-perpendicular squares, (3) time-parallel squares (oriented in a diamond-like fashion with respect to time).
There are two types of vertices:
(1) ``Horizontal'' vertices that are adjacent to a time-perpendicular square, and (2) ``vertical'' vertices that are not.
Horizontal vertices have half-integer $t$-coordinates $\{\frac12 k\}_{k\in \zz}$  with respect to the coordinate system in Fig.~\ref{fig:bcc_100_bulk}, and vertical vertices have $t$-coordinates $\{\frac14+\frac12 k\}_{k\in\zz}$.
The circuit consists of the following operators and measurements:
\begin{itemize}
\item One $M_{XX}$ measurement at every hexagon,
\item one $M_{Z^\otimes 4}$ measurement at every time-perpendicular square,
\item one $CX$ gate and one $M_{ZZ}$ measurement at every horizontal vertex,
\item four $CX$ gates at every vertical vertex,
\item and one $T^{\sgn_p}$ gate at every vertex $p$.
\end{itemize}
Let us now describe how the tensors of the $ZX$ path integral are reshaped to give rise to the operators above.
The $X$-tensor at a hexagon is traversed by two qubit worldlines, is split up like in Eq.~\eqref{eq:time_diagonal_hexagon_splitting}.
It yields the $M_{XX}$ measurement of the hexagon, and contributes the $X$-tensors of the $CX$ gates at the two adjacent vertical vertices.
The $X$-tensor at a time-perpendicular square is not traversed by any qubit worldlines, and yields the central $X$-tensor of the $M_{Z^{\otimes 4}}$ measurement at the square.
The $X$-tensor at a time-parallel square is traversed by one qubit worldline.
It contributes the $X$-tensor of the two $CX$ gates at the adjacent horizontal vertices.
The $Z$-tensor at a horizontal vertex is traversed by two qubit worldlines.
It is split up like in Eq.~\eqref{eq:hexagon_ztensor_splitting}, and gives rise to the $M_{ZZ}$ measurement at the vertex, one $Z$-tensor of the $M_{Z^{\otimes 4}}$ measurement at the adjacent time-perpendicular square, the $Z$-tensor of the $CX$ gate at the adjacent time-parallel square, and the $T^{\sgn_p}$ gate at the vertex.
The $Z$-tensor at a vertical vertex is traversed by one qubit worldline.
It gives rise to the $Z$-tensor of the four $CX$ gates at the vertex, and the $T^{\sgn_p}$ gate at the vertex.

\begin{figure}
\includegraphics[width=0.85\linewidth]{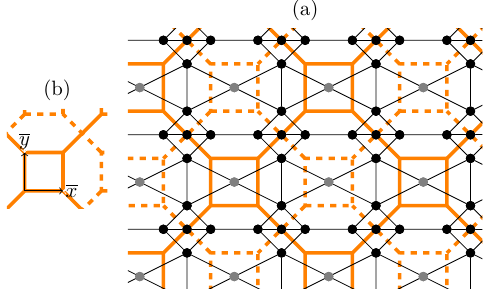}
\caption{
(a) Projection of the 3-colex along the $t$ direction in orange:
It consists of two 4-8-8 lattices, one of which we draw dashed.
Qubits and their connectivity are depicted in black.
Gray dots represent ancilla qubits for performing the weight-4 Pauli measurements.
(b) Coordinate system.
}
\label{fig:bcc_100_qubits}
\end{figure}

\begin{figure}
\includegraphics[width=0.9\linewidth]{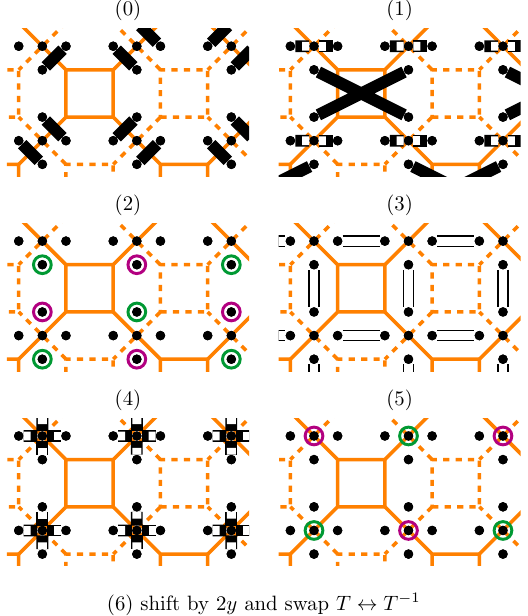}
\caption{
Twisted color circuit derived from the 3-colex, time direction, and qubit worldlines shown in Fig.~\ref{fig:bcc_100_bulk}.
Symbols for gates are the same as in Fig.~\ref{fig:bulk_circuit}.
Some steps contain multiple operations which overlap on qubits, but commute.
}
\label{fig:bcc_100_circuit}
\end{figure}

With this, we can write down the circuit by reading off from Fig.~\ref{fig:bcc_100_bulk} in which order the gates and measurements above act.
The resulting circuit is shown in Fig.~\ref{fig:bcc_100_circuit} and the required qubit connectivity is shown in Fig.~\ref{fig:bcc_100_qubits}.

It suffices to write down the circuit for one spacetime unit cell.
We choose this unit cell to be spanned by $4\ovl x$, $2\ovl x+2\ovl y$, and $\frac12 t+2\ovl x$.
Note that a shift by $\frac12 t+2\ovl x$ flips the vertex signs, which is why we need to swap $T$ and $T^{-1}$ in step (6) of Fig.~\ref{fig:bcc_100_circuit}.

\subsection{Chamfered cubic lattice}
\begin{figure}
\includegraphics[width=5cm]{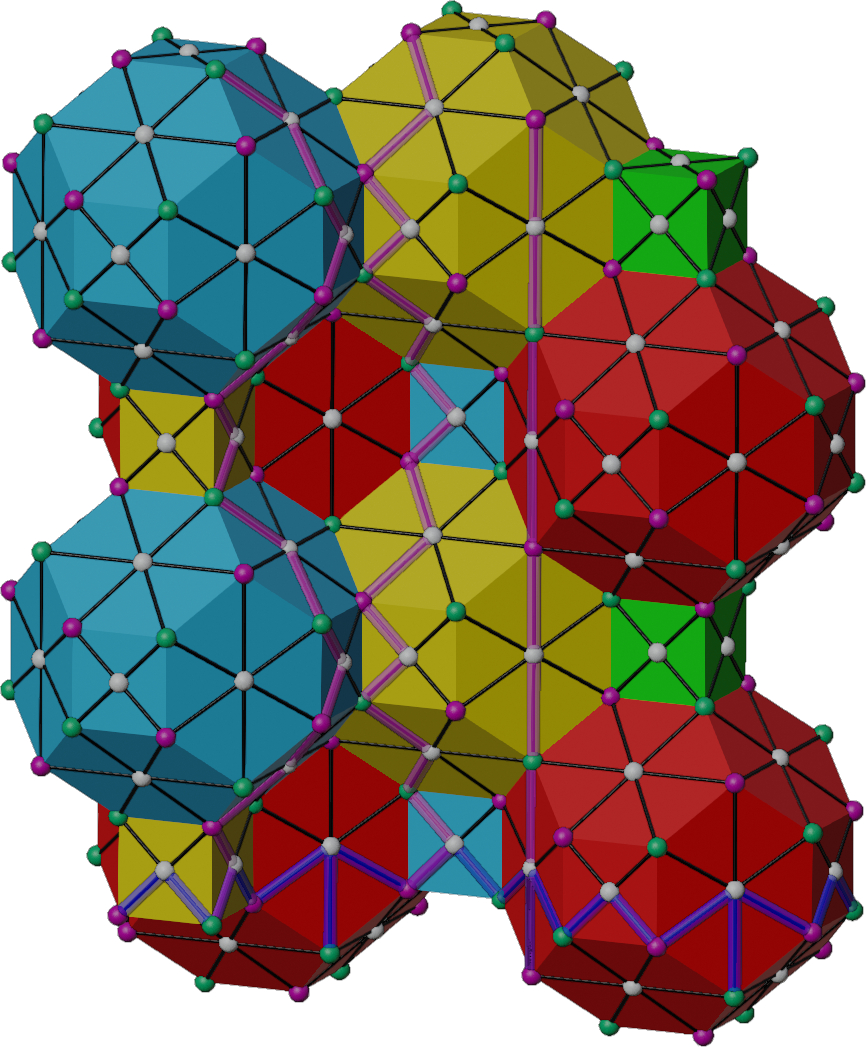}
\caption{
$ZX$ color path integral on a snippet of the chamfered cubic 3-colex.
Time goes upward.
Two representative qubit worldlines are marked in purple, and one representative qubit timeslice is marked in dark blue.
}
\label{fig:mixcube_zx}
\end{figure}

In this section, we briefly consider a combinatorially different 3-colex.
It consists of two kinds of volumes:
(1) Smaller Cubes, and (2) larger cubes whose edges have been truncated, yielding a polyhedron with square and hexagon faces called a \emph{chamfered cube}.
The 3-colex with the associated $ZX$ color path integral is shown in Fig.~\ref{fig:mixcube_zx}.
The spatial projection and qubit distribution is combinatorially equivalent to the one shown in Fig.~\ref{fig:bcc_100_qubits}.
The circuit is similar to Fig.~\ref{fig:bcc_100_circuit}, with the difference being that steps (1) to (3) are repeated a second time (with $T\leftrightarrow T^{-1}$ exchanged).

\subsection{Circuits for other boundaries}

\begin{figure*}
\centering
\begin{tikzpicture}
\node[inner sep=0] (x0) at (0,0){
\includegraphics[width=5cm]{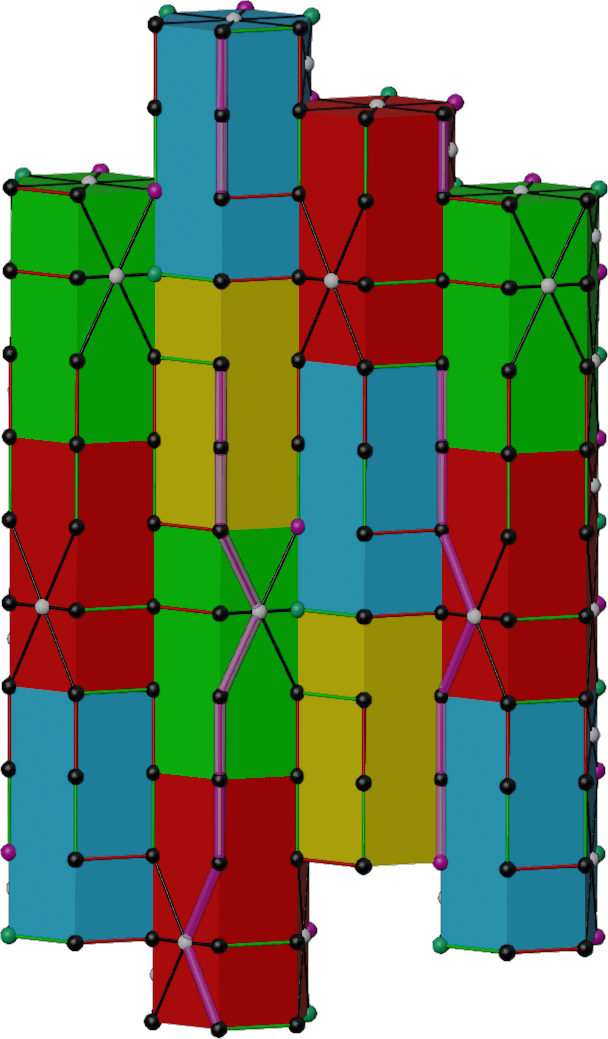}
};
\node[inner sep=0] (x1) at (8,0){
\includegraphics[width=5cm]{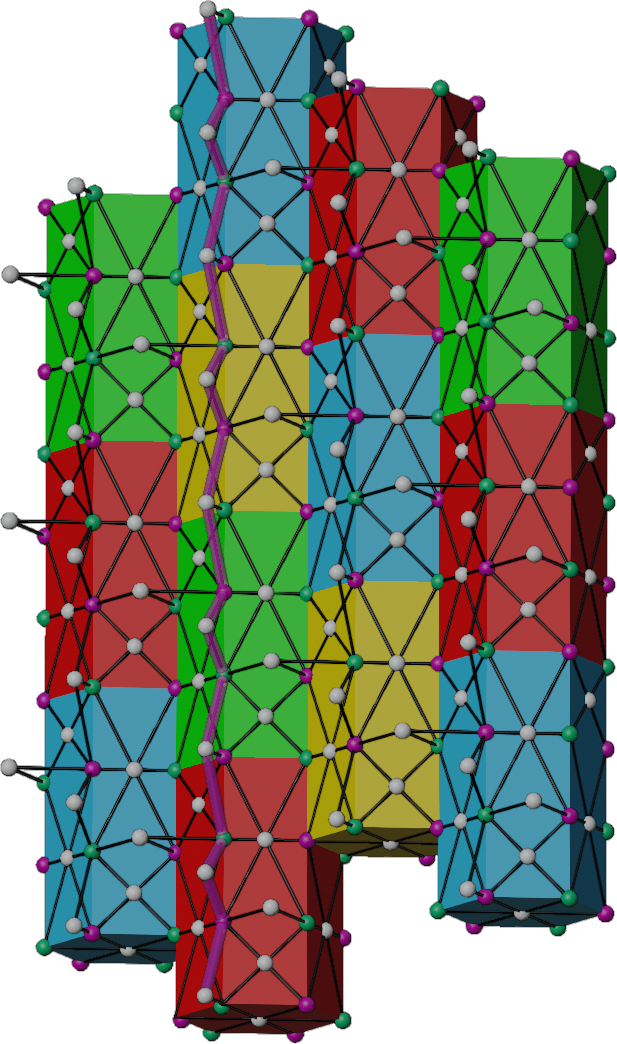}
};
\node[anchor=east] at (x0.west){(a)};
\node[anchor=east] at (x1.west){(b)};
\end{tikzpicture}
\caption{
$ZX$ diagrams for the double-color boundary (a) and all-rough boundary (b) of the color path integral.
The $Z$ and $X$-tensors correspond the generators in $X_1$ and $X_2$ of the according cohomology theories in Eq.~\eqref{eq:colorcohomology_doublecolor_boundary} for the double-color boundary and Eq.~\eqref{eq:colorcohomology_allrough_boundary} for the all-rough boundary.
The bonds correspond to the coboundary map $d_1$.
Note that in (b) we should have put a 2-index $X$-tensor at every $\ovl r$ and $\ovl g$-colored edge, but a 2-index $X$-tensor is the same as a bond without tensor.
One representative qubit worldline has been marked in purple in each (a) and (b).
}
\label{fig:other_boundaries}
\end{figure*}

In the previous examples, we have restricted ourselves to the bulk circuits for simplicity.
Here, we will describe additional boundary conditions for twisted color circuits, for the geometry considered in Section~\ref{sec:simple_example}.
In Section~\ref{sec:boundaries_easy}, we have only shown how to implement color boundaries of the color path integral.
Let us now briefly describe how to implement the other types of boundary conditions discussed in Section~\ref{sec:boundaries}, namely the double-color and all-rough boundaries.
Fig.~\ref{fig:other_boundaries} shows the according $ZX$ diagrams, along with representative qubit worldlines.
Deriving the according circuits is straight-forward using the methods developed in this paper.

Using the 2-color boundary, we can also obtain a gate for a unitary $T$ gate (instead of a $\ket T$-state measurement).
The according spacetime diagram (c.f. Ref.~\cite{Davydova2025}) for the global protocol is as follows:
\begin{align}
\label{eq:logical_tgate}
\begin{tikzpicture}[scale=0.8]
\draw[gray] (-0.3,2.3)edge[->]node[midway,rotate=90,anchor=south]{time} (-0.3,4.7);
\draw (0,0)rectangle++(3,2) (0,5)rectangle++(3,2) (0,2)--++(0,3) (3,2)--++(0,3) (0,7)--++(2,0.7)--(3,7);
\draw[dashed] (0,0)--++(2,0.7)--++(0,7) (2,0.7)--(3,0) (0,2)--++(2,0.7)--(3,2) (0,5)--++(2,0.7)--(3,5);
\draw[gray,align=left] (1.7,3.7)--++(-10:2.2)node[anchor=west]{twisted color circuit};
\draw[gray] (1.7,1.2)--++(-10:2.2)node[anchor=west]{2D color code};
\draw[gray] (1.7,6.2)--++(-10:2.2)node[anchor=west]{2D color code};
\node[colr] at (1.5,3.5){$r$};
\node[colb] at (1,1.35){$b$};
\node[colg] at (2.5,1.35){$g$};
\node[colr] at (1.5,1){$r$};
\node at (1,3.85){$\textcolor{colr}{r}\textcolor{colg}{g}$};
\node[colg] at (2.5,3.85){$g$};
\node[colr] at (1.5,6){$r$};
\node[colb] at (1,6.35){$b$};
\node[colg] at (2.5,6.35){$g$};
\node[colb] at (1.8,2.3){$b$};
\node[coly] at (1.8,5.3){$y$};
\end{tikzpicture}
\end{align}
The twisted color circuit in the middle has two color boundaries, and also one double-color boundary.
The input and output domain walls couple the 2D color code with two different color boundaries.
It should be noted that at the bottom domain wall, the 3-colex restricts to a 2-colex with colors $r$, $g$, and $y$, which we have to identify with $r,g,b$.

\section{Discussion}
\myparagraph{Summary}
We have constructed \emph{twisted color circuits}, a new class of fault-tolerant scalable planar circuits that realize logical non-Clifford gates on the 2D color code.
The circuits are natively described in spacetime by a \emph{color path integral}, and can be derived from an arbitrary 3-colex by traversing it in an arbitrary time direction.
We have constructed physical boundaries of the twisted color circuits, and domain walls with the 2D color code, to realize various logical non-Clifford gates, using spacetime geometries presented in Ref.~\cite{Davydova2025}.
These include a logical $\ket T$-state preparation via $XS$ measurement, a logical unitary $T$ gate, and a logical $CCZ$ gate.

We have given a formulation of the color path integral as a 2+1D lattice gauge theory, which is based on a cohomology theory on 3-colexes that we call \emph{color cohomology}.
We have directly proven the gauge invariance of the underlying action, and its equivalence to a known Dijkgraaf-Witten theory, or twisted quantum double.
This establishes that the path integral represents a non-Abelian topological phase, which coincides with that of the (untwisted) Kitaev quantum double for the dihedral group $D_4$.
We have related the flux and charge defects that are used for the decoding of our circuits to the anyons of the non-Abelian phase.

Further, we have described the decoding of our circuits in detail, consisting of a \emph{just-in-time} decoder for the $X$ errors and a global decoder for the $Z$ errors.
We have shown how to concretely implement the ideas in Ref.~\cite{Bombin2018} for just-in-time decoding by using two global decoders at every time step.
We have made extensive use of color cohomology to construct such efficient, matching-based global decoders, extending previous work on color-code decoders~\cite{Kubica2019}.
In particular, we have proposed a new decoder for the $X$ errors, which could also be used for single-shot decoding of the 3D color code.
We have also described how knowledge of the just-in-time $X$ corrections can be used to optimize the global decoding for $Z$ errors.

\myparagraph{Outlook}
The most important remaining open task is to benchmark our circuits, in order to obtain a better estimate for realistic fault-tolerance thresholds of our protocols under circuit-level noise.
While this is beyond the scope of this paper, we have established the essential theoretical tools for simulating the logical performance under local Pauli-$X$ and $Z$ noise:
Our analysis of flux and charge defects in the color path integral allows us to correctly sample syndrome configurations and to determine the correctness of the decoder, despite the non-Clifford $T$ gates in the circuit.
We believe that the steps we have taken have potential for greatly improving the modest logical performance of the protocols found in simulations so far~\cite{Scruby2020, Scruby2024}.
Two reasons that indicate the possibility of an improved performance are (1) the comparatively low complexity of our circuits, (2) our proposed matching-based decoder which is more fine-tuned to the underlying lattice microscopics than previously used RG-based decoders as in Refs.~\cite{Brown2019,Scruby2020}.

A second open question is to which extent the methods in this paper carry over to non-geometrically local qLDPC codes.
There is a variety of recent constructions for transversal $CCZ$ gates in qLDPC codes based on cup products or triple-intersection forms~\cite{Breuckmann2024,Zhu2025,Hsin2024,Lin2024}.
The equivalence that we establish between color and cellular cohomology can be used to translate between the transversal $CCZ$ gate of three 3D toric codes and the transversal $T$ gate of the 3D color code.
It will be interesting to see if this cohomological perspective can be used to find transversal logical $T$ gates in more general qLDPC codes.

\subsubsection*{Acknowledgments}
We would like to thank M.~Davydova, D.~Williamson, M.~Webster, and B.~Brown for collaboration on Ref.~\cite{Davydova2025}, and B.~Brown and M.~Vasmer for feedback on an earlier draft of this manuscript.
A.B. was supported by the U.S. Army Research Laboratory and the U.S. Army Research Office under contract/grant number W911NF2310255, and by the U.S. Department of Energy, Office of Science, National Quantum Information Science Research Centers, and the Co-design Center for Quantum Advantage (C2QA) under contract number DE-SC0012704.
J.M. is supported by the DFG (CRC 183).

\appendix

\section{Path integral boundaries: Derivation}
\label{sec:boundaries_derivation}
In this appendix, we derive the boundaries stated in Section~\ref{sec:boundaries} in a detailed and pedagogical way.
We start with boundaries of the 2D color path integral and then generalize to three dimensions.
The boundaries of the color path integral from which we construct the twisted color circuits only make use of the coboundary map $d_1$ of color cohomology.
However, if we want to describe the decoding problem in terms of color cohomology we have to also define $d_0$ and $d_2$ in the presence of boundaries.
The boundary maps $d_{-1}$ and $d_3$ are less natural (and also not locally exact), especially in the presence of boundaries, so we will not define these.

\myparagraph{2D boundaries}
We start by discussing some 1D physical boundaries for the 2D color path integral in Eq.~\eqref{eq:2d_color_path_integral}.
These provide a useful exercise to get some intuition for the construction, and are related to the boundaries of the 2D color code used in Section~\ref{sec:input_ouput_domain}.
Geometrically, the boundary is defined along any path of edges of the 2-colex.
We will use the following notation:
\begin{itemize}
\item $\Delta_i$ denotes the set of $i$-cells of the 2-colex including its boundary edges and vertices.
\item $\Delta_0^{\mbd}\subset \Delta_0$ ($\Delta_1^{\mbd,c}\subset \Delta_1$) denote the sets of vertices and edges (of color $c$) that form the boundary.
\item $\Delta_2^{\bdin}\subset \Delta_2$ denotes the set of faces adjacent to some boundary edge.
\item $\Delta_2^{\bdout}\not\subset \Delta_2$ denotes the ``phantom'' faces that are adjacent to a boundary edge on the outside of the boundary.
\item For every color $c$, let $\Delta_0^{\bdout,c}$ ($\Delta_0^{\bdin,c}$) denote the set of boundary vertices that are adjacent to one single face in $\Delta_2^{\bdout}$ ($\Delta_2^{\bdin}$), which has of color $c$.
\end{itemize}

\myparagraph{2D state boundary}
The most naive way to extend 2D color cohomology to the boundary is by restricting the generator sets and coboundary maps of some larger 2-colex-without-boundary containing the 2-colex-with-boundary,
\begin{equation}
\label{eq:2d_colorcohomology_open_boundary}
X_0\coloneqq \Delta_2\;,\quad
X_1\coloneqq \Delta_0\;,\quad
X_2\coloneqq\Delta_2\;.
\end{equation}
This does not define a physical boundary for the path integral (or, at least not a topological one), since it lacks the local exactness condition discussed in Section~\ref{eq:color_cohomology_definition}.
That is, there are locally supported 1-cocycles that are not 1-coboundaries, namely one for every face in $\Delta_2^{\bdout}$.
Rather than a physical boundary, the above can be viewed as a state boundary of the path integral as discussed in Section~\ref{sec:path_integrals_general}.
It also does not define a topological boundary for the 2D color code, where the lack of local exactness implies low-weight logical operators.

\myparagraph{2D all-smooth boundary}
Local exactness can be regained by trivializing each local 1-cocycle by adding an according generator to $X_0$,
\begin{equation}
\label{eq:2d_colorcohomology_free_boundary}
\begin{aligned}
X_0&\coloneqq \Delta_2+\Delta_2^{\bdout}\;,&
X_1&\coloneqq \Delta_0\;,\\
X_2&\coloneqq\Delta_2\;,&&
\end{aligned}
\end{equation}
where $+$ denotes set union.
The coboundary maps look the same as for the bulk.

The chain complex does define a topological boundary for the 2D untwisted color path integral.
However, it does not yield a boundary for the 2D (twisted) color path integral, since there is no way to extend the action in Eq.~\eqref{eq:2d_color_action} to the boundary in a gauge-invariant way.
One can see this by mapping the boundary to a boundary for cellular cohomology as shown at the end of this section:
It corresponds to the full subgroup of $\zz_2^2$, and the action is a group 2-cocycle that is non-trivial on this subgroup.
The chain complex also yields a boundary for the 2D color code, which represents the same boundary topological phase as the Pauli-$X$ boundary~\cite{Kesselring2018}, though it is defined in a geometrically slightly different way.
In analogy to the path-integral action being incompatible with this boundary, the transversal $\ovl S$ gate of the 2D color code cannot be extended to this boundary.

\myparagraph{2D all-rough boundary}
Instead of adding generators to $X_0$, we can also achieve local exactness by adding generators to $X_2$:
\begin{equation}
\label{eq:2d_colorcohomology_rough_boundary}
\begin{aligned}
X_0&\coloneqq \Delta_2\;,&
X_1&\coloneqq \Delta_0\;,\\
X_2&\coloneqq\Delta_2+\Delta_2^{\bdout}\;.&&
\end{aligned}
\end{equation}
This chain complex defines a boundary for both the untwisted and twisted 2D color path integral.
The action looks the same as in Eq.~\eqref{eq:2d_color_action}, where the summation runs over all vertices $p$ including these inside the boundary.
The analog 2D-color code boundary is the Pauli-$Z$ boundary~\cite{Kesselring2018}.

\myparagraph{2D color boundary}
There is one additional boundary cohomology that is in between the all-smooth and all-rough boundaries:
We can define this boundary for each color $c$, but will set $c=r$ without loss of generality below:
\begin{equation}
\label{eq:2d_color_boundary}
\begin{aligned}
X_0&\coloneqq \Delta_2-\Delta_2^{\bdin,r} +\Delta_2^{\bdout,g}+\Delta_2^{\bdout,b}\;,\\
X_1&\coloneqq \Delta_0\;,\\
X_2&\coloneqq \Delta_2 + \Delta_1^{\mbd,\ovl r}\;,
\end{aligned}
\end{equation}
where $-$ denotes set subtraction.
$d_1^T$ maps every edge $e\in \Delta_1^{\mbd,\ovl r}$ to both of its endpoints, otherwise the coboundary maps are like in the bulk.
Note that after adding $\Delta_1^{\mbd,\ovl r}$ to $X_2$ we must exclude $\Delta_2^{\bdin,r}$ from $X_0$, otherwise we would have $(d_1d_0)_{ef}=1$ for an $r$ boundary face $f\in X_0$ and an $\ovl r$ boundary edge $e\in X_2$ sharing a vertex.

The above chain complex yields a boundary for the untwisted as well as for the twisted color path integral.
In the twisted case, the action in Eq.~\eqref{eq:2d_color_action} is extended to the boundary as follows:
\begin{equation}
\label{eq:2d_color_boundary_action}
\begin{aligned}
S_{\text{2D}}^{\mbd,r}[A]
&= \frac14 \sum_{p\in \Delta_0-\Delta_0^{\bdin,r}} \sgn_p \ovl{A_p}\\
&= \frac14 \sum_{p\in \Delta_0-\Delta_0^{\mbd}+\Delta_0^{\bdout,r}} \sgn_p \ovl{A_p}
\;.
\end{aligned}
\end{equation}
The second equation holds because $(dA)_e=0$ for every edge $e\in \Delta_1^{\mbd,\ovl r}$, so we have $A_{p_0}=A_{p_1}$ for the two endpoints $p_0$ and $p_1$ of $e$, and their action cancels, $\frac14(\sgn_{p_0} \ovl A_{p_0} + \sgn_{p_1} \ovl A_{p_1})=0$.
It suffices to test gauge invariance for $\alpha$ consisting of a single $X_0$ generator $f$.
For faces $f\in \Delta_2^{\bdin,g}$ or $f\in \Delta_2^{\bdin,b}$, the action on the adjacent vertices is the same as for any bulk face, if we use the first line in Eq.~\eqref{eq:2d_color_boundary_action}.
For faces $f\in \Delta_2^{\bdout,g}$ or $f\in \Delta_2^{\bdout,b}$, the action is zero if we use the first line in Eq.~\eqref{eq:2d_color_boundary_action}.

The chain complex also defines a boundary for the 2D color code, which we use in Section~\ref{sec:simple_example} in Fig.~\ref{fig:color_code}.
This boundary is in the same phase as the \emph{color boundary}~\cite{Bombin2006b,Kesselring2018}, though defined in a microscopically slightly different way.
Further, the phases in Eq.~\eqref{eq:2d_color_boundary_action} of the form $e^{2\pi i\frac14\sgn_p \ovl A_p}$ define a transversal $\ovl S$ gate for the 2D color-code boundary.

\myparagraph{3D boundaries}
Let us continue with boundaries of the 3D color path integral.
We recall the notations like $\Delta_3^{\bdout}$ or $\Delta_2^{\mbd,gb}$ from Section~\ref{sec:boundaries}.

\myparagraph{3D state boundary}
As in 2D, simply restricting the bulk cohomology,
\begin{equation}
\label{eq:colorcohomology_state_boundary}
\begin{aligned}
X_0 &\coloneqq \Delta_3\;,&
X_1&\coloneqq \Delta_0\;,\\
X_2&\coloneqq\Delta_2\;,&
X_3&\coloneqq \bigsqcup_{c\in\col_1} \Delta_3^c\times \ovl c\;,
\end{aligned}
\end{equation}
is not locally exact and yields a state boundary rather than a physical (and topological) boundary.
Concretely, there are locally supported 1-cocycles that are not 1-coboundaries, namely one for every volume in $\Delta_3^{\bdout}$.

\myparagraph{3D all-smooth boundary}
Local exactness can be fixed by adding the one generator to $X_0$ for each local 1-cocycle:%
\footnote{In the language of subsystem codes this is similar to \textit{gauge fixing} an associated (spacetime) logical operator.}
\begin{equation}
\label{eq:colorcohomology_free_boundary}
\begin{aligned}
X_0 &\coloneqq \Delta_3+\Delta_3^{\bdout}\;,&
X_1&\coloneqq \Delta_0\;,\\
X_2&\coloneqq\Delta_2\;,&
X_3&\coloneqq \bigsqcup_{c\in\col_1} \Delta_3^c\times \ovl c\;.
\end{aligned}
\end{equation}
The coboundary maps are still restrictions of the bulk ones in Eq.~\eqref{eq:color_cohomology_boundaries}.
This chain complex defines a boundary of the untwisted color path integral.
We call it ``all-smooth'', since it corresponds to a stack of three smooth boundary conditions for the equivalent cellular cohomology.
However, the chain complex does \emph{not} yield a boundary for the (twisted) color path integral, as there is no way to extend the action of Eq.~\eqref{eq:twisted_action} in a gauge-invariant way.
The chain complex also yields a boundary for the 3D color code, which represents the same topological phase as the Pauli-$X$ boundary~\cite{Song2024}.
The fact that the action is not gauge invariant on the boundary is analogous to the fact that this color-code boundary does not admit a transversal $\ovl T$ gate.%
\footnote{Even though the same chain complexes can be used to define boundaries for both the color path integral and the 3D color code, their classifications are fundamentally different:
In general, the boundaries of the 3D color code correspond to the boundaries of a 4D path integral.}

\myparagraph{3D all-rough boundary}
To obtain a different boundary type we can, instead of adding generators to $X_0$, regain local exactness by adding generators to $X_2$ and $X_3$.
This yields the cohomology theory shown in Eq.~\eqref{eq:colorcohomology_allrough_boundary}.
In this case, the resulting chain complex does define a boundary for the untwisted as well as the twisted color path integral.
The boundary action looks the same as the bulk one in Eq.~\eqref{eq:twisted_action}, where the summation runs over all vertices $p\in \Delta_0$ including those on the boundary.
Gauge invariance follows directly, since it is equivalent to the bulk gauge invariance for the special case where all vertices $p$ beyond the boundary are set to $A_p=0$.
The chain complex also defines a boundary for the 3D color code, which is equivalent to a Pauli-$Z$ boundary~\cite{Song2024}.

\myparagraph{3D color boundary}
Instead of adding all locally supported generators to either $X_0$ or to $X_2$ and $X_3$, we can add or remove some generators for both.
There are two such families of boundaries.
The first boundary can be defined for every color $c$, where we set $c=r$ (red) without loss of generality.
The cohomology theory is discussed around Eq.~\eqref{eq:colorcohomology_color_boundary}.
The chain complex obtained in this way defines a boundary for the untwisted as well as the twisted color path integral.
The resulting gauge-invariant action extends from Eq.~\eqref{eq:twisted_action} to the boundary as shown in Eq.~\eqref{eq:color_boundary_action}.
As in the 2D case, we can rewrite the action as
\begin{equation}
\label{eq:color_boundary_action_appendix}
\begin{aligned}
S^{\mbd,r}[A] &= \frac18 \sum_{p\in \Delta_0-\Delta_0^{\bdin,r}} \sgn_p \ovl{A_p}\\
&= \frac18 \sum_{p\in \Delta_0-\Delta_0^{\mbd}+\Delta_0^{\bdout,r}} \sgn_p \ovl{A_p}\;,
\end{aligned}
\end{equation}
since $(dA)_e=0$ for any $e\in \Delta_1^{\mbd,\ovl r}$.
It suffices to verify the gauge invariance $S^{\mbd,r}[A+\alpha]=S^{\mbd,r}[A]$ for $\alpha$ equal to a single generator $v\in X_0$.
For volumes $v\in \Delta_3^{\bdin,g}$, the action on the adjacent vertices is the same as for any bulk volume if we use the first line in Eq.~\eqref{eq:color_boundary_action_appendix}, so gauge invariance follows as in the bulk.
For volumes $v\in \Delta_3^{\bdout,g}$, the action on the adjacent vertices is zero if we use the second line in Eq.~\eqref{eq:color_boundary_action_appendix}, so gauge invariance follows trivially.
The analogous holds for $b$ or $y$ instead of $g$.

The chain complex above also yields a boundary for the 3D color code, which is in the same phase as the \emph{color boundary}~\cite{Bombin2006}.
Conventionally, the color boundary is defined on a 3-colex whose boundary forms the surface of a single large $r$ volume on the outside, which is a 2-colex.
In this case, we have $\Delta_1^{\mbd,\ovl r}=\varnothing$, $\Delta_3^{\bdin,r}=\varnothing$, and $\Delta_0^{\bdout,r}=\varnothing$, and the chain complex and action are slightly simplified.
The color boundaries we have defined generalize these boundaries, by defining them on arbitrary surfaces in the 3-colex.
The action in Eq.~\eqref{eq:color_boundary_action_appendix} is analogous to the phases introduced by a transversal $\ovl T$ gate acting on the 3D color code and its color boundary.

Finally, we note that an alternative form of the chain complex is given by exchanging
\begin{equation}
\label{eq:colorcohomology_color_boundary_alt}
\begin{aligned}
X_2&\coloneqq\Delta_2+\Delta_1^{\mbd,\ovl r}\;,\\
X_3&\coloneqq\bigsqcup_{c\in\col_1} \Delta_3^c\times \ovl c +\Delta_2^{\mbd,gb}+\Delta_2^{\mbd,gy}+\Delta_2^{\mbd,by}\;.
\end{aligned}
\end{equation}
That is, instead of removing faces from $X_2$, we add the same faces to $X_3$.

The $ZX$ tensor network corresponding to the $r$ color boundary is shown in Section~\ref{sec:simple_example} in Fig.~\ref{fig:r_boundary_zx}.
In fact, Fig.~\ref{fig:r_boundary_zx} corresponds to a mix of Eq.~\eqref{eq:colorcohomology_color_boundary} and Eq.~\eqref{eq:colorcohomology_color_boundary_alt}, where some of the faces of $\Delta_2^{\mbd,gb}+\Delta_2^{\mbd,gy}+\Delta_2^{\mbd,by}$ have been removed from $X_2$ but not all.
The $Z$-tensor phases in Fig.~\ref{fig:r_boundary_zx} correspond to the second line in Eq.~\eqref{eq:color_boundary_action_appendix}.

\myparagraph{3D double-color boundary}
The second type of boundary that is neither all-smooth nor all-rough can be defined for every color pair $\kappa$.
Without loss of generality, we choose $\kappa=rg$.
The cohomology theory is discussed around Eq.~\eqref{eq:colorcohomology_doublecolor_boundary}.
Compared to the all-smooth boundary in Eq.~\eqref{eq:colorcohomology_free_boundary}, we have removed all boundary faces from $X_2$, and added both the $\ovl r$ and $\ovl g$-colored boundary edges.
Again, this chain complex defines a boundary of both the untwisted and twisted color path integral, and the gauge-invariant boundary action looks the same as in Eq.~\eqref{eq:color_boundary_action_appendix}.

In total, we have found 11 distinct boundaries of the 3D twisted color path integral corresponding to three families, namely 1 all-rough, 4 color, and 6 double-color boundaries.

\myparagraph{Boundaries of the twisted cellular path integral}
In order to show that boundaries defined above capture all boundary topological phases, we relate them to well-studied boundaries of the twisted cellular path integral, which we briefly review as follows.
There are two possibilities of extending $\zz_2$-valued 3D cellular cohomology to a boundary in a locally exact way:
The \emph{smooth boundary cohomology} is defined on a 3-cellulation with boundary, and looks the same as in Eq.~\eqref{eq:cellular_cohomology}, where $X_i$ also includes the $i$-cells inside the boundary.
In contrast, for the \emph{rough boundary cohomology}, $X_i$ does not include the $i$-cells inside the boundary.
The rough boundary cohomology is equivalent to the smooth boundary cohomology on the Poincar\'e dual lattice.
When we picture 1-cocycles and 2-cocycles as closed-membrane and closed-loop patterns, respectively, then the smooth boundary corresponds to one where these membrane and loops can freely terminate at the boundary.

If we have three copies of $\zz_2$-valued cellular cohomology, each copy may either have a rough or a smooth boundary, yielding 8 different boundaries.
However, there are additional boundaries that couple the different copies.
In general, there is one boundary cohomology for every subgroup $H\subset \zz_2^3$:
It is defined by restricting any $i$-(co-)chain to $H$ on all $i$-cells that are part of the boundary.
Not all boundary cohomology theories yield boundaries of the twisted cellular path integral in Eq.~\eqref{eq:3d_cellular_path_integral}, since there may be no gauge-invariant way of extending the action in Eq.~\eqref{eq:3d_cellular_action} to the boundary.
For example, this is not possible for the triple-smooth boundary corresponding to $H=\zz_2^3$.
In the intuitive closed-membrane picture discussed in Section~\ref{sec:path_integral_equivalence}, this can be shown as follows:
Consider a point on the boundary where the termination lines of the membranes $A^r$ and $A^g$ intersect.
There is a gauge transformation that adds a small semi-sphere to the closed-membrane pattern $A^b$ around this point on the boundary.
Such a gauge transformation adds a triple-intersection between the $A^r$, $A^g$, and $A^b$ membranes, and thus adds $\frac12$ to the action in Eq.~\eqref{eq:3d_cellular_action}.%
\footnote{Note that the cellular ``twist'' $(-1)^{A^r\cup A^g\cup A^b}$ is described by a group 3-cocycle $\omega\in H^3(B\zz_2^3,U(1))$, and a boundary is described by a subgroup $H$ and a group 2-cocycle $\psi\in H^2(B\zz_2^3, U(1))$ such that $d\psi=\omega|_H$.
So we only get boundaries if $\omega|_H$ is a cohomologically trivial group 3-cocycle, which is the case for subgroups $H\subset \zz_2^3$ that do not contain $(1,1,1)\in\zz_2\times\zz_2\times\zz_2$.
Note that for $H\sim\zz_2^2$, there are two distinct group 2-cohomology classes corresponding to two distinct boundaries, but these distinct boundaries turn out to belong to the same boundary topological phase nonetheless~\cite{Davydova2025}.}
The subgroups $H$ that allow for a gauge-invariant action are given by
\begin{itemize}
\item the trivial subgroup $\langle \varnothing \rangle$,
\item the 6 $\zz_2$ subgroups generated by
\begin{equation}
\begin{gathered}
\langle (1,0,0)\rangle,\quad  \langle (0,1,0)\rangle,\quad  \langle (0,0,1)\rangle,\\  \langle (1,1,0)\rangle,\quad  \langle (0,1,1)\rangle,\quad \langle (1,0,1)\rangle\;,
\end{gathered}
\end{equation}
\item and the 4 $\zz_2^2$ subgroups generated by
\begin{equation}
\begin{gathered}
\langle (1,0,0), (0,1,0)\rangle,\quad  \langle (1,0,0), (0,0,1)\rangle,\\  \langle (0,1,0), (0,0,1)\rangle,\quad \langle (1,1,0), (0,1,1)\rangle\;.
\end{gathered}
\end{equation}
\end{itemize}
All in all, we find that there are 11 distinct boundary phases for the type-III twisted cellular path integral~\cite{Bullivant2017,Magdalena2023}.

\myparagraph{Equivalence to cellular boundaries}
We have found 11 different boundaries for the 3D color path integral, and also 11 different boundaries of the twisted cellular path integral.
Indeed, if we fix one way to map between the cellular and the color path integral, then there is a one-to-one correspondence between the two sets of boundaries.
If we use the mappings $M^r$, $M^g$, and $M^b$ from Section~\ref{sec:color_cohomology_equivalence}, then the one-to-one correspondence is shown in Fig.~\ref{fig:boundary_equivalence}.
We will not give full chain maps for all the equivalences but instead illustrate how color 1-cocycles $A$ are mapped to cellular 1-cocycles $A^r\coloneqq M^rA$, $A^g\coloneqq M^gA$, and $A^b\coloneqq M^bA$.
The $r$-color boundary adds the $\ovl r$-colored edges to $X_2$, which forces $A^r=0$ on the boundary as $A^r$ is defined on these edges.
$A^g$ and $A^b$ on the other hand are unrestricted, so this corresponds to the cellular boundary with $H=\langle (0,1,0), (0,0,1)\rangle$.
The analogous holds for the $g$ and $b$-color boundaries.
Since the sum $M^r+M^g+M^b+M^y$ is trivial, the $y$-color boundary corresponds to one where $A^y=A^r+A^g+A^b=0$ at the boundary, corresponding to $H=\langle (1,1,0), (0,1,1)\rangle$.
The $rg$-color boundary adds both the $\ovl r$ and $\ovl g$ edges to $X_2$, which forces $A^r=0$ and $A^g=0$ on the boundary, corresponding to $H=\langle (0,0,1)\rangle$.
The $ry$-color boundary forces $A^r=0$ and $A^y=A^r+A^g+A^b=0$ on the boundary, corresponding to $H=\langle (0,1,1)\rangle$.
The other double-color boundaries are analogous.
The equivalence establishes that we have found path integrals for all 11 boundary phases of the color path integral.

\begin{figure}
\begin{tabular}{l|l}
Color boundary & Cellular boundary\\
\hline
All-rough boundary & $H=\langle\varnothing\rangle$\\
$r$-color boundary & $H=\langle(0,1,0),(0,0,1)\rangle$\\
$g$-color boundary & $H=\langle(1,0,0),(0,0,1)\rangle$\\
$b$-color boundary & $H=\langle(1,0,0),(0,1,0)\rangle$\\
$y$-color boundary & $H=\langle(1,1,0),(0,1,1)\rangle$\\
$rg$-double-color boundary & $H=\langle(0,0,1)\rangle$\\
$rb$-double-color boundary & $H=\langle(0,1,0)\rangle$\\
$gb$-double-color boundary & $H=\langle(1,0,0)\rangle$\\
$ry$-double-color boundary & $H=\langle(0,1,1)\rangle$\\
$gy$-double-color boundary & $H=\langle(1,0,1)\rangle$\\
$by$-double-color boundary & $H=\langle(1,1,0)\rangle$
\end{tabular}
\caption{
Equivalence between boundaries of the color path integral (left) and boundaries of the type-III twisted path integral (right), defined by subgroup $H\subset \zz_2^3$.}
\label{fig:boundary_equivalence}
\end{figure}

\section{Fault-tolerance for arbitrary local noise}
\label{app:local_fault_tolerance}
In this appendix, we will present a fault-tolerance argument for arbitrary $p$-bounded local noise in the untwisted color circuits.
Such an argument can be made \emph{without} expressing the noisy circuit in terms of Pauli errors, following the proof in Ref.~\cite{twisted_double_code}.
We assume a general noise model where the unitary channel or measurement instrument $C_0^{(l)}$ at the spacetime location $l\in \mathcal L$ in the circuit may be replaced with one of a list of faulty channels or instruments $\{C_x^{(l)}\}_{1\leq x<n}$:
Let $\vec x$ be an error configuration, that is, at each location $l\in \mathcal L$, we apply $C_{\vec x_l}^{(l)}$ instead of the noise-free channel $C_0^{(l)}$.
In other words, all locations with $\vec x_l\neq 0$ experience an error.
The noisy circuit is given by applying the error configuration $\vec x$ with probability $P(\vec x)$.
We assume that $P$ is \emph{$p$-bounded}, meaning that
\begin{equation}
\label{eq:error_probabilities}
\sum_{\vec x: \vec x_l\neq 0\;\forall l\in L} P(\vec x) < p^{|L|} \qquad \forall L\subset \mathcal L\;.
\end{equation}
for some small enough $p$.
Note that $p$-bounded noise includes on-site circuit-level noise, but also includes correlated noise to some extent.

Now, the conditions in Eqs.~\eqref{eq:flux_gauss_law} and \eqref{eq:charge_gauss_law} imply that the measured flux and charge configuration always obeys $(db)_{(v,c)}= 0$ or $(d^Tc)_v= 0$ on every volume $v$ that is separated from any error location $l$ ($\vec x_l\neq 0$) by more than some constant lattice distance.
That is, if $R_{\vec x}$ denotes the region obtained by padding the set of error locations $\{l: \vec x_l\neq 0\}$ by some constant-size error-free margins, then $db$ and $d^Tc$ are only supported on $R_{\vec x}$.
This is because outside of $R_{\vec x}$, the circuit equals the path integral, which is zero if $(db)_{(v,c)}\neq 0$ or $(d^Tc)_v\neq 0$, and so the probability for measuring such a $b$ or $c$ is zero as well.
This allows us to write down sufficient conditions for when a noisy circuit does not introduce a logical error:
\begin{enumerate}
\item $R_{\vec x}$ decomposes into topologically trivial components each contained inside a 3-ball.
\item $\widetilde b$ ($\widetilde c$) are homologically equal to a color 1-cochain $b'$ (1-chain $c'$) supported only inside $R_{\vec x}$.
\end{enumerate}
This is because the fixed-point condition in Eq.~\eqref{eq:fixed_point_property} of the underlying path integral together with (1) implies that the circuit with errors is equal to the circuit without errors, with measurement outcomes given by $b+b'$ ($c+c'$).
(2) then implies that $\widetilde b+b+b_c$ ($\widetilde c+c+c_c$) has the same (co)homology class as $b'+b+b_c$ ($c'+c+c_c$), which by construction is the trivial one.
Due to the invariance in Eqs.~\eqref{eq:flux_invariance} and \eqref{eq:charge_invariance}, the post-selected circuit with corrections is thus equal to the clean, error-free, path integral without any flux or charge configuration.

It is left to be shown that the configurations of $\vec x$, $b$, and $\widetilde b$ that violate the conditions (1) or (2) are exponentially unlikely in the lattice size $L$.
Here we assume that the hypergraph matching algorithm really yields the exact minimum-weight configuration, but the argument can be adapted if this is only approximately the case.
Assume that we have a configuration of $\vec x$, $b$, and $c$ violating (1) or (2).
Then, $b'+\widetilde b$ ($c'+\widetilde c$) is a 2-cocycle (1-cycle) of non-trivial (co)homology such that $|b'+\widetilde b|\cap R_{\vec x} < \frac12 |b'+\widetilde b|$, where $|\bullet|$ denotes the weight.
This is because otherwise, $b'$ ($c'$) has a lower weight than $\widetilde b$ ($\widetilde c$) which contradicts the assumption that $\widetilde b$ ($\widetilde c$) is of minimum weight.
The cohomologically non-trivial configuration $b'+\widetilde b$ ($c'+\widetilde c$) can be pictured as a non-contractible loop configuration.
The number of closed loops of length $L$ is at most $\alpha^l$ for some constant $\alpha$, and the probability for a loop that is covered half by errors scales like $p^{l/\eta}$ for some constant $\eta$, due to Eq.~\eqref{eq:error_probabilities}.
So the overall probability of a length-$l$ loop scales like $(p^{1/\eta}\alpha)^l$.
Since the size of a non-contractible loop scales like the lattice size, $l\sim L$, the probability of a logical error vanishes exponentially quickly in $L$ if $p^{1/\eta}\alpha<1$.

\bibliography{fourcolor_references}{}

%apsrev4-2.bst 2019-01-14 (MD) hand-edited version of apsrev4-1.bst
%Control: key (0)
%Control: author (8) initials jnrlst
%Control: editor formatted (1) identically to author
%Control: production of article title (0) allowed
%Control: page (0) single
%Control: year (1) truncated
%Control: production of eprint (0) enabled
\begin{thebibliography}{78}%
\makeatletter
\providecommand \@ifxundefined [1]{%
 \@ifx{#1\undefined}
}%
\providecommand \@ifnum [1]{%
 \ifnum #1\expandafter \@firstoftwo
 \else \expandafter \@secondoftwo
 \fi
}%
\providecommand \@ifx [1]{%
 \ifx #1\expandafter \@firstoftwo
 \else \expandafter \@secondoftwo
 \fi
}%
\providecommand \natexlab [1]{#1}%
\providecommand \enquote  [1]{``#1''}%
\providecommand \bibnamefont  [1]{#1}%
\providecommand \bibfnamefont [1]{#1}%
\providecommand \citenamefont [1]{#1}%
\providecommand \href@noop [0]{\@secondoftwo}%
\providecommand \href [0]{\begingroup \@sanitize@url \@href}%
\providecommand \@href[1]{\@@startlink{#1}\@@href}%
\providecommand \@@href[1]{\endgroup#1\@@endlink}%
\providecommand \@sanitize@url [0]{\catcode `\\12\catcode `\$12\catcode
  `\&12\catcode `\#12\catcode `\^12\catcode `\_12\catcode `\%12\relax}%
\providecommand \@@startlink[1]{}%
\providecommand \@@endlink[0]{}%
\providecommand \url  [0]{\begingroup\@sanitize@url \@url }%
\providecommand \@url [1]{\endgroup\@href {#1}{\urlprefix }}%
\providecommand \urlprefix  [0]{URL }%
\providecommand \Eprint [0]{\href }%
\providecommand \doibase [0]{https://doi.org/}%
\providecommand \selectlanguage [0]{\@gobble}%
\providecommand \bibinfo  [0]{\@secondoftwo}%
\providecommand \bibfield  [0]{\@secondoftwo}%
\providecommand \translation [1]{[#1]}%
\providecommand \BibitemOpen [0]{}%
\providecommand \bibitemStop [0]{}%
\providecommand \bibitemNoStop [0]{.\EOS\space}%
\providecommand \EOS [0]{\spacefactor3000\relax}%
\providecommand \BibitemShut  [1]{\csname bibitem#1\endcsname}%
\let\auto@bib@innerbib\@empty
%</preamble>
\bibitem [{\citenamefont {Kitaev}(2003)}]{Kitaev1997}%
  \BibitemOpen
  \bibfield  {author} {\bibinfo {author} {\bibfnamefont {A.~Y.}\ \bibnamefont
  {Kitaev}},\ }\bibfield  {title} {\bibinfo {title} {Fault-tolerant quantum
  computation by anyons},\ }\href
  {https://doi.org/10.1016/S0003-4916(02)00018-0} {\bibfield  {journal}
  {\bibinfo  {journal} {Ann. Phys.}\ }\textbf {\bibinfo {volume} {303}},\
  \bibinfo {pages} {2 } (\bibinfo {year} {2003})},\ \Eprint
  {https://arxiv.org/abs/quant-ph/9707021} {arXiv:quant-ph/9707021}
  \BibitemShut {NoStop}%
\bibitem [{\citenamefont {Dennis}\ \emph {et~al.}(2002)\citenamefont {Dennis},
  \citenamefont {Kitaev}, \citenamefont {Landahl},\ and\ \citenamefont
  {Preskill}}]{Dennis2001}%
  \BibitemOpen
  \bibfield  {author} {\bibinfo {author} {\bibfnamefont {E.}~\bibnamefont
  {Dennis}}, \bibinfo {author} {\bibfnamefont {A.}~\bibnamefont {Kitaev}},
  \bibinfo {author} {\bibfnamefont {A.}~\bibnamefont {Landahl}},\ and\ \bibinfo
  {author} {\bibfnamefont {J.}~\bibnamefont {Preskill}},\ }\bibfield  {title}
  {\bibinfo {title} {Topological quantum memory},\ }\href
  {https://doi.org/10.1063/1.1499754} {\bibfield  {journal} {\bibinfo
  {journal} {J. Math. Phys.}\ }\textbf {\bibinfo {volume} {43}},\ \bibinfo
  {pages} {4452} (\bibinfo {year} {2002})},\ \Eprint
  {https://arxiv.org/abs/quant-ph/0110143} {arXiv:quant-ph/0110143}
  \BibitemShut {NoStop}%
\bibitem [{\citenamefont {Bravyi}\ and\ \citenamefont
  {Kitaev}(1998)}]{Bravyi1998}%
  \BibitemOpen
  \bibfield  {author} {\bibinfo {author} {\bibfnamefont {S.~B.}\ \bibnamefont
  {Bravyi}}\ and\ \bibinfo {author} {\bibfnamefont {A.~Y.}\ \bibnamefont
  {Kitaev}},\ }\href@noop {} {\bibinfo {title} {Quantum codes on a lattice with
  boundary}} (\bibinfo {year} {1998}),\ \Eprint
  {https://arxiv.org/abs/quant-ph/9811052} {arXiv:quant-ph/9811052}
  \BibitemShut {NoStop}%
\bibitem [{\citenamefont {Bombin}\ and\ \citenamefont
  {Martin-Delgado}(2006)}]{Bombin2006b}%
  \BibitemOpen
  \bibfield  {author} {\bibinfo {author} {\bibfnamefont {H.}~\bibnamefont
  {Bombin}}\ and\ \bibinfo {author} {\bibfnamefont {M.~A.}\ \bibnamefont
  {Martin-Delgado}},\ }\bibfield  {title} {\bibinfo {title} {Topological
  quantum distillation},\ }\href
  {https://doi.org/10.1103/PhysRevLett.97.180501} {\bibfield  {journal}
  {\bibinfo  {journal} {Phys. Rev. Lett.}\ }\textbf {\bibinfo {volume} {97}},\
  \bibinfo {pages} {180501} (\bibinfo {year} {2006})},\ \Eprint
  {https://arxiv.org/abs/quant-ph/0605138} {arXiv:quant-ph/0605138}
  \BibitemShut {NoStop}%
\bibitem [{\citenamefont {et~al.}(2024)}]{Lacroix2024}%
  \BibitemOpen
  \bibfield  {author} {\bibinfo {author} {\bibfnamefont {N.~L.}\ \bibnamefont
  {et~al.}},\ }\href@noop {} {\bibinfo {title} {Scaling and logic in the color
  code on a superconducting quantum processor}} (\bibinfo {year} {2024}),\
  \Eprint {https://arxiv.org/abs/2412.14256} {arXiv:2412.14256} \BibitemShut
  {NoStop}%
\bibitem [{\citenamefont {Krinner}\ \emph {et~al.}(2022)\citenamefont
  {Krinner}, \citenamefont {Lacroix}, \citenamefont {Remm}, \citenamefont
  {Di~Paolo}, \citenamefont {Genois}, \citenamefont {Leroux}, \citenamefont
  {Hellings}, \citenamefont {Lazar}, \citenamefont {Swiadek}, \citenamefont
  {Herrmann}, \citenamefont {Norris}, \citenamefont {Andersen}, \citenamefont
  {M{\ifmmode\ddot{u}\else\"{u}\fi}ller}, \citenamefont {Blais}, \citenamefont
  {Eichler},\ and\ \citenamefont {Wallraff}}]{Krinner2022}%
  \BibitemOpen
  \bibfield  {author} {\bibinfo {author} {\bibfnamefont {S.}~\bibnamefont
  {Krinner}}, \bibinfo {author} {\bibfnamefont {N.}~\bibnamefont {Lacroix}},
  \bibinfo {author} {\bibfnamefont {A.}~\bibnamefont {Remm}}, \bibinfo {author}
  {\bibfnamefont {A.}~\bibnamefont {Di~Paolo}}, \bibinfo {author}
  {\bibfnamefont {E.}~\bibnamefont {Genois}}, \bibinfo {author} {\bibfnamefont
  {C.}~\bibnamefont {Leroux}}, \bibinfo {author} {\bibfnamefont
  {C.}~\bibnamefont {Hellings}}, \bibinfo {author} {\bibfnamefont
  {S.}~\bibnamefont {Lazar}}, \bibinfo {author} {\bibfnamefont
  {F.}~\bibnamefont {Swiadek}}, \bibinfo {author} {\bibfnamefont
  {J.}~\bibnamefont {Herrmann}}, \bibinfo {author} {\bibfnamefont {G.~J.}\
  \bibnamefont {Norris}}, \bibinfo {author} {\bibfnamefont {C.~K.}\
  \bibnamefont {Andersen}}, \bibinfo {author} {\bibfnamefont {M.}~\bibnamefont
  {M{\ifmmode\ddot{u}\else\"{u}\fi}ller}}, \bibinfo {author} {\bibfnamefont
  {A.}~\bibnamefont {Blais}}, \bibinfo {author} {\bibfnamefont
  {C.}~\bibnamefont {Eichler}},\ and\ \bibinfo {author} {\bibfnamefont
  {A.}~\bibnamefont {Wallraff}},\ }\bibfield  {title} {\bibinfo {title}
  {Realizing repeated quantum error correction in a distance-three surface
  code},\ }\href {https://doi.org/10.1038/s41586-022-04566-8} {\bibfield
  {journal} {\bibinfo  {journal} {Nature}\ }\textbf {\bibinfo {volume} {605}},\
  \bibinfo {pages} {669} (\bibinfo {year} {2022})}\BibitemShut {NoStop}%
\bibitem [{\citenamefont {et~al}(2024)}]{Acharya2024}%
  \BibitemOpen
  \bibfield  {author} {\bibinfo {author} {\bibfnamefont {R.~A.}\ \bibnamefont
  {et~al}},\ }\href@noop {} {\bibinfo {title} {Quantum error correction below
  the surface code threshold}} (\bibinfo {year} {2024}),\ \Eprint
  {https://arxiv.org/abs/2408.13687} {arXiv:2408.13687} \BibitemShut {NoStop}%
\bibitem [{\citenamefont {Horsman}\ \emph {et~al.}(2012)\citenamefont
  {Horsman}, \citenamefont {Fowler}, \citenamefont {Devitt},\ and\
  \citenamefont {Meter}}]{Horsman2011}%
  \BibitemOpen
  \bibfield  {author} {\bibinfo {author} {\bibfnamefont {D.}~\bibnamefont
  {Horsman}}, \bibinfo {author} {\bibfnamefont {A.~G.}\ \bibnamefont {Fowler}},
  \bibinfo {author} {\bibfnamefont {S.}~\bibnamefont {Devitt}},\ and\ \bibinfo
  {author} {\bibfnamefont {R.~V.}\ \bibnamefont {Meter}},\ }\bibfield  {title}
  {\bibinfo {title} {Surface code quantum computing by lattice surgery},\
  }\href {https://doi.org/10.1088/1367-2630/14/12/123011} {\bibfield  {journal}
  {\bibinfo  {journal} {New J. Phys.}\ }\textbf {\bibinfo {volume} {14}},\
  \bibinfo {pages} {123011} (\bibinfo {year} {2012})},\ \Eprint
  {https://arxiv.org/abs/1111.4022} {arXiv:1111.4022} \BibitemShut {NoStop}%
\bibitem [{\citenamefont {Litinski}(2019{\natexlab{a}})}]{Litinski2018}%
  \BibitemOpen
  \bibfield  {author} {\bibinfo {author} {\bibfnamefont {D.}~\bibnamefont
  {Litinski}},\ }\bibfield  {title} {\bibinfo {title} {A game of surface codes:
  Large-scale quantum computing with lattice surgery},\ }\href
  {https://doi.org/10.22331/q-2019-03-05-128} {\bibfield  {journal} {\bibinfo
  {journal} {Quantum}\ }\textbf {\bibinfo {volume} {3}},\ \bibinfo {pages}
  {128} (\bibinfo {year} {2019}{\natexlab{a}})},\ \Eprint
  {https://arxiv.org/abs/arXiv:1808.02892} {arXiv:1808.02892} \BibitemShut
  {NoStop}%
\bibitem [{\citenamefont {Landahl}\ and\ \citenamefont
  {Ryan-Anderson}(2014)}]{Landahl2014}%
  \BibitemOpen
  \bibfield  {author} {\bibinfo {author} {\bibfnamefont {A.~J.}\ \bibnamefont
  {Landahl}}\ and\ \bibinfo {author} {\bibfnamefont {C.}~\bibnamefont
  {Ryan-Anderson}},\ }\href@noop {} {\bibinfo {title} {Quantum computing by
  color-code lattice surgery}} (\bibinfo {year} {2014}),\ \Eprint
  {https://arxiv.org/abs/1407.5103} {arXiv:1407.5103} \BibitemShut {NoStop}%
\bibitem [{\citenamefont {Thomsen}\ \emph {et~al.}(2024)\citenamefont
  {Thomsen}, \citenamefont {Kesselring}, \citenamefont {Bartlett},\ and\
  \citenamefont {Brown}}]{Thomsen2022}%
  \BibitemOpen
  \bibfield  {author} {\bibinfo {author} {\bibfnamefont {F.}~\bibnamefont
  {Thomsen}}, \bibinfo {author} {\bibfnamefont {M.~S.}\ \bibnamefont
  {Kesselring}}, \bibinfo {author} {\bibfnamefont {S.~D.}\ \bibnamefont
  {Bartlett}},\ and\ \bibinfo {author} {\bibfnamefont {B.~J.}\ \bibnamefont
  {Brown}},\ }\bibfield  {title} {\bibinfo {title} {Low-overhead quantum
  computing with the color code},\ }\href
  {https://doi.org/10.1103/PhysRevResearch.6.043125} {\bibfield  {journal}
  {\bibinfo  {journal} {Phys. Rev. Research}\ }\textbf {\bibinfo {volume}
  {6}},\ \bibinfo {pages} {043125} (\bibinfo {year} {2024})},\ \Eprint
  {https://arxiv.org/abs/2201.07806} {2201.07806} \BibitemShut {NoStop}%
\bibitem [{\citenamefont {Ryan-Anderson}\ \emph {et~al.}(2022)\citenamefont
  {Ryan-Anderson}, \citenamefont {Brown}, \citenamefont {Allman}, \citenamefont
  {Arkin}, \citenamefont {Asa-Attuah}, \citenamefont {Baldwin}, \citenamefont
  {Berg}, \citenamefont {Bohnet}, \citenamefont {Braxton}, \citenamefont
  {Burdick}, \citenamefont {Campora}, \citenamefont {Chernoguzov},
  \citenamefont {Esposito}, \citenamefont {Evans}, \citenamefont {Francois},
  \citenamefont {Gaebler}, \citenamefont {Gatterman}, \citenamefont {Gerber},
  \citenamefont {Gilmore}, \citenamefont {Gresh}, \citenamefont {Hall},
  \citenamefont {Hankin}, \citenamefont {Hostetter}, \citenamefont {Lucchetti},
  \citenamefont {Mayer}, \citenamefont {Myers}, \citenamefont {Neyenhuis},
  \citenamefont {Santiago}, \citenamefont {Sedlacek}, \citenamefont {Skripka},
  \citenamefont {Slattery}, \citenamefont {Stutz}, \citenamefont {Tait},
  \citenamefont {Tobey}, \citenamefont {Vittorini}, \citenamefont {Walker},\
  and\ \citenamefont {Hayes}}]{Ryan2022}%
  \BibitemOpen
  \bibfield  {author} {\bibinfo {author} {\bibfnamefont {C.}~\bibnamefont
  {Ryan-Anderson}}, \bibinfo {author} {\bibfnamefont {N.~C.}\ \bibnamefont
  {Brown}}, \bibinfo {author} {\bibfnamefont {M.~S.}\ \bibnamefont {Allman}},
  \bibinfo {author} {\bibfnamefont {B.}~\bibnamefont {Arkin}}, \bibinfo
  {author} {\bibfnamefont {G.}~\bibnamefont {Asa-Attuah}}, \bibinfo {author}
  {\bibfnamefont {C.}~\bibnamefont {Baldwin}}, \bibinfo {author} {\bibfnamefont
  {J.}~\bibnamefont {Berg}}, \bibinfo {author} {\bibfnamefont {J.~G.}\
  \bibnamefont {Bohnet}}, \bibinfo {author} {\bibfnamefont {S.}~\bibnamefont
  {Braxton}}, \bibinfo {author} {\bibfnamefont {N.}~\bibnamefont {Burdick}},
  \bibinfo {author} {\bibfnamefont {J.~P.}\ \bibnamefont {Campora}}, \bibinfo
  {author} {\bibfnamefont {A.}~\bibnamefont {Chernoguzov}}, \bibinfo {author}
  {\bibfnamefont {J.}~\bibnamefont {Esposito}}, \bibinfo {author}
  {\bibfnamefont {B.}~\bibnamefont {Evans}}, \bibinfo {author} {\bibfnamefont
  {D.}~\bibnamefont {Francois}}, \bibinfo {author} {\bibfnamefont {J.~P.}\
  \bibnamefont {Gaebler}}, \bibinfo {author} {\bibfnamefont {T.~M.}\
  \bibnamefont {Gatterman}}, \bibinfo {author} {\bibfnamefont {J.}~\bibnamefont
  {Gerber}}, \bibinfo {author} {\bibfnamefont {K.}~\bibnamefont {Gilmore}},
  \bibinfo {author} {\bibfnamefont {D.}~\bibnamefont {Gresh}}, \bibinfo
  {author} {\bibfnamefont {A.}~\bibnamefont {Hall}}, \bibinfo {author}
  {\bibfnamefont {A.}~\bibnamefont {Hankin}}, \bibinfo {author} {\bibfnamefont
  {J.}~\bibnamefont {Hostetter}}, \bibinfo {author} {\bibfnamefont
  {D.}~\bibnamefont {Lucchetti}}, \bibinfo {author} {\bibfnamefont
  {K.}~\bibnamefont {Mayer}}, \bibinfo {author} {\bibfnamefont
  {J.}~\bibnamefont {Myers}}, \bibinfo {author} {\bibfnamefont
  {B.}~\bibnamefont {Neyenhuis}}, \bibinfo {author} {\bibfnamefont
  {J.}~\bibnamefont {Santiago}}, \bibinfo {author} {\bibfnamefont
  {J.}~\bibnamefont {Sedlacek}}, \bibinfo {author} {\bibfnamefont
  {T.}~\bibnamefont {Skripka}}, \bibinfo {author} {\bibfnamefont
  {A.}~\bibnamefont {Slattery}}, \bibinfo {author} {\bibfnamefont {R.~P.}\
  \bibnamefont {Stutz}}, \bibinfo {author} {\bibfnamefont {J.}~\bibnamefont
  {Tait}}, \bibinfo {author} {\bibfnamefont {R.}~\bibnamefont {Tobey}},
  \bibinfo {author} {\bibfnamefont {G.}~\bibnamefont {Vittorini}}, \bibinfo
  {author} {\bibfnamefont {J.}~\bibnamefont {Walker}},\ and\ \bibinfo {author}
  {\bibfnamefont {D.}~\bibnamefont {Hayes}},\ }\href@noop {} {\bibinfo {title}
  {Implementing fault-tolerant entangling gates on the five-qubit code and the
  color code}} (\bibinfo {year} {2022}),\ \Eprint
  {https://arxiv.org/abs/2208.01863} {arXiv:2208.01863} \BibitemShut {NoStop}%
\bibitem [{\citenamefont {Bluvstein}\ \emph {et~al.}(2024)\citenamefont
  {Bluvstein}, \citenamefont {Evered}, \citenamefont {Geim}, \citenamefont
  {Li}, \citenamefont {Zhou}, \citenamefont {Manovitz}, \citenamefont {Ebadi},
  \citenamefont {Cain}, \citenamefont {Kalinowski}, \citenamefont {Hangleiter},
  \citenamefont {Bonilla~Ataides}, \citenamefont {Maskara}, \citenamefont
  {Cong}, \citenamefont {Gao}, \citenamefont {Sales~Rodriguez}, \citenamefont
  {Karolyshyn}, \citenamefont {Semeghini}, \citenamefont {Gullans},
  \citenamefont {Greiner}, \citenamefont
  {Vuleti{\ifmmode\acute{c}\else\'{c}\fi}},\ and\ \citenamefont
  {Lukin}}]{Bluvstein2024}%
  \BibitemOpen
  \bibfield  {author} {\bibinfo {author} {\bibfnamefont {D.}~\bibnamefont
  {Bluvstein}}, \bibinfo {author} {\bibfnamefont {S.~J.}\ \bibnamefont
  {Evered}}, \bibinfo {author} {\bibfnamefont {A.~A.}\ \bibnamefont {Geim}},
  \bibinfo {author} {\bibfnamefont {S.~H.}\ \bibnamefont {Li}}, \bibinfo
  {author} {\bibfnamefont {H.}~\bibnamefont {Zhou}}, \bibinfo {author}
  {\bibfnamefont {T.}~\bibnamefont {Manovitz}}, \bibinfo {author}
  {\bibfnamefont {S.}~\bibnamefont {Ebadi}}, \bibinfo {author} {\bibfnamefont
  {M.}~\bibnamefont {Cain}}, \bibinfo {author} {\bibfnamefont {M.}~\bibnamefont
  {Kalinowski}}, \bibinfo {author} {\bibfnamefont {D.}~\bibnamefont
  {Hangleiter}}, \bibinfo {author} {\bibfnamefont {J.~P.}\ \bibnamefont
  {Bonilla~Ataides}}, \bibinfo {author} {\bibfnamefont {N.}~\bibnamefont
  {Maskara}}, \bibinfo {author} {\bibfnamefont {I.}~\bibnamefont {Cong}},
  \bibinfo {author} {\bibfnamefont {X.}~\bibnamefont {Gao}}, \bibinfo {author}
  {\bibfnamefont {P.}~\bibnamefont {Sales~Rodriguez}}, \bibinfo {author}
  {\bibfnamefont {T.}~\bibnamefont {Karolyshyn}}, \bibinfo {author}
  {\bibfnamefont {G.}~\bibnamefont {Semeghini}}, \bibinfo {author}
  {\bibfnamefont {M.~J.}\ \bibnamefont {Gullans}}, \bibinfo {author}
  {\bibfnamefont {M.}~\bibnamefont {Greiner}}, \bibinfo {author} {\bibfnamefont
  {V.}~\bibnamefont {Vuleti{\ifmmode\acute{c}\else\'{c}\fi}}},\ and\ \bibinfo
  {author} {\bibfnamefont {M.~D.}\ \bibnamefont {Lukin}},\ }\bibfield  {title}
  {\bibinfo {title} {{Logical quantum processor based on reconfigurable atom
  arrays}},\ }\href {https://doi.org/10.1038/s41586-023-06927-3} {\bibfield
  {journal} {\bibinfo  {journal} {Nature}\ }\textbf {\bibinfo {volume} {626}},\
  \bibinfo {pages} {58} (\bibinfo {year} {2024})}\BibitemShut {NoStop}%
\bibitem [{\citenamefont {Bravyi}\ and\ \citenamefont
  {Kitaev}(2005)}]{Bravyi2004}%
  \BibitemOpen
  \bibfield  {author} {\bibinfo {author} {\bibfnamefont {S.}~\bibnamefont
  {Bravyi}}\ and\ \bibinfo {author} {\bibfnamefont {A.}~\bibnamefont
  {Kitaev}},\ }\bibfield  {title} {\bibinfo {title} {Universal quantum
  computation with ideal clifford gates and noisy ancillas},\ }\href
  {https://doi.org/10.1103/PhysRevA.71.022316} {\bibfield  {journal} {\bibinfo
  {journal} {Phys. Rev. A}\ }\textbf {\bibinfo {volume} {71}},\ \bibinfo
  {pages} {022316} (\bibinfo {year} {2005})},\ \Eprint
  {https://arxiv.org/abs/quant-ph/0403025} {arXiv:quant-ph/0403025}
  \BibitemShut {NoStop}%
\bibitem [{\citenamefont {Bravyi}\ and\ \citenamefont
  {Haah}(2012)}]{Bravyi2012a}%
  \BibitemOpen
  \bibfield  {author} {\bibinfo {author} {\bibfnamefont {S.}~\bibnamefont
  {Bravyi}}\ and\ \bibinfo {author} {\bibfnamefont {J.}~\bibnamefont {Haah}},\
  }\bibfield  {title} {\bibinfo {title} {Magic state distillation with low
  overhead},\ }\href {https://doi.org/10.1103/PhysRevA.86.052329} {\bibfield
  {journal} {\bibinfo  {journal} {Phys. Rev. A}\ }\textbf {\bibinfo {volume}
  {86}},\ \bibinfo {pages} {052329} (\bibinfo {year} {2012})},\ \Eprint
  {https://arxiv.org/abs/1209.2426} {arXiv:1209.2426} \BibitemShut {NoStop}%
\bibitem [{\citenamefont {Litinski}(2019{\natexlab{b}})}]{Litinski2019}%
  \BibitemOpen
  \bibfield  {author} {\bibinfo {author} {\bibfnamefont {D.}~\bibnamefont
  {Litinski}},\ }\bibfield  {title} {\bibinfo {title} {Magic state
  distillation: Not as costly as you think},\ }\href
  {https://doi.org/10.22331/q-2019-12-02-205} {\bibfield  {journal} {\bibinfo
  {journal} {Quantum}\ }\textbf {\bibinfo {volume} {3}},\ \bibinfo {pages}
  {205} (\bibinfo {year} {2019}{\natexlab{b}})},\ \Eprint
  {https://arxiv.org/abs/1905.06903} {arXiv:1905.06903} \BibitemShut {NoStop}%
\bibitem [{\citenamefont {Davydova}\ \emph {et~al.}(2025)\citenamefont
  {Davydova}, \citenamefont {Bauer}, \citenamefont {Magdalena de~la Fuente},
  \citenamefont {Webster}, \citenamefont {Williamson},\ and\ \citenamefont
  {Brown}}]{Davydova2025}%
  \BibitemOpen
  \bibfield  {author} {\bibinfo {author} {\bibfnamefont {M.}~\bibnamefont
  {Davydova}}, \bibinfo {author} {\bibfnamefont {A.}~\bibnamefont {Bauer}},
  \bibinfo {author} {\bibfnamefont {J.~C.}\ \bibnamefont {Magdalena de~la
  Fuente}}, \bibinfo {author} {\bibfnamefont {M.}~\bibnamefont {Webster}},
  \bibinfo {author} {\bibfnamefont {D.~J.}\ \bibnamefont {Williamson}},\ and\
  \bibinfo {author} {\bibfnamefont {B.~J.}\ \bibnamefont {Brown}},\ }\href@noop
  {} {\bibinfo {title} {Universal fault tolerant quantum computing in 2d
  without getting tied in knots}} (\bibinfo {year} {2025}),\ \Eprint
  {https://arxiv.org/abs/2503.15751} {arXiv:2503.15751} \BibitemShut {NoStop}%
\bibitem [{\citenamefont {Bombin}(2016)}]{Bombin2014}%
  \BibitemOpen
  \bibfield  {author} {\bibinfo {author} {\bibfnamefont {H.}~\bibnamefont
  {Bombin}},\ }\bibfield  {title} {\bibinfo {title} {Dimensional jump in
  quantum error correction},\ }\href@noop {} {\bibfield  {journal} {\bibinfo
  {journal} {New Journal of Physics}\ }\textbf {\bibinfo {volume} {18}},\
  \bibinfo {pages} {043038} (\bibinfo {year} {2016})},\ \Eprint
  {https://arxiv.org/abs/1412.5079} {arXiv:1412.5079} \BibitemShut {NoStop}%
\bibitem [{\citenamefont {Beverland}\ \emph {et~al.}(2021)\citenamefont
  {Beverland}, \citenamefont {Kubica},\ and\ \citenamefont
  {Svore}}]{Beverland2021}%
  \BibitemOpen
  \bibfield  {author} {\bibinfo {author} {\bibfnamefont {M.~E.}\ \bibnamefont
  {Beverland}}, \bibinfo {author} {\bibfnamefont {A.}~\bibnamefont {Kubica}},\
  and\ \bibinfo {author} {\bibfnamefont {K.~M.}\ \bibnamefont {Svore}},\
  }\bibfield  {title} {\bibinfo {title} {The cost of universality: A
  comparative study of the overhead of state distillation and code switching
  with color codes},\ }\href {https://doi.org/10.1103/PRXQuantum.2.020341}
  {\bibfield  {journal} {\bibinfo  {journal} {PRX Quantum}\ }\textbf {\bibinfo
  {volume} {2}},\ \bibinfo {pages} {020341} (\bibinfo {year} {2021})},\ \Eprint
  {https://arxiv.org/abs/2101.02211} {arXiv:2101.02211} \BibitemShut {NoStop}%
\bibitem [{\citenamefont {Bombin}\ and\ \citenamefont
  {Martin-Delgado}(2007{\natexlab{a}})}]{Bombin2006}%
  \BibitemOpen
  \bibfield  {author} {\bibinfo {author} {\bibfnamefont {H.}~\bibnamefont
  {Bombin}}\ and\ \bibinfo {author} {\bibfnamefont {M.~A.}\ \bibnamefont
  {Martin-Delgado}},\ }\bibfield  {title} {\bibinfo {title} {Topological
  computation without braiding},\ }\href
  {https://doi.org/10.1103/PhysRevLett.98.160502} {\bibfield  {journal}
  {\bibinfo  {journal} {Phys.Rev.Lett.}\ }\textbf {\bibinfo {volume} {98}},\
  \bibinfo {pages} {160502} (\bibinfo {year} {2007}{\natexlab{a}})},\ \Eprint
  {https://arxiv.org/abs/arXiv:quant-ph/0610024} {arXiv:quant-ph/0610024}
  \BibitemShut {NoStop}%
\bibitem [{\citenamefont {Vasmer}\ and\ \citenamefont
  {Browne}(2019)}]{Vasmer2018}%
  \BibitemOpen
  \bibfield  {author} {\bibinfo {author} {\bibfnamefont {M.}~\bibnamefont
  {Vasmer}}\ and\ \bibinfo {author} {\bibfnamefont {D.~E.}\ \bibnamefont
  {Browne}},\ }\bibfield  {title} {\bibinfo {title} {Three-dimensional surface
  codes: Transversal gates and fault-tolerant architectures},\ }\href
  {https://doi.org/10.1103/PhysRevA.100.012312} {\bibfield  {journal} {\bibinfo
   {journal} {Phys. Rev. A}\ }\textbf {\bibinfo {volume} {100}},\ \bibinfo
  {pages} {012312} (\bibinfo {year} {2019})},\ \Eprint
  {https://arxiv.org/abs/1801.04255} {arXiv:1801.04255} \BibitemShut {NoStop}%
\bibitem [{\citenamefont {Bombin}(2018{\natexlab{a}})}]{Bombin2018}%
  \BibitemOpen
  \bibfield  {author} {\bibinfo {author} {\bibfnamefont {H.}~\bibnamefont
  {Bombin}},\ }\href@noop {} {\bibinfo {title} {2d quantum computation with 3d
  topological codes}} (\bibinfo {year} {2018}{\natexlab{a}}),\ \Eprint
  {https://arxiv.org/abs/1810.09571} {arXiv:1810.09571} \BibitemShut {NoStop}%
\bibitem [{\citenamefont {Brown}(2020)}]{Brown2019}%
  \BibitemOpen
  \bibfield  {author} {\bibinfo {author} {\bibfnamefont {B.~J.}\ \bibnamefont
  {Brown}},\ }\bibfield  {title} {\bibinfo {title} {A fault-tolerant
  non-clifford gate for the surface code in two dimensions},\ }\href
  {https://doi.org/10.1126/sciadv.aay4929} {\bibfield  {journal} {\bibinfo
  {journal} {Sci. Adv.}\ }\textbf {\bibinfo {volume} {6}},\ \bibinfo {pages}
  {eaay4929} (\bibinfo {year} {2020})},\ \Eprint
  {https://arxiv.org/abs/1903.11634} {arXiv:1903.11634} \BibitemShut {NoStop}%
\bibitem [{\citenamefont {Scruby}\ \emph {et~al.}(2022)\citenamefont {Scruby},
  \citenamefont {Browne}, \citenamefont {Webster},\ and\ \citenamefont
  {Vasmer}}]{Scruby2020}%
  \BibitemOpen
  \bibfield  {author} {\bibinfo {author} {\bibfnamefont {T.~R.}\ \bibnamefont
  {Scruby}}, \bibinfo {author} {\bibfnamefont {D.~E.}\ \bibnamefont {Browne}},
  \bibinfo {author} {\bibfnamefont {P.}~\bibnamefont {Webster}},\ and\ \bibinfo
  {author} {\bibfnamefont {M.}~\bibnamefont {Vasmer}},\ }\bibfield  {title}
  {\bibinfo {title} {Numerical implementation of just-in-time decoding in novel
  lattice slices through the three-dimensional surface code},\ }\href
  {https://doi.org/10.22331/q-2022-05-24-721} {\bibfield  {journal} {\bibinfo
  {journal} {Quantum}\ }\textbf {\bibinfo {volume} {6}},\ \bibinfo {pages}
  {721} (\bibinfo {year} {2022})},\ \Eprint {https://arxiv.org/abs/2012.08536}
  {arXiv:2012.08536} \BibitemShut {NoStop}%
\bibitem [{\citenamefont {Bombin}\ \emph {et~al.}(2023)\citenamefont {Bombin},
  \citenamefont {Litinski}, \citenamefont {Nickerson}, \citenamefont
  {Pastawski},\ and\ \citenamefont {Roberts}}]{Bombin2023}%
  \BibitemOpen
  \bibfield  {author} {\bibinfo {author} {\bibfnamefont {H.}~\bibnamefont
  {Bombin}}, \bibinfo {author} {\bibfnamefont {D.}~\bibnamefont {Litinski}},
  \bibinfo {author} {\bibfnamefont {N.}~\bibnamefont {Nickerson}}, \bibinfo
  {author} {\bibfnamefont {F.}~\bibnamefont {Pastawski}},\ and\ \bibinfo
  {author} {\bibfnamefont {S.}~\bibnamefont {Roberts}},\ }\href@noop {}
  {\bibinfo {title} {Unifying flavors of fault tolerance with the zx calculus}}
  (\bibinfo {year} {2023}),\ \Eprint {https://arxiv.org/abs/2303.08829}
  {arXiv:2303.08829} \BibitemShut {NoStop}%
\bibitem [{\citenamefont {Townsend-Teague}\ \emph {et~al.}(2023)\citenamefont
  {Townsend-Teague}, \citenamefont {de~la Fuente},\ and\ \citenamefont
  {Kesselring}}]{Teague2023}%
  \BibitemOpen
  \bibfield  {author} {\bibinfo {author} {\bibfnamefont {A.}~\bibnamefont
  {Townsend-Teague}}, \bibinfo {author} {\bibfnamefont {J.~M.}\ \bibnamefont
  {de~la Fuente}},\ and\ \bibinfo {author} {\bibfnamefont {M.}~\bibnamefont
  {Kesselring}},\ }\bibfield  {title} {\bibinfo {title} {Floquetifying the
  colour code},\ }\href {https://doi.org/10.4204/EPTCS.384.14} {\bibfield
  {journal} {\bibinfo  {journal} {EPTCS}\ }\textbf {\bibinfo {volume} {384}},\
  \bibinfo {pages} {265} (\bibinfo {year} {2023})},\ \Eprint
  {https://arxiv.org/abs/2307.11136} {arXiv:2307.11136} \BibitemShut {NoStop}%
\bibitem [{\citenamefont {Bauer}(2024)}]{path_integral_qec}%
  \BibitemOpen
  \bibfield  {author} {\bibinfo {author} {\bibfnamefont {A.}~\bibnamefont
  {Bauer}},\ }\bibfield  {title} {\bibinfo {title} {Topological error
  correcting processes from fixed-point path integrals},\ }\href
  {https://doi.org/10.22331/q-2024-03-20-1288} {\bibfield  {journal} {\bibinfo
  {journal} {Quantum 8}\ ,\ \bibinfo {pages} {1288}} (\bibinfo {year}
  {2024})},\ \Eprint {https://arxiv.org/abs/2303.16405} {arXiv:2303.16405}
  \BibitemShut {NoStop}%
\bibitem [{\citenamefont {Bauer}(2025{\natexlab{a}})}]{twisted_double_code}%
  \BibitemOpen
  \bibfield  {author} {\bibinfo {author} {\bibfnamefont {A.}~\bibnamefont
  {Bauer}},\ }\bibfield  {title} {\bibinfo {title} {Low-overhead non-clifford
  topological fault-tolerant circuits for all non-chiral abelian topological
  phases},\ }\href {https://doi.org/10.22331/q-2025-03-25-1673} {\bibfield
  {journal} {\bibinfo  {journal} {Quantum 9}\ ,\ \bibinfo {pages} {1673}}
  (\bibinfo {year} {2025}{\natexlab{a}})},\ \Eprint
  {https://arxiv.org/abs/arXiv:2403.12119} {arXiv:2403.12119} \BibitemShut
  {NoStop}%
\bibitem [{\citenamefont {Bauer}(2025{\natexlab{b}})}]{xy_floquet}%
  \BibitemOpen
  \bibfield  {author} {\bibinfo {author} {\bibfnamefont {A.}~\bibnamefont
  {Bauer}},\ }\bibfield  {title} {\bibinfo {title} {The x+y floquet code: A
  simple example for topological quantum computation in the path integral
  approach},\ }\href@noop {} {\bibfield  {journal} {\bibinfo  {journal} {Phys.
  Rev. A}\ }\textbf {\bibinfo {volume} {111}},\ \bibinfo {pages} {032413}
  (\bibinfo {year} {2025}{\natexlab{b}})},\ \Eprint
  {https://arxiv.org/abs/2408.07265} {arXiv:2408.07265} \BibitemShut {NoStop}%
\bibitem [{\citenamefont {Kubica}\ and\ \citenamefont
  {Delfosse}(2023)}]{Kubica2019}%
  \BibitemOpen
  \bibfield  {author} {\bibinfo {author} {\bibfnamefont {A.}~\bibnamefont
  {Kubica}}\ and\ \bibinfo {author} {\bibfnamefont {N.}~\bibnamefont
  {Delfosse}},\ }\bibfield  {title} {\bibinfo {title} {Efficient color code
  decoders in $d\geq 2$ dimensions from toric code decoders},\ }\bibfield
  {journal} {\bibinfo  {journal} {Quantum 7}\ }\textbf {\bibinfo {volume}
  {929}},\ \href {https://doi.org/10.22331/q-2023-02-21-929}
  {10.22331/q-2023-02-21-929} (\bibinfo {year} {2023}),\ \Eprint
  {https://arxiv.org/abs/1905.07393} {arXiv:1905.07393} \BibitemShut {NoStop}%
\bibitem [{\citenamefont {Bombin}\ and\ \citenamefont
  {Martin-Delgado}(2007{\natexlab{b}})}]{Bombin2006a}%
  \BibitemOpen
  \bibfield  {author} {\bibinfo {author} {\bibfnamefont {H.}~\bibnamefont
  {Bombin}}\ and\ \bibinfo {author} {\bibfnamefont {M.~A.}\ \bibnamefont
  {Martin-Delgado}},\ }\bibfield  {title} {\bibinfo {title} {Exact topological
  quantum order in d=3 and beyond: Branyons and brane-net condensates},\ }\href
  {https://doi.org/10.1103/PhysRevB.75.075103} {\bibfield  {journal} {\bibinfo
  {journal} {Phys.Rev.B}\ }\textbf {\bibinfo {volume} {75}},\ \bibinfo {pages}
  {075103} (\bibinfo {year} {2007}{\natexlab{b}})},\ \Eprint
  {https://arxiv.org/abs/arXiv:cond-mat/0607736} {arXiv:cond-mat/0607736}
  \BibitemShut {NoStop}%
\bibitem [{\citenamefont {Gidney}(2023)}]{Gidney2022}%
  \BibitemOpen
  \bibfield  {author} {\bibinfo {author} {\bibfnamefont {C.}~\bibnamefont
  {Gidney}},\ }\bibfield  {title} {\bibinfo {title} {A pair measurement surface
  code on pentagons},\ }\href {https://doi.org/10.22331/q-2023-10-25-1156}
  {\bibfield  {journal} {\bibinfo  {journal} {Quantum}\ }\textbf {\bibinfo
  {volume} {7}},\ \bibinfo {pages} {1156} (\bibinfo {year} {2023})},\ \Eprint
  {https://arxiv.org/abs/2206.12780} {arXiv:2206.12780} \BibitemShut {NoStop}%
\bibitem [{\citenamefont {Kissinger}(2022)}]{Kissinger2022}%
  \BibitemOpen
  \bibfield  {author} {\bibinfo {author} {\bibfnamefont {A.}~\bibnamefont
  {Kissinger}},\ }\href@noop {} {\bibinfo {title} {Phase-free zx diagrams are
  css codes (...or how to graphically grok the surface code)}} (\bibinfo {year}
  {2022}),\ \Eprint {https://arxiv.org/abs/2204.14038} {arXiv:2204.14038}
  \BibitemShut {NoStop}%
\bibitem [{\citenamefont {Magdalena de~la Fuente}\ \emph
  {et~al.}(2025)\citenamefont {Magdalena de~la Fuente}, \citenamefont {Old},
  \citenamefont {Townsend-Teague}, \citenamefont {Rispler}, \citenamefont
  {Eisert},\ and\ \citenamefont {M\"uller}}]{xyzrubycode}%
  \BibitemOpen
  \bibfield  {author} {\bibinfo {author} {\bibfnamefont {J.~C.}\ \bibnamefont
  {Magdalena de~la Fuente}}, \bibinfo {author} {\bibfnamefont {J.}~\bibnamefont
  {Old}}, \bibinfo {author} {\bibfnamefont {A.}~\bibnamefont
  {Townsend-Teague}}, \bibinfo {author} {\bibfnamefont {M.}~\bibnamefont
  {Rispler}}, \bibinfo {author} {\bibfnamefont {J.}~\bibnamefont {Eisert}},\
  and\ \bibinfo {author} {\bibfnamefont {M.}~\bibnamefont {M\"uller}},\
  }\bibfield  {title} {\bibinfo {title} {$\mathrm{XYZ}$ ruby code: Making a
  case for a three-colored graphical calculus for quantum error correction in
  spacetime},\ }\href {https://doi.org/10.1103/PRXQuantum.6.010360} {\bibfield
  {journal} {\bibinfo  {journal} {PRX Quantum}\ }\textbf {\bibinfo {volume}
  {6}},\ \bibinfo {pages} {010360} (\bibinfo {year} {2025})}\BibitemShut
  {NoStop}%
\bibitem [{\citenamefont {Raussendorf}\ \emph {et~al.}(2007)\citenamefont
  {Raussendorf}, \citenamefont {Harrington},\ and\ \citenamefont
  {Goyal}}]{Raussendorf2007}%
  \BibitemOpen
  \bibfield  {author} {\bibinfo {author} {\bibfnamefont {R.}~\bibnamefont
  {Raussendorf}}, \bibinfo {author} {\bibfnamefont {J.}~\bibnamefont
  {Harrington}},\ and\ \bibinfo {author} {\bibfnamefont {K.}~\bibnamefont
  {Goyal}},\ }\bibfield  {title} {\bibinfo {title} {Topological fault-tolerance
  in cluster state quantum computation},\ }\href
  {https://doi.org/10.1088/1367-2630/9/6/199} {\bibfield  {journal} {\bibinfo
  {journal} {New Journal of Physics}\ }\textbf {\bibinfo {volume} {9}},\
  \bibinfo {pages} {199} (\bibinfo {year} {2007})},\ \Eprint
  {https://arxiv.org/abs/quant-ph/0703143} {arXiv:quant-ph/0703143}
  \BibitemShut {NoStop}%
\bibitem [{\citenamefont {Nickerson}\ and\ \citenamefont
  {Bombín}(2018)}]{Nickerson2018}%
  \BibitemOpen
  \bibfield  {author} {\bibinfo {author} {\bibfnamefont {N.}~\bibnamefont
  {Nickerson}}\ and\ \bibinfo {author} {\bibfnamefont {H.}~\bibnamefont
  {Bombín}},\ }\href@noop {} {\bibinfo {title} {Measurement based fault
  tolerance beyond foliation}} (\bibinfo {year} {2018}),\ \Eprint
  {https://arxiv.org/abs/1810.09621} {arXiv:1810.09621} \BibitemShut {NoStop}%
\bibitem [{\citenamefont {Wikipedia}()}]{Wiki_bitrunc}%
  \BibitemOpen
  \bibfield  {author} {\bibinfo {author} {\bibnamefont {Wikipedia}},\ }\href
  {https://en.wikipedia.org/wiki/Bitruncated_cubic_honeycomb} {\bibinfo {title}
  {Bitruncated cubic honeycomb}}\BibitemShut {NoStop}%
\bibitem [{\citenamefont {Coecke}\ and\ \citenamefont
  {Kissinger}(2017)}]{Coecke2017}%
  \BibitemOpen
  \bibfield  {author} {\bibinfo {author} {\bibfnamefont {B.}~\bibnamefont
  {Coecke}}\ and\ \bibinfo {author} {\bibfnamefont {A.}~\bibnamefont
  {Kissinger}},\ }\href {https://doi.org/10.1017/9781316219317} {\emph
  {\bibinfo {title} {Picturing Quantum Processes: A First Course in Quantum
  Theory and Diagrammatic Reasoning}}}\ (\bibinfo  {publisher} {Cambridge
  University Press},\ \bibinfo {year} {2017})\BibitemShut {NoStop}%
\bibitem [{\citenamefont {van~de Wetering}(2020)}]{Wetering2020}%
  \BibitemOpen
  \bibfield  {author} {\bibinfo {author} {\bibfnamefont {J.}~\bibnamefont
  {van~de Wetering}},\ }\href@noop {} {\bibinfo {title} {Zx-calculus for the
  working quantum computer scientist}} (\bibinfo {year} {2020}),\ \Eprint
  {https://arxiv.org/abs/2012.13966} {arXiv:2012.13966} \BibitemShut {NoStop}%
\bibitem [{\citenamefont {Chamberland}\ and\ \citenamefont
  {Beverland}(2018)}]{Chamberland2018}%
  \BibitemOpen
  \bibfield  {author} {\bibinfo {author} {\bibfnamefont {C.}~\bibnamefont
  {Chamberland}}\ and\ \bibinfo {author} {\bibfnamefont {M.~E.}\ \bibnamefont
  {Beverland}},\ }\bibfield  {title} {\bibinfo {title} {Flag fault-tolerant
  error correction with arbitrary distance codes},\ }\href
  {https://doi.org/10.22331/q-2018-02-08-53} {\bibfield  {journal} {\bibinfo
  {journal} {Quantum}\ }\textbf {\bibinfo {volume} {2}},\ \bibinfo {pages} {53}
  (\bibinfo {year} {2018})}\BibitemShut {NoStop}%
\bibitem [{\citenamefont {Magdalena de~la Fuente}(2024)}]{Magdalena2024}%
  \BibitemOpen
  \bibfield  {author} {\bibinfo {author} {\bibfnamefont {J.~C.}\ \bibnamefont
  {Magdalena de~la Fuente}},\ }\href {https://arxiv.org/abs/2410.12527}
  {\bibinfo {title} {Dynamical weight reduction of pauli measurements}}
  (\bibinfo {year} {2024}),\ \Eprint {https://arxiv.org/abs/2410.12527}
  {arXiv:2410.12527 [quant-ph]} \BibitemShut {NoStop}%
\bibitem [{\citenamefont {Rodatz}\ \emph {et~al.}(2024)\citenamefont {Rodatz},
  \citenamefont {Poór},\ and\ \citenamefont {Kissinger}}]{rodatz2024}%
  \BibitemOpen
  \bibfield  {author} {\bibinfo {author} {\bibfnamefont {B.}~\bibnamefont
  {Rodatz}}, \bibinfo {author} {\bibfnamefont {B.}~\bibnamefont {Poór}},\ and\
  \bibinfo {author} {\bibfnamefont {A.}~\bibnamefont {Kissinger}},\ }\href
  {https://arxiv.org/abs/2410.17240} {\bibinfo {title} {Floquetifying
  stabiliser codes with distance-preserving rewrites}} (\bibinfo {year}
  {2024}),\ \Eprint {https://arxiv.org/abs/2410.17240} {arXiv:2410.17240
  [quant-ph]} \BibitemShut {NoStop}%
\bibitem [{\citenamefont {Song}\ and\ \citenamefont {Zhu}(2024)}]{Song2024}%
  \BibitemOpen
  \bibfield  {author} {\bibinfo {author} {\bibfnamefont {Z.}~\bibnamefont
  {Song}}\ and\ \bibinfo {author} {\bibfnamefont {G.}~\bibnamefont {Zhu}},\
  }\href@noop {} {\bibinfo {title} {Magic boundaries of 3d color codes}}
  (\bibinfo {year} {2024}),\ \Eprint {https://arxiv.org/abs/arXiv:2404.05033}
  {arXiv:2404.05033} \BibitemShut {NoStop}%
\bibitem [{\citenamefont {Bombin}(2015)}]{Bombin2014a}%
  \BibitemOpen
  \bibfield  {author} {\bibinfo {author} {\bibfnamefont {H.}~\bibnamefont
  {Bombin}},\ }\bibfield  {title} {\bibinfo {title} {Single-shot fault-tolerant
  quantum error correction},\ }\href
  {https://doi.org/10.1103/PhysRevX.5.031043} {\bibfield  {journal} {\bibinfo
  {journal} {Phys. Rev. X}\ }\textbf {\bibinfo {volume} {5}},\ \bibinfo {pages}
  {031043} (\bibinfo {year} {2015})},\ \Eprint
  {https://arxiv.org/abs/1404.5504} {arXiv:1404.5504} \BibitemShut {NoStop}%
\bibitem [{\citenamefont {Bombin}(2018{\natexlab{b}})}]{Bombin2018a}%
  \BibitemOpen
  \bibfield  {author} {\bibinfo {author} {\bibfnamefont {H.}~\bibnamefont
  {Bombin}},\ }\href@noop {} {\bibinfo {title} {Transversal gates and error
  propagation in 3d topological codes}} (\bibinfo {year}
  {2018}{\natexlab{b}}),\ \Eprint {https://arxiv.org/abs/1810.09575}
  {arXiv:1810.09575} \BibitemShut {NoStop}%
\bibitem [{\citenamefont {Delfosse}(2014)}]{Delfosse2013}%
  \BibitemOpen
  \bibfield  {author} {\bibinfo {author} {\bibfnamefont {N.}~\bibnamefont
  {Delfosse}},\ }\bibfield  {title} {\bibinfo {title} {Decoding color codes by
  projection onto surface codes},\ }\href
  {https://doi.org/10.1103/PhysRevA.89.012317} {\bibfield  {journal} {\bibinfo
  {journal} {Phys. Rev. A}\ }\textbf {\bibinfo {volume} {89}},\ \bibinfo
  {pages} {012317} (\bibinfo {year} {2014})},\ \Eprint
  {https://arxiv.org/abs/1308.6207} {arXiv:1308.6207} \BibitemShut {NoStop}%
\bibitem [{\citenamefont {Turaev}\ and\ \citenamefont
  {Viro}(1992)}]{Turaev1992}%
  \BibitemOpen
  \bibfield  {author} {\bibinfo {author} {\bibfnamefont {V.~G.}\ \bibnamefont
  {Turaev}}\ and\ \bibinfo {author} {\bibfnamefont {O.~Y.}\ \bibnamefont
  {Viro}},\ }\bibfield  {title} {\bibinfo {title} {State sum invariants of
  3-manifolds and quantum 6j-symbols},\ }\href
  {https://doi.org/10.1016/0040-9383(92)90015-A} {\bibfield  {journal}
  {\bibinfo  {journal} {Topology}\ }\textbf {\bibinfo {volume} {31}},\ \bibinfo
  {pages} {865} (\bibinfo {year} {1992})}\BibitemShut {NoStop}%
\bibitem [{\citenamefont {Barrett}\ and\ \citenamefont
  {Westbury}(1996)}]{Barrett1993}%
  \BibitemOpen
  \bibfield  {author} {\bibinfo {author} {\bibfnamefont {J.~W.}\ \bibnamefont
  {Barrett}}\ and\ \bibinfo {author} {\bibfnamefont {B.~W.}\ \bibnamefont
  {Westbury}},\ }\bibfield  {title} {\bibinfo {title} {Invariants of
  piecewise-linear 3-manifolds},\ }\href
  {https://doi.org/10.1090/S0002-9947-96-01660-1} {\bibfield  {journal}
  {\bibinfo  {journal} {Trans. Amer. Math. Soc.}\ }\textbf {\bibinfo {volume}
  {348}},\ \bibinfo {pages} {3997} (\bibinfo {year} {1996})},\ \Eprint
  {https://arxiv.org/abs/hep-th/9311155} {arXiv:hep-th/9311155} \BibitemShut
  {NoStop}%
\bibitem [{\citenamefont {Fukuma}\ \emph {et~al.}(1994)\citenamefont {Fukuma},
  \citenamefont {Hosono},\ and\ \citenamefont {Kawai}}]{Fukuma1992}%
  \BibitemOpen
  \bibfield  {author} {\bibinfo {author} {\bibfnamefont {M.}~\bibnamefont
  {Fukuma}}, \bibinfo {author} {\bibfnamefont {S.}~\bibnamefont {Hosono}},\
  and\ \bibinfo {author} {\bibfnamefont {H.}~\bibnamefont {Kawai}},\ }\bibfield
   {title} {\bibinfo {title} {Lattice topological field theory in two
  dimensions},\ }\href {https://doi.org/10.1007/BF02099416} {\bibfield
  {journal} {\bibinfo  {journal} {Commun. Math. Phys.}\ }\textbf {\bibinfo
  {volume} {161}},\ \bibinfo {pages} {157} (\bibinfo {year} {1994})},\ \Eprint
  {https://arxiv.org/abs/hep-th/9212154} {arXiv:hep-th/9212154} \BibitemShut
  {NoStop}%
\bibitem [{\citenamefont {Dijkgraaf}\ and\ \citenamefont
  {Witten}(1990)}]{Dijkgraaf1990}%
  \BibitemOpen
  \bibfield  {author} {\bibinfo {author} {\bibfnamefont {R.}~\bibnamefont
  {Dijkgraaf}}\ and\ \bibinfo {author} {\bibfnamefont {E.}~\bibnamefont
  {Witten}},\ }\bibfield  {title} {\bibinfo {title} {Topological gauge theories
  and group cohomology},\ }\href {https://doi.org/10.1007/BF02096988}
  {\bibfield  {journal} {\bibinfo  {journal} {Commun. Math. Phys.}\ }\textbf
  {\bibinfo {volume} {129}},\ \bibinfo {pages} {393} (\bibinfo {year}
  {1990})}\BibitemShut {NoStop}%
\bibitem [{\citenamefont {Crane}\ and\ \citenamefont
  {Yetter}(1993)}]{Crane1993}%
  \BibitemOpen
  \bibfield  {author} {\bibinfo {author} {\bibfnamefont {L.}~\bibnamefont
  {Crane}}\ and\ \bibinfo {author} {\bibfnamefont {D.~N.}\ \bibnamefont
  {Yetter}},\ }\bibfield  {title} {\bibinfo {title} {A categorical construction
  of 4d tqfts},\ }in\ \href {https://doi.org/10.1142/9789812796387_0005} {\emph
  {\bibinfo {booktitle} {Quantum Topology}}},\ \bibinfo {editor} {edited by\
  \bibinfo {editor} {\bibfnamefont {L.}~\bibnamefont {Kauffman}}\ and\ \bibinfo
  {editor} {\bibfnamefont {R.}~\bibnamefont {Baadhio}}}\ (\bibinfo  {publisher}
  {World Scientific},\ \bibinfo {address} {Singapore},\ \bibinfo {year}
  {1993})\ \Eprint {https://arxiv.org/abs/hep-th/9301062}
  {arXiv:hep-th/9301062} \BibitemShut {NoStop}%
\bibitem [{\citenamefont {Bauer}\ \emph {et~al.}(2022)\citenamefont {Bauer},
  \citenamefont {Eisert},\ and\ \citenamefont {Wille}}]{liquid_intro}%
  \BibitemOpen
  \bibfield  {author} {\bibinfo {author} {\bibfnamefont {A.}~\bibnamefont
  {Bauer}}, \bibinfo {author} {\bibfnamefont {J.}~\bibnamefont {Eisert}},\ and\
  \bibinfo {author} {\bibfnamefont {C.}~\bibnamefont {Wille}},\ }\bibfield
  {title} {\bibinfo {title} {A unified diagrammatic approach to topological
  fixed point models},\ }\href
  {https://doi.org/10.21468/SciPostPhysCore.5.3.038} {\bibfield  {journal}
  {\bibinfo  {journal} {SciPost Phys. Core}\ }\textbf {\bibinfo {volume} {5}},\
  \bibinfo {pages} {38} (\bibinfo {year} {2022})},\ \Eprint
  {https://arxiv.org/abs/2011.12064} {arXiv:2011.12064} \BibitemShut {NoStop}%
\bibitem [{\citenamefont {Bauer}(2023)}]{thesis}%
  \BibitemOpen
  \bibfield  {author} {\bibinfo {author} {\bibfnamefont {A.}~\bibnamefont
  {Bauer}},\ }\href
  {https://refubium.fu-berlin.de/bitstream/handle/fub188/45937/Dissertation_Andreas_Bauer.pdf?sequence=4}
  {\bibinfo {title} {Topological phases, fixed-point models, extended tqft, and
  fault-tolerant quantum computation-tensors in spacetime}} (\bibinfo {year}
  {2023}),\ \bibinfo {note} {phD thesis, Freie Universit\"at
  Berlin}\BibitemShut {NoStop}%
\bibitem [{\citenamefont {Propitius}(1995)}]{Propitius1995}%
  \BibitemOpen
  \bibfield  {author} {\bibinfo {author} {\bibfnamefont {M.~d.~W.}\
  \bibnamefont {Propitius}},\ }\emph {\bibinfo {title} {Topological
  interactions in broken gauge theories}},\ \href
  {http://arxiv.org/abs/hep-th/9511195} {Ph.D. thesis},\ \bibinfo  {school}
  {University of Amsterdam} (\bibinfo {year} {1995}),\ \Eprint
  {https://arxiv.org/abs/hep-th/9511195} {arXiv:hep-th/9511195} \BibitemShut
  {NoStop}%
\bibitem [{\citenamefont {Steenrod}(1947)}]{Steenrod1947}%
  \BibitemOpen
  \bibfield  {author} {\bibinfo {author} {\bibfnamefont {N.~E.}\ \bibnamefont
  {Steenrod}},\ }\bibfield  {title} {\bibinfo {title} {Products of cocycles and
  extensions of mappings},\ }\href {https://doi.org/10.2307/1969172} {\bibfield
   {journal} {\bibinfo  {journal} {Ann. Math.}\ }\textbf {\bibinfo {volume}
  {48}},\ \bibinfo {pages} {290} (\bibinfo {year} {1947})}\BibitemShut
  {NoStop}%
\bibitem [{\citenamefont {Chen}\ and\ \citenamefont {Tata}(2023)}]{Chen2021}%
  \BibitemOpen
  \bibfield  {author} {\bibinfo {author} {\bibfnamefont {Y.-A.}\ \bibnamefont
  {Chen}}\ and\ \bibinfo {author} {\bibfnamefont {S.}~\bibnamefont {Tata}},\
  }\bibfield  {title} {\bibinfo {title} {Higher cup products on hypercubic
  lattices: application to lattice models of topological phases},\ }\href
  {https://doi.org/10.1063/5.0095189} {\bibfield  {journal} {\bibinfo
  {journal} {J. Math. Phys.}\ }\textbf {\bibinfo {volume} {64}},\ \bibinfo
  {pages} {091902} (\bibinfo {year} {2023})},\ \Eprint
  {https://arxiv.org/abs/2106.05274} {arXiv:2106.05274} \BibitemShut {NoStop}%
\bibitem [{\citenamefont {Roberts}\ and\ \citenamefont
  {Williamson}(2020)}]{Roberts2020}%
  \BibitemOpen
  \bibfield  {author} {\bibinfo {author} {\bibfnamefont {S.}~\bibnamefont
  {Roberts}}\ and\ \bibinfo {author} {\bibfnamefont {D.~J.}\ \bibnamefont
  {Williamson}},\ }\href@noop {} {\bibinfo {title} {3-fermion topological
  quantum computation}} (\bibinfo {year} {2020}),\ \Eprint
  {https://arxiv.org/abs/2011.04693} {arXiv:2011.04693} \BibitemShut {NoStop}%
\bibitem [{\citenamefont {Bulmash}\ and\ \citenamefont
  {Barkeshli}(2020)}]{Bulmash2020}%
  \BibitemOpen
  \bibfield  {author} {\bibinfo {author} {\bibfnamefont {D.}~\bibnamefont
  {Bulmash}}\ and\ \bibinfo {author} {\bibfnamefont {M.}~\bibnamefont
  {Barkeshli}},\ }\bibfield  {title} {\bibinfo {title} {Absolute anomalies in
  (2+1)d symmetry-enriched topological states and exact (3+1)d constructions},\
  }\href {https://doi.org/10.1103/PhysRevResearch.2.043033} {\bibfield
  {journal} {\bibinfo  {journal} {Phys. Rev. Research}\ }\textbf {\bibinfo
  {volume} {2}},\ \bibinfo {pages} {043033} (\bibinfo {year} {2020})},\ \Eprint
  {https://arxiv.org/abs/2003.11553} {arXiv:2003.11553} \BibitemShut {NoStop}%
\bibitem [{\citenamefont {McEwen}\ \emph {et~al.}(2023)\citenamefont {McEwen},
  \citenamefont {Bacon},\ and\ \citenamefont {Gidney}}]{McEven2023}%
  \BibitemOpen
  \bibfield  {author} {\bibinfo {author} {\bibfnamefont {M.}~\bibnamefont
  {McEwen}}, \bibinfo {author} {\bibfnamefont {D.}~\bibnamefont {Bacon}},\ and\
  \bibinfo {author} {\bibfnamefont {C.}~\bibnamefont {Gidney}},\ }\bibfield
  {title} {\bibinfo {title} {Relaxing hardware requirements for surface code
  circuits using time-dynamics},\ }\href
  {https://doi.org/10.22331/q-2023-11-07-1172} {\bibfield  {journal} {\bibinfo
  {journal} {Quantum 7}\ ,\ \bibinfo {pages} {1172}} (\bibinfo {year}
  {2023})},\ \Eprint {https://arxiv.org/abs/arXiv:2302.02192}
  {arXiv:2302.02192} \BibitemShut {NoStop}%
\bibitem [{\citenamefont {nCat Lab}()}]{Quadratic_refinement_nlab}%
  \BibitemOpen
  \bibfield  {author} {\bibinfo {author} {\bibnamefont {nCat Lab}},\ }\href
  {https://ncatlab.org/nlab/show/quadratic+refinement} {\bibinfo {title}
  {Quadratic refinement}}\BibitemShut {NoStop}%
\bibitem [{\citenamefont {Sahay}\ and\ \citenamefont
  {Brown}(2022)}]{Sahay2021}%
  \BibitemOpen
  \bibfield  {author} {\bibinfo {author} {\bibfnamefont {K.}~\bibnamefont
  {Sahay}}\ and\ \bibinfo {author} {\bibfnamefont {B.~J.}\ \bibnamefont
  {Brown}},\ }\bibfield  {title} {\bibinfo {title} {A decoder for the
  triangular color code by matching on a möbius strip},\ }\href
  {https://doi.org/10.1103/PRXQuantum.3.010310} {\bibfield  {journal} {\bibinfo
   {journal} {PRX Quantum}\ }\textbf {\bibinfo {volume} {3}},\ \bibinfo {pages}
  {010310} (\bibinfo {year} {2022})},\ \Eprint
  {https://arxiv.org/abs/arXiv:2108.11395} {arXiv:2108.11395} \BibitemShut
  {NoStop}%
\bibitem [{\citenamefont {Gidney}\ and\ \citenamefont
  {Jones}(2023)}]{Gidney2023}%
  \BibitemOpen
  \bibfield  {author} {\bibinfo {author} {\bibfnamefont {C.}~\bibnamefont
  {Gidney}}\ and\ \bibinfo {author} {\bibfnamefont {C.}~\bibnamefont {Jones}},\
  }\href@noop {} {\bibinfo {title} {New circuits and an open source decoder for
  the color code}} (\bibinfo {year} {2023}),\ \Eprint
  {https://arxiv.org/abs/2312.08813} {arXiv:2312.08813} \BibitemShut {NoStop}%
\bibitem [{\citenamefont {Hastings}\ and\ \citenamefont
  {Haah}(2021)}]{Hastings2021}%
  \BibitemOpen
  \bibfield  {author} {\bibinfo {author} {\bibfnamefont {M.~B.}\ \bibnamefont
  {Hastings}}\ and\ \bibinfo {author} {\bibfnamefont {J.}~\bibnamefont
  {Haah}},\ }\bibfield  {title} {\bibinfo {title} {Dynamically generated
  logical qubits},\ }\href {https://doi.org/10.22331/q-2021-10-19-564}
  {\bibfield  {journal} {\bibinfo  {journal} {Quantum}\ }\textbf {\bibinfo
  {volume} {5}},\ \bibinfo {pages} {564} (\bibinfo {year} {2021})},\ \Eprint
  {https://arxiv.org/abs/2107.02194} {arXiv:2107.02194} \BibitemShut {NoStop}%
\bibitem [{\citenamefont {Paetznick}\ \emph {et~al.}(2023)\citenamefont
  {Paetznick}, \citenamefont {Knapp}, \citenamefont {Delfosse}, \citenamefont
  {Bauer}, \citenamefont {Haah}, \citenamefont {Hastings},\ and\ \citenamefont
  {da~Silva}}]{Paetznick2022}%
  \BibitemOpen
  \bibfield  {author} {\bibinfo {author} {\bibfnamefont {A.}~\bibnamefont
  {Paetznick}}, \bibinfo {author} {\bibfnamefont {C.}~\bibnamefont {Knapp}},
  \bibinfo {author} {\bibfnamefont {N.}~\bibnamefont {Delfosse}}, \bibinfo
  {author} {\bibfnamefont {B.}~\bibnamefont {Bauer}}, \bibinfo {author}
  {\bibfnamefont {J.}~\bibnamefont {Haah}}, \bibinfo {author} {\bibfnamefont
  {M.~B.}\ \bibnamefont {Hastings}},\ and\ \bibinfo {author} {\bibfnamefont
  {M.~P.}\ \bibnamefont {da~Silva}},\ }\bibfield  {title} {\bibinfo {title}
  {Performance of planar floquet codes with majorana-based qubits},\ }\href
  {https://doi.org/10.1103/PRXQuantum.4.010310} {\bibfield  {journal} {\bibinfo
   {journal} {PRX Quantum}\ }\textbf {\bibinfo {volume} {4}},\ \bibinfo {pages}
  {010310} (\bibinfo {year} {2023})},\ \Eprint
  {https://arxiv.org/abs/2202.11829} {arXiv:2202.11829} \BibitemShut {NoStop}%
\bibitem [{\citenamefont {Bombín}\ \emph {et~al.}(2023)\citenamefont
  {Bombín}, \citenamefont {Dawson}, \citenamefont {Liu}, \citenamefont
  {Nickerson}, \citenamefont {Pastawski},\ and\ \citenamefont
  {Roberts}}]{Bombin2023a}%
  \BibitemOpen
  \bibfield  {author} {\bibinfo {author} {\bibfnamefont {H.}~\bibnamefont
  {Bombín}}, \bibinfo {author} {\bibfnamefont {C.}~\bibnamefont {Dawson}},
  \bibinfo {author} {\bibfnamefont {Y.-H.}\ \bibnamefont {Liu}}, \bibinfo
  {author} {\bibfnamefont {N.}~\bibnamefont {Nickerson}}, \bibinfo {author}
  {\bibfnamefont {F.}~\bibnamefont {Pastawski}},\ and\ \bibinfo {author}
  {\bibfnamefont {S.}~\bibnamefont {Roberts}},\ }\href@noop {} {\bibinfo
  {title} {Modular decoding: parallelizable real-time decoding for quantum
  computers}} (\bibinfo {year} {2023}),\ \Eprint
  {https://arxiv.org/abs/2303.04846} {arXiv:2303.04846} \BibitemShut {NoStop}%
\bibitem [{\citenamefont {Edmonds}(1965)}]{Edmonds1965}%
  \BibitemOpen
  \bibfield  {author} {\bibinfo {author} {\bibfnamefont {J.}~\bibnamefont
  {Edmonds}},\ }\bibfield  {title} {\bibinfo {title} {Paths, trees, and
  flowers},\ }\href {https://doi.org/10.4153/CJM-1965-045-4} {\bibfield
  {journal} {\bibinfo  {journal} {Canadian Journal of Mathematics}\ }\textbf
  {\bibinfo {volume} {17}},\ \bibinfo {pages} {449–467} (\bibinfo {year}
  {1965})}\BibitemShut {NoStop}%
\bibitem [{\citenamefont {Higgott}\ and\ \citenamefont
  {Gidney}(2025)}]{Higgott2023}%
  \BibitemOpen
  \bibfield  {author} {\bibinfo {author} {\bibfnamefont {O.}~\bibnamefont
  {Higgott}}\ and\ \bibinfo {author} {\bibfnamefont {C.}~\bibnamefont
  {Gidney}},\ }\bibfield  {title} {\bibinfo {title} {Sparse blossom: correcting
  a million errors per core second with minimum-weight matching},\ }\href
  {https://doi.org/10.22331/q-2025-01-20-1600} {\bibfield  {journal} {\bibinfo
  {journal} {Quantum}\ }\textbf {\bibinfo {volume} {9}},\ \bibinfo {pages}
  {1600} (\bibinfo {year} {2025})},\ \Eprint {https://arxiv.org/abs/2303.15933}
  {arXiv:2303.15933} \BibitemShut {NoStop}%
\bibitem [{\citenamefont {Kubica}\ \emph {et~al.}(2018)\citenamefont {Kubica},
  \citenamefont {Beverland}, \citenamefont {Brandao}, \citenamefont
  {Preskill},\ and\ \citenamefont {Svore}}]{Kubica2017}%
  \BibitemOpen
  \bibfield  {author} {\bibinfo {author} {\bibfnamefont {A.}~\bibnamefont
  {Kubica}}, \bibinfo {author} {\bibfnamefont {M.~E.}\ \bibnamefont
  {Beverland}}, \bibinfo {author} {\bibfnamefont {F.}~\bibnamefont {Brandao}},
  \bibinfo {author} {\bibfnamefont {J.}~\bibnamefont {Preskill}},\ and\
  \bibinfo {author} {\bibfnamefont {K.~M.}\ \bibnamefont {Svore}},\ }\bibfield
  {title} {\bibinfo {title} {Three-dimensional color code thresholds via
  statistical-mechanical mapping},\ }\href
  {https://doi.org/10.1103/PhysRevLett.120.180501} {\bibfield  {journal}
  {\bibinfo  {journal} {Phys. Rev. Lett.}\ }\textbf {\bibinfo {volume} {120}},\
  \bibinfo {pages} {180501} (\bibinfo {year} {2018})},\ \Eprint
  {https://arxiv.org/abs/1708.07131} {arXiv:1708.07131} \BibitemShut {NoStop}%
\bibitem [{\citenamefont {Brown}\ \emph {et~al.}(2016)\citenamefont {Brown},
  \citenamefont {Nickerson},\ and\ \citenamefont {Browne}}]{Brown2015}%
  \BibitemOpen
  \bibfield  {author} {\bibinfo {author} {\bibfnamefont {B.~J.}\ \bibnamefont
  {Brown}}, \bibinfo {author} {\bibfnamefont {N.~H.}\ \bibnamefont
  {Nickerson}},\ and\ \bibinfo {author} {\bibfnamefont {D.~E.}\ \bibnamefont
  {Browne}},\ }\bibfield  {title} {\bibinfo {title} {Fault-tolerant error
  correction with the gauge color code},\ }\href
  {https://doi.org/10.1038/ncomms12302} {\bibfield  {journal} {\bibinfo
  {journal} {Nat. Commun.}\ }\textbf {\bibinfo {volume} {7}},\ \bibinfo {pages}
  {12302} (\bibinfo {year} {2016})},\ \Eprint
  {https://arxiv.org/abs/1503.08217} {arXiv:1503.08217} \BibitemShut {NoStop}%
\bibitem [{\citenamefont {Katzgraber}\ \emph {et~al.}(2009)\citenamefont
  {Katzgraber}, \citenamefont {Bombin},\ and\ \citenamefont
  {Martin-Delgado}}]{Katzgraber2009}%
  \BibitemOpen
  \bibfield  {author} {\bibinfo {author} {\bibfnamefont {H.~G.}\ \bibnamefont
  {Katzgraber}}, \bibinfo {author} {\bibfnamefont {H.}~\bibnamefont {Bombin}},\
  and\ \bibinfo {author} {\bibfnamefont {M.~A.}\ \bibnamefont
  {Martin-Delgado}},\ }\bibfield  {title} {\bibinfo {title} {Error threshold
  for color codes and random three-body ising models},\ }\bibfield  {journal}
  {\bibinfo  {journal} {Phys. Rev. Lett.}\ }\textbf {\bibinfo {volume} {103}},\
  \href {https://doi.org/10.1103/physrevlett.103.090501}
  {10.1103/physrevlett.103.090501} (\bibinfo {year} {2009}),\ \Eprint
  {https://arxiv.org/abs/0902.4845} {arXiv:0902.4845} \BibitemShut {NoStop}%
\bibitem [{\citenamefont {Scruby}\ and\ \citenamefont
  {Cai}(2024)}]{Scruby2024}%
  \BibitemOpen
  \bibfield  {author} {\bibinfo {author} {\bibfnamefont {T.~R.}\ \bibnamefont
  {Scruby}}\ and\ \bibinfo {author} {\bibfnamefont {Z.}~\bibnamefont {Cai}},\
  }\href@noop {} {\bibinfo {title} {Fault-tolerant quantum computation without
  distillation on a 2d device}} (\bibinfo {year} {2024}),\ \Eprint
  {https://arxiv.org/abs/2412.12529} {arXiv:2412.12529} \BibitemShut {NoStop}%
\bibitem [{\citenamefont {Breuckmann}\ \emph {et~al.}(2024)\citenamefont
  {Breuckmann}, \citenamefont {Davydova}, \citenamefont {Eberhardt},\ and\
  \citenamefont {Tantivasadakarn}}]{Breuckmann2024}%
  \BibitemOpen
  \bibfield  {author} {\bibinfo {author} {\bibfnamefont {N.~P.}\ \bibnamefont
  {Breuckmann}}, \bibinfo {author} {\bibfnamefont {M.}~\bibnamefont
  {Davydova}}, \bibinfo {author} {\bibfnamefont {J.~N.}\ \bibnamefont
  {Eberhardt}},\ and\ \bibinfo {author} {\bibfnamefont {N.}~\bibnamefont
  {Tantivasadakarn}},\ }\href@noop {} {\bibinfo {title} {Cups and gates i:
  Cohomology invariants and logical quantum operations}} (\bibinfo {year}
  {2024}),\ \Eprint {https://arxiv.org/abs/2410.16250} {arXiv:2410.16250}
  \BibitemShut {NoStop}%
\bibitem [{\citenamefont {Zhu}(2025)}]{Zhu2025}%
  \BibitemOpen
  \bibfield  {author} {\bibinfo {author} {\bibfnamefont {G.}~\bibnamefont
  {Zhu}},\ }\href@noop {} {\bibinfo {title} {A topological theory for qldpc:
  non-clifford gates and magic state fountain on homological product codes with
  constant rate and beyond the $n^{1/3}$ distance barrier}} (\bibinfo {year}
  {2025}),\ \Eprint {https://arxiv.org/abs/2501.19375} {arXiv:2501.19375}
  \BibitemShut {NoStop}%
\bibitem [{\citenamefont {Hsin}\ \emph {et~al.}(2024)\citenamefont {Hsin},
  \citenamefont {Kobayashi},\ and\ \citenamefont {Zhu}}]{Hsin2024}%
  \BibitemOpen
  \bibfield  {author} {\bibinfo {author} {\bibfnamefont {P.-S.}\ \bibnamefont
  {Hsin}}, \bibinfo {author} {\bibfnamefont {R.}~\bibnamefont {Kobayashi}},\
  and\ \bibinfo {author} {\bibfnamefont {G.}~\bibnamefont {Zhu}},\ }\href@noop
  {} {\bibinfo {title} {Classifying logical gates in quantum codes via
  cohomology operations and symmetry}} (\bibinfo {year} {2024}),\ \Eprint
  {https://arxiv.org/abs/2411.15848} {arXiv:2411.15848} \BibitemShut {NoStop}%
\bibitem [{\citenamefont {Lin}(2024)}]{Lin2024}%
  \BibitemOpen
  \bibfield  {author} {\bibinfo {author} {\bibfnamefont {T.-C.}\ \bibnamefont
  {Lin}},\ }\href@noop {} {\bibinfo {title} {Transversal non-clifford gates for
  quantum ldpc codes on sheaves}} (\bibinfo {year} {2024}),\ \Eprint
  {https://arxiv.org/abs/2410.14631} {arXiv:2410.14631} \BibitemShut {NoStop}%
\bibitem [{\citenamefont {Kesselring}\ \emph {et~al.}(2018)\citenamefont
  {Kesselring}, \citenamefont {Pastawski}, \citenamefont {Eisert},\ and\
  \citenamefont {Brown}}]{Kesselring2018}%
  \BibitemOpen
  \bibfield  {author} {\bibinfo {author} {\bibfnamefont {M.~S.}\ \bibnamefont
  {Kesselring}}, \bibinfo {author} {\bibfnamefont {F.}~\bibnamefont
  {Pastawski}}, \bibinfo {author} {\bibfnamefont {J.}~\bibnamefont {Eisert}},\
  and\ \bibinfo {author} {\bibfnamefont {B.~J.}\ \bibnamefont {Brown}},\
  }\bibfield  {title} {\bibinfo {title} {The boundaries and twist defects of
  the color code and their applications to topological quantum computation},\
  }\href {https://doi.org/10.22331/q-2018-10-19-101} {\bibfield  {journal}
  {\bibinfo  {journal} {Quantum}\ }\textbf {\bibinfo {volume} {2}},\ \bibinfo
  {pages} {101} (\bibinfo {year} {2018})},\ \Eprint
  {https://arxiv.org/abs/1806.02820} {arXiv:1806.02820} \BibitemShut {NoStop}%
\bibitem [{\citenamefont {Bullivant}\ \emph {et~al.}(2017)\citenamefont
  {Bullivant}, \citenamefont {Hu},\ and\ \citenamefont {Wan}}]{Bullivant2017}%
  \BibitemOpen
  \bibfield  {author} {\bibinfo {author} {\bibfnamefont {A.}~\bibnamefont
  {Bullivant}}, \bibinfo {author} {\bibfnamefont {Y.}~\bibnamefont {Hu}},\ and\
  \bibinfo {author} {\bibfnamefont {Y.}~\bibnamefont {Wan}},\ }\bibfield
  {title} {\bibinfo {title} {Twisted quantum double model of topological orders
  with boundaries},\ }\href {https://doi.org/10.1103/PhysRevB.96.165138}
  {\bibfield  {journal} {\bibinfo  {journal} {Phys. Rev. B}\ }\textbf {\bibinfo
  {volume} {96}},\ \bibinfo {pages} {165138} (\bibinfo {year} {2017})},\
  \Eprint {https://arxiv.org/abs/1706.03611} {arXiv:1706.03611} \BibitemShut
  {NoStop}%
\bibitem [{\citenamefont {Magdalena de~la Fuente}\ \emph
  {et~al.}(2023)\citenamefont {Magdalena de~la Fuente}, \citenamefont
  {Eisert},\ and\ \citenamefont {Bauer}}]{Magdalena2023}%
  \BibitemOpen
  \bibfield  {author} {\bibinfo {author} {\bibfnamefont {J.~C.}\ \bibnamefont
  {Magdalena de~la Fuente}}, \bibinfo {author} {\bibfnamefont {J.}~\bibnamefont
  {Eisert}},\ and\ \bibinfo {author} {\bibfnamefont {A.}~\bibnamefont
  {Bauer}},\ }\bibfield  {title} {\bibinfo {title} {Bulk-to-boundary anyon
  fusion from microscopic models},\ }\bibfield  {journal} {\bibinfo  {journal}
  {J. Math. Phys.}\ }\textbf {\bibinfo {volume} {64}},\ \href
  {https://doi.org/10.1063/5.0147335} {10.1063/5.0147335} (\bibinfo {year}
  {2023})\BibitemShut {NoStop}%
\end{thebibliography}%

\end{document}